\documentclass[rmp,aps,twocolumn,nofootinbib]{revtex4}
\usepackage{greekctr}

\usepackage[pdftex]{graphicx}
\usepackage{epstopdf}
\usepackage{psfrag}
\usepackage{epsfig}
\usepackage{pstricks}
\usepackage{pstcol,pst-fill,pst-grad}
\usepackage{pdfpages}

\usepackage{calligra}
\usepackage{frcursive}

\usepackage{lettrine}

\usepackage{amssymb}

\usepackage{amsfonts}
\usepackage{bm}
\usepackage{amsmath}
\usepackage{mathrsfs}
\usepackage{color}
\usepackage{boxedminipage}
\usepackage{wallpaper}
\usepackage{booktabs}
\usepackage{enumerate}

\DeclareMathAlphabet{\mathpzc}{OT1}{pzc}{m}{it}

\newcommand{\mc}[1]{\mathcal{#1}}

\renewcommand{\vec}[1]{{\mathbf #1}}
\newcommand{\be}{\begin{equation}}
\newcommand{\ee}{\end{equation}}

\newcommand{\bra}[1]{\langle#1|}
\newcommand{\ket}[1]{|#1\rangle}
\newcommand{\scp}[2]{\langle#1|#2\rangle}
\newcommand{\braket}[3]{\langle#1|#2|#3\rangle}
\newcommand{\ketbra}[2]{|#1\rangle\langle#2|}

\newcommand{\moy}[1]{\langle #1\rangle}

\newcommand{\comment}[1]{} 

\begin{document}
\title{Euclidean random matrices and their applications in physics}

\author{A. Goetschy}
\email{Arthur.Goetschy@yale.edu}
\affiliation{Department of Applied Physics, Yale University, New Haven, Connecticut 06520, USA}

\author{S.E. Skipetrov}
\email{Sergey.Skipetrov@grenoble.cnrs.fr}
\affiliation{Universit\'{e} Grenoble 1/CNRS, Laboratoire de Physique et Mod\'elisation des
Milieux Condens\'es UMR 5493,\\ B.P. 166, 38042 Grenoble, France}

\begin{abstract}
We review the state of the art of the theory of Euclidean random matrices, focusing on the density of their eigenvalues. Both Hermitian and non-Hermitian matrices are considered and links with simpler, standard random matrix ensembles are established. We discuss applications of Euclidean random matrices to contemporary problems in condensed matter physics, optics, and quantum chaos.
\end{abstract}

\date{\today}
\maketitle
\tableofcontents

\section{Introduction}
\label{Introduction}

Random matrix theory (RMT) is a powerful tool of modern theoretical physics \cite{mehta04}. Its main goal is to calculate the statistical properties of eigenvalues and eigenvectors of large matrices with random elements. First introduced by \citet{wishart28} and then used by \citet{wigner55} to describe the statistics of energy levels in complex nuclei, random matrices are nowadays omnipresent in physics \cite{brody81,beenakker97,guhr98,tulino04}. The majority of works---including the seminal papers by \citet{wigner55} and \citet{dyson62a, dyson62b, dyson62c}---deal with Hermitian matrices. Hermitian matrices are of special importance in physics because of the Hermiticity of operators associated with observables in quantum mechanics.
On the other hand, non-Hermitian random matrices have also attracted considerable attention \cite{hatano96, janik97a, janik97b, jarosz06, feinberg97a, feinberg97b, feinberg99, feinberg99b, feinberg06}. They can be used to model such physical phenomena as scattering in dissipative or open systems \cite{haake92, fyodorov97, fyodorov03, fyodorov05}, dynamics of neural networks \cite{sommers88, rajan06}, diffusion in random velocity fields \cite{chalker97}, or chiral symmetry breaking of the QCD Dirac operator \cite{stephanov96, verbaarschot00}.

A special class of random matrices are the so-called Euclidean random matrices (ERMs) \cite{mezard99}. The elements $A_{ij}$ of a $N \times N$ Euclidean random matrix $A$ are given by a deterministic function $f$ of positions of pairs of points that are randomly distributed in a finite region $V$ of Euclidean space: $A_{ij} = f(\vec{r}_i, \vec{r}_j)$, $i = 1, \ldots, N$. Hermitian ERM models play an important role in the theoretical description of vibrations in topologically disordered systems \cite{mezard99, grigera01a, grigera01b, grigera02, grigera03,ganter11,grigera11} and relaxation in glasses \cite{amir08,amir10,amir11}. They have been used to study Anderson localization \cite{bogomolny03,ciliberti05,huang09}. A number of analytic approaches were developed to deal with Hermitian ERMs \cite{mezard99, grigera01a, grigera01b,grigera02,grigera03,ganter11,grigera11,chamon01,amir10,bogomolny03, ciliberti05,skipetrov11}.
In contrast, an analytic approach to non-Hermitian ERMs was proposed only recently \cite{goetschy11a,goetschy11b}. A particular non-Hermitian ERM---the so-called Green's matrix---was a subject of extensive numerical studies \cite{rusek00,pinheiro04,antezza10,skipetrov11}.
The principal difficulties that one encounters when trying to develop a theory for ERMs  stem from the nontrivial statistics of their elements and the correlations between them. Both are not known analytically and are often difficult to calculate. This is in sharp contrast with standard approaches \cite{mehta04,beenakker97,guhr98} in which the joint probability distribution of the elements of the random matrix under study is the starting point of analysis.

The purpose of this paper is to review the existing analytical results for the eigenvalue density of large ERMs ($N \gg 1$) and to discuss their applications to modern problems in condensed matter physics, optics, and quantum chaos. Analytical results will be compared with numerical diagonalization and special attention will be given to the range of validity of the former. To derive the analytical results, we will use both the diagrammatic techniques which are more familiar to a theoretical physicist, and the recently introduced free probability theory \citep{voiculescu92} which is a mathematically involved but powerful technique. Arriving at the same result in two different ways will help to understand which approximations were essential and which were not and will, we hope, promote the use of the new mathematical tool---free probability theory---by the community of physicists.

Because ERMs are a relatively new class of random matrices, we believe that it is essential to make links with more `standard' random matrix ensembles, such as the matrices with independent, identically distributed elements or Wishart matrices \citep{mehta04}. We will show that for some of the ERMs under study and in a restricted range of parameters, the eigenvalue density is very close to that of the known, simpler matrix ensembles: Wigner semi-circle \citep{mehta04}, Marchenko-Pastur law \citep{marchenko67}, Ginibre's disk \citep{ginibre65, girko85}. These well-known distributions will be introduced and discussed as well, to ensure a self-contained presentation.

Finally, we will discuss what is known about the properties of eigenvectors of Euclidean random matrices. The study of eigenvectors is a much more difficult problem, which explains our very poor understanding of their statistical properties. We will focus on the most interesting question of spatial (Anderson) localization of eigenvectors.

The paper contains three main sections: Sec.\ \ref{HermitianERM} is devoted to Hermitian matrices, Sec.\ \ref{NonHermitianERM}---to non-Hermitian ERMs, and, finally, Sec.\ \ref{Applications} discusses applications of ERMs in physics. Sections \ref{HermitianERM} and \ref{NonHermitianERM} are self-contained, even though the latter makes frequent references to the former to help the reader understand the more complicated case of non-Hermitian matrices by analogy with the simpler Hermitian one. This review is supposed to be accessible to everyone with a general theoretical physics background and does not require knowledge of RMT to be understood. It can serve as an introduction to RMT, even though the view it provides is biased by its main focus on ERM ensembles which are only a small fraction of many important ensembles of RMT. For a more general and unbiased view of the modern RMT, we invite the interested reader to consult the book of \citet{mehta04} or the review papers by \citet{guhr98} and \citet{stephanov01}. Excellent accounts of more specific issues of RMT and of their applications in physics can be found in review papers by    \citet{brody81,beenakker97,mitchell10}.

\section{Hermitian Euclidean random matrix theory}
\label{HermitianERM}

In this section, we start by introducing a number of well-known ensembles of Hermitian random matrices (Sec.\ \ref{RMEnsInterest}), including the Euclidean random matrix ensemble that will be of interest for us (Sec.\ \ref{SubsecERM}). We then review the main approaches to study the statistical properties of eigenvalues of Hermitian random matrices in general and of Euclidean matrices in particular
%(Sec.\ \ref{GBR, MappingDyson, Fieldreprersentation, SecDiagrammaticApproach, FreeProba, Renorm})%
(Secs.\ \ref{GBR}--\ref{Renorm}) and apply them to analyze the eigenvalue distributions of several specific Euclidean random matrices (Sec.\ \ref{Applherm}).

\subsection{Some standard random matrix ensembles}
\label{RMEnsInterest}

\subsubsection{Gaussian matrices}
\label{SecGaussianMatrices}
The best known random matrix ensembles are probably the Gaussian ensembles. They are ensembles of $N \times N$ Hermitian matrices $A=A^\dagger$, that have independent and identically distributed (i.i.d.) zero-mean Gaussian entries. The probability distribution of $A$ is
\be
\label{GaussianProba}
P(A)=C_Ne^{-\frac{\beta N}{4}\textrm{Tr}A^2},
\ee
where $C_N$ is a normalization constant, and $\beta$ is the symmetry index, that counts the number of degrees of freedom in the matrix elements.

For our purpose, it is sufficient to consider matrices $A$ with entries being either real or complex numbers ($\beta=1$ or $2$). Let us first analyze the case of complex elements, for which $\beta=2$. Since the transformation $A\to UAU^{-1}$, with $U$ unitary, leaves $P(A)$ invariant, the ensemble is called `Gaussian unitary ensemble' (GUE). From Eq.~(\ref{GaussianProba}), we easily verify that the second moments of $A_{ij}$ are
\be
\label{ContractionGUE}
\begin{array}{lcr}
\moy{A_{ij}A_{kl}}=\frac{1}{N}\delta_{il}\delta_{jk}, & &\textrm{GUE}\;\;(\beta=2).
\end{array}
\ee

On the other hand, if elements of $A$ are real numbers, $\beta=1$, and the transformation  $A\to UAU^{-1}$ leaves $P(A)$ invariant for $U$ orthogonal. The ensemble is called `Gaussian orthogonal ensemble' (GOE), and the second moments are given by
\be
\label{ContractionGOE}
\begin{array}{lcr}
\moy{A_{ij}A_{kl}}=\frac{1}{N}(\delta_{il}\delta_{jk}+\delta_{ik}\delta_{jl}),  & &\textrm{GOE}\;\;(\beta=1).
\end{array}
\ee

As we shall see later, the density of eigenvalues
of a Gaussian matrix $A$ converges, in the limit $N\to \infty$, to the so-called `semicircle' law, first discovered by \citet{wigner55}.

\subsubsection{Wishart matrices}

Another ensemble of particular interest for us is the Wishart ensemble, that is as old as RMT itself \cite{wishart28}. It is useful in many contexts, such as neural networks, image processing, or wireless communications, where Wishart matrices naturally arise to characterize the singular values of `channel matrices' \cite{tulino04}. A $N\times N$ Wishart matrix $A$ is of the form
\be
\label{WishartMatrix}
A=HH^\dagger,
\ee
where $H$ is a rectangular $N\times M$ matrix, with columns that are zero-mean independent real or complex Gaussian random vectors with covariance matrix $\vec{\Sigma}$ \cite{tulino04}. In this section, we will work with $H$ complex and $\vec{\Sigma}$ proportional to the identity matrix $I_N$. In this case, the entries of $H$ are zero-mean i.i.d. complex Gaussian random numbers. The probability distribution of the non-Hermitian matrix $H$ is
\be
\label{GaussProbDistr}
P(H)=C_{N,M}e^{-N\textrm{Tr}HH^\dagger},
\ee
so that the second-moments of elements of $H$ obey
\be
\label{Contraction0}
\moy{H_{i\alpha}H^\dagger_{\beta j}}=\frac{1}{N}\delta_{ij}\delta_{\alpha \beta}=\moy{H^\dagger_{\alpha i}H_{j \beta }}.
\ee

For $c=N/M<1$, \citet{wishart28} showed that the probability distribution of (\ref{WishartMatrix}) is given by
\be
\label{PAWishart}
P(A)=C'_{N,M}\textrm{det}A^{M-N}e^{-N\textrm{Tr}A},
\ee
see also \cite{tulino04}. Quite surprisingly, no such explicit formula was known for $c>1$ (`anti-Wishart case') until recently \cite{janik03}. However, as far as the eigenvalue distribution of $A$ is concerned, it is straightforward to obtain the result for $c>1$ from the one for $c<1$ (see Sec.\ \ref{SecExApplDyson}).

In Sec.\ \ref{SecExApplDyson}, we will see that the eigenvalue distribution of $A=HH^\dagger$ converges, in the limit $N, M \to \infty$ with $c=N/M$ fixed, to the so-called Marchenko-Pastur law. It was first established by \citet{marchenko67}, and then rediscovered several times \cite{tulino04}.

\subsubsection{Euclidean random matrices}
\label{SubsecERM}

As explained in the introduction, ERMs are matrices with elements $A_{ij}$ defined with the help of some deterministic function $f$ of positions of pairs of points:
\be
\label{DefinitionERM}
A_{ij}=f(\vec{r}_i, \vec{r}_j)=\braket{\vec{r}_i}{\hat{A}}{\vec{r}_j},
\ee
where we introduced an operator $\hat{A}$ associated with the matrix $A$.
We assume that the $N$ points $\vec{r}_i$ are randomly distributed inside some region $V$ of $d$-dimensional space with a uniform density $\rho = N/V$. In this review, we will mainly focus on $d = 3$. Most of the analysis that will be presented below can be extended to the case of correlated positions of points by replacing $f(\vec{r_i}, \vec{r_j})$ with $f(\vec{r_i}, \vec{r_j}) g(\vec{r_i}-\vec{r_j})$, where $g(\vec{r})$ is the pair correlation function \cite{martin01}.

Contrary to Gaussian or Wishart matrices, the probability distribution $P(A)$ is not known analytically. When computing the statistical properties of $A$, averaging is performed not with respect to $P(A)$ but with respect to the probability density of points $\{ \vec{r}_i \}$ in $V$.

Depending on a particular physical problem under study, the matrix $A$ defined by Eq.\ (\ref{DefinitionERM}) may be subject to additional constraints. For ERMs that arise in the study of vibrational modes of an amorphous solid, instantaneous normal modes of a liquid, or random master equations \cite{mezard99}, a condition $\sum_j A_{ij}=0$ expresses the global translational invariance (conservation of momentum in the case of propagating excitations). Typically, it is obeyed by using Eq.\ (\ref{DefinitionERM}) only for off-diagonal elements of $A$ and adopting a different definition for diagonal elements, $A_{ii} = -\sum_{j \ne i} A_{ij}$. Both cases can be combined in a single definition \cite{mezard99}
\be
\label{DefinitionERMu}
A_{ij}=f(\vec{r}_i, \vec{r}_j) - u \delta_{ij} \sum\limits_{k =1}^{N} f(\vec{r}_i, \vec{r}_k),
\ee
where $u = 0$ or 1.

A useful trick to study statistical properties of matrices defined by Eq.\ (\ref{DefinitionERM}) is to change the basis from $\{\vec{r}_i\}$ to $\{\psi_\alpha\}$, which is orthogonal in $V$ \cite{skipetrov11}. Inserting the closure relation $\hat{\vec{1}}=\sum_\alpha \ketbra{\psi_\alpha}{\psi_\alpha}$ into Eq.~(\ref{DefinitionERM}), we obtain for arbitrary $V$:
\be
\label{AHTH}
A=HTH^\dagger,
\ee
where
\begin{align}
\label{DefHinA}
H_{i\alpha}&=\frac{1}{\sqrt{\rho}}
\scp{\vec{r}_i}{\psi_\alpha},
\\
\label{DefTinA}
T_{\alpha\beta}&=\rho\,\bra{\psi_\alpha}\hat{A}\ket{\psi_\beta}.
\end{align}
In Eq.~(\ref{DefHinA}) and (\ref{DefTinA}), the prefactor $\rho$ is introduced for later convenience. In a rectangular box, for example, $\ket{\psi_\alpha} = \ket{\vec{k}_\alpha}$ with
$\scp{\vec{r}}{\vec{k}_{\alpha}} = \exp(i \vec{k}_{\alpha} \vec{r})/\sqrt{V}$, so that $T_{\alpha\beta}$ are simply the Fourier coefficients of $f(\vec{r}_i, \vec{r}_j)$:
\be
\label{TSerriesCoeff}
T_{\alpha\beta} =N
\iint_V \frac{\textrm{d}^d \vec{r}_i}{V} \frac{\textrm{d}^d \vec{r}_j}{V} f(\vec{r}_i, \vec{r}_j)
e^{-\textrm{i}(\vec{k}_\alpha \cdot \vec{r}_i-
\vec{k}_\beta \cdot \vec{r}_j)}.
\ee
The advantage of the representation (\ref{AHTH}) lies in the separation of two different sources of complexity: the matrix $H$ is random but independent of the function $f$, whereas the matrix $T$ depends on $f$ but is not random.

Furthermore, we assume that
\be
\int_{V} \textrm{d}^d \mathbf{r}\; \psi_\alpha(\mathbf{r}) = 0,
\ee
which in a box is obeyed for all $\alpha$ except when $\vec{k}_{\alpha} = 0$. We readily find that $H_{i\alpha}$ are identically distributed random variables with zero mean and variance equal to $1/N$:
\begin{align}
\langle H_{i\alpha} \rangle &= \frac{1}{\sqrt{\rho}} \int_{V} \frac{\textrm{d}^d  \mathbf{r}_i }{V} \psi_\alpha(\mathbf{r}_i) = 0,
\label{matrixh1}
\\
\langle H_{i\alpha} H_{j \beta}^* \rangle &= \frac{1}{\rho}\iint_{V}  \frac{\textrm{d}^d  \mathbf{r}_i }{V} \frac{\textrm{d}^d  \mathbf{r}_j }{V} \psi_\alpha(\mathbf{r}_i) \psi_\beta^*(\mathbf{r}_j)
\nonumber
\\
&= \langle H_{i\alpha} \rangle \langle H_{j \beta}^* \rangle = 0 \;\;\; (i \neq j),
\label{matrixh2}
\\
\langle H_{i \alpha} H_{i \beta}^* \rangle &=  \frac{1}{\rho} \int_{V}\frac{\textrm{d}^d  \mathbf{r}_i }{V}  \psi_\alpha(\mathbf{r}_i) \psi_ \beta^*(\mathbf{r}_i) = \frac{1}{N}\delta_{\alpha  \beta}.
\label{matrixh3}
\end{align}
Equations (\ref{matrixh2}) and (\ref{matrixh3}) show that $H$ satisfies Eq.\ (\ref{Contraction0}), i.e. that the covariance matrix of the columns of $H$ is $\vec{\Sigma}=I_N/N$.
If $H_{i\alpha}$ were Gaussian random variables, then this would be sufficient to conclude that $H_{i\alpha}$ are independent. However, they are not Gaussian and hence not necessarily independent. For example, the cumulant $\moy{A_{ij} A_{ji}A_{ij} }_c$ is not zero. It turns out that neglecting these complications and assuming $H_{i\alpha}$ Gaussian i.i.d.  amounts to disregarding a class of `dependent scattering' events  corresponding to the formation of `cavities' by clusters of points $\vec{r}_i$.

In Sec.\ \ref{SecDiagrammaticApproach}, we will explicitly assume that $H_{i\alpha}$ are independent Gaussian random variables. This assumption largely simplifies calculations but may limit the applicability of results, at least for certain types of Euclidean matrices, as we will see later. Within this assumption, the only but crucial difference between an ERM (\ref{AHTH}) and a Wishart matrix (\ref{WishartMatrix}) is the matrix $T$ that contains all information about the function $f$ defining the ERM. It can modify the eigenvalue distribution in a non-trivial way and lead to transitions between topologically different supports $\mc{D}$ of the eigenvalue density. Illustrations of such transitions are given by the examples considered in Secs.\ \ref{ERMSinc} and \ref{ERMCosc}.

\subsection{Resolvent, Blue function, and $\mathcal{R}$-transform}
\label{GBR}

Eigenvalues $\Lambda_n$ of a $N \times N$ Hermitian matrix $A$ are real. Their density,
\be
p(\Lambda)=\frac{1}{N}\left\langle \sum_{n=1}^N\delta(\Lambda-\Lambda_n)
 \right\rangle,
\ee
can be obtained from the (one-point) resolvent
\be
\label{Defresolvent}
g(z) = \frac{1}{N} \left\langle \textrm{Tr} \frac{1}{z-A} \right\rangle
= \frac{1}{N} \left\langle \sum_{n=1}^N \frac{1}{z-\Lambda_n} \right\rangle.
\ee
Using the standard relation $\lim_{\epsilon\to0^+}1/(\Lambda+\textrm{i}\epsilon)=\textrm{P}1/\Lambda-\textrm{i}\pi\delta(\Lambda)$ (P denotes the Cauchy principal value), we rewrite Eq.\ (\ref{Defresolvent}) as
\be
g(\Lambda+\textrm{i}\epsilon)=\textrm{P}\int_{-\infty}^\infty\textrm{d}\Lambda'\frac{p(\Lambda')}{\Lambda-\Lambda'}-\textrm{i}\pi p(\Lambda),
\ee
so that $p(\Lambda)$ may be reconstructed from either the imaginary or the real part of $g(\Lambda+\textrm{i}\epsilon)$:\footnote{Physically, $g(\Lambda+\textrm{i}\epsilon)$ is the Fourier transform of the causal propagator $e^{-\textrm{i}At}\Theta(t)$ [with $\Theta(t)$ the Heaviside step function], and therefore its real and imaginary parts obey Kramers-Kronig relations.}
\begin{align}
\label{LinkpImG}
&p(\Lambda) = - \frac{1}{\pi} \lim_{\epsilon \to 0^+} \mathrm{Im}
g(\Lambda + \textrm{i} \epsilon),
\\
\label{LinkpReG}
&\textrm{P}\int_{-\infty}^\infty   \textrm{d}\Lambda' \frac{p(\Lambda')}{\Lambda-\Lambda'} = \textrm{Re}\,g(\Lambda + \textrm{i} \epsilon).
\end{align}
Useless to say, Eq.\ (\ref{LinkpImG}) provides a much more direct way to compute $p(\Lambda)$ than the Fredholm integral equation of the first kind (\ref{LinkpReG}). However, the inversion of the latter may sometimes yield the solution in a very efficient manner. Indeed, if $p(\Lambda)$ has a finite support $[a,b]$, the solution of Eq.~(\ref{LinkpReG}) is given by Tricomi's theorem \cite{tricomi57}:
\begin{align}
\label{pVsReG}
p(\Lambda)&=\frac{1}{\pi^2\sqrt{(\Lambda-a)(b-\Lambda)}}
\\
&\times
\left[ \pi - \textrm{P}\int_a^b  \textrm{d}\Lambda'  \frac{\sqrt{(\Lambda'-a)(b-\Lambda')}}{\Lambda'-\Lambda}
\textrm{Re}g(\Lambda' + \textrm{i} \epsilon)
\right].
\nonumber
\end{align}
Such an expression for $p(\Lambda)$ turns out to be particularly useful within the framework of the Dyson gas model (see Sec.\ \ref{MappingDyson}).

In order to compute $g(z)$, we can rewrite Eq.\ (\ref{Defresolvent}) in various forms. Each of them is the starting point of a specific analysis developed in the following sections. First, we note that
\be
\sum_{n=1}^N\frac{1}{z-\Lambda_n}=\partial_z\ln\left[\prod_{n=1}^N(z-\Lambda_n)\right],
\ee
and express the resolvent as
\begin{align}
g(z)&=\frac{1}{N}\partial_z \left\langle
\textrm{ln det}(z-A)
 \right\rangle.
 \label{GasDet}
\end{align}
This expression will be used in the field-theoretical approach presented in Sec.\ \ref{Fieldreprersentation}. Another interesting expression for $g(z)$ is a decomposition in terms of the moments of $p(\Lambda)$,
\be
\label{MomentsTrA}
\moy{\Lambda^{n}}=\int_{-\infty}^\infty \textrm{d}\Lambda\; p(\Lambda) \Lambda^n=\frac{1}{N} \left\langle
\textrm{Tr}A^n
 \right\rangle.
\ee
For Hermitian matrices, $g(z)$ is a holomorphic function of $z\in \mathbb{C}$ except for some cuts along the real axis where eigenvalues of $A$ are concentrated. Therefore, we can reconstruct $g(z)$ for all $z$ by analytic continuation of its series expansion
\be
\label{gSumMoments}
g(z)= \sum_{n=0}^\infty\frac{\moy{\Lambda^{n}}}{z^{n+1}},
\ee
which, in general, converges only in the vicinity of $\vert z\vert \to \infty$. We will work with the representation (\ref{gSumMoments}) in Sec.\ \ref{SecDiagrammaticApproach} to perform a diagrammatic computation of $g(z)$. In this perspective, it is also convenient to define the self-energy $\sigma(z)$, that contains all irreducible diagrams in Eq.~(\ref{gSumMoments}):
\be
\label{SelfenergyRMT}
g(z)=\frac{1}{z-\sigma(z)}.
\ee

Other important objects for us are the functional inverse of $g(z)$, also called the Blue function, and the $\mathcal{R}$-transform:
\begin{align}
\label{DefBlue}
\mc{B}(z)&=g^{-1}(z),
\\
\label{DefRed}
\mathcal{R}(z)&=\mc{B}(z)-\frac{1}{z}.
\end{align}
Both of them are fundamental objects of the free random variable theory, discussed in Sec.\ \ref{FreeProba}. In particular, $\mathcal{R}(z)$ is the generating function of the `free cumulants' (see Sec.\ \ref{FreeProba} for more details).
According to Eq.~(\ref{SelfenergyRMT}), $\mc{B}(z)$ and $\mathcal{R}(z)$ are related to the self-energy $\sigma(z)$ by
\begin{align}
\label{LinksigmaR}
\sigma(z)&=\mathcal{R}[g(z)],
\\
\mc{B}(z)&=\frac{1}{z}+\sigma[\mc{B}(z)].
\end{align}

Let us now mention a couple of properties useful for further analysis. The functions $g(z)$, $\mc{B}(z)$, and $\mc{R}(z)$ obey the following scaling relations:
\begin{align}
g_{\alpha A}(z)&=\frac{1}{\alpha}g_A(z/\alpha),
\nonumber
\\
\label{scalingPpty}
\mc{B}_{\alpha A}(z)&=\alpha \mc{B}_A(\alpha z),
\\
\mc{R}_{\alpha A}(z)&=\alpha \mc{R}_A(\alpha z),
\nonumber
\end{align}
where $\alpha \in \mathbb{C}^*$.
Besides, the moments $\moy{\Lambda^{n}}$ can be obtained from $g(z)$, $\mc{B}(z)$, and $\mathcal{R}(z)$. Using Eqs.~(\ref{gSumMoments}), (\ref{DefBlue}), and (\ref{DefRed}), we readily show that
\begin{align}
\label{momentg}
\langle \Lambda^n \rangle& =
\left. \frac{1}{(n+1)!} \frac{\mathrm{d}^{n+1} g(z)}{\mathrm{d}(1/z)^{n+1}}
\right|_{z \rightarrow \infty},
\\
\label{momentb}
\langle \Lambda^n \rangle &=
\left. \frac{1}{(n+1)!}
\left[ -\frac{\mc{B}^2(z)}{\mc{B}'(z)} \frac{\mathrm{d}}{\mathrm{d}z} \right]^n
\left[ -\frac{\mc{B}^2(z)}{\mc{B}'(z)} \right]
\right|_{z \rightarrow 0}.
\end{align}
In particular,
\begin{align}
\langle \Lambda \rangle&=\mathcal{R}(0),
\\
\label{VarianceLambda}
\mathrm{var} \Lambda &= \langle (\Lambda - \langle \Lambda \rangle)^2 \rangle = \mathcal{R}'(z)|_{z \rightarrow 0},
\end{align}
where $\mc{B}'(z) = \textrm{d}\mc{B}(z)/\textrm{d}z$ and $\mathcal{R}'(z) = \textrm{d}\mathcal{R}(z)/\textrm{d}z$.
Finally, we note that the boundaries $\Lambda_*$ of the domain of existence of eigenvalues, $p(\Lambda_*)=0$,  are given by the following simple relations \cite{zee96}:
\begin{align}
  \label{gboundaries}
 g'(\Lambda_*)&\to\infty,
 \\
 \label{Bboundaries}
 \mc{B}'(\Lambda_*)&=0.
 \end{align}

\subsection{Mapping to the Dyson gas}
\label{MappingDyson}

\subsubsection{Dyson gas picture}
Observing that the electric field created by a point charge in two dimensions is inversely proportional to the distance from the charge, we can interpret the resolvent (\ref{Defresolvent}) as the electric field created at a point $z$ in the complex plane by charges $q = 1$ situated at positions $\Lambda_n$ on the real axis. This suggests an analogy between the statistical properties of random matrices and those of a gas of charged particles restricted to move in one dimension, the so-called Dyson gas \cite{dyson62a, dyson62b, dyson62c, mehta04}.

For a large class of random matrices $A$, the distribution of the eigenvalues $\Lambda_n$ can be seen as the equilibrium distribution of fictitious point charges repealing each other by Coulomb interaction, and submitted to an external one-body potential determined by the precise form of the probability distribution $P(A)$. In particular, this statement is true for the Wigner-Dyson ensemble defined as
\be
\label{WignerDyson}
P(A)=C_Ne^{-\beta N\textrm{Tr}V^g(A)},
\ee
where the `potential' $V^g(A)$ is arbitrary, provided the existence of the partition function $C_N^{-1}$. If $V^g$ is quadratic, we recover the Gaussian ensemble (\ref{GaussianProba}). To justify the Dyson gas picture, it is sufficient to consider the (joint) probability distribution of the eigenvalues (for the proof, see Sec.\ \ref{SubsecDysonBrownian}):
\begin{align}
\label{GibbsDG}
P(\{\Lambda_n\})&=C'_Ne^{- \beta H^g(\{ \Lambda_n\})},
\\
\label{HgDG}
H^g(\{\Lambda_n\})&=N\sum_{n=1}^NV^g(\Lambda_n)-\sum_{n<m}\ln|\Lambda_n-\Lambda_m|.
\end{align}
We recognize the Boltzmann-Gibbs distribution of a classical gas in thermal equilibrium at a temperature $T=1/\beta$. The logarithmic pairwise repulsion
\be
\label{VCoulRep}
V^{\mathrm{int}}(z)=-\sum_{n=1}^N\ln|z- \Lambda_n|
\ee
is the Coulomb interaction in 2D, associated with the electric field $\vec{g}=(\textrm{Re}g, \textrm{Im}g)$ represented by the resolvent (\ref{Defresolvent}):
\begin{align}
\label{gAsElecField}
N\mathbf{g}(z=x+\textrm{i}y) &= -\mathbf{\nabla}_{x,y}V^{\mathrm{int}}
\nonumber \\
&=\sum_{n=1}^N\left(\frac{x-\textrm{Re}\Lambda_n}{|z-\Lambda_n|^2},\frac{y-\textrm{Im}\Lambda_n}{|z-\Lambda_n|^2}\right).
\end{align}
For Hermitian matrices, the Dyson gas is a two-dimensional Coulomb gas, experiencing the one-body potential $V^g$, with the kinematic restriction that the charges move along the line $\textrm{Im}\Lambda_n=0$.\footnote{This kinematic restriction is suppressed for non-Hermitian matrices (see Sec.\ \ref{NonHermitianERM}).}

The main advantage of the Dyson gas picture is that it allows to apply methods of statistical mechanics to calculate distributions and correlations of eigenvalues, giving therefore a physical intuition about the statistical properties of eigenvalues. In particular, it is clear that the shape of the overall density will strongly depend on the one-body potential $V^g$,  while the correlations in the relative positions of eigenvalues are affected by the interaction $V^{\mathrm{int}}$ and are much less sensitive to $V^g$.

\subsubsection{Brownian motion of eigenvalues}
\label{SubsecDysonBrownian}

Before exploiting further the Dyson gas picture, let us justify Eq.\ (\ref{GibbsDG}) for $P(\{\Lambda_n\})$ in two different ways. Mathematically, Eq.\ (\ref{GibbsDG}) can be obtained from Eq.\ (\ref{WignerDyson}) by changing variables from the matrix elements of $A$ to parameters related to eigenvalues and eigenvectors of $A$. The Jacobian of the transformation contains, in particular, a factor $\vert \mc{V}(\{\Lambda_n\})\vert^\beta$, where
\be
\mc{V}(\{\Lambda_n\})=\prod_{n<m}(\Lambda_n-\Lambda_m)
\ee
is a Vandermonde determinant. $\mc{V}(\{\Lambda_n\})$ is the source of the logarithmic repulsion in $H^g$. Integrating over the parameters related to the eigenvectors, one obtains Eqs.~(\ref{GibbsDG}) and (\ref{HgDG}) \cite{mehta04}.

Equation (\ref{GibbsDG}) can also be proved elegantly using physical arguments. Interpreting  Eq.~(\ref{WignerDyson}) as the stationary solution of a Fokker-Planck equation \cite{dyson72, mehta04}, it is easy to infer the associated Langevin equation that controls the fictitious dynamics, parametrized by the fictitious time $\tau$, of the independent matrix elements $A_{\eta}(\tau)$,\footnote{ $\eta$ labels independent elements of $A$. Alternatively, we can write $A_{\eta}= A_{mn}^{(\mu)}$, with $\mu=0, \dots, \beta-1$, see Eq.~(\ref{Perturbdev}).} as well as the drift and diffusion coefficients of the matrix elements $A_{\eta}$:
\begin{align}
\label{KramerscoeffM1}
M_1(A_{\eta})&=\lim_{\Delta\tau\to0}\frac{\langle \Delta A_{\eta}\rangle }{\Delta\tau}=-N \frac{\partial V^{g}(A_{\eta})}{\partial A_{\eta}},
\\
\label{KramerscoeffM2}
M_2(A_{\eta})&=\lim_{\Delta\tau\to0}\frac{\langle \Delta A_{\eta}^2\rangle}{2\Delta\tau}=\frac{1}{2\beta}\left[1+\delta_{\eta,(m,m)}\right],
\end{align}
where $\langle \ldots \rangle$ denotes the ensemble average over the fictitious Markov processes. This averaging must not be confused with averaging over different realizations of the random matrix $A$. The key point now is that we can calculate, by a second-order perturbative expansion in time, how the eigenvalues are modified during a time interval $\Delta\tau$ :
\be
\label{Perturbdev}
\Delta\Lambda _{n}=\Delta A_{nn}+\sum_{m\neq n}
\sum_{\mu=0}^{\beta-1}
\frac{\Delta A_{mn}^{(\mu)2}}{\Lambda _{m}-\Lambda _{n}}.
\ee
Averaging this equation using Eqs.~(\ref{KramerscoeffM1}) and (\ref{KramerscoeffM2}), and keeping only the terms $\sim \mc{O}(\Delta\tau)$, we find $\left<\Delta\Lambda _{n}\right>$ and $\left<\Delta\Lambda _{n}^2\right>$, and the related drift and diffusion coefficients for the eigenvalues:
\begin{align}
\label{KramerscoeffM1ev}
M_1(\Lambda _{n})&=-N \frac{\partial V^g(\Lambda_{n})}{\partial \Lambda_{n}} +\sum_{m\neq n}\frac{1}{\Lambda_{m}-\Lambda_{n}},
\\
\label{KramerscoeffM2ev}
M_2(\Lambda _{n})&=\frac{1}{\beta}.
\end{align}
We recognize in the drift coefficient (\ref{KramerscoeffM1ev}) the deterministic force driving the point charge located at $\Lambda _{n}$. In particular, we understand in a new way the origin of the electrostatic repulsion (\ref{VCoulRep}), since in the present context it arises from the second-order perturbative term in Eq.~(\ref{Perturbdev}). Finally, from the coefficients (\ref{KramerscoeffM1ev}) and (\ref{KramerscoeffM2ev}), it is straightforward to reconstruct the Fokker-Planck equation obeyed by the fictitious time-dependent joint probability density of the eigenvalues, and its stationnary solution is precisely the desired result (\ref{GibbsDG}).

\subsubsection{Mean-field approximation}
\label{SecMeanField}

Once the probability distribution $P(\{\Lambda_n\})$ is known, the density of eigenvalues $p(\Lambda)$ can formally be recovered by integrating it $(N-1)$ times. Luckily, we can avoid this cumbersome calculation by taking advantage of the Dyson gas picture. In a naive mean-field approach, the distribution of charges at equilibrium is found by minimizing the energy $H^g$ given by Eq.\ (\ref{HgDG}). This leads to
\be
\label{LinkVintVg}
-\partial_\Lambda V^{\mathrm{int}}(\Lambda)=N\partial_\Lambda V^g(\Lambda).
\ee
Furthermore, since for Hermitian matrices $\textrm{Im}\Lambda_n=0$, Eqs.~(\ref{VCoulRep}) and (\ref{gAsElecField}) yield
\be
\label{LinkRegVint}
N\,\textrm{Re}g(\Lambda + \textrm{i} \epsilon)=\sum_{n=1}^N\frac{1}{\Lambda-\Lambda_n}=-\partial_{\Lambda}V^{\mathrm{int}}(\Lambda),
\ee
so that the combination of Eqs.~(\ref{LinkVintVg}) and (\ref{LinkRegVint}) allows to relate $\textrm{Re}g(\Lambda' + \textrm{i} \epsilon)$ with the one-body potential $V^g$. Inserting the result into Eq.~(\ref{pVsReG}), we obtain
\begin{align}
\label{pVsVg}
p(\Lambda) &= \frac{1}{\pi^2\sqrt{(\Lambda-a)(b-\Lambda)}}
\\
&\times \left[
\pi-\textrm{P}\int_a^b  \textrm{d}\Lambda'  \frac{\sqrt{(\Lambda'-a)(b-\Lambda')}}{\Lambda'-\Lambda}
\partial_{\Lambda'} V^g(\Lambda')
\right].
\nonumber
\end{align}

Let us now justify this mean-field result in a different way. In the large $N$ limit, we can coarse-grain the energy functional $H^g$:
\begin{align}
\label{energyCG}
H^g(p)&\simeq N^2 \int_{-\infty}^\infty \textrm{d}\Lambda p(\Lambda)V^g(\Lambda)
\nonumber \\
&-\frac{N^2}{2}\iint_{-\infty}^\infty \textrm{d}\Lambda \textrm{d}\Lambda' p(\Lambda)p(\Lambda')\ln|\Lambda-\Lambda'|.
\end{align}
Rigorously, when changing the integration variables from $\{ \Lambda_n\}$ to the density `field'  $p$ in the partition function, a Jacobian appears, which physically takes into account the entropy associated with $p$. We neglect all corresponding sub-leading terms of order $\ln N/N$ \cite{dyson72}.\footnote{This is justified when the confining potential $V^g$ is `strong'. \citet{tierz03} gives explicit examples of failure of Eq.\ (\ref{energyCG}) for a `weak' confining potential.} The equilibrium of the Dyson gas is given by the extremum of this functional. Note that we also have to take into account the normalization constraint on $p$, which can be done by introducing a Lagrange multiplier $c$. We thus find:
\be
\label{Wignerequation}
V^g(\Lambda)-\int_{-\infty}^\infty \textrm{d}\Lambda'p(\Lambda')\ln|\Lambda-\Lambda'|+c=0.
\ee
Differentiating Eq.~(\ref{Wignerequation}) with respect to $\Lambda$ we get
\be
\label{WigneDiffEq}
\textrm{P}\int_{-\infty}^\infty   \textrm{d}\Lambda' \frac{p(\Lambda')}{\Lambda-\Lambda'} =\partial_\Lambda V^g(\Lambda),
\ee
which admits the solution (\ref{pVsVg}) for $p$ defined on the compact support $[a, b]$, as expected.

The mean-field approach used to infer the eigenvalue distribution $p(\Lambda)$ from the joint probability distribution $P(\{\Lambda_n\})$ is general and can be applied to any ensemble, provided that $P(\{\Lambda_n\})$ is known. Actually,  $P(\{\Lambda_n\})$ can be found for a larger class of matrices than the Wigner-Dyson ensemble (\ref{WignerDyson}). It is straightforward for any distribution $P(A)$ that is simply expressed in terms of the eigenvalues of $A$, e.g. through $\textrm{Tr}A$ or $\textrm{det}A$: $P(\{\Lambda_n\})$ is then obtained by multiplying $P(A)$ by the Vandermonde-type Jacobian $\vert \mc{V}(\{\Lambda_n\})\vert^\beta$ responsible for the logarithmic repulsion between eigenvalues.

\subsubsection{Application to specific random matrix ensembles}
\label{SecExApplDyson}

\paragraph{Gaussian ensemble.}

We start by considering the Gaussian ensemble (\ref{GaussianProba}) corresponding to $V^g(x)=x^2/4$ in Eq.\ (\ref{WignerDyson}). Using Eq.~(\ref{pVsVg}) with $a=-b$ found by the normalization condition $\int_{-b}^b \textrm{d}\Lambda p(\Lambda)=1$, we readily obtain the celebrated Wigner semicircle law \cite{wigner55}:\footnote{Note that if the quadratic $V^g$ is multiplied by an arbitrary constant $\alpha$, the eigenvalue density is found by a simple rescaling of variables: $p_{\alpha}(\Lambda)=\sqrt{\alpha} p_{\alpha=1}(\sqrt{\alpha}  \Lambda)$.}
\be
\label{semi-circle}
p(\Lambda)=\frac{1}{2\pi}\left(4-\Lambda^2\right)^{1/2}.
\ee
It states that for large $N$ and on average, the $N$ eigenvalues lie within a finite interval $[-2, 2]$, sometimes referred to as the `Wigner sea'. Within this sea, the eigenvalue distribution has a semicircular form.

\paragraph{Wishart ensemble.}

Our second example is the Wishart ensemble defined by Eqs.\ (\ref{WishartMatrix}) and (\ref{GaussProbDistr}). Let us focus on $P(A)$ given by Eq.~(\ref{PAWishart}) that corresponds to $c=N/M<1$. As explained above, $P(\{\Lambda_n\})$ follows by multiplying Eq.\ (\ref{PAWishart}) by the Jacobian $\vert \mc{V}(\{\Lambda_n\})\vert^2$:
\begin{align}
\label{GibbsWishart}
P(\{\Lambda_n\})&=C''_{N,M}e^{- 2 H^g(\{ \Lambda_n\})},
\\
\label{HgWishart}
H^g(\{\Lambda_n\})&=\frac{1}{2}\sum_{n=1}^N \left[ N\Lambda_n-(M-N)\ln \Lambda_n
\right]
\nonumber \\
&-\sum_{n<m}\ln|\Lambda_n-\Lambda_m|.
\end{align}
This result has the same form as Eqs.~(\ref{GibbsDG}) and (\ref{HgDG}), with the one-body potential
\be
\label{VgWishart}
V^g(x)=\frac{1}{2}\left[x-\left(\frac{1}{c}-1\right)\ln x \right],
\ee
which is repulsive in the limit $x\to 0^+$.
The linear and logarithmic contributions come from $\textrm{Tr}A$ and $\textrm{det}A$ in Eq.~(\ref{PAWishart}), respectively. Note the difference with $H$ entering in the definition of $A=HH^\dagger$, for which $V^g$ is harmonic due to the term $\textrm{Tr}HH^\dagger$ in Eq.~(\ref{GaussProbDistr}). Inserting the potential (\ref{VgWishart}) into Eq.~(\ref{pVsVg}), we find
\be
\label{MPfunction}
p(\Lambda)=\frac{1}{2\pi \Lambda}\sqrt{(\Lambda_+-\Lambda)(\Lambda-\Lambda_-)},
\ee
which is defined on the compact support $[\Lambda_-, \Lambda_+]$ with
\be
\label{Lambdapm}
 \Lambda_{\pm}=\left(\frac{1}{\sqrt{c}}\pm1\right)^2.
 \ee
This result was derived for $c<1$. It is easy to find the solution for $c>1$, by noting that, according to its definition (\ref{Defresolvent}),  $g$ is the average of
\be
\textrm{Tr}_{(N)}\frac{1}{z-HH^\dagger}=\textrm{Tr}_{(M)}\frac{1}{z-H^\dagger H}+\frac{N-M}{z},
\ee
where we used the cyclic permutation of the trace operator. From Eq.~(\ref{LinkpImG}), it is thus clear that the case $c>1$ is obtained by adding $N-M$ zero eigenvalues to $p(\Lambda)$. For arbitrary $c$, $p(\Lambda)$ has the generic form\footnote{If Eq.\ (\ref{GaussProbDistr}) is modified into
$P_\alpha(H)=C_{N,M}e^{-\alpha N\textrm{Tr}HH^\dagger}$, a rescaling of variables shows that $p_{\alpha}(\Lambda)=\alpha\, p_{\alpha=1}(\alpha  \Lambda)$.}
\be
\label{MarchenkoPastur}
p(\Lambda)=\left(1-\frac{1}{c}\right)^+\delta(\Lambda)+\frac{1}{2\pi \Lambda}\sqrt{(\Lambda_+-\Lambda)^+(\Lambda-\Lambda_-)^+},
\ee
where $x^+=\textrm{max}(x, 0)$. The result (\ref{MarchenkoPastur}) is the famous Marchenko-Pastur law \cite{marchenko67, tulino04}.

\paragraph{Euclidean random matrices?}

It would be tempting to apply the Dyson gas picture to ERMs. This requires to find $P(\{\Lambda_n\})$ in a form similar to Eqs.~(\ref{GibbsWishart}) and (\ref{HgWishart}). The problem is that, in order to derive $P(\{\Lambda_n\})$ with standard tools of RMT, we need $P(A)$, which is unfortunately unknown for ERMs. However, as we discussed in Sec.\ \ref{SubsecERM}, an ERM $A$ can be rewritten as $A=HTH^\dagger$, with entries $H_{i\alpha}$ that approximately behave as i.i.d. Gaussian random variables. The probability distribution of $H$ is then given by Eq.~(\ref{GaussProbDistr}). Hence, following the original Wishart's idea \cite{wishart28}, we expect $P(A)$ to be of the form
\be
\label{PAERM}
P(A)=C_{N,M}(T)\textrm{det}A^{M-N}e^{-N\textrm{Tr}(H\,T^{-1}H^\dagger)},
\ee
where the size $M$ of the matrix $T$ can be arbitrary, and in fact it will be infinite for the majority of functions $f(\vec{r}_i, \vec{r}_j)$. In Eq.~(\ref{PAERM}), we assume $N<M$  and $C_{N,M}(T)$ is a normalization coefficient that depends on the matrix $T$. For $T=I_M$, we recover the Wishart case (\ref{PAWishart}). This shows that the eigenvalue density of the ERM associated with the simplest matrix $T$ yields already a non-trivial result, the Marchenko-Pastur law (\ref{MarchenkoPastur}). An explicit example of ERM that obeys this law for a certain range of parameters is given in Sec.\ \ref{ERMSinc}. For arbitrary $T$, inferring $P(\{\Lambda_n\})$ from Eq.~(\ref{PAERM}) is \textit{a priori} not easy, inasmuch as the argument $\textrm{Tr}(H\,T^{-1}H^\dagger)$ cannot be expressed in terms of the eigenvalues of $A$. Therefore, integration over the independent parameters related to the eigenvectors of $A$ may be complicated. We believe, however, that the Dyson gas picture may be promising for ERMs, in particular when one is interested in more complicated quantities than just the density of eigenvalues, such as, e.g., eigenvalue correlations.

\subsection{Euclidean random matrices: heuristic approach}
\label{ERMHeuristic}

Before introducing advanced techniques to describe the spectal properties of ERMs, it might be useful to present simple arguments that apply for ERMs of the form $A_{ij}=f(\vec{r}_i - \vec{r}_j)$, in the situation where the function $f(\vec{r})$ decays fast enough for large $r$ so that, in the limit $V\to \infty$, the eigenvalue density $p(\Lambda)$ depends on the density $\rho=N/V$ only. We note that this is far from being the case for all ERMs; a couple of explicit counterexamples is given in Sec.\ \ref{ERMSinc} and \ref{ERMCosc}.

\subsubsection{High density expansion}
\label{HeuristicHigh}

For any ERM $A$, we can always formally write $\sum_{j=1}^NA_{ij}\Phi_j(\mathbf{k})=\Lambda_i(\mathbf{k})
\Phi_i(\mathbf{k})$  with $\Phi_i(\mathbf{k})=e^{-\textrm{i}\mathbf{k} \cdot \mathbf{r}_i}$ and
\be
\label{Pseudoev}
\Lambda_i(\mathbf{k})=\sum_{j=1}^Ne^{\textrm{i}\mathbf{k} \cdot (\mathbf{r}_i-\mathbf{r}_j)}f(\mathbf{r}_i-\mathbf{r}_j).
\ee
Suppose now that the density is large enough for the phase $\textrm{i}\mathbf{k}\cdot(\mathbf{r}_i-\mathbf{r}_j)$ to vary only weakly between neighboring points $\mathbf{r}_i$ and $\mathbf{r}_j$. In $d$ dimensions, this is roughly satisfied for $\rho^{1/d} \gg k$.  The sum in Eq.~(\ref{Pseudoev}) can then be approximated by an integral, so that $ \Lambda_i(\mathbf{k})$ does not depend anymore on $i$, becoming an eigenvalue of $A$, $\Lambda(\mathbf{k})=\rho f_0(\mathbf{k})$,
where
\be
\label{Deff0k}
 f_0(\vec{k})=\int \textrm{d}^d\vec{r}f(\vec{r})e^{\textrm{i}\vec{k}\cdot\vec{r}}
\ee
is the Fourier transform of $f(\vec{r})$. This eigenvalue is associated with the eigenvector $(e^{-\textrm{i}\mathbf{k} \cdot \mathbf{r}_1}, \ldots, e^{-\textrm{i}\mathbf{k} \cdot \mathbf{r}_N})$. Summing over the different eigenvalues labelled by $\mathbf{k}$, we obtain the eigenvalue density \cite{mezard99}:
\be
\label{pHighDensity1}
p(\Lambda)=\frac{1}{\rho}\int \frac{\textrm{d}^d\mathbf{k}}{(2\pi)^d}\;\delta\left[\Lambda-\rho f_0(\mathbf{k})\right].
\ee

For ERMs of the form (\ref{DefinitionERMu}) with $u = 1$, Eq.(\ref{Pseudoev}) is replaced by
\be
\label{Pseudoev2}
\Lambda_i(\mathbf{k})=\sum_{j=1}^N\left[e^{\textrm{i}\mathbf{k} \cdot (\mathbf{r}_i-\mathbf{r}_j)}-1\right] f(\mathbf{r}_i-\mathbf{r}_j)
\ee
and the eigenvalue density reads
\be
\label{pHighDensity11}
p(\Lambda)=\frac{1}{\rho}\int \frac{\textrm{d}^d\mathbf{k}}{(2\pi)^d}\;\delta\left[\Lambda-\rho \left[f_0(\mathbf{k})-f_0(\vec{0})\right]\right].
\ee

\subsubsection{Low density expansion}
\label{HeuristicLow}
In the low density limit $\rho\to 0$, for a rapidly decaying function $f(\vec{r}_i -\vec{r}_j)$, the matrix elements $A_{ij}$ are sizable only if the points $\vec{r}_i$ and $\vec{r}_j$ are nearest neighbors. In this case, the matrix $A$ can be approximated by a block diagonal matrix with $N/2$ blocks of size $2\times 2$. Each block describes the coupling between two nearest neighbors $i$ and $j$. Its eigenvalues are $\Lambda=f(\vec{0})\pm f(\mathbf{r}_i-\mathbf{r}_j)$. Hence, the eigenvalue density of $A$ is
\begin{eqnarray}
\label{pLowDensity1}
p(\Lambda) &=& \frac{1}{2}\int \textrm{d}^d \Delta r\,
p_{nn}( \Delta r) \left\{
\delta \left[\Lambda- f(0) - f(\Delta r) \right]
\right.
\nonumber\\
&+& \left.\delta\left[\Lambda -f(0) + f(\Delta r) \right]
\right\},
\end{eqnarray}
where $ p_{nn}( \Delta r)$ is the probability to find two nearest neighbors separated by a distance $\Delta r$:
\be
p_{nn}( \Delta r)=d\,\mc{V}\,\rho\,(\Delta r)^{d-1}e^{-\mc{V}\rho (\Delta r)^d}.
\ee
In this expression, $\mc{V}=\pi^{d/2}/\Gamma(d/2+1)$ is the volume of a $d$-dimensional unit sphere. If $f(r)$ is a monotonically decreasing function decaying to $0$ in the limit $r\to\infty$, the cumulative $C(\Lambda)=\int_{\Lambda}^{\infty} \textrm{d}\Lambda'\, p(\Lambda')$ takes a simple form:
\be
C(\Lambda)=\frac{1}{2}
\textrm{sgn}[f(0)-\lambda]e^{-\mc{V}\rho \{f^{-1}[\vert f(0)-\Lambda \vert]\}^d} +\frac{1}{2}
\label{cumul}
\ee
for $\Lambda \in [0,2f(0)]$, $C(\Lambda)=1$ for $\Lambda<0$ and $C(\Lambda)=0$ for $\Lambda>2f(0)$. $p(\Lambda)$ can be obtained as the negative derivative of $C(\Lambda)$.

For an ERM defined by Eq.\ (\ref{DefinitionERMu}) with $u = 1$, the two eigenvalues of the $2\times 2$ blocks become $-2f(\mathbf{r}_i-\mathbf{r}_j)$ and $0$. Therefore, the eigenvalue density reads
\be
\label{pLowDensity11}
p(\Lambda)=\frac{1}{2}\int \textrm{d}^d \Delta r\,  p_{nn}( \Delta r)
\delta \left[\Lambda+ 2f(\Delta r) \right] +
\frac{1}{2}\delta(\Lambda),
\ee
and for a monotonically decreasing function $f(r)$ decaying to $0$, the cumulative reduces to
\be
C(\Lambda)=\frac{1}{2}
e^{-\mc{V}\rho \{f^{-1}[-\Lambda/2]\}^d}+\frac{1}{2}
\ee
for $\Lambda \in [-2f(0),0]$, $C(\Lambda)=1$ for $\Lambda<-2f(0)$ and $C(\Lambda)=0$ for $\Lambda>0$.
Equation (\ref{pLowDensity11}) shows that half of the eigenvalues are zero. This unphysical result may come from the fact that the condition $A_{ii}=-\sum_{j\neq i}A_{ij}$ cannot be applied for every block independently. A simple way to get rid of those zero eigenvalues is to suppress the delta contribution as well as the prefactor $1/2$ in Eq. (\ref{pLowDensity11}). The fact that half of the $N$ points contribute to the remaining term can be taken into account by replacing $\rho$ by $\rho/2$ in $ p_{nn}( \Delta r)$. The corresponding cumulative
 \be
\label{Cumul1}
C(\Lambda)=
e^{-\mc{V}\rho \{f^{-1}[-\Lambda/2]\}^d/2}
\ee
was used by \citet{amir10} to study the eigenvalue density of the exponential ERM. We review this study in more detail in Sec.\ \ref{expERM}.

\subsection{Field representation}
\label{Fieldreprersentation}

In this section we discuss a field-theoretical representation of the resolvent $g(z)$. The starting point is Eq.\ (\ref{GasDet}), that we rewrite as
\be
\label{GasDet2}
g(z)= -\frac{2}{N}\partial_z \left\langle\textrm{ln det}(z-A)^{-1/2}\right\rangle.
\ee
The determinant $\textrm{det}(z-A)^{-1/2}$ can be represented as a canonical partition function
\begin{align}
\label{partition1}
\mc{Z}(z) &= \textrm{det}(z-A)^{-1/2}
\nonumber \\
&= \int\frac{\textrm{d}\phi_1}{\sqrt{2\pi}}...\frac{\textrm{d}\phi_N}{\sqrt{2\pi}}\exp{\left[-\frac{1}{2}\Phi^T\left(zI_N-A\right)\Phi\right]},
\end{align}
where $\Phi^T$ is the transpose of the vector $\Phi=(\phi_1,...,\phi_N)$. In this way, we recast the calculation of the resolvent $g(z)$ into a statistical mechanics problem of $N$ interacting particles $\phi_i$ with a Hamiltonian
\be
\mathcal{H}(\Phi,z)=\frac{z}{2}\sum_{i=1}^N\phi_i^2-\frac{1}{2}\sum_{i\neq j=1}^NA_{ij}\phi_i\phi_j.
\ee
The corresponding Boltzmann-Gibbs distribution is
\be
\label{GibbsField}
P(\Phi,z)=\frac{1}{\mathcal{Z}(z)}e^{-\mathcal{H}(\Phi,z)},
\ee
so that the resolvent (\ref{GasDet2}) is proportional to the derivative of the average thermodynamic free energy, $-\ln \mc{Z}(z)$:
\be
\label{GAsLnZ}
g(z)=-\frac{2}{N}\partial_z \left<\textrm{ln}\mathcal{Z}(z) \right>=-\frac{1}{N}\left<\sum_{i=1}^N\left<\phi_i^2\right>_z\right>,
\ee
where $\langle \ldots \rangle_z$ denotes the field average with respect to $P(\Phi,z)$ defined by Eq.~(\ref{GibbsField}).
In order to compute $\left<\textrm{ln}\mathcal{Z}(z) \right>$, we apply the replica method based on a smart use of the identity
\begin{equation}
\textrm{ln}\;x=\lim_{n\to 0}\frac{x^n-1}{n}.
\end{equation}
The idea is to compute the right-hand side for finite and integer $n$ and then perform the analytic continuation to $n\to0$.\footnote{For some models, the analytic continuation may not be unique, and the replica trick may fail. In a more rigorous treatment, we have to use the supersymmetric approach \cite{efetov97, haake10}. } Eq.~(\ref{GAsLnZ}) becomes
\begin{equation}\label{Green vs partition}
g(z)=-\frac{2}{N}\partial_z \left[\lim_{n\to 0}\frac{1}{n}\left<\mathcal{Z}^n(z)\right> \right].
\end{equation}
The quantity that we now want to evaluate is $\left<\mathcal{Z}^n(z)\right> $. It contains $n$ copies (replicas) of the original system (\ref{partition1}):
\begin{align}
\label{partition2}
\left<\mathcal{Z}^n(z)\right> &=\left(\frac{1}{2\pi}\right)^{\frac{Nn}{2}}
\int(\textrm{d}\phi_1^1 \ldots \textrm{d}\phi_1^n) \ldots (\textrm{d}\phi_N^1 \ldots \textrm{d}\phi_N^n)
\nonumber \\
&\left<\exp{\left[-\frac{1}{2}\sum_{\alpha=1}^n\Phi^{\alpha T}\left(zI_N-A\right)\Phi^{\alpha}\right]}\right>.
\end{align}

The averaging $\langle \ldots \rangle$ in this equation can be performed in different ways, depending on what we know about $A$. In the standard RMT, the averaging is performed over the distribution $P(A)$ that is known. Without entering into details, let us describe the way to proceed with  Eq.\ (\ref{partition2}) in this case. First, we perform two algebraic manipulations: we integrate over the matrix elements (which is possible, in practice, for Gaussian-like distributions), and we introduce auxiliary fields such that integration over replica variables can be carried out. We thus get a new integral form that depends only on these new fields. Second, in the $N\to \infty$ limit, we find the relevant values of these fields by making a saddle point approximation. This method was originally applied to the Gaussian ensemble (\ref{GaussianProba}) by \citet{edwards79} who rederived the semicircle law (\ref{semi-circle}). More recently, it was also applied to Wishart matrices (\ref{WishartMatrix}) (with arbitrary covariance matrix), and the Marchenko-Pastur law (\ref{MarchenkoPastur}) was recovered \cite{sengupta99}.

For Hermitian ERMs of the form $f(\vec{r}_i, \vec{r}_j)=f(\vec{r}_i - \vec{r}_j)$, the field-theoretical approach was first proposed by \citet{mezard99}. Let us review some details of their calculation. For $A_{ij}=f(\vec{r}_i-\vec{r}_j)$, Eq.~(\ref{partition2}) becomes:
\begin{align}
 \label{partition3}
\left<\mathcal{Z}^n(z)\right> \varpropto &\int(\textrm{d}\phi_1^1\dots \textrm{d}\phi_1^n)\dots(\textrm{d}\phi_N^1\dots \textrm{d}\phi_N^n)
\nonumber
\\
&\int\frac{\textrm{d}^d\vec{r}_1}{V}\dots \frac{\textrm{d}^d\vec{r}_N}{V}
\exp\left[-\frac{z}{2}\sum_{\alpha=1}^n
\sum_{i=1}^N(\phi_i^{\alpha})^2
\right.
\nonumber \\
&+ \left. \frac{1}{2}\sum_{\alpha=1}^n\sum_{i,j=1}^Nf(\vec{r}_i-\vec{r}_j)
\phi_i^{\alpha}\phi_j^{\alpha}\right].
\end{align}
As explained just above, in order to perform the Gaussian integration over the replica fields, we introduce two sets of auxiliary (bosonic) fields $\psi^{\alpha}$ and $\hat{\psi}^{\alpha}$, i.e. we insert into Eq.~(\ref{partition3}) the relation
\begin{equation}
\int\prod_{\alpha=1}^n \textrm{D}[\psi^{\alpha}]\;\delta_F\left[\psi^{\alpha}(\vec{r})-\sum_{i=1}^N\phi_i^{\alpha}\delta(\vec{r}-\vec{r}_i)\right],
\end{equation}
where $\delta_F$ stands for the functional Dirac delta:
\begin{equation}
\delta_F[\psi]=\int  \textrm{D} [\hat{\psi}]\exp\left[ \textrm{i} \int \textrm{d}^d\vec{r}\;\psi(\vec{r})\hat{\psi}(\vec{r})\right].
\end{equation}
We then integrate out the $\phi$ variables to obtain a field representation of the partition function
\be
\left<\mathcal{Z}^n(z)\right>=\frac{1}{z^{Nn/2}}\int D[\psi^{\alpha},\hat{\psi}^{\alpha}]{\cal A}^N e^{\mathcal{S}_0},
\ee
where
\begin{align}
{\cal A} &= \int\textrm{d}^d\vec{r}\exp\left[
-\frac{1}{2z}\sum_{\alpha=1}^n\hat{\psi}^{\alpha}(\vec{r})^2\right] ,\nonumber
\\
\mathcal{S}_0 &= \textrm{i} \sum_{\alpha=1}^n\int \textrm{d}^d\vec{r}\;
\psi^{\alpha}(\vec{r})\hat{\psi}^{\alpha}(\vec{r})
\nonumber \\
&+ \frac{1}{2}\sum_{\alpha=1}^n\int \textrm{d}^d\vec{r}\; \textrm{d}^d\vec{r}'\psi^{\alpha}(\vec{r})f(\vec{r}-\vec{r}')\psi^{\alpha}(\vec{r}').
\end{align}
Finally, integrating out the $\psi$ fields, we get an expression which is a good starting point for different approximations:
\begin{equation}\label{partition4}
\left<\mathcal{Z}^n(z)\right>=\int D[\hat{\psi}^{\alpha}]e^{\mathcal{S}_1},
\end{equation}
with
\begin{align}
\label{actionS1}
\mathcal{S}_1 &= N\textrm{ln}\left\{z^{-n/2}\int \textrm{d}^d\vec{r}
\exp\left[-\frac{1}{2z}\sum_{\alpha=1}^n\hat{\psi}^{\alpha}(\vec{r})^2 \right] \right\}
\nonumber \\
&+ \frac{1}{2}\sum_{\alpha=1}^n\int \textrm{d}^d\vec{r}\textrm{d}^d\vec{r}'\hat{\psi}^{\alpha}(\vec{r})f^{-1}(\vec{r}-\vec{r}')\hat{\psi}^{\alpha}(\vec{r}').
\end{align}
Here $f^{-1}$ is the operator inverse of $f$ considered as an integral operator:
\be
\int \textrm{d}^d\vec{r}''f^{-1}(\vec{r}-\vec{r}'')f(\vec{r}''-\vec{r}')=\delta(\vec{r}-\vec{r}').
\ee

Suppose now that we can expand the logarithmic term in Eq.~(\ref{actionS1}). We omit terms independent of $\psi$ and apply the Wick rotation $\hat{\psi}\to \mathrm{i} \hat{\psi}$, so that the action $\mathcal{S}_1$ becomes:
\begin{align}
\label{actionS1approx1}
\mathcal{S}_1 &\simeq \rho z^{-n/2}\int \textrm{d}^d\vec{r}\exp\left[\frac{1}{2z}
\sum_{\alpha=1}^n\hat{\psi}^{\alpha}(\vec{r})^2\right]
\nonumber \\
&- \frac{1}{2}\sum_{\alpha=1}^n\int \textrm{d}^d\vec{r}\textrm{d}^d\vec{r}'\hat{\psi}^{\alpha}(r)f^{-1}(\vec{r}-\vec{r}')\hat{\psi}^{\alpha}(\vec{r}').
\end{align}
In the high density limit $\rho=N/V\to \infty$, \citet{mezard99} proposed to expand the exponential term of the action (\ref{actionS1approx1}), at $z/\rho$ fixed. Inserting the result into Eq.~(\ref{partition4}), we obtain:
\be
\label{GreenHighDensity1}
g(z)=\frac{1}{\rho}\int \frac{\textrm{d}^d\mathbf{k}}{(2\pi)^d} \frac{1}{z-\rho f_0(\mathbf{k})},
\ee
where $f_0(\vec{k})$ is the Fourier transform of $f(\vec{r})$, defined in Eq. (\ref{Deff0k}). The corresponding density of eigenvalues (\ref{LinkpImG}) coincides with the result (\ref{pHighDensity1}) obtained from heuristic arguments.

In order to obtain an expression for the resolvent $g(z)$ valid beyond the high density regime, \citet{mezard99} looked for the best quadratic action $\mathcal{S}_v$ that approximates the full problem (\ref{actionS1approx1}):
\be
\mathcal{S}_v=-\frac{1}{2}\int \textrm{d}^d\vec{r}\textrm{d}^d\vec{r}'\hat{\Psi}^{T}(\vec{r})K^{-1}(\vec{r},\vec{r}')\hat{\Psi}(\vec{r}'),
\ee
where $\hat{\Psi}^{T}=(\hat{\psi}^1,...,\hat{\psi}^n)$. The $n\times n$ matrix $K^{-1}(\vec{r},\vec{r}')$ is obtained by minimizing the variational free energy $F_v=\left<\mathcal{S}_1\right>_v-\textrm{ln}\mathcal{Z}_v$, where $\mathcal{Z}_v=\int  \textrm{D} [\hat{\Psi}]e^{\mathcal{S}_v}$ and $\left<...\right>_v$ is defined with respect to the measure $P_v=e^{\mathcal{S}_v}/\mathcal{Z}_v$. This leads to the following self-consistent equations for the resolvent $g(z)$:\footnote{Although Eqs.~(\ref{SelfconsistentMezard}) and (\ref{SelfconsistentMezard2}) do not appear explicitly in \cite{mezard99}, it is straightforward to obtain them from the results presented in that paper.}
\begin{align}
\label{SelfconsistentMezard}
g(z)&=\frac{1}{z-\sigma(z)},
\\
\label{SelfconsistentMezard2}
\sigma(z)&=\int\frac{\textrm{d}^d\mathbf{k}}{(2\pi)^d} \frac{f_0(\mathbf{k})}{1-\rho f_0(\mathbf{k})g(z)}.
\end{align}
These equation implicitly assume that the function $f(\vec{r})$ decays fast enough for large $r$, so that the translational invariance is restored in the limit $V \to \infty$ at fixed density $\rho=N/V$. Consequently, the resolvent $g(z)$ and the density of eigenvalues $p(\Lambda)$ depend only on the density $\rho$.

\subsection{Diagrammatic approach}
\label{SecDiagrammaticApproach}

\subsubsection{From Gaussian and Wishart ensembles to Euclidean random matrices}

Before discussing in details the diagrammatic treatment of Hermitian ERMs, we briefly review the results for Gaussian and Wishart matrices. The starting point of a diagrammatic computation of the resolvent (\ref{Defresolvent}) is its series expansion (\ref{gSumMoments}). For Gaussian-like ensembles, the result of averaging can be expressed through pairwise contractions, such as Eq.\ (\ref{ContractionGUE}). The different terms (diagrams) arising from this calculation are conveniently collected in the self-energy $\sigma(z)$ defined by Eq.~(\ref{SelfenergyRMT}). By construction, $\sigma(z)$ is the sum of all irreducible diagrams contained in the expansion of $g(z)$, i.e. those that cannot be separated into two independent diagrams linked by the propagator $1/z$. We do not detail the diagrammatic representation of $\sigma(z)$ for Gaussian and Wishart ensembles because these ensembles can be considered as special cases of ERMs, for which a diagrammatic calculation is given below.

It is easy to show, using the pairwise contractions (\ref{ContractionGUE}) for GUE and (\ref{ContractionGOE}) for GOE, that the self-energy $\sigma(z)$ of the Gaussian ensemble (\ref{GaussianProba}) is given by \cite{jurkiewicz08}\footnote{\citet{jurkiewicz08} treat GUE. The GOE case is slightly more involved since Eq.\ (\ref{ContractionGOE}) generates two types of diagrams rather than one for Eq.\ (\ref{ContractionGUE}). However, in the large $N$ limit, the second term of (\ref{ContractionGOE}) does not contribute to $\sigma(z)$ because it gives rise to non-planar diagrams only (for the definition of these diagrams, see Sec.\ \ref{SecERMSelfConsistent}).}
\be
\label{SelfenergyGaussian}
\sigma(z)=g(z).
\ee
Inserting Eq.\ (\ref{SelfenergyGaussian}) into Eq.\ (\ref{SelfenergyRMT}), we find the resolvent
\be
g(z)=\frac{1}{2}\left(z-\sqrt{z^2-4}\right),
\ee
which leads, via Eq.~(\ref{LinkpImG}), to the semicircle law (\ref{semi-circle}).

The self-energy $\sigma(z)$ of Wishart random matrix ensemble (\ref{WishartMatrix}) is obtained in a similar way. The main difference with the Gaussian case is that we now have to distinguish, when manipulating pairwise contractions (\ref{Contraction0}), indices $i=1,\dots, N$ and $\alpha=1, \dots, M$. For $c=N/M<1$, the self-energy is \cite{jurkiewicz08}:
\be
\label{SelfenergyWishart}
\sigma(z)=\frac{1}{c}\frac{1}{1-g(z)}.
\ee
Equations (\ref{SelfenergyRMT}) and (\ref{SelfenergyWishart}) lead to a quadratic equation for $g(z)$, that has a normalizable solution
\be
\label{ResolventMP}
g(z)=\frac{1}{2z}\left[z+1-\frac{1}{c}-\sqrt{(z-\Lambda_+)(z-\Lambda_-)}
\right],
\ee
with $\Lambda_\pm$ given by Eq.~(\ref{Lambdapm}). From Eq.~(\ref{LinkpImG}), we recover the Marchenko-Pastur function (\ref{MPfunction}).

Historically, neither the Wigner semicircle law (\ref{semi-circle}) nor the Marchenko-Pastur law (\ref{MarchenkoPastur}) were derived by calculating diagrammatically the self-energy $\sigma(z)$. Wigner's original proof \cite{wigner58} is based on an explicit calculation of the moments $\moy{\Lambda^n}$ that appear in the series expansion (\ref{gSumMoments}) of the resolvent. This is somewhat surprising inasmuch as the counting procedure required to evaluate the moments is more complicated than the direct evaluation of the self-energy (\ref{SelfenergyGaussian}). Odd moments of the symmetric semicircle law are zero, and even moments are the Catalan numbers:
\be
\moy{\Lambda^{2p}}=\frac{(2p)!}{p!(p+1)!}.
\ee
A calculation of the Marchenko-Pastur law from its moments can also be performed \cite{bai99}. The procedure is quite tricky, as we can imagine by looking at the result for the moments:
\be
\moy{\Lambda^n}=\frac{1}{c^n}\sum_{k=0}^n\frac{n!(n-1)!}{(n+1-k)!(n-k)!\left[(k-1)!\right]^2}\frac{c^{k-1}}{k}.
\ee

Undoubtedly, if we are interested in the full distribution $p(\Lambda)$, the counting procedure for evaluating the moments is less appropriate than the diagrammatic self-consistent calculation of the self-energy. The same remark holds for ERMs, as we will see shortly.

The main object in the diagrammatic analysis of a Hermitian ERM $A$ is the operator
 \be
 \label{DefOA2}
\hat{O}(z)=\left\langle
\sum_{i=1}^N\sum_{j=1}^N
\left[
\frac{1}{z-A}
\right]_{ij}
\ket{\vec{r}_i}\bra{\vec{r}_j}
\right\rangle.
 \ee
For later convenience, we also define
\begin{align}
g_\vec{k}(z)&=\frac{1}{\rho}\braket{\vec{k}}{\hat{O}(z)}{\vec{k}},
\nonumber
\\
\label{Defgk}
&=\frac{1}{N}\left<\sum_{i=1}^N\sum_{j=1}^N e^{\textrm{i}\vec{k}\cdot(\vec{r}_i-\vec{r}_j)}\left[\frac{1}{z-A}\right]_{i,j}\right>.
\end{align}
Since in the limit $k\to\infty$ only terms $i=j$ contribute significantly in Eq.~(\ref{Defgk}), $g_\vec{k}(z)$ is related to the resolvent (\ref{Defresolvent}) by
\be
\label{LinkggkERM}
 g(z)=\lim_{k\to\infty}g_\vec{k}(z).
\ee

Similarly to $g(z)$, $g_\vec{k}(z)$ admits a series expansion in its holomorphic domain:
\begin{align}
\label{gasMomentskERM}
g_\vec{k}(z)&= \sum_{n=0}^\infty\frac{\moy{\Lambda^{n}}_\vec{k}}{z^{n+1}},
\\
\label{MomentskERM}
\moy{\Lambda^{n}}_\vec{k}&=\frac{1}{N}\left<\sum_{i=1}^N\sum_{j=1}^Ne^{\textrm{i}\vec{k}\cdot(\vec{r}_i-\vec{r}_j)}\left[A^n\right]_{ij}\right>.
\end{align}

\subsubsection{Euclidean random matrices: high density expansion}
\label{SecERMHighDensity}

In this section, inspired by the work of \citet{grigera01b, grigera11}, we will present a perturbative calculation of the resolvent $(\ref{Defgk})$ by direct evaluation of moments (\ref{MomentskERM}) for ERMs of the form  $A_{ij}=f(\vec{r}_i, \vec{r}_j)=f(\vec{r}_i - \vec{r}_j)$. The moments
\begin{align}
\label{explicitFouriermoment}
\moy{\Lambda^{n}}_\vec{k} &=
\frac{1}{N}\left<\sum_{i_1=1}^N \dots \sum_{i_{n+1}=1}^N e^{\textrm{i}\mathbf{k} \cdot(\mathbf{r}_{i_1}-\mathbf{r}_{i_{n+1}})}
\right.
\nonumber \\
&\times \left. \vphantom{\sum_{i_{n+1}=1}^N} A_{i_1,i_2}A_{i_2,i_3} \dots A_{i_{n-1},i_n}A_{i_n,i_{n+1}}\right>
\end{align}
can be expressed as sums of $n$ terms characterized by the number of repeating indices. The term with all indices different is
\begin{align}
\label{momentkERM}
&\moy{\Lambda^{n}}^{(n)}_\vec{k} = N^{n}\int_V\frac{\textrm{d}^d\vec{r}_1}{V} \dots \frac{\textrm{d}^d\vec{r}_{n+1}}{V}
e^{\textrm{i}\mathbf{k} \cdot(\mathbf{r}_{i_1}-\mathbf{r}_{i_{n+1}})}
\nonumber \\
&\times f(\vec{r}_1-\vec{r}_2)\dots f(\vec{r}_{n}-\vec{r}_{n+1})f(\vec{r}_{n+1}-\vec{r}_1).
\end{align}
Assuming translational invariance, we can eliminate one integral in Eq.~(\ref{momentkERM}), thus showing that $\moy{\Lambda^{n}}^{(n)}_\vec{k}\sim \rho^{n}$. When two indices are equal in Eq.~(\ref{explicitFouriermoment}), we have a missing $N$ factor from the sum and a missing $1/V$ factor from the average, leading to $\moy{\Lambda^{n}}^{(n-1)}_\vec{k}\sim \rho^{n-1}$. Therefore, $\moy{\Lambda^{n}}_\vec{k}$ has the following density expansion:
\be
\label{DensityExpansionMomk}
\moy{\Lambda^{n}}_\vec{k}=\sum_{i=1}^n
\moy{\Lambda^{n}}^{(i)}_\vec{k}
\;\;\;\textrm{with}\;\;\;
\moy{\Lambda^{n}}^{(i)}_\vec{k} \sim \rho^{i}.
\ee

Let us compute explicitly the two first leading terms in the high density regime, $\moy{\Lambda^{n}}^{(n)}_\vec{k}$ and $\moy{\Lambda^{n}}^{(n-1)}_\vec{k}$. We replace all functions $f(\vec{r}_i-\vec{r}_j)$ in Eq.~(\ref{momentkERM}) by $\int \textrm{d}^d\vec{k}f_0(\vec{k})e^{-\textrm{i}\vec{k}\cdot(\vec{r}_i-\vec{r}_j)}/(2\pi)^d$, and perform the $n$ spatial integrations. For points $\vec{r}_i$  in a box of side $L$ and volume $V=L^d$, $\moy{\Lambda^{n}}^{(n)}_\vec{k}$ becomes
\begin{align}
&\moy{\Lambda^{n}}^{(n)}_\vec{k} = \frac{N^n}{(2\pi)^{nd}}\int \textrm{d}^d\mathbf{k}_1 \dots \textrm{d}^d\mathbf{k}_n\,
\textrm{sinc}\left[\frac{\mathbf{k}-\mathbf{k}_1}{2/L}\right]\dots
\nonumber
\\
&\times \textrm{sinc}\left[\frac{\mathbf{k}_{n-1}-\mathbf{k}_n}{2/L}\right]\textrm{sinc}\left[\frac{\mathbf{k}_n-\mathbf{k}}{2/L}\right]   f_0(\mathbf{k}_1) \dots f_0(\mathbf{k}_n),
\end{align}
where, for $d=3$, $\textrm{sinc}\left[\mathbf{k}\right]=
\textrm{sinc}\left[k^x\right]\textrm{sinc}\left[k^y\right]
\textrm{sinc}\left[k^z\right]$. Assuming that $f_0(\vec{k})$ is centered around, say, $\vec{k}_{a}$, with a width $\Delta k_{a}$ such that $k_{a}L \gg 1$ and $\Delta k_{a}L \ll 1$, we use $\textrm{sinc}[(\mathbf{k}_i-\mathbf{k}_j)L/2]\simeq (2\pi)^d\delta(\mathbf{k}_i-\mathbf{k}_j)/L^d$ and obtain
\be
\label{Gamma Fourier high rho}
\moy{\Lambda^{n}}^{(n)}_\vec{k}=\left[\rho f_0(\mathbf{k})\right]^n.
\ee
Inserting this into Eq.~(\ref{gasMomentskERM}), we obtain the crudest approximation $g^0_\vec{k}(z)$ for the resolvent $g_\vec{k}(z)$:
\be
g^0_\vec{k}(z)=\frac{1}{z-\rho f_0(\mathbf{k})}.
\ee
This result is the `bare' propagator that does not capture any fluctuations of $A$. Using $g_\vec{k}(z) = g^0_\vec{k}(z)$ leads to a trivial result $g(z)=\lim_{k\to \infty} g_\vec{k}(z)=1/z$. We thus calculate the next-order contribution $\moy{\Lambda^{n}}^{(n-1)}_\vec{k}$, which contains two equal indices. There are two differences with the calculation of $\moy{\Lambda^{n}}^{(n)}_\vec{k}$. First, we can choose the two positions of the equal indices. Second, for given positions such that we have $\beta+2$ functions $f$ between the two equal indices,\footnote{$\beta\in[0, n-2]$ is an integer that should not be confused with the symmetry index defined in Sec.\ \ref{SecGaussianMatrices}.} we replace $\beta +1$ sinc terms by $\delta$-functions. The result reads
\begin{align}
\label{GammaFourierhighrho2}
\moy{\Lambda^{n}}^{(n-1)}_\vec{k} &= \frac{1}{\rho}\sum_{\alpha+\beta+\gamma=n-2}\left[\rho f_0(\mathbf{k})\right]^\alpha
\nonumber \\
&\times \left[\int \frac{\textrm{d}^d\mathbf{q}}{(2\pi)^d} \left[\rho f_0(\mathbf{q})\right]^{(\beta+2)}\right]\left[\rho f_0(\mathbf{k})\right]^\gamma.
\end{align}
Summing over $n$ to get the corresponding resolvent (\ref{gasMomentskERM}),  $g_\vec{k}(z)\simeq g^1_\vec{k}(z)$, suppresses the restriction imposed on $\alpha$, $\beta$ and $\gamma$:
\begin{align}
g^1_\vec{k}(z) &=
\left[\frac{1}{z-\rho f_0(\mathbf{k})}\right]
\left[\frac{1}{\rho}\int\frac{\textrm{d}^d\mathbf{q}}{(2\pi)^d} \frac{1}{z-\rho f_0(\mathbf{q})}\left[\rho f_0(\mathbf{q})\right]^2\right]
\nonumber \\
&\times
\left[\frac{1}{z-\rho f_0(\mathbf{k})}\right],
\end{align}
which is of the form $g^0_\vec{k}(z)\sigma^1 (z)g^0_\vec{k}(z)$. The first irreducible diagram contained in the self-energy $\sigma_\vec{k}(z)=1/g^0_\vec{k}(z)-1/g_\vec{k}(z)$ is therefore independent of $\mathbf{k}$ and reads
\begin{align}
\label{self1}
\sigma^1(z) & =  \frac{1}{\rho}\int\frac{\textrm{d}^d\mathbf{q}}{(2\pi)^d}\left[\rho f_0(\mathbf{q})\right]^2g^0_\vec{q}(z).
\end{align}

If now we restrict the density expansion of the self-energy to the first order (\ref{self1}), $\sigma_\vec{k}(z)\simeq\sigma^1(z)$, the resolvent (\ref{Defgk}) is given by
\be
\label{Approxgksigma1}
g_\vec{k}(z)=\frac{1}{z-\rho f_0(\mathbf{k})-\sigma^1(z)},
\ee
and, from Eqs.~(\ref{LinkpImG}) and (\ref{LinkggkERM}), the density of eigenvalues takes the form
\be
p(\Lambda) = \frac{\textrm{Im}\sigma^1(\Lambda+\textrm{i}\epsilon)}{\left[\Lambda-\textrm{Re}\sigma^1(\Lambda+\textrm{i}\epsilon)\right]^2+\left[\textrm{Im}\sigma^1(\Lambda+\textrm{i}\epsilon)\right]^2}.
\ee
For $|\Lambda|\gg |\textrm{Re}\sigma^1(\Lambda+\textrm{i}\epsilon)|,|\textrm{Im}\sigma^1(\Lambda+\textrm{i}\epsilon)|$, $p(\Lambda)\simeq\textrm{Im}\sigma^1(\Lambda+\textrm{i}\epsilon)/\Lambda^2$. Using the explicit form (\ref{self1}) of $\sigma_1$, we recover the result (\ref{pHighDensity1}).\footnote{Another way to recover Eq.~(\ref{pHighDensity1}) is to compute the series (\ref{gSumMoments}) with $\moy{\Lambda^{n}}\simeq \moy{\Lambda^{n}}^{(n)}$ calculated following the same procedure as for $\moy{\Lambda^{n}}^{(n)}_\vec{k}$.} This indicates that the more diagrams we take into account in $\sigma_\vec{k}(z)$, the more accurate is $p(\Lambda)$ at small  $|\Lambda|$. Applying essentially the same procedure as for the calculation of $\sigma^1(z)$, it is also possible to compute higher-order contributions $\sigma_\vec{k}^i$ ($i>1$) of order $1/\rho^i$  to the self-energy $\sigma_\vec{k}(z)=\sigma^1(z)+\sigma_\vec{k}^2(z)+\dots$, even though the combinatorial rules presented in the recent literature \cite{ganter11, grigera11} are quite involved.  A simple way to improve Eq.\ (\ref{self1}) is to replace the bare propagator $g^0_\vec{q}(z)$ in the integrand by $g_\vec{q}(z)$, thus obtaining a system of self-consistent equations for $\sigma(z)$ and $g_\vec{k}(z)$.

With minor modifications, the analysis of this section can be repeated for ERMs defined by Eq.\ (\ref{DefinitionERMu}) with $u = 1$. An important complication in this case is that the self-energy $\sigma^1$ becomes $\vec{k}$-dependent. The self-consistent equations for $g_\vec{k}(z)$ and $\sigma_{\vec{k}}(z)$ read \cite{grigera01a}:
\begin{eqnarray}
\label{sc1u1}
g_\vec{k}(z) &=& \frac{1}{z-\rho [f_0(\mathbf{k})-f_0(\mathbf{0})]-\sigma_{\vec{k}}(z)},
\\
\label{sc2u1}
\sigma_{\vec{k}}(z) &=&  \frac{1}{\rho}\int\frac{\textrm{d}^d\vec{q}}{(2\pi)^d}\left\{ \rho \left[ f_0(\vec{q}) - f_0(\vec{k}-\vec{q}) \right] \right\}^2 g_{\vec{q}}(z).\hspace*{8mm}
\end{eqnarray}

\subsubsection{Euclidean random matrices: self-consistent equations}
\label{SecERMSelfConsistent}

%%%%%FIG%%%%%
\begin{figure}[t!]
%\vspace{3mm}
\centering{
\includegraphics[angle=0,width=\columnwidth]{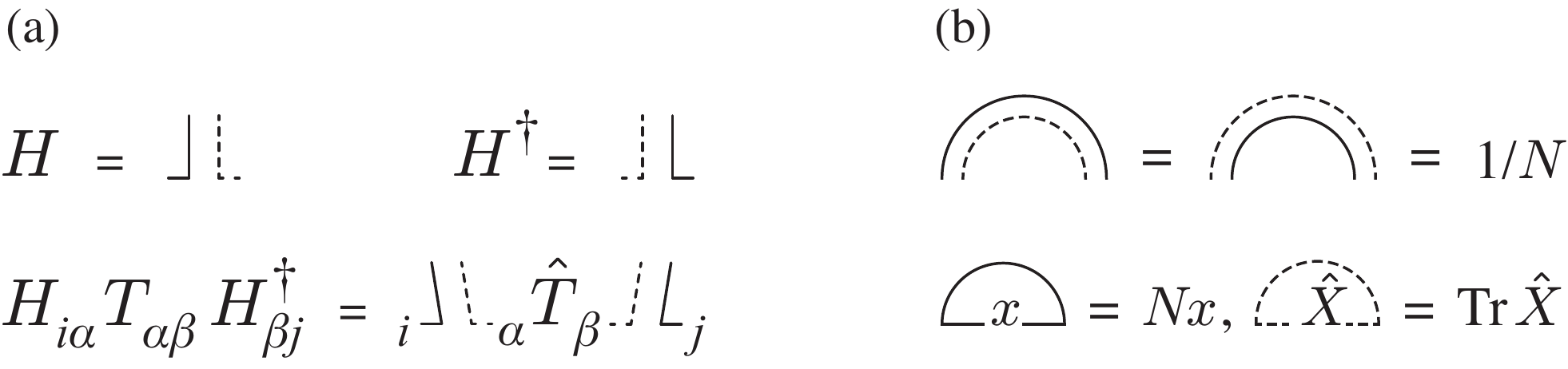}
\caption{
\label{HHAandTracediag}
(a) Diagrammatic representations of the matrices $H$, $H^{\dagger}$, and $A=HTH^{\dagger}$. Full and dashed lines propagate in the bases $\{\mathbf{r}_i\}$ and $\{\psi_\alpha\}$ defined in Sec.\ \ref{SubsecERM}, respectively; $\hat{T}=\rho\hat{A}$.
(b) Diagrammatic notation for pairwise contractions (\ref{Contraction0}) and loop diagrams for any scalar $x$ in the basis $\{\mathbf{r}_i\}$ and for any operator $\hat{X}$ in an arbitrary basis $\{\psi_\alpha\}$.}}
\end{figure}
%%%%%%%%%%%%%

In this section, we derive self-consistent equations for the operator (\ref{DefOA2}), using the representation (\ref{AHTH}) for an ERM $A_{ij}=f(\vec{r}_i, \vec{r}_j)=\braket{\vec{r}_i}{\hat{A}}{\vec{r}_j}$. We recall that the matrix $H$ is random but independent of the function $f$, whereas the matrix $T$ depends on $f$ but is not random (see Sec.\ \ref{SubsecERM}).

\paragraph{Diagrammatic calculation.}

We start by expanding the resolvent (\ref{Defresolvent}) in series in $1/z$:
\be
\label{SeriesgERM}
g(z)=\frac{1}{N}\left\langle
\textrm{Tr}
\left[
\frac{1}{z}+\frac{1}{z}A\frac{1}{z}+\frac{1}{z}A\frac{1}{z}A\frac{1}{z}+\dots
\right]
\right\rangle,
\ee
where averaging $\left<\dots\right>$ is performed over the ensemble of matrices $H$. As explained in Sec.\ \ref{SubsecERM}, we assume that $H$ has i.i.d. complex entries distributed according to the Gaussian law (\ref{GaussProbDistr}). Using the properties of Gaussian random variables (such as the Wick's theorem), the result of averaging in Eq.~(\ref{SeriesgERM}) can be expressed through pairwise contractions (\ref{Contraction0}). To evaluate efficiently the weight of different terms that arise in the calculation, it is convenient to introduce diagrammatic notations. The matrices $H$, $H^\dagger$, and $A$ will be represented as shown in Fig.\ \ref{HHAandTracediag}(a).

\begin{figure}
\vspace{2mm}
\centering{
\includegraphics[angle=0,width=\columnwidth]{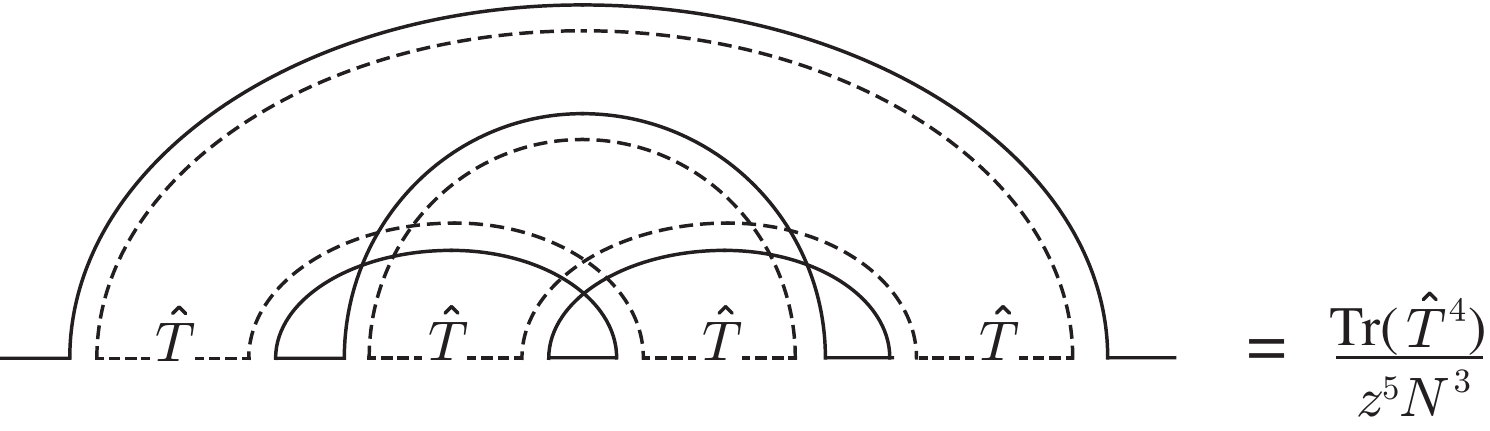}
\caption{
\label{nonplanardiag}
A typical non-planar diagram appearing in the expansion of the resolvent $g(z)$. Its value follows after application of `Feynman' rules defined in Fig.\ \ref{HHAandTracediag}(b). It does not survive in the limit $N \rightarrow \infty$.
}
}
\end{figure}

Each contraction (\ref{Contraction0}) brings a factor $1/N$, and each loop corresponding to taking the trace of a matrix brings a factor $N$, see Fig.\ \ref{HHAandTracediag}(b). In the limit $N \rightarrow \infty$, only the diagrams that contain as many loops as contractions will survive. These diagrams are those where full and dashed lines do not cross.
%%%%%FIG%%%%%
\begin{figure}
\vspace{2mm}
\centering{
\includegraphics[angle=0,width=\columnwidth]{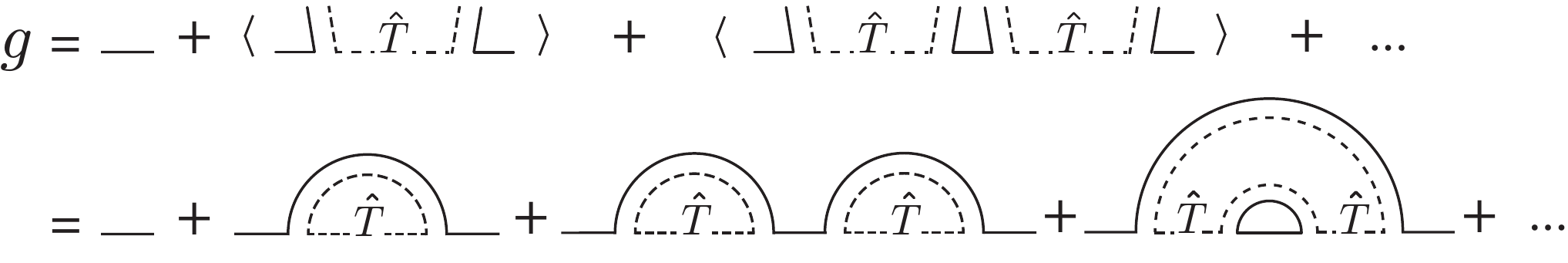}
\caption{
\label{gdiag}
Diagrammatic expansion of the resolvent $g(z)$. A horizontal straight line reprepresents the propagator $1/z$.
}
}
\end{figure}
%%%%%%%%%%%%%
Therefore, the leading order expansion of the resolvent (\ref{SeriesgERM}) involves only diagrams which are planar and look like rainbows, see Fig.\ \ref{gdiag} where we show the beginning of the expansion of $g(z)$. Note that the prefactor $1/N$ of Eq.~(\ref{SeriesgERM}) does not appear in Fig.\ \ref{gdiag} because it is compensated by the external trace. An example of a non-planar diagram is represented in Fig.\ \ref{nonplanardiag}. It vanishes in the limit $N\to\infty$.
%%%%%FIG%%%%%
%%%%%%%%%%%%%

%%%%%FIG%%%%%
\begin{figure}
\vspace{2mm}
\centering{
\includegraphics[angle=0,width=\columnwidth]{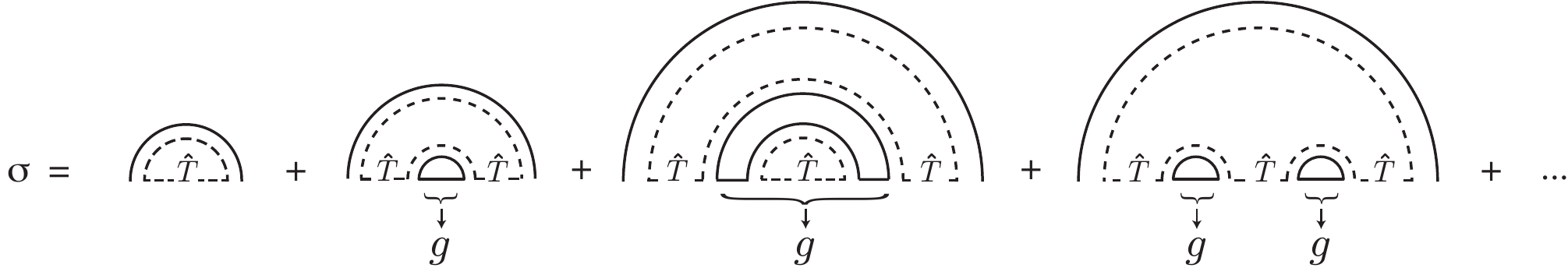}
\caption{
\label{sigmadiag}
Diagrammatic expansion of the self-energy $\sigma(z)$. Braces with arrows denote parts of diagrams that are the beginning of the diagrammatic expansion of the resolvent $g(z)$. }
}
\end{figure}
%%%%%%%%%%%%%

%%%%%FIG%%%%%
\begin{figure*}
\vspace{2mm}
\centering{
\includegraphics[angle=0,width=0.8\textwidth]{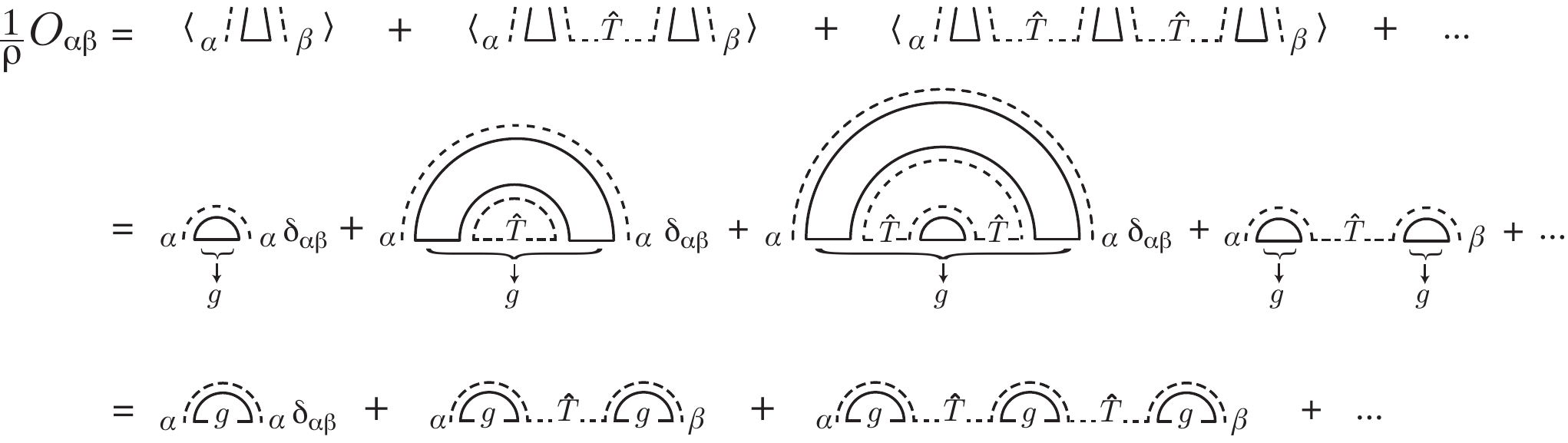}
\caption{
\label{Odiag}
Diagrammatic expansion of $O_{\alpha \beta}/\rho$. Braces with arrows denote parts of diagrams that are the beginning of the  expansion depicted in Fig.\ \ref{gdiag}. }
}
\end{figure*}
%%%%%%%%%%%%%

The self-energy $\sigma(z)$ is the sum of all one-particle irreducible diagrams contained in $zg(z)z$. The first dominant terms that appear in the expansion of $\sigma(z)$ are represented in Fig.\ \ref{sigmadiag}. Under a pairwise contraction, we recognize $g(z)$ depicted in Fig.\ \ref{gdiag}. After summation of all planar rainbow diagrams in the expansion of Fig.\ \ref{sigmadiag} and application of `Feynman' rules defined in Fig.\ \ref{HHAandTracediag}(b), the self-energy becomes
\begin{align}
\label{sigmaScERM}
\sigma(z)&=\frac{1}{N}\textrm{Tr}
\left[
\frac{\hat{T}}{1-g(z)\hat{T}}
\right]
\\
\label{sigmaScERM2}
&=\frac{\textrm{Tr}\hat{T}}{N}+\frac{g(z)}{N}\textrm{Tr}\frac{\hat{T}^2}{1-g(z)\hat{T}}.
\end{align}
where $\hat{T}=\rho\hat{A}$, and $\textrm{Tr}$ denotes the trace of an operator. Inserting Eq.~(\ref{sigmaScERM}) into Eq.~(\ref{SelfenergyRMT}), we obtain:
\be
\label{SolgERM}
z=\frac{1}{g(z)}+\frac{1}{N}\textrm{Tr}
\left[
\frac{\hat{T}}{1-g(z)\hat{T}}
\right]
\ee
that is a closed equation for $g(z)$. Noting that\footnote{From here on, we denote by $\textrm{Tr}_{N}$ the trace of a $N\times N$ matrix when confusion is possible with the trace of an operator.}
\be
\textrm{Tr}\hat{T}=\rho \textrm{Tr}\hat{A}=\moy{ \textrm{Tr}_{N}A}=N\moy{\Lambda},
\ee
we conclude that $\textrm{Tr}\hat{T}/N$ in Eq.~(\ref{sigmaScERM2}) leads to a shift in the distribution of eigenvalues $p(\Lambda)$.

Before commenting on the result (\ref{sigmaScERM}), let us see how the operator (\ref{DefOA2}) can be expressed through the solution $g(z)$ and $\hat{T}$. In the basis $\{\psi_\alpha\}$, Eq.\ (\ref{DefOA2}) reads
\begin{eqnarray}
O_{\alpha \beta} &=& \braket{\psi_\alpha}{\hat{O}}{\psi_\beta}
= \rho\sum_{i=1}^N\sum_{j=1}^N
H^\dagger_{\alpha i}
\nonumber \\
&\times& \left[
\frac{1}{z}+\frac{1}{z}A\frac{1}{z}+\frac{1}{z}A\frac{1}{z}A\frac{1}{z}+\dots
\right]_{ij}
H_{j \beta},
\end{eqnarray}
where we used the definition (\ref{DefHinA}) of the matrix $H$. In Fig.\ \ref{Odiag}, we represent the beginning of the expansion of  $O_{\alpha \beta}/\rho$ with the diagrammatic notations of Fig.\ \ref{HHAandTracediag}(a).
Note that all diagrams in Fig.\ \ref{Odiag} are irreducible. As it was the case for $\sigma(z)$, we recognize the expansion of $g(z)$ under pairwise contractions. After summation of planar diagrams, the operator $\hat{O}(z)$  is finally given by:
\begin{eqnarray}
\label{LinkOAg2}
\hat{O}(z) = \frac{\rho g(z)}{1-g(z)\hat{T}}
%% \label{LinkOAg22}
= \frac{\rho}{z-\hat{T}-\sigma(z)}.
\end{eqnarray}

\paragraph{Low density limit.}

Without loss of generality, let us assume that the diagonal elements of the matrix $A$ are all equal, $A_{ii}=\moy{\Lambda}$. At low density $\rho\to0$, an approximation of the self-energy (\ref{sigmaScERM}) can be obtained by neglecting the term $g(z)\hat{T}$ in the denominator:
\be
\label{ApproxSigmaLowDens}
\sigma(z)\simeq \frac{\textrm{Tr}(\hat{T})}{N}+\frac{\textrm{Tr}(\hat{T}^2)}{N}g(z)=\moy{\Lambda}+\textrm{Var}\Lambda\, g(z).
\ee
The last equality of Eq.~(\ref{ApproxSigmaLowDens}) follows from
\begin{align}
\label{Moment2HERM}
\textrm{Tr}(\hat{T}^2) &= \rho^2 \iint_{V}  \mathrm{d}^d\vec{r}\; \mathrm{d}^d\vec{r'} \left|f(\vec{r},\vec{r'})\right|^2
\nonumber \\
&= \moy{\textrm{Tr}_N(A^2)}-\frac{\moy{\textrm{Tr}_N(A)}^2}{N}=N\textrm{Var}\Lambda.
\end{align}

The implication of Eq.~(\ref{ApproxSigmaLowDens}) is that in the limit of low density $\rho$, the approach described in this section yields for all ERMs an eigenvalue density identical to that of a Gaussian matrix:
\be
\label{WignerERM}
p(\Lambda)=\frac{1}{2\pi \textrm{Var}\Lambda}
\sqrt{4\textrm{Var}\Lambda-(\Lambda-\moy{\Lambda})^2},
\ee
where the variance $\textrm{Var} \Lambda$ is given by Eq.~(\ref{Moment2HERM}). This result holds for ERMs for which  the representation (\ref{AHTH}) with matrices $H$ having i.i.d. elements and $T$  having a sufficiently large number of non-zero eigenvalues, is a good approximation. When considering specific examples of ERMs in Sec.\ \ref{Applherm}, we will see that both matrices that fall into this category (Sec.\ \ref{ERMSinc} and \ref{ERMCosc}) as well as those that do not (Sec.\ \ref{gaussERM}) exist. The low-density limit of the latter can be understood using the heuristic analysis of Sec.\ \ref{ERMHeuristic}.

\paragraph{Relation to previous results.}

Finally, let us show how the various approximations for $g(z)$, $\sigma(z)$, and $g_\vec{k}(z)$ derived in the previous sections can be recovered from Eqs.~(\ref{sigmaScERM}) and (\ref{LinkOAg2}). We need to assume that
\begin{align}
\label{Deffkk}
f(\vec{k}, \vec{k'}) &= \braket{\vec{k}}{\hat{A}}{\vec{k}'}
\nonumber \\
&= \frac{1}{V}\iint_{V}  \mathrm{d}^d\vec{r}\; \mathrm{d}^d\vec{r'} e^{-\textrm{i}(\vec{k}\cdot\vec{r}-\vec{k}'\cdot \vec{r}')}f(\vec{r}-\vec{r'})
\end{align}
is diagonal: $f(\vec{k}, \vec{k'})\simeq\braket{\vec{k}}{\hat{A}}{\vec{k}}\delta_{\vec{k}\vec{k}'}\equiv f(\vec{k})\delta_{\vec{k}\vec{k}'}$, which is not exact in a finite volume $V$. In the momentum representation, Eqs.~(\ref{sigmaScERM}) and (\ref{Defgk}) read now
\begin{align}
\label{sigmaApprox1}
\sigma(z)&\simeq
\int\frac{\textrm{d}^d\mathbf{k}}{(2\pi)^d} \frac{f(\mathbf{k})}{1-\rho f(\mathbf{k})g(z)},
\\
\label{gkApprox1}
\frac{1}{\rho} \braket{\vec{k}}{\hat{O}(z)}{\vec{k}}&=g_\vec{k}(z)\simeq\frac{1}{z-\rho f(\mathbf{k}) - \sigma(z)},
\end{align}
where $f(\vec{k})=\braket{\vec{k}}{\hat{A}}{\vec{k}}$ can be further approximated by $f_0(\vec{k})$ defined in Eq.~(\ref{Deff0k}). Hence, Eq.~(\ref{sigmaApprox1}) becomes identical to Eq.~(\ref{SelfconsistentMezard2}). If the integrand of Eq.~(\ref{sigmaApprox1}) is expanded in series in $\rho$, Eq.~(\ref{gkApprox1}) becomes consistent with Eqs.~(\ref{self1}) and (\ref{Approxgksigma1}). This means that the approximate self-energy (\ref{self1}) corresponds to a truncation of the expansion depicted in Fig.\ \ref{sigmadiag} after the second diagram.

\paragraph{Solving self-consistent equations in practice.}
\label{SolvingPractice}

The solution of Eq.~(\ref{SolgERM}) for a given matrix $A$ can be greatly facilitated by a suitable choice of the basis in which the trace appearing in this equation is expressed. In addition to $\{ \vec{r} \}$ and $\{ \vec{k}_{\alpha} \}$, a basis of   eigenvectors $\ket{\mathcal{R}_{\alpha}}$ of $\hat{T}$ can be quite convenient. The eigenvector $\ket{\mathcal{R}_{\alpha}}$ obeys
\be
\label{rightEv}
\bra{\vec{r}} \hat{T} \ket{\mathcal{R}_{\alpha}} =
\rho \int_V\textrm{d}^d \vec{r'}f(\vec{r}, \vec{r'}) \mathcal{R}_\alpha(\vec{r'}) = \mu_\alpha \mathcal{R}_\alpha(\vec{r}),
\ee
where $\mu_{\alpha}$ is the eigenvalue corresponding to the eigenvector $\ket{\mathcal{R}_{\alpha}}$. In this basis, Eq.~(\ref{SolgERM}) becomes
\be
\label{SolgERMEv}
z=\frac{1}{g(z)}+
\frac{1}{N} \sum_{\alpha} \frac{\mu_\alpha}{1-g(z)\mu_\alpha}.
\ee

\subsection{Free probability theory}
\label{FreeProba}

The term `free probability theory' designates a discipline founded by \citet{voiculescu83} [see also \cite{voiculescu92}] in order to solve the following problem: can we say anything about the spectral properties of the sum of two matrices $X_1+X_2$ when the spectral properties of the summands $X_1$ and $X_2$ are known? Unless the two matrices commute, knowing their eigenvalues is, in general, not enough to find the eigenvalues of the sum. However, the free probability theory identifies a certain sufficient condition, called asymptotic freeness, under which this problem can be tackled without involving the eigenvectors of the matrices. The notion of asymptotic freeness is equivalent to the notion of statistical independence that we are familiar with for random variables. It is a generalization of the latter to the case where the variables---here, the matrices---do not commute.

\subsubsection{Theoretical framework}

Let us briefly recall basic properties of independent variables. We denote by $p_x$ the probability density of the variable $x$, by  $g_x(z)\equiv\left<e^{zx}\right>=\sum_{n\geqslant0}\left<x^n\right>z^n/n!$ its characteristic function, and by $r_x(z)\equiv\textrm{ln}g_x(z)=\sum_{n\geqslant0}c_{x,n}z^n$ its cumulant generating function. For two independent real random variables $x_1$ and $x_2$, the following relations hold: \begin{align}
\left<x_1x_2\right>&=\left<x_1\right>\left<x_2\right>,
\\
\label{distrib convolution}
p_{x_1+x_2}&=p_{x_1}*p_{x_2},
\\
\label{cumulant}
r_{x_1+x_2}&=r_{x_1}+r_{x_2},
\end{align}
where `$*$' denotes the convolution. We will see that these relations find their equivalents for asymptotically free matrices.

By definition, two Hermitian matrices $X_1$ and $X_2$ are asymptotically free if for all $l\in \mathbb{N}$ and for all polynomials $p_i$ and $q_i$ ($1\leq i \leq l$), we have \cite{tulino04}
\begin{align}
\label{freeness}
&\left<p_i(X_1)\right>_{\Lambda}=\left<q_i(X_2)\right>_{\Lambda}=0 \nonumber \\
&\Rightarrow\;\;\left<p_1(X_1)q_1(X_2) \dots p_l(X_1)q_l(X_2)\right>_{\Lambda}=0,
\end{align}
where the expectation value $\left<...\right>_{\Lambda}$ is defined as
\be
\left<X\right>_{\Lambda}=\frac{1}{N} \left<\textrm{Tr}X\right>.
\ee
The interpretation of the formal definition (\ref{freeness}) is the following: two matrices are asymptotically free if their eigenbases are related to one another by a random rotation, or said differently, if their eigenvectors are almost surely orthogonal.

From the definition (\ref{freeness}), it is easy to compute various mixed moments of $X_1$ and $X_2$. Considering $\tilde{X}_i= X_i-\left<X_i\right>_{\Lambda}$ that obey $\langle\tilde{X}_1\rangle_{\Lambda}=\langle\tilde{X}_2
\rangle_{\Lambda}=0$, we obtain from Eq.~(\ref{freeness}):
\be
 \left<X_1X_2\right>_{\Lambda}=\left<X_1\right>_{\Lambda}\left<X_2\right>_{\Lambda}.
\ee
Note that this last condition is not enough to define asymptotic freeness, since matrices do not commute. For example, from Eq.~(\ref{freeness}), forth moments read
\begin{align}
\left<X_1X_1X_2X_2\right>_{\Lambda}
&= \left<X_1^2\right>_{\Lambda}\left<X_2^2\right>_{\Lambda} ,
\nonumber
\\
\left<X_1X_2X_1X_2\right>_{\Lambda}
&= \left<X_1^2\right>_{\Lambda}\left<X_2\right>_{\Lambda} ^2+\left<X_1\right>_{\Lambda}^2\left<X_2^2\right>_{\Lambda}
\nonumber \\
&- \left<X_1\right>^2_{\Lambda}\left<X_2\right>^2_{\Lambda} .
\end{align}

Free cumulants are defined such that the sum property (\ref{cumulant}) is preserved for the generating function of the free cumulants, the so-called $\mc{R}$-transform \cite{tulino04, jarosz06}. Interestingly, the $\mc{R}$-transform is simply related to the Blue function (\ref{DefBlue})---the functional inverse of the resolvent $g(z)$---by Eq.~(\ref{DefRed}).\footnote{ Note that $g(z)$ plays the role of a free characteristic function, see Eqs.~(\ref{MomentsTrA}) and  (\ref{gSumMoments}). \citet{jarosz06} provide an insightful discussion about the relation between the free cumulants and the moments $\moy{\Lambda^n}$.}
The $\mc{R}$-transform of the sum of two asymptotically free matrices $X_1$ and $X_2$ obeys:
\be
\label{Rsum}
\mc{R}_{X_1+X_2}(z)=\mc{R}_{X_1}(z)+\mc{R}_{X_2}(z).
\ee
Hence, the problem of finding the eigenvalue distribution of the sum of two free random matrices is straightforward. Applying successively Eqs.~(\ref{DefBlue}), (\ref{DefRed}), and (\ref{Rsum}), one readily infers $g_{X_1+X_2}$ from $g_{X_1}$ and  $g_{X_2}$. The steps of the algorithm are as follows:
\begin{align}
\label{algosum}
g_{X_i} &\to \mc{B}_{X_i} \to \mc{R}_{X_i} \to \mc{R}_{X_1+X_2} \nonumber \\
&\to \mc{B}_{X_1+X_2} \to g_{X_1+X_2}.
\end{align}

There is an analogous result for the product of free matrices, which involves the so-called $\mc{S}$-transform \cite{tulino04}.
If we define $\chi(z)$ as a solution of
\be
\label{DefchiRMT}
\frac{1}{\chi(z)}g\left[ \frac{1}{\chi(z)} \right]-1=z,
\ee
then the $\mc{S}$-transform is
\be
\label{DefSRMT}
\mc{S}(z)=\frac{1+z}{z}\chi(z).
\ee
Equations (\ref{DefchiRMT}) and (\ref{DefSRMT}) are equivalent to the following implicit equation for $\mc{S}(z)$:
\be
\label{Simplicit}
\mc{S}(z)\mc{R}[z\mc{S}(z)]=1.
\ee
The $\mc{S}$-transform of the product of two asymptotically free matrices $X_1$ and $X_2$ satisfies \cite{tulino04}
\be
\label{Sproduct}
\mc{S}_{X_1X_2}(z)=\mc{S}_{X_1}(z)\mc{S}_{X_2}(z).
\ee
Therefore, the $S$-transform plays a role analogous to the $\mc{R}$-transform for products (instead of sums) of free matrices. The recipe to find the eigenvalue density of $X_1X_2$ is analogous to Eq.\ (\ref{algosum}):
\be
\label{algoproduct}
g_{X_i} \to \chi_{X_i} \to \mc{S}_{X_i} \to \mc{S}_{X_1X_2} \to \chi_{X_1X_2} \to g_{X_1X_2}.
\ee

\subsubsection{Gaussian and Wishart ensembles revisited}
\label{revisited}

A good attitude when searching for the eigenvalue density of a given matrix, is to look for a possible decomposition of the latter in a sum or product of free matrices for which resolvents are known. Let us apply this idea to recover in a new way the now familiar semicircle and Marchenko-Pastur  laws.

Let us first consider a matrix $A$ from the Gaussian orthogonal ensemble (GOE), with the probability distribution $P(A)=C_N e^{-\frac{N}{4}\textrm{Tr}(A^2)}$. From Eq.~(\ref{distrib convolution}), it is clear that the distribution of the variable $x_1+x_2$, where $x_1$ and $x_2$ are independent Gaussian random variables with variances $\sigma^2$, is still Gaussian with a variance $2\sigma^2$. We can therefore decompose any Gaussian matrix $A$ in a sum of two independent rescaled matrices $A_1$ and $A_2$ that have the same probability density $P$: $A=\frac{1}{\sqrt{2}}(A_1+A_2)$. In addition, two independent Gaussian matrices are asymptotically free. Indeed,  since the measure $P(A)$ is invariant under orthogonal transformation, rotation matrices $O_1$ and $O_2$, diagonalizing $A_1$ and $A_2$, respectively, are random rotations over the orthogonal group. This means that the rotation $O_1^\dagger O_2$ from the eigenbasis of $A_1$  to that of $A_2$ is also random, which is precisely the intuitive definition of asymptotic freeness [\citet{tulino04} give a formal proof of this statement]. The additive property of the $\mc{R}$-transform and the scaling property (\ref{scalingPpty}) yield:
\be
\mc{R}_A(z)=\mc{R}_{\frac{A_1}{\sqrt{2}}}(z)+\mc{R}_{\frac{A_2}{\sqrt{2}}}(z)
= \sqrt{2} \mc{R}_A\left( \frac{z}{\sqrt{2}} \right).
\ee
A solution of this equation is $\mc{R}_A(z)\varpropto z$. According to Eq.~(\ref{VarianceLambda}), $\mc{R}'(0)=\moy{\Lambda^2}=\moy{\textrm{Tr}A^2}/N=1$, so that
\be
\mc{R}(z)= z.
\ee
As could be expected from Eqs.\ (\ref{LinksigmaR}) and (\ref{SelfenergyGaussian}), this is the $\mc{R}$-transform of the semicircle law. Thus, we can claim that the semicircle law is the free counterpart of the Gaussian distribution in the classical probability theory.

In order to use the powerful arsenal of free probability for Wishart matrices, we decompose the $N\times N$ matrix $A=HH^{\dagger}$ as
\be
\label{HHasSum}
HH^{\dagger}=
\sum_{\alpha=1}^M
\mathbf{h}^{(\alpha)\dagger}
\mathbf{h}^{(\alpha)}
\;\;\;\textrm{with}\;\;\;
\mathbf{h}^{(\alpha)}=(H^*_{1\alpha},\dots,H^*_{N \alpha}).
\ee
The spectrum of each matrix $\mathbf{h}^{(\alpha)\dagger}\mathbf{h}^{(\alpha)}$ is simple because it has only one non-zero eigenvalue $\Lambda_\alpha=\|\mathbf{h}^{(\alpha)}\|^2=\sum_{i=1}^N\vert H_{i \alpha}\vert^2$, associated with an eigenvector $\mathbf{h}^{(\alpha)*}$. The $(N-1)$ other eigenvectors associated with zero eigenvalue form the basis in the hyperplane perpendicular to the vector $\mathbf{h}^{(\alpha)*}$.  Since the vectors $\mathbf{h}^{(\alpha)}$ are uncorrelated, we can replace the resolvent of the matrix  $\mathbf{h}^{(\alpha)\dagger}\mathbf{h}^{(\alpha)}$ by
\be
g_{\mathbf{h}^{(\alpha)\dagger}\mathbf{h}^{(\alpha)}}(z)=\frac{1}{N}\left[\frac{N-1}{z}+\frac{1}{z-1}\right],
\ee
where we used $\left<\Lambda_0\right>=1$ ($\langle \vert H_{i \alpha}\vert^2\rangle=1/N$).
Inverting this relation gives
\begin{align}
\mc{R}_{\mathbf{h}^{(\alpha)\dagger}\mathbf{h}^{(\alpha)}}(z)&=\frac{1}{2z}\left(z-1-\sqrt{(z-1)^2+\frac{4 z}{N}}\right)
\nonumber
\\
\label{Rofhh}
&=\frac{1}{N}\;\frac{1}{1- z}+\mc{O}\left(\frac{1}{N^2}\right).
\end{align}
For independent vectors $\mathbf{h}^{(\alpha)}$, that have independent entries with variances equal to $1/N$ and identical means, it can be shown that the matrices $\mathbf{h}^{(\alpha)\dagger}\mathbf{h}^{(\alpha)}$ are asymptotically free \cite{tulino04}. Thus,
\begin{align}
\mc{R}_{HH^\dagger}(z)&=\sum_{\alpha=1}^M\mc{R}_{\mathbf{h}^{(\alpha)\dagger}\mathbf{h}^{(\alpha)}}(z)
\\
\label{RWishart}
&=\frac{1}{c}\frac{1}{1-z},
\end{align}
where $c=N/M$. This is the $\mc{R}$-transform of the Marchenko-Pastur law, as could also be obtained from Eqs.\ (\ref{LinksigmaR}) and (\ref{SelfenergyWishart}). It is interesting to note that if we took the $N$th classical convolution (by inverting the sum of cumulant-generating functions) of the distributions of the variables $\Lambda_\alpha$, we would obtain asymptotically ($N, M\to \infty$ at fixed $c=N/M$) the Poisson distribution. However, the distribution that we obtain by taking the $N$th free convolution (by inverting the sum of $\mc{R}$-transforms) is the Marchenko-Pastur law. The latter is therefore the free analog of the Poisson law in the classical probability theory \cite{tulino04}. Another simple proof of this law was given by \citet{janik97a} using the product decomposition of the matrix $HH^\dagger$ and the properties of the  $\mc{S}$-transform.

\subsubsection{Application to Euclidean random matrices}
\label{SecFRVERM}

By analogy with results for Wishart ensemble, it is straightforward to apply the toolbox of free probability to ERMs. We start with the decomposition $A=HTH^\dagger$, where the basis $\{\psi_\alpha\}$, that defines $H_{i\alpha}$ through Eq.~(\ref{DefHinA}), is assumed to be the eigenbasis of the operator $\hat{T}$: $\hat{T}\ket{\psi_\alpha}=\mu_\alpha\ket{\psi_\alpha}$. The matrix $A$ is conveniently rewritten as
\be
\label{DecompoHTH}
HTH^\dagger=\sum_{\alpha=1}^M
\mu_\alpha
\mathbf{h}^{(\alpha)\dagger}
\mathbf{h}^{(\alpha)},
\ee
where $\mathbf{h}^{(\alpha)}$ is defined in Eq.~(\ref{HHasSum}). As explained above, the $M$ matrices $\mathbf{h}^{(\alpha)\dagger}\mathbf{h}^{(\alpha)}$ are asymptotically free, as long as the vectors $\mathbf{h}^{(\alpha)}$ are independent. Hence,
\begin{align}
\label{HTHasSum1}
\mc{R}_{HTH^\dagger}(z) &=
\sum_{\alpha=1}^M
\mc{R}_{\mu_\alpha \mathbf{h}^{(\alpha)\dagger}\mathbf{h}^{(\alpha)}}(z)
\nonumber \\
&=\sum_{\alpha=1}^M
\mu_\alpha\mc{R}_{\mathbf{h}^{(\alpha)\dagger}\mathbf{h}^{(\alpha)}}(\mu_\alpha z)
\\
\label{HTHasSum2}
&=\frac{1}{N}\sum_{\alpha=1}^M
\frac{\mu_\alpha}{1-\mu_\alpha z}
=\frac{1}{N}\textrm{Tr}_{M}\left[
\frac{T}{1-zT}
\right]
\\
\label{HTHasSum3}
&=\frac{1}{N}\textrm{Tr}\left[
\frac{\hat{T}}{1-z\hat{T}}
\right]
\\
\label{HTHasSum4}
&=\frac{1}{cz}
\left[
\frac{1}{z}g_T\left(\frac{1}{z}\right)-1
\right].
\end{align}
Equation (\ref{HTHasSum1}) follows from Eqs.\ (\ref{Rsum}) and (\ref{scalingPpty}), whereas Eq.~(\ref{HTHasSum2})---from Eq.\ (\ref{Rofhh}), and Eq.~(\ref{HTHasSum4})---from $g_T(z)=\sum_{\alpha=1}^M1/(z-\mu_\alpha)M$. Using the definition (\ref{DefSRMT}) of the $\mc{S}$-transform, one also easily shows that Eq.~(\ref{HTHasSum4}) is equivalent to
\be
\label{SHTH}
\mc{S}_{HTH^\dagger}(z)=\frac{1}{z+1/c}\mc{S}_T(cz).
\ee

For completeness, we now derive Eq.\ (\ref{SHTH}) from Eq.\ (\ref{Sproduct}). From the definitions of the resolvent $g(z)$ and the $\mc{S}$-transform, one can check that, for arbitrary matrices $A$ and $B$ of size $N\times M$ and $M \times N$, respectively,
\be
\mc{S}_{AB}(z)=\frac{z+1}{z+1/c}\mc{S}_{BA}(cz).
\ee
Applying this result to $A=HT$ and $B=H^\dagger$, we obtain
\begin{align}
\mc{S}_{HTH^\dagger}(z)&=\frac{z+1}{z+1/c}\mc{S}_{H^\dagger H T}(cz),
\nonumber
\\
\label{SHTHProd}
&=\frac{z+1}{z+1/c}\mc{S}_{H^\dagger H}(cz)\mc{S}_T(cz).
\end{align}
Here we made use of the fact that the deterministic matrix $T$ and the random matrix $H^\dagger H$ are asymptotically free. Besides, the combination of Eq.~(\ref{Simplicit}) with $\mc{R}_{H^\dagger H}(z) = c\mc{R}_{HH^\dagger}(z/c) = c/(c-z)$ gives
\be
\label{SHH2}
\mc{S}_{H^\dagger H}(z)=\frac{c}{c+z}.
\ee
From Eqs.~(\ref{SHTHProd}) and (\ref{SHH2}), we finally recover Eq.\ (\ref{SHTH}).

The result (\ref{SHTH}), or equivalently its operator form (\ref{HTHasSum3}), is in perfect agreement with the solution obtained by a diagrammatic approach in Sec.\ \ref{SecERMSelfConsistent}. Indeed, the self-energy $\sigma(z)=\mc{R}[g(z)]$ inferred from Eq.~(\ref{HTHasSum3}) is exactly the result (\ref{sigmaScERM}). It is worth recalling that Eq.\ (\ref{HTHasSum3}) was obtained from the asymptotic freeness of the matrices $\mathbf{h}^{(\alpha)\dagger}\mathbf{h}^{(\alpha)}$, that holds as long as the elements $H_{i\alpha}$ are i.i.d. with a finite second moment \cite{tulino04}. In particular, it means that Eq.\ (\ref{sigmaScERM}) is valid even if $H_{i\alpha}$ are not Gaussian variables. Therefore, Gaussian hypothesis, that largely simplified diagrammatic calculations in Sec.\ \ref{SecERMSelfConsistent}, turns out to be not essential.\footnote{A rigorous diagrammatic proof of Eq.\ (\ref{sigmaScERM}) with the finiteness of the second moment of $H_{i\alpha}$ as the only assumption is nontrivial.} In particular, this remark holds for the Wigner semicircle and the Marchenko-Pastur laws,\footnote{We are not aware of diagrammatic proofs of the Wigner semicircle and the Marchenko-Pastur laws that would not invoke the Gaussian assumption.} and justifies their large degree of universality.

As far as ERMs are concerned, we conclude that the only assumption that may limit the applicability of Eq.\ (\ref{sigmaScERM}) is the independence of the vectors $\mathbf{h}^{(\alpha)}$. We know that their covariance matrix is proportional to the identity (see Sec.\ \ref{SubsecERM}), but this is not enough to insure their independence, precisely because $H_{i\alpha}$ are not Gaussian random variables. We will see how correlations between $H_{i\alpha}$ limit the applicability of Eq.\ (\ref{sigmaScERM}) by considering specific examples of ERMs in Sec.\ \ref{Applherm}.

\subsection{Renormalization group approach}
\label{Renorm}

Renormalization group (RG) approaches are powerful methods to identify phase transitions and study universal properties in their vicinity. Most RG procedures that are applied to both ordered and disordered systems renormalize the space in a homogenous way. This choice, natural for translationally invariant systems, is not the only possible one when we deal with disordered systems \cite{igloi05}. Indeed, for the latter, space can also be renormalized in an inhomogeneous way in order to `adapt' to local fluctuations due to disorder. Such real-space RG was first developed in the context of quantum spin chains \cite{ma79, dasgupta80, fisher95} and then was successfully applied to various quantum and classical systems; see the review by \citet{igloi05} for works before 2004 and \cite{monthus11} for more recent references.

To illustrate the application of RG approach in ERM theory, let us consider a matrix ensemble defined by Eq.\ (\ref{DefinitionERMu}) with $u = 1$ and $f(\vec{r}_i-\vec{r}_j)$ a positive function decaying to $0$ at large distances. Following \citet{amir10} and  \citet{monthus11}, we  analyze the eigenproblem $AR_n=\Lambda_nR_n$  with the RG approach making use of the following mechanical analogy. Imagine a network of unit masses connected by springs, such that $A_{ij}$ is the random spring constant between masses $i \ne j$. Then, the displacement $x_i$ of the mass $i$ obeys $\textrm{d}^2x_i/\textrm{d}t^2=\sum_{j}A_{ij} x_j$. Therefore, the eigenvalues $\Lambda_n$ of $A$ are related to the frequencies $\omega_n$ of vibrational modes by $\Lambda_n=-\omega_n^2$.  The frequencies $\omega_n$ can be found by eliminating iteratively the modes with highest frequencies. Initially, if the density is low enough, the highest frequency corresponds to the vibrational mode associated with the pair of the two closest masses: $\omega = [\max(A_{ij})/\mu]^{1/2}$, where $\mu=m_1m_2/(m_1+m_2)=1/2$ is the reduced mass. Provided that $\omega$ is high, this mode will not be coupled to the lower-frequency modes of the system and can be decimated: the relative displacement of particles $i$ and $j$ is set to $0$ and the pair is assumed to behave as a single mass $\tilde{m}_G = m_1 + m_2$ connected by springs to the $N - 2$ other masses $\alpha$, with spring constants $\tilde{A}_{G \alpha} = A_{1 \alpha} + A_{2 \alpha}$. This decimation procedure can be repeated iteratively. At step $k$, where we have $N-k+1$ masses $\tilde{m}_i$ connected by springs of strength $\tilde{A}_{ij}$, the mode associated with the renormalized frequency scale
\be
\label{FreqRG}
\omega=\max{\sqrt{\tilde{A}_{ij}(1/\tilde{m}_i+1/\tilde{m}_j})},
\ee
is decimated and the two corresponding masses $\tilde{m}_{i_0}$ and $\tilde{m}_{j_0}$ are replaced by a single mass
\be
\label{MassRG}
\tilde{m}_{G}=\tilde{m}_{i_0}+\tilde{m}_{j_0}
\ee
with the new spring constants
\be
\label{SpringRG}
\tilde{A}_{G \alpha}=\tilde{A}_{i_0 \alpha}+\tilde{A}_{j_0 \alpha}.
\ee
This renormalization scheme is consistent only if the new generated frequencies at each step are smaller than the decimated frequency, so that $\omega$ decreases along the RG flow. This restricts the analysis to the regime of low density in which the distribution of couplings $A_{ij}$ is sufficiently broad \cite{fisher95, amir10}. It is also worth noting that alternative or more sophisticated RG rules can be considered \cite{monthus11}. In particular, such rules may be needed to take into account the modes associated with individual masses and following the slow motion of their neighbors adiabatically. Such modes are neglected in the RG scheme described above.

As the RG flow progresses, the eigenvalues $\Lambda=-\omega^2$ of $A$ are obtained in an ascending order in the range from $-2f(\vec{0})$ to $0$. After the step $k$, the flow has reached the typical value $\Lambda$ given by the relation
\be
\label{LinkKLambda}
k=N\int_{-2}^{\Lambda}\textrm{d}\Lambda'p(\Lambda').
\ee
On the other hand, the typical size $m_c$ of the eigenvector associated with $\Lambda$ is given by the the typical mass of the clusters of points at step $k$. By construction, this mass grows along the flow, starting from pairs of points for the smallest eigenvalue and converging to clusters of $N$ points for the largest one. If we assume that the formation of clusters is a random process independent of the choice of the frequency scale (\ref{FreqRG}), then it is easy to verify that $m_c=N/(N-k)$ after the step $k$ \cite{amir10}. Combining this result with  Eq.\ (\ref{LinkKLambda}) yields
\be
\label{VectorRG}
m_c(\Lambda)\simeq\frac{1}{C(\Lambda)},
\ee
where $C(\Lambda)=\int_{\Lambda}^{0} \textrm{d}z\, p(z)$ is the cumulative eigenvalue density. In the regime of low density, it is given by Eq. (\ref{Cumul1}).

The RG rules (\ref{FreqRG}), (\ref{MassRG}) and (\ref{SpringRG}) are explicitly implemented in Sec.\ \ref{expERM} where we apply them to study the eigenvalue density of an ERM defined through an exponentially decaying function $f(\vec{r}_i-\vec{r}_j)$. We stress that these rules only apply for ERMs defined by Eq.\ (\ref{DefinitionERMu}) with $u = 1$ and cannot be used at $u=0$ because in this case the spring analogy does not hold anymore.

\subsection{Application of the general theory to specific Euclidean random matrix ensembles}
\label{Applherm}

\subsubsection{Gaussian Euclidean random matrix}
\label{gaussERM}

Historically, the Euclidean matrix defined through the Gaussian function $f(\vec{r}_i, \vec{r}_j) = f(\vec{r}_i-\vec{r}_j) = \exp(-| \vec{r}_i - \vec{r}_j |^2/2 a^2)$ was the first to be considered in the literature \cite{mezard99}. Assume that the $N$ points that determine the particular realization of our random matrix are randomly chosen in a cube of side $L \gg a$ centered at the origin. In the low density limit $\rho a^3 \to 0$ (with $\rho = N/V$ and $V = L^3$), we can apply heuristic arguments of Sec.\ \ref{ERMHeuristic} and calculate the eigenvalue density by differentiating Eq.\ (\ref{cumul}). This yields a sharp peak at $\Lambda = 1$. To obtain $p(\Lambda)$ at higher densities, we will follow the approach developed in Sec.\ \ref{SolvingPractice}. First, we analyze Eq.\ (\ref{rightEv}) for the eigenvalues $\mu_{\alpha}$ and eigenvectors ${\cal R}_{\alpha}(\vec{r})$ of the operator ${\hat T}$. Because the function $f(\vec{r}_i, \vec{r}_j)$ is separable, i.e. it can be expressed as $f(\vec{r}_i, \vec{r}_j) = f(x_i, x_j) f(y_i, y_j) f(z_i, z_j)$, $\mu_{\alpha}$ and ${\cal R}_{\alpha}(\vec{r})$ obeying Eq.\ (\ref{rightEv}) can be written as products of corresponding eigenvalues $\mu_{\alpha_i}$ and eigenvectors ${\cal R}_{\alpha_i}(\vec{r})$ of a one-dimensional problem:
$\mu_{\alpha} = \rho \mu_{\alpha_i} \mu_{\alpha_j} \mu_{\alpha_k}$ and ${\cal R}_{\alpha}(\vec{r}) = {\cal R}_{\alpha_i}(x) {\cal R}_{\alpha_j}(y) {\cal R}_{\alpha_k}(z)$, where $\mu_{\alpha_i}$ and ${\cal R}_{\alpha_i}(\vec{r})$ obey
\begin{eqnarray}
\label{rightEv1d}
\int_{-L/2}^{L/2} \textrm{d}x' f(x, x') \mathcal{R}_{\alpha_i}(x') = \mu_{\alpha_i} \mathcal{R}_{\alpha_i}(x)
\end{eqnarray}
and the index $\alpha$ has now three components: $\alpha = \{\alpha_i, \alpha_j, \alpha_k \}$. Equation (\ref{SolgERMEv}) for the resolvent $g(z)$ can be then rewritten as
\begin{eqnarray}
\label{SolgERMEv1d}
z=\frac{1}{g(z)}+
\frac{1}{N} \sum_{i, j, k} \frac{\rho \mu_{\alpha_i} \mu_{\alpha_j} \mu_{\alpha_k}}{1-\rho \mu_{\alpha_i} \mu_{\alpha_j} \mu_{\alpha_k} g(z)}.
\end{eqnarray}

If now we discretize Eq.\ (\ref{rightEv1d}) on a grid of $M$ equidistant points with a spacing $\Delta x$, we see that $\lambda_{\alpha_i} = \mu_{\alpha_i}/\Delta x$ are the eigenvalues of the Euclidean $M \times M$ matrix $B$ with elements $B_{ij} = f(i \Delta x, j \Delta x) = \exp(-\Delta x^2 |i-j|^2/2 a^2)$. The matrices with elements $B_{ij}$ that depend on the difference $i-j$ only are called Toeplitz matrices \cite{grenander58,gray06}. Of the whole arsenal of powerful analytical tools developed for Toeplitz matrices, we will make use of the so-called fundamental eigenvalue distribution theorem of Szeg{\"{o}}: under suitable assumptions and for $M \to \infty$, the average of $F(\lambda_{\alpha_i})$ over all eigenvalues $\lambda_{\alpha_i}$ converges to the average of $F[{\tilde f}(\xi)]$ over $\xi \in [0, 2\pi)$,  where
\begin{eqnarray}
\label{fourierfk}
{\tilde f}(\xi) = \sum_{k=-\infty}^{\infty}
f(\Delta x k) \exp(i k \xi).
\end{eqnarray}
Expressing $\mu_{\alpha_i}$ through $\lambda_{\alpha_i}$ in Eq.\ (\ref{SolgERMEv1d}) and applying the above theorem to the sum over eigenvalues, we obtain
\begin{eqnarray}
\label{gauss1}
z &=& \frac{1}{g(z)}+
\frac{L^3}{(2 \pi)^3 N}
\nonumber \\
&\times& \iiint_0^{2 \pi} \frac{{\rm d} \xi_1 {\rm d} \xi_2 {\rm d} \xi_3\; \rho {\tilde f}(\xi_1) {\tilde f}(\xi_2) {\tilde f}(\xi_3)}{1- \Delta x^3 \rho {\tilde f}(\xi_1) {\tilde f}(\xi_2) {\tilde f}(\xi_3) g(z)}.
\end{eqnarray}
Without giving technical details of derivations, we note now that the series in Eq.\ (\ref{fourierfk}) can be summed up, leading to an expression for ${\tilde f}(\xi)$ involving special functions, and then the integrals in Eq.\ (\ref{gauss1}) can be carried out in the limit of $\Delta x \to 0$, which corresponds to the continuous limit that we had initially in Eq.\ (\ref{rightEv1d}). The resulting equation for the resolvent $g(z)$ is
\begin{eqnarray}
\label{gauss2}
&&z = \frac{1}{g(z)}+
\frac{1}{(2 \pi)^{3/2} \rho a^3 g(z)}
\mathrm{Li}_{3/2}[(2 \pi)^{3/2} \rho a^3 g(z)],\hspace*{1cm}
\end{eqnarray}
where $\mathrm{Li}_{3/2}(z)$ is the polylogarithm function \cite{prudnikov90}.
An important feature of this equation is that it depends on a single parameter characterizing the system---the dimensionless density of points $\rho a^3$---but is independent of the total number of points $N$ and the system size $L$. This is a consequence of the fast decay of $f(\vec{r}_i, \vec{r}_j)$ with $|\vec{r}_i-\vec{r}_j|$ in conjunction with the assumption of $L \gg a$.

\begin{figure}[t]
\centering{
\includegraphics[angle=0,width=0.9\columnwidth]{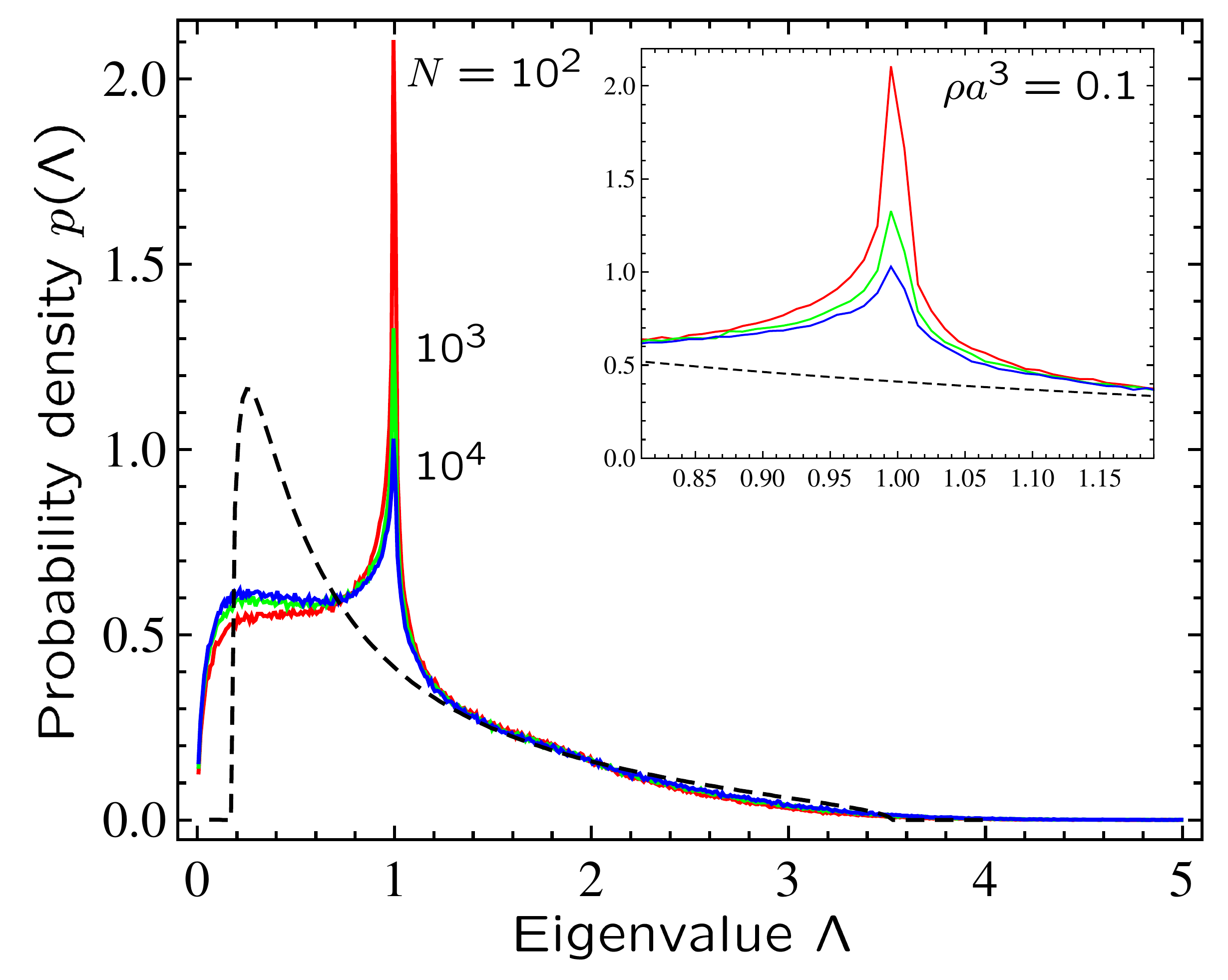}
\caption{
\label{figgauss1}
Eigenvalue density of $N \times N$ Euclidean matrix $A$ with elements $A_{ij} = \exp(-|\vec{r}_i - \vec{r}_j|^2/2 a^2)$ at an intermediate density of points $\rho a^3 = 0.1$. The analytic result obtained by solving Eq.\ (\ref{gauss2}) (dashed line) is compared to numerical diagonalization (solid lines). The inset is a zoom onto a part of the main plot around $\Lambda = 1$.
}}
\end{figure}

\begin{figure}[t]
\centering{
%\hspace*{-8mm}
\includegraphics[angle=0,width=0.9\columnwidth]{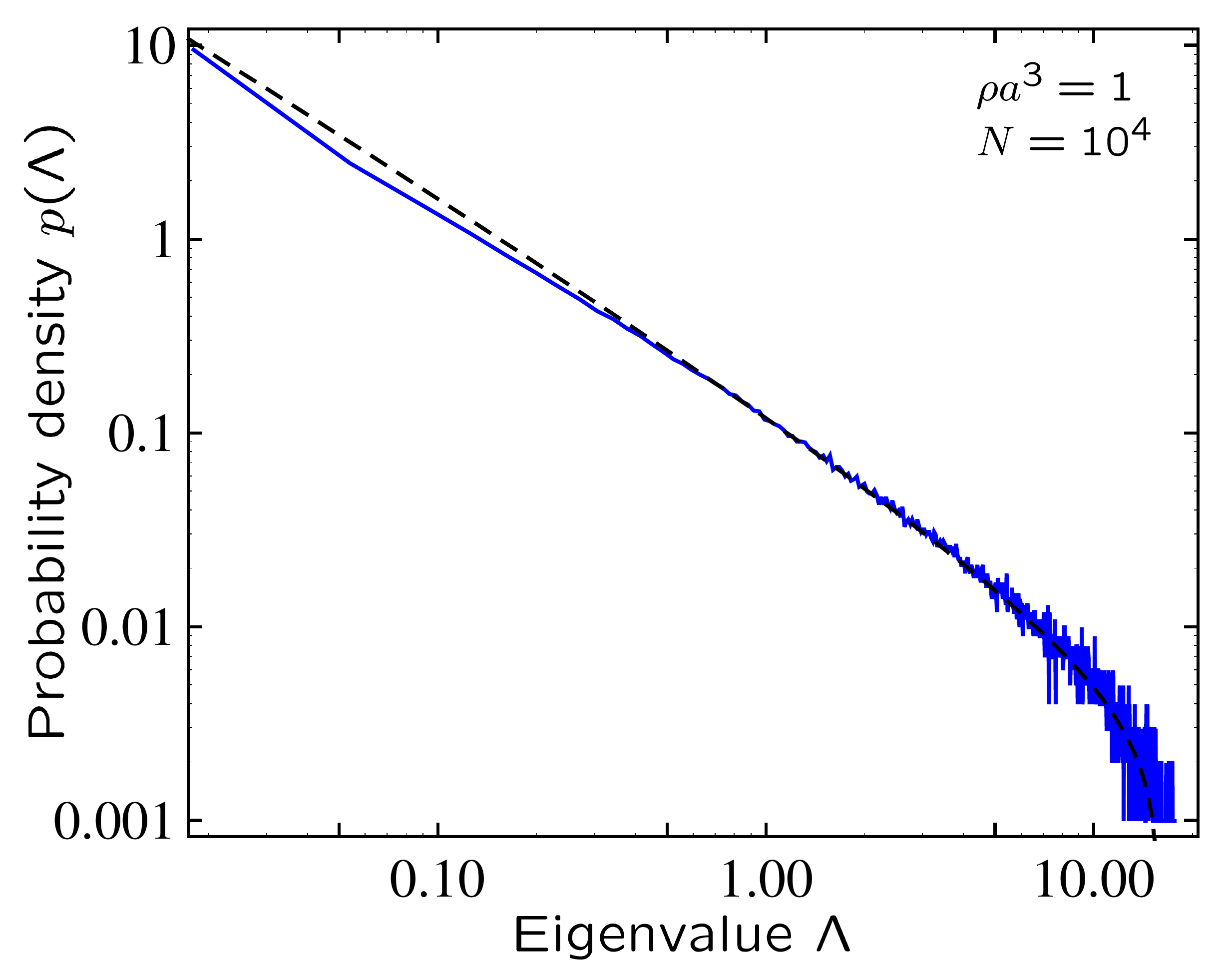}
\caption{
\label{figgauss2}
Eigenvalue density of $N \times N$ Euclidean matrix $A$ with elements $A_{ij} = \exp(-\frac12 |\vec{r}_i - \vec{r}_j|^2)$ at a high density of points $\rho a^3 = 1$. The approximate Eq.\ (\ref{gauss3}) (dashed line) is compared to numerical diagonalization (solid lines).
}}
\end{figure}

In the limit of low density $\rho a^3 \to 0$, the solution of Eq.\ (\ref{gauss2}) is $g(z) = (z-1)^{-1}$ and hence $p(\Lambda) = \delta(\Lambda-1)$. A better approximation is obtained from Eq.\ (\ref{cumul}). When the density $\rho a^3$ is increased, the eigenvalue distribution widens, but a narrow peak at $\Lambda = 1$ resists much longer than predicted by Eq.\ (\ref{gauss2}), as can be seen from Fig.\ \ref{figgauss1}, where we compare the eigenvalue density following from Eq.\ (\ref{gauss2}) with the results of numerical diagonalization at $\rho a^3 = 0.1$ and three different numbers of points $N$. Even though the height of the peak at $\Lambda = 1$ decreases with $N$, it is clear that the theory fails to describe $p(\Lambda)$ in this regime \cite{mezard99}. At higher densities $\rho a^3 \gtrsim 1$, the peak disappears and a satisfactory approximation to $p(\Lambda)$ is obtained by taking the high-density limit of Eq.\ (\ref{gauss2}) that is very close to the result obtained by \citet{mezard99} in the high-density approximation [see also Sec.\ \ref{HeuristicHigh} and, in particular, Eq.\ (\ref{pHighDensity1})]:\footnote{This approximation does not describe the rounding of $p(\Lambda)$ at small $\Lambda$ and does not obey the normalization of probability.}
\begin{eqnarray}
\label{gauss3}
p(\Lambda) \simeq \frac{1}{\sqrt{2} \pi^2 \rho a^3 \Lambda}
\ln \left[  \frac{(2\pi)^{3/2} \rho a^3}{\Lambda} \right]^{1/2}.
\end{eqnarray}
We compare this equation with the results of numerical simulations in Fig.\ \ref{figgauss2}.

In conclusion, the heuristic analysis of Sec.\ \ref{ERMHeuristic} correctly describes $p(\Lambda)$ in the limit of low density, when it is peaked around $\Lambda = 1$. In contrast, the self-consistent equations of Sec.\ \ref{SecERMSelfConsistent} apply at higher densities. None of the two approaches correctly follow the transition between the two regimes. In particular, the reason of failure of Eq.\ (\ref{gauss2}) resides in the assumption of independence of elements $H_{i \alpha}$ of the matrix $H$ in the representation (\ref{AHTH}) of an arbitrary ERM. We checked that if, instead of taking $H_{i \alpha} = \exp(i \vec{k}_{\alpha} \vec{r}_i)/\sqrt{N}$, we generate random matrices $A = HTH^{\dagger}$ numerically using matrices $H$ with independent elements,\footnote{We tried different statistical distributions of $H_{i \alpha}$, like, e.g., the circular Gaussian distribution and $H_{i \alpha} = \exp(i \varphi_{i \alpha})/\sqrt{N}$, with independent phases $\varphi_{i \alpha}$ uniformly distributed in $[0, 2\pi]$.} the eigenvalue density of resulting random matrices nicely follows Eq.\ (\ref{gauss2}). As we will see in the following, the assumption of mutual independence of $H_{i \alpha}$ will limit the applicability of the analysis relying on Eq.\ (\ref{SolgERM}) to other ERMs too, although in a less important way.

\subsubsection{Exponential Euclidean random matrix}
\label{expERM}

%%%%%%%%%%%%%FIG%%%%%%%%%%%
\begin{figure*}
\centering{
\includegraphics[angle=0,width=0.8\textwidth]{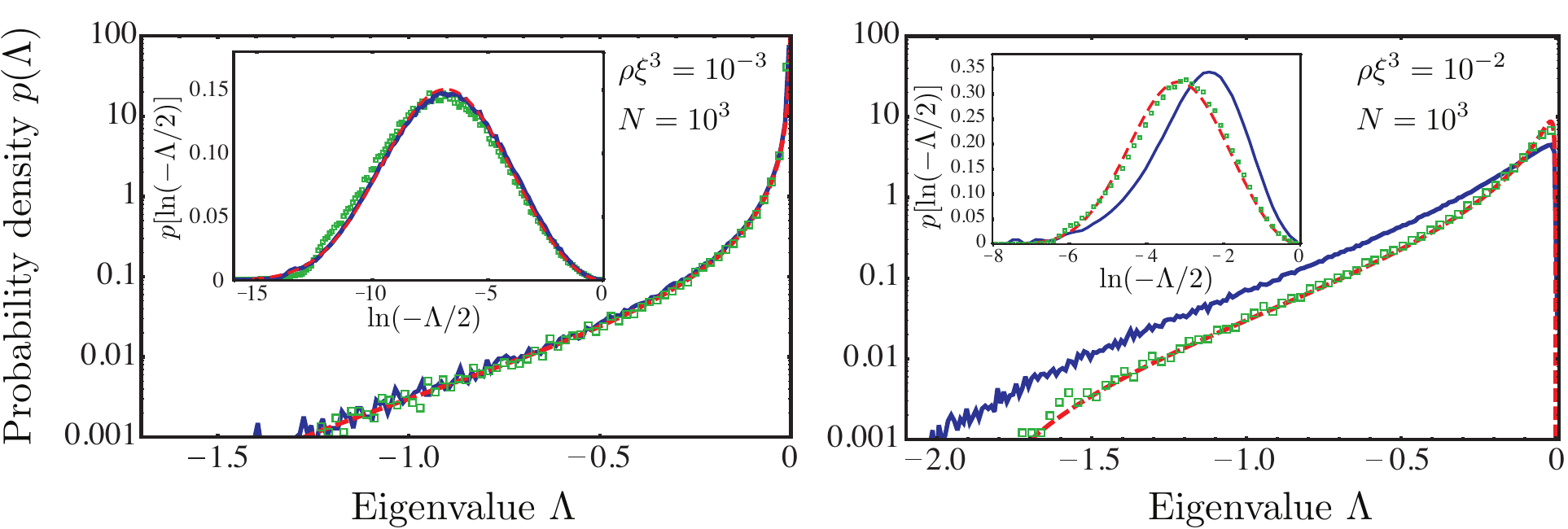}
\caption{
\label{figexp}
Eigenvalue density of $N \times N$ Euclidean matrix $A$ with elements $A_{ij} = \exp(-|\vec{r}_i - \vec{r}_j|/\xi)-\delta_{ij}\sum_{k=1}^{N} \exp(-|\vec{r}_i - \vec{r}_k|/\xi)$ at densities $\rho \xi^3 = 10^{-3}$ (left panel) and $10^{-2}$ (right panel) for $N = 10^3$ points $\mathbf{r}_i$ randomly distributed inside a cube.
The analytic result (\ref{PexpLowDensity}) (red dashed line) is compared to the numerical diagonalization (blue solid lines) and to the RG approach (green squares). The insets show $p[\ln(-\Lambda/2)]$ to emphasize the deviation from $p(\Lambda) \sim 1/|\Lambda|$ that would correspond to $p[\ln(-\Lambda/2)] = \mathrm{const}$.
}}
\end{figure*}
%%%%%%%%%%%%%%%%%%%%%%%%%%

Another example of ERM defined using a rapidly decaying function $f(\vec{r}_i-\vec{r}_j)$ can be obtained by taking $f(\vec{r}_i-\vec{r}_j) = \exp(-|\vec{r}_i-\vec{r}_j|/\xi)$. Let us study the eigenvalue density of the matrix $A$ defined by this function and Eq.\ (\ref{DefinitionERMu}) with $u = 1$. Such matrices appear in various contexts; as an example, in Sec.\ \ref{ElectronGlass} we will use them to study the electron glass dynamics. If we assume that $A_{ij}$ represent the hopping rates between randomly distributed, exponentially localized states $i$, the probability $p_i$ to occupy the state $i$ obeys the master equation $\mathrm{d}p_i/\mathrm{d}t = \sum A_{ij} p_j$. We emphasize that the symmetry constraint $\sum_j A_{ij} = 0$ combined with the fact that $A_{ij} > 0$ has a profound impact on the eigenvalues $\Lambda$ of $A$: they are all negative, whereas they would be positive is the diagonal elements of $A$ were defined as $A_{ii} = f(\vec{0}) = 1$ [i.e., for $u = 0$ in Eq.\ (\ref{DefinitionERMu})] \cite{mezard99}.

\citet{amir10} studied the eigenvalue density $p(\Lambda)$ of the exponential ERM $A$ in the limit of low density $\rho\xi^d\ll1$. In the space of arbitrary dimensionality $d$, their main result can be rederived using the heuristic approach of Sec.\ \ref{HeuristicLow}. Inside its support $[-2,0]$, $p(\Lambda)$ is readily found by differentiating Eq.\ (\ref{Cumul1}):
\be
\label{PexpLowDensity}
p(\Lambda)=\frac{-d\mathcal{V}\rho\xi^d[-\textrm{ln}(-\Lambda/2)]^{d-1}e^{-\mathcal{V}\rho\xi^d[-\textrm{ln}(-\Lambda/2)]^{d}/2}}{2\Lambda}.
\ee
This equation shows that $p(\Lambda)\sim -1/\Lambda$ for all $d$ over a broad range of $\Lambda$'s. We compare it with the result of numerical diagonalization in Fig.\ \ref{figexp} for $d = 3$. Good agreement is found as long as $\rho\xi^3 \lesssim 10^{-3}$. At larger densities, a more sophisticated theory is required. This theory should include diagrams similar to those taken into account in Sec.\ \ref{SecERMSelfConsistent} and corresponding to elementary excitations that involve clusters of many points. Unfortunately, the results of Sec.\ \ref{SecERMSelfConsistent} cannot be used as such, since they were obtained for ERMs defined by Eq.\ (\ref{DefinitionERMu}) with $u=0$ instead of $u = 1$.

In addition to Eq.\ (\ref{Cumul1}), $p(\Lambda)$ of the exponential ERM can be studied with the help of the real-space RG approach presented in Sec.\ \ref{Renorm}. There, we argued that RG gives quantitative results in the low density limit $\rho\xi^d\ll1$ only. This is confirmed in Fig.\ \ref{figexp} where the symbols represent the results of the RG flow following from Eqs.\ (\ref{FreqRG}), (\ref{MassRG}) and (\ref{SpringRG}). These results are in agreement with those of \citet{amir10} who used a renormalization rule that is slightly different from Eq.\ (\ref{FreqRG}). At low densities $\rho\xi^3 \lesssim 10^{-3}$, when both Eq.\ (\ref{PexpLowDensity}) and the RG prediction are good estimates of the eigenvalue density, the typical size $m_c(\Lambda)$ of the eigenvector associated with the eigenvalue $\Lambda$ is, according to Eq.\ (\ref{VectorRG}), given by the inverse of the cumulative of (\ref{PexpLowDensity}) \cite{amir10}:
\be
m_c(\Lambda)\simeq e^{2\rho\xi^3 [\textrm{ln}(-\Lambda/2)]^3/3}.
\ee
This indicates that the eigenvectors are localized with a spatial extent $m_c(\Lambda)$ that diverges when $\Lambda$ goes to zero. By means of a more refined analysis based on the percolation theory, \citet{amir13} show that a small fraction $(\rho\xi^3)^{\nu}$ of delocalized eigenvectors survive close to $\Lambda=0$, with $\nu$ the critical exponent of the percolation transition.

\subsubsection{Cardinal sine Euclidean random matrix}
\label{ERMSinc}

%%%%%%%%%%%%%FIG%%%%%%%%%%%
\begin{figure*}
\centering{
\includegraphics[angle=0,width=0.8\textwidth]{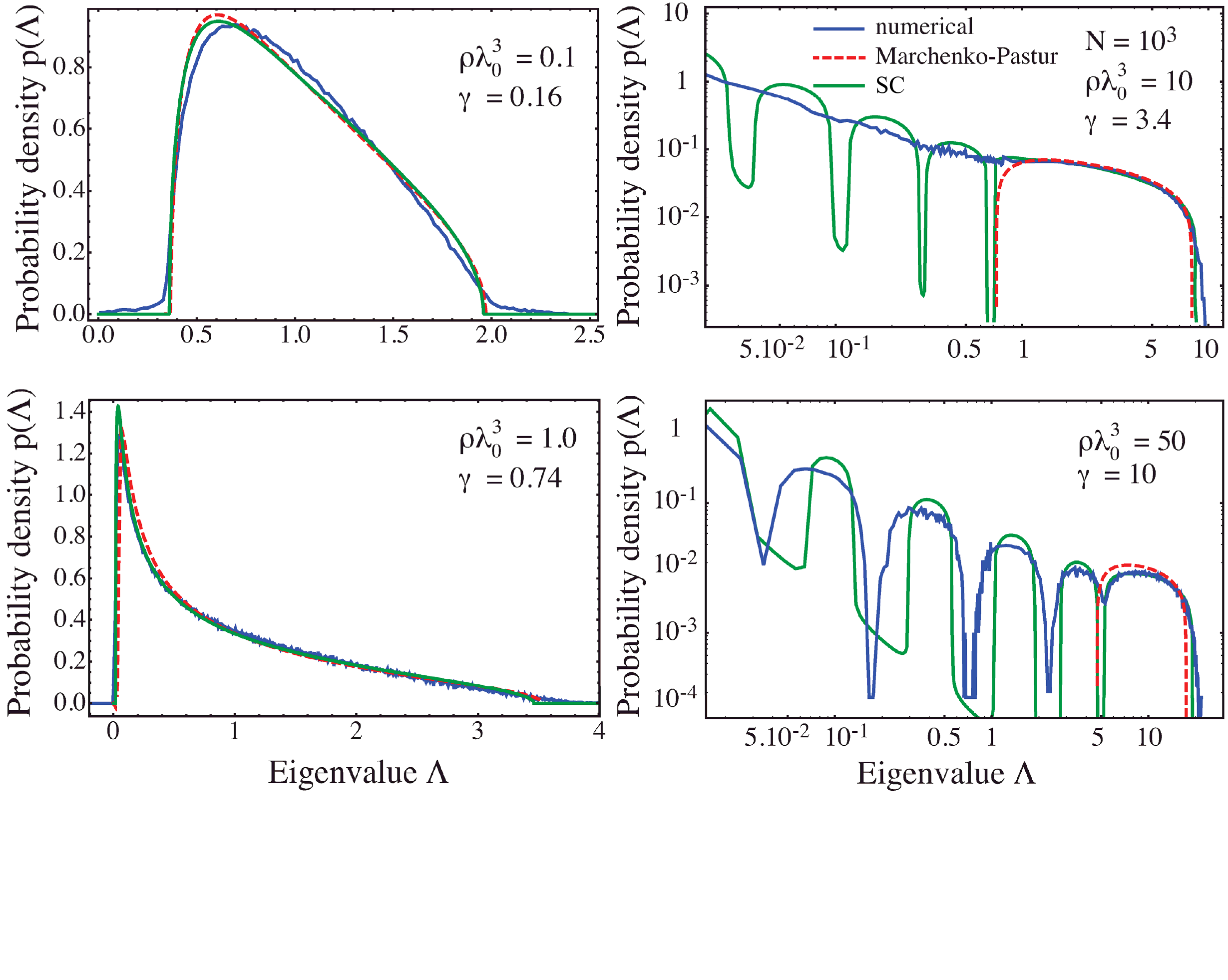}
\vspace*{-2cm}
\caption{
\label{figsincSphere1}
Eigenvalue density of the  $N \times N$ ERM $S$ defined by Eq.\ (\ref{sincERM}), where the $N$ points $\mathbf{r}_i$ are randomly chosen inside a sphere of radius $R$. Numerical results (blue solid line) are compared to the Marchenko-Pastur law (\ref{pmp}) (red dashed lines), and to Eq.\ (\ref{SolgSincEv}) (SC, green solid lines) at 4 different densities $\rho \lambda_0^3$ of points ($\lambda_0 = 2\pi/k_0$).}}
\end{figure*}
%%%%%%%%%%%%%%%%%%%%%%%%%%

A common feature of the Gaussian and exponential ERMs considered in the two previous subsections is the fast decay of their generating function $f(\vec{r}_i, \vec{r}_j)$ with $|\vec{r}_i - \vec{r}_j|$. As a consequence, at a fixed density $\rho$ of points $\vec{r}_i$ and in the limit of large $N$, the density of eigenvalues $p(\Lambda)$ becomes independent of $N$ and is controlled by a single parameter $\rho$. In this section we consider an example of ERM generated by a function $f$ that decays slowly and oscillates in space, so that $p(\Lambda)$ always depends on two parameters (e.g., $\rho$ and $N$), even in the limit of $N \to \infty$.

Consider a matrix $S$ with elements
\be
\label{sincERM}
S_{ij} = f( \mathbf{r}_i- \mathbf{r}_j ) = \frac{\sin(k_0 |\mathbf{r}_i - \mathbf{r}_j|)}{
k_0 |\mathbf{r}_i - \mathbf{r}_j|},
\ee
with a wavenumber $k_0 = 2\pi/\lambda_0$ and a wavelength $\lambda_0$ (this terminology anticipates the application of the matrix $S$ to problems of wave propagation, see Secs.\ \ref{collective}, \ref{laser}, and \ref{SecAndersonGreen}). The matrix $S$ was studied by \citet{akkermans08,svidzinsky08,scully09,svidzinsky09,svidzinsky10} in the context of light scattering and emission by large ensembles of atoms.

An approximate solution for the eigenvalue density of the ERM (\ref{sincERM}) for points $\vec{r}_i$ randomly distributed in a box of side $L$ can be found by calculating the matrix $T$ in the limit of large $k_0 L$ using Eq.\ (\ref{TSerriesCoeff}) and then applying the free probability theory as described in Sec.\ \ref{SecFRVERM} \cite{skipetrov11}. $T$ can be approximated by a diagonal matrix: $T \simeq (N/M) I_{M}$ with $M = (k_0 L)^2/2.8$. This readily leads to the Marchenko-Pastur law for the eigenvalues $\Lambda$ of the matrix $S$:
\be
\label{pmp}
p(\Lambda) = \left(1 - \frac{1}{\gamma} \right)^+ \delta(\Lambda)
+ \frac{\sqrt{(\Lambda_{+}-
\Lambda)^+(\Lambda - \Lambda_{-})^+}}{2\pi \gamma \Lambda},
\ee
where $\Lambda_{\pm} =(1 \pm \sqrt{\gamma})^2$. This distribution is parameterized by a single parameter $\gamma = \langle \Lambda^2 \rangle - 1 \simeq 2.8 N/(k_0 L)^2$ equal to its variance. It is easy to show that the same result holds for arbitrary shape of the volume $V$, provided that $\gamma$ is calculated accordingly. In Fig.\ \ref{figsincSphere1} we compare Eq.\ (\ref{pmp}) with numerical simulations in a sphere of radius $R$, where $\gamma = 9N/8(k_0 R)^2$. The agreement is good at low densities and $\gamma \lesssim 1$, but Eq.\ (\ref{pmp}) fails to describe numerical results at high densities and $\gamma > 1$. A better solution for $p(\Lambda)$ is therefore required.

Let us calculate the eigenvalue density of the matrix $S$ using the self-consistent equation (\ref{SolgERM}) derived in Sec.\ \ref{SecERMSelfConsistent}. As explained in Sec.\ \ref{SolvingPractice}, a general way to solve this equation is to express the latter in the eigenbasis of the operator $\hat{T}$. In order to solve the eigenvalue equation (\ref{rightEv}), it is convenient to decompose its kernel in spherical harmonics \cite{gradshteyn80}:
\begin{align}
\label{DecompoSinc}
\frac{\sin(k_0 |\mathbf{r} - \mathbf{r}'|)}{
k_0 |\mathbf{r} - \mathbf{r}'|} &= 4\pi \sum_{l=0}^\infty\sum_{m=-l}^l
j_l(k_0r)j_l(k_0r')
\nonumber \\
&\times Y_{lm}(\theta,\phi)Y_{lm}(\theta',\phi')^*,
\end{align}
where $\theta$ and $\phi$ are the polar and azimuthal angles of the vector $\vec{r}$, respectively, $j_l$ are spherical Bessel functions of the first kind, and $Y_{lm}$ are spherical harmonics. Inserting this decomposition into Eq.~(\ref{rightEv}), we readily find that
\begin{align}
\mc{R}_\alpha(\vec{r})&=\mc{R}_{lm}(\vec{r})=\mc{A}_{l}j_l(k_0r)Y_{lm}(\theta,\phi),
\\
\label{mulSinc}
\mu_\alpha&=\mu_{l}=4\pi\rho\int_0^R\textrm{d}r'j_l(k_0r')^2r'^2
\nonumber
\\
&
=\frac{3}{2}N\left[j_l(k_0R)^2-j_{l-1}(k_0R)j_{l+1}(k_0R)\right],
\end{align}
where $\mc{A}_{l}$ are normalization coefficients and $\alpha=\{l, m\}$. Eigenvalues $\mu_{l}$ are $(2l+1)$-times degenerate ($m\in[-l, l]$). Eq.~(\ref{SolgERMEv}) then becomes
\be
\label{SolgSincEv}
z = \frac{1}{g(z)}+
\frac{1}{N} \sum_{l} \frac{(2l+1)\mu_l}{1-g(z)\mu_l}.
\ee
Once this equation is solved for $g(z)$, $p(\Lambda)$ can be obtained from Eq.\ (\ref{LinkpImG}). In contrast to Eq.\ (\ref{pmp}), $p(\Lambda)$ following from Eq.\ (\ref{SolgSincEv}) depends on two parameters $\gamma$ and $\rho \lambda_0^3$.

As can be seen from Fig.\ \ref{figsincSphere1}, $p(\Lambda)$ following from Eq.\ (\ref{SolgSincEv}) is in good agreement with numerical results at all densities $\rho \lambda_0^3$ and for all values of $\gamma$. It even describes the splitting of the support of $p(\Lambda)$ is separate `islands' for $R \lesssim \lambda_0$, even though the agreement with numerics is slightly worse in this regime, probably due to the finite value of $N$ in numerical calculations [the limit $N \to \infty$ is taken in the analytic result (\ref{SolgSincEv})].

\subsubsection{Cardinal cosine Euclidean random matrix}
\label{ERMCosc}

%%%%%%%%%%%%%FIG%%%%%%%%%%%
\begin{figure*}[t]
\centering{
\includegraphics[angle=0,width=0.8\textwidth]{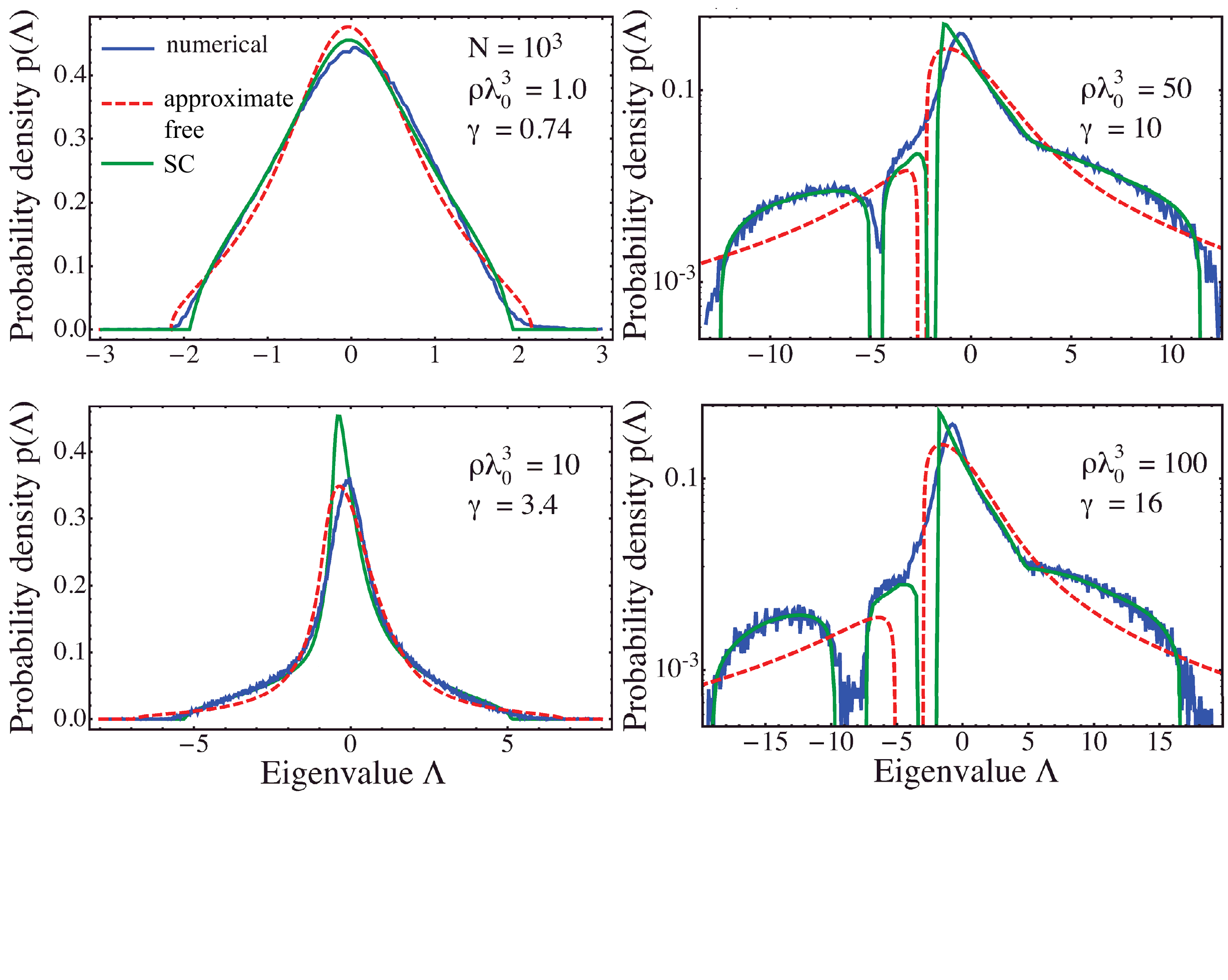}
\vspace*{-2cm}
\caption{
\label{figcoscSphere1}
Eigenvalue density of the  $N \times N$ ERM $C$ defined by Eq.\ (\ref{CoscERM}), where the $N$ points $\mathbf{r}_i$ are randomly chosen inside a sphere of radius $R$. Numerical results (blue solid lines) are compared to Eq.~(\ref{RCoscBox1}) (red dashed lines), and to Eq.~(\ref{SolgCoscEv}) (SC, green solid lines) at 4 different densities $\rho \lambda_0^3$ of points ($\lambda_0 = 2\pi/k_0$).
}}
\end{figure*}
%%%%%%%%%%%%%%%%%%%%%%%%%%

Another example of ERM generated by a slowly decaying and oscillating function $f(\vec{r}_i, \vec{r}_j)$ is the matrix $C$ with elements
\be
\label{CoscERM}
C_{ij} = f( \mathbf{r}_i - \mathbf{r}_j ) =
(1 - \delta_{ij})
\frac{\cos(k_0 |\mathbf{r}_i - \mathbf{r}_j|)}{
k_0 |\mathbf{r}_i - \mathbf{r}_j|}.
\ee
This matrix is relevant, for example, for understanding the collective Lamb shift in the problem of light interaction with clouds of cold atoms (see Sec.\ \ref{collective}). In contrast to Eq.\ (\ref{sincERM}), the Fourier transform of Eq.\ (\ref{CoscERM}) is not always positive, and hence, the spectrum of $C$ is not bounded from below.

In a complete analogy with Sec.\ \ref{ERMSinc}, we compute the matrix $T$ [see Eq.\ (\ref{TSerriesCoeff})] for the ERM $C$ and then apply the free probability theory of Sec.\ \ref{SecFRVERM} \cite{skipetrov11}. This time, however, the calculation can be performed analytically only for the ${\cal R}$-transform of the probability density:
\begin{align}
&\mc{R}(z) =
- \frac{2}{\pi} \mathrm{arccoth} \frac{4 \pi^3 \gamma}{\rho \lambda_0^3} - \frac{2}{\pi}
\sqrt{-1 - \frac{\rho \lambda_0^3}{2\pi^2} z}
\label{RCoscBox1}
\\
&\times
\left[ \frac{\pi}{2}
- \arctan \frac{1 + \frac{\rho \lambda_0^3}{2 \pi^3 \gamma}}{\sqrt{-1 - \frac{\rho \lambda_0^3}{2\pi^2} z}} +
\arctan \frac{1 - \frac{\rho \lambda_0^3}{2 \pi^3 \gamma}}{\sqrt{-1 - \frac{\rho \lambda_0^3 }{2\pi^2} z}}
\right].
\nonumber
\end{align}
The resolvent $g(z)$ has to be found numerically by inverting this equation.

As can be seen from Fig.\ \ref{figcoscSphere1}, Eq.\ (\ref{RCoscBox1}) is not sufficient at high densities and a better solution is needed. We will search for it using the self-consistent equation (\ref{SolgERM}) of Sec.\ \ref{SecERMSelfConsistent}. The eigenvalue equation (\ref{rightEv}) is solved by using the expansion \cite{gradshteyn80}
\begin{align}
\label{DecompoCosc}
\frac{\cos(k_0 |\mathbf{r} - \mathbf{r}'|)}{
k_0 |\mathbf{r} - \mathbf{r}'|} &= -4\pi \sum_{l=0}^\infty\sum_{m=-l}^l
j_l\left[k_0\textrm{min}(r,r')\right]
\\
&\times n_l\left[k_0\textrm{max}(r,r')\right]
Y_{lm}(\theta,\phi)Y_{lm}(\theta',\phi')^*,
\nonumber
\end{align}
where $\theta$ and $\phi$ are the polar and azimuthal angles of the vector $\vec{r}$, respectively, $Y_{lm}$ are spherical harmonics, and $j_l$ and $n_l$ are spherical Bessel functions of the first and second kind, respectively. Inserting Eq.~(\ref{DecompoCosc}) into Eq.~(\ref{rightEv}), and using standard properties of spherical harmonics and spherical Bessel functions \cite{morse53}, it is easy to show that the eigenvectors of $\hat{T}$ are necessarily of the form
\be
\mc{R}_\alpha(\vec{r})=\mc{R}_{lmp}(\vec{r})=\mc{A}_{lp}j_l(\kappa_{lp}r)Y_{lm}(\theta,\phi),
\ee
where the coefficients $\kappa_{lp}$ obey
\be
\label{modesCosc}
\frac{\kappa_{lp}}{k_0}=\frac{j_l(\kappa_{lp}R)}{j_{l-1}(\kappa_{lp}R)}\frac{n_{l-1}(k_0R)}{n_l(k_0R)}.
\ee
Integer $p$ labels the different solutions of this equation for a given $l$. $\kappa_{lp}$ are either real or imaginary numbers, and the corresponding eigenvalues
\be
\label{muCosc}
\mu_{\alpha}=\mu_{lp}=\frac{\rho \lambda_0^3}{2\pi^2}\frac{1}{\left(\kappa_{lp}/k_0\right)^2-1}
\ee
are $(2l+1)$-times degenerate ($m\in[-l, l]$). In terms of the solutions $\mu_{lp}$ of Eqs.~(\ref{modesCosc}) and (\ref{muCosc}), Eq.~(\ref{SolgERMEv}) reads finally
\be
\label{SolgCoscEv}
z=\frac{1}{g(z)}+
\frac{g(z)}{N} \sum_{l} \sum_p \frac{(2l+1)\mu_{lp}^2}{1-g(z)\mu_{lp}}.
\ee

We solve Eq.\ (\ref{modesCosc}) numerically for $\kappa_{lp}$, then compute $\mu_{lp}$ using Eq.\ (\ref{muCosc}), solve Eq.\ (\ref{SolgCoscEv}) numerically to find $g(z)$ and, finally, obtain $p(\Lambda)$ from the imaginary part of $g(z)$ using Eq.\ (\ref{LinkpImG}). The results are shown in Fig.\ \ref{figcoscSphere1}. We see that in contrast to Eq.\ (\ref{RCoscBox1}), Eq.\ (\ref{SolgCoscEv}) describes the eigenvalue distribution quite nicely even at high densities, but both equation fail for negative eigenvalues close to zero, where they predict a gap in the eigenvalue density, whereas numerical results do not show evidence for such a gap. Detailed analysis reveals that states corresponding to the eigenvalues in the gap are localized in space, suggesting a possible link between the failure of Eqs.\ (\ref{RCoscBox1}) and (\ref{SolgCoscEv}) and Anderson localization \cite{goetschy11c}. However, no definite conclusion can be made about this issue at the time of writing.

\section{Non-Hermitian Euclidean random matrix theory}
\label{NonHermitianERM}

In contrast to Hermitian matrices, the eigenvalues of non-Hermitian matrices are not constrained to lie on the real axis and  may invade the complex plane. Consequently, various methods developed for Hermitian matrices and heavily exploiting the analytic function theory are no longer applicable and require non-trivial modifications \cite{janik97b,janik97a,jarosz06,feinberg97a,feinberg97b,feinberg06}.

Most of the literature on random non-Hermitian matrices focus on Gaussian randomness. A paradigmatic example is the ensemble of $N\times N$ matrices $A$ with the probability distribution
\be
\label{ComplexGaussianNH}
P(A)=C_Ne^{- N\textrm{Tr}AA^\dagger}.
\ee
\citet{ginibre65} showed that in the limit $N\to \infty$, the eigenvalues of $A$ are uniformly distributed within a disk of radius unity on the complex plane. Twenty years later, \citet{girko85} generalized this result to matrix elements $A_{ij}$ that are i.i.d. with zero mean and variance $1/N$. This is commonly referred to as Girko's law.
Here we would like to tackle the problem of computing the density of eigenvalues of matrices that break away from this law: the non-Hermitian ERMs. Non-Hermitian ERMs appear in such important physical problems as Anderson localization of light \cite{rusek00, pinheiro04} and matter waves \cite{massignan06, antezza10}, random lasing \cite{pinheiro06,goetschy11b}, propagation of light in nonlinear disordered media \cite{gremaud10}, and collective spontaneous emission of atomic systems \cite{ernst69,akkermans08,svidzinsky10,skipetrov11}. However, no analytic theory was available to deal with these matrices until very recently, and our knowledge about their statistical properties was based exclusively on large-scale numerical simulations \cite{rusek00, pinheiro04, massignan06, antezza10, pinheiro06, gremaud10, skipetrov11}.

Our review of non-Hermitian ERMs is organized as follows. In Sec.\ \ref{SecDefnhERM} we introduce new mathematical objects that allow to generalize the methods developed for Hermitian matrices to the non-Hermitian case. A diagrammatic theory for the density of eigenvalues of an arbitrary non-Hermitian ERM in the limit of large matrix size  ($N\to \infty$) is developed in Sec.\ \ref{SecDiagnhERM}. Alternative approaches are also briefly discussed (Sec.\ \ref{SecOthernhERM}). We illustrate the theory by applying it to two specific non-Hermitian ERMs in Sec.\ \ref{applnonherm}.

\subsection{Foundations of the non-Hermitian random matrix theory}
\label{SecDefnhERM}

This section is devoted to the introduction of basic definitions and relations useful in the study of non-Hermitian matrices.

\subsubsection{Eigenvalue density and Hermitization}

Eigenvalues $\Lambda_n$ of a $N \times N$ non-Hermitian matrix $A$ are, in general, complex. Their density on the complex plane is
\be
\label{DefpNH}
p(\Lambda)=\frac{1}{N}\left\langle \sum_{n=1}^N\delta^{(2)}(\Lambda-\Lambda_n)
 \right\rangle,
\ee
where we use a shorthand notation $\delta^{(2)}(\Lambda-\Lambda_n)=\delta(\textrm{Re}\Lambda-\textrm{Re}\Lambda_n)\delta(\textrm{Im}\Lambda-\textrm{Im}\Lambda_n)$. The relation between $p(\Lambda)$ and the resolvent
\be
\label{Defresolvent2}
g(z) = \frac{1}{N} \left\langle \textrm{Tr} \frac{1}{z-A} \right\rangle
= \frac{1}{N} \left\langle \sum_{n=1}^N \frac{1}{z-\Lambda_n} \right\rangle
\ee
can be found using $\partial_{z^*}(1/z)=\pi \delta(x)\delta(y)$, with the standard notation $\partial_{z^*}=\frac{1}{2}(\partial_{x}+\textrm{i}\partial_{y})$ for $z=x+\textrm{i}y$. We obtain:
\begin{align}
\label{LinkpgNH}
p(\Lambda)&= \left. \frac{1}{\pi} \partial_{z^*} g(z) \right|_{z=\Lambda}
\\
\label{LinkpgNH2}
&=\left.
\frac{1}{2\pi}\left[
\partial_x \textrm{Re}g(z)
-\partial_y \textrm{Im}g(z)
\right]
\right|_{z=\Lambda}.
\end{align}
Note that $\partial_y \textrm{Re}g(z)=-\partial_x \textrm{Im}g(z)$ because $p(\Lambda)$ is real. The right-hand side of Eq.~(\ref{LinkpgNH2}) vanishes if $g(z)$ obeys the Cauchy-Riemann conditions, i.e. if it is an analytic function of the complex variable $z$. In general, the eigenvalues $\Lambda_n$ occupy, on average, a two-dimensional domain $\mc{D}$ on the complex plane where $g(z)$ is nonanalytic, and $p(\Lambda)$ describes the location and the amount of this nonanalyticity.

We now recall that the resolvent $g(z)$ can be interpreted as the electric field  $\vec{g}(z)$ created at a point $z$ on the complex plane by charges $q = 1$ situated at positions $\Lambda_n$, see Eq.~(\ref{gAsElecField}). Equation (\ref{LinkpgNH2}) can thus be seen as the Gauss law $p(z)=\mathbf{\nabla}_{x,y} \cdot \vec{g}(z)/2\pi$. Hence, we readily obtain a new relation between $p(\Lambda)$ and the logarithmic pairwise repulsion $V^{\mathrm{int}}(z)$ defined by $\vec{g}(z)=-\mathbf{\nabla}_{x,y}V^{\mathrm{int}}(z)$ [see Eq.~(\ref{VCoulRep})]:
\be
\label{LinkpVSVintNH}
p(\Lambda)=- \left. \frac{1}{2\pi N}\Delta_{x,y}V^{\mathrm{int}}(z)\right|_{z=\Lambda},
\ee
where $\Delta_{x,y}=4\partial_z\partial_{z^*}$ is the Laplacian in the coordinates $x$ and $y$. Clearly, Eq.~(\ref{LinkpVSVintNH}) may be particularly useful within the framework of the Dyson gas model where $V^{\mathrm{int}}$ may be related to the one-body potential determined by the probability distribution $P(A)$ (see Sec.\ \ref{MappingDyson} and \ref{SecDysonNH}).

Inserting the explicit expression (\ref{VCoulRep}) of $V^{\mathrm{int}}(z)$ into Eq.~(\ref{LinkpVSVintNH}), we express $p(\Lambda)$ in an alternative form:
\begin{align}
\label{pAAdagger}
p(\Lambda)&=\left. \frac{1}{\pi N}\partial_z\partial_{z^*}
\left\langle
\textrm{Tr}\,
\textrm{ln}(z-A)(z^*-A^\dagger)
\right\rangle
\right|_{z=\Lambda}
\\
\label{pAAdagger2}
&=\left. \frac{1}{\pi N}\partial_z\partial_{z^*}
\left\langle
\textrm{ln}\,
\textrm{det}
(z-A)(z^*-A^\dagger)
\right\rangle
\right|_{z=\Lambda}
\\
\label{pAAdagger3}
&=\left. \frac{1}{\pi N}
\lim_{\epsilon\to 0}
\partial_z\partial_{z^*}
\left\langle
\textrm{ln}\,
\textrm{det}
\left[
\mc{H}_A(z)
-i\epsilon I_{2N}
\right]
\right\rangle
\right|_{z=\Lambda},
\end{align}
where $I_{2N}$ is the $2 N \times 2N$ identity matrix and $\mc{H}_A$ is the $2 N \times 2N$ chiral Hermitian matrix
\be
\label{DefHA}
\mc{H}_A(z)=
\left( \begin{array}{cc}
0& A-z  \\
A^\dagger - z^*& 0
\end{array} \right).
\ee
Note that Eq.~(\ref{pAAdagger}) can also be derived from Eq.~(\ref{DefpNH}) using $\partial_z\partial_{z^*}\textrm{ln}zz^*=\pi\delta(x)\delta(y)$. The representation (\ref{pAAdagger3}) is generally used in field-theoretical approaches (Sec.\ \ref{SecFieldNH}). In addition, since the matrix (\ref{DefHA}) is Hermitian, one can compute its resolvent with well-established Hermitian techniques, from which it is still possible to recover the eigenvalue density of $A$ \cite{feinberg97b, feinberg06}. This is the so-called `Hermitization method'. In the following, we will use an alternative method which has various advantages: it is technically slightly simpler, it reveals a relation between $g(z)$ and the correlator of right and left eigenvectors of $A$, and, finally, it allows for a generalization of the free probability calculus.

\subsubsection{Quaternions and the eigenvector correlator}

If $A$ is Hermitian, its eigenvalues $\Lambda_n$ lie, on average, on some intervals (cuts) of the real axis. Therefore, it is possible to reconstruct $g(z)$ by analytic continuation of its series expansion (\ref{gSumMoments}) performed in the vicinity of $\vert z\vert \to \infty$. The eigenvalue distribution $p(\Lambda)$ follows from the discontinuities of $g(z)$ on the real axis [see Eqs.~(\ref{LinkpImG}) and (\ref{LinkpgNH})]. For a non-Hermitian matrix $A$, however, $g(z)$ loses its analyticity inside a two-dimensional domain $\mc{D}$ where $\Lambda_n$ are concentrated, meaning that $g(z)$ for $z\in \mc{D}$ cannot be simply assessed by analytic continuation of its series expansion. A way to circumvent this problem is based on the algebra of quaternions:  while $p(\Lambda)$ for an Hermitian $A$ is obtained  by approaching the real axis from orthogonal directions (in the complex plane), $p(\Lambda)$ for a  non-Hermitian $A$ can be found by approaching two sides of $\mc{D}$ from directions `orthogonal' to the complex plane in the quaternion space \cite{jarosz06}. Doubling the size of the matrix under study, we now work with a new $2N \times 2N$ matrix
\be
\label{Adupplicate}
A^D=
\left( \begin{array}{cc}
A & 0  \\
0& A^{\dagger}
\end{array} \right)
\ee
and a quaternion resolvent matrix
\be
\label{resolventQuaternion}
G(Q)=\frac{1}{N} \left\langle \textrm{Tr}_N \frac{1}{Q\otimes I_N - A^D} \right\rangle.
\ee
The $2\times 2$ matrix $Q$ is an arbitrary quaternion in the matrix representation:
\be
\label{DefQuaternion}
Q=
\left( \begin{array}{cc}
a& \mathrm{i} b^*  \\
\mathrm{i} b& a^*
\end{array} \right)=x_0I_2+\textrm{i}\vec{x}\cdot \bm{\sigma},
\ee
where $\vec{x}=(x_1, x_2, x_3)$, $ \bm{\sigma}$ is the triplet of Pauli matrices, $a=x_0+\textrm{i}x_3$, and $b=x_1+\textrm{i}x_2$.
$\textrm{Tr}_N$ in Eq.~(\ref{resolventQuaternion}) denotes the block trace of an arbitrary $2N \times 2N$ matrix $X$. It is defined by separating $X$ in four $N \times N$ blocks $X_{11}$, $X_{12}$, $X_{21}$, $X_{22}$ and taking the trace of each of the latter separately:
\begin{align}
\mathrm{Tr}_N X &= \mathrm{Tr}_N \left( \begin{array}{cc}
X_{11} & X_{12}  \\
X_{21} & X_{22}
\end{array} \right)
\nonumber
\\
\label{btr}
&= \left( \begin{array}{cc}
\mathrm{Tr} X_{11} & \mathrm{Tr} X_{12}  \\
\mathrm{Tr} X_{21} & \mathrm{Tr} X_{22}
\end{array} \right).
\end{align}
Algebraic properties of the quaternions are useful to generalize the free probability theory to non-Hermitan matrices (see Sec.\ \ref{SecFreeNH}). However, if we wish to compute $g(z)$ by a diagrammatic approach, it is sufficient to consider the quaternion $Q=Z_{\epsilon}$, where
\be
\label{ze}
Z_{\epsilon}=
\left( \begin{array}{cc}
z& \mathrm{i} \epsilon  \\
\mathrm{i} \epsilon& z^*
\end{array} \right).
\ee
The generalized resolvent matrix $G(Z_{\epsilon})$ is then safely equal to its series expansion in $1/Z_{\epsilon}$ \cite{janik97b, janik01, jarosz06}. By evaluating the block trace in Eq.~(\ref{resolventQuaternion}) explicitly, one readily finds that
\be
\label{Gas2el}
G(Z_{\epsilon})
=\left( \begin{array}{cc}
G_{11}^\epsilon & G_{12}^\epsilon  \\
G_{12}^\epsilon & G_{11}^{\epsilon *}
\end{array} \right)
\ee
with
\begin{align}
\label{G11explicit}
G_{11}^\epsilon&=\frac{1}{N} \left\langle \textrm{Tr} \frac{z^*-A^\dagger}{(z-A)(z^*-A^\dagger)+\epsilon^2} \right\rangle,
\\
\label{G12explicit}
G_{12}^\epsilon&=-\frac{\mathrm{i} \epsilon}{N} \left\langle \textrm{Tr}\frac{1}{(z-A)(z^*-A^\dagger)+\epsilon^2} \right\rangle,
\end{align}
so that
\be
\label{LimitResolvent}
\lim_{\epsilon \to 0}G(Z_{\epsilon})=
\left[ \begin{array}{cc}
g(z) &c(z)   \\
c(z)& g(z)^*
\end{array} \right].
\ee
Interestingly, the off-diagonal elements $c(z)=\lim_{\epsilon \to 0}G_{12}^\epsilon$ yield the correlator of right $\ket{R_n}$ and left $\ket{L_n}$ eigenvectors of $A$ \cite{chalker98, janik99}:
\begin{align}
\label{EVcorrelator}
\mathcal{C}(z) &=
-\frac{\pi}{N} \left\langle \sum_{n=1}^N \scp{L_n}{L_n} \scp{R_n}{R_n} \delta^{(2)}(z-\Lambda_n) \right\rangle
\nonumber \\
&= N c(z)^2.
\end{align}
This shows that $c(z)$ must vanish on the boundary $\delta\mc{D}$ of the support of the eigenvalue density $\mc{D}$. In order to obtain $p(\Lambda)$, one should compute $G(Z_{\epsilon})$ at finite $\epsilon\in \mathbb{R}$ (by a diagrammatic or any other approach), then take the limit $\epsilon\to 0$ to extract $g(z)$ from the diagonal elements of (\ref{LimitResolvent}), and finally apply Eq.~(\ref{LinkpgNH}).

\subsubsection{Biorthogonal basis of left and right eigenvectors}

For the sake of completeness, we recall here basic properties of right $\ket{R_n}$ and left $\ket{L_n}$ eigenvectors of a non-Hermitian matrix $A$ (or operator $\hat{A}$). By definition,
\begin{align}
A\ket{R_n}&=\Lambda_n\ket{R_n},
\\
\bra{L_n}A&=\Lambda_n\bra{L_n} \iff A^\dagger \ket{L_n}=\Lambda_n^*\ket{L_n},
\end{align}
meaning that $\ket{L_n}$ are the right eigenvectors of $A^\dagger$. Obviously, $A$ and $A^\dagger$ have  complex conjugated eigenvalues for $\textrm{det}(A-\Lambda_nI_n)=0=\textrm{det}(A^\dagger-\Lambda_n^*I_n)$. Besides, $\ket{L_n}$ and $\ket{R_m}$ are necessarily orthogonal because $\braket{L_n}{A}{R_m}=\Lambda_n\scp{L_n}{R_m}=\Lambda_m\scp{L_n}{R_m}$. Assuming that the eigenvalues $\Lambda_n$ are not degenerate, we normalize $\ket{R_n}$ and $\ket{L_n}$ such that
\be
\scp{L_n}{R_m}=\sum_{i=1}^N L_n^{i*}R_m ^i=\delta_{nm}.
\ee
Note that $\scp{R_n}{R_m}\neq \delta_{nm}$. Finally, the following properties hold:
\begin{align}
I_N&=\sum_n\ket{R_n}\bra{L_n}=\sum_n\ket{L_n}\bra{R_n},
\\
\textrm{Tr}X&=\sum_{n}\braket{L_n}{X}{R_n},
\end{align}
where $X$ is an arbitrary matrix.

\subsection{Diagrammatic approach for non-Hermitian Euclidean random matrices}
\label{SecDiagnhERM}

Our goal is to derive equations for the resolvent $g(z)$ and the correlator $c(z)$ of an arbitrary $N\times N$ non-Hermitian ERM $A$ with elements $A_{ij}=f(\vec{r}_i, \vec{r}_j)=\braket{\vec{r}_i}{\hat{A}}{\vec{r}_j}$ in the limit of $N\to \infty$. For this purpose, we follow \citet{goetschy11a} and make use of the representation $A=HTH^\dagger$ introduced in Sec.\ \ref{SubsecERM}, with the assumption that $H$ has i.i.d. complex Gaussian entries satisfying Eq.\ (\ref{Contraction0}). The assumption of Gaussian statistics for the elements of $H$ simplifies diagrammatic calculations but is not essential, contrary to the assumption of independence of different elements that may limit the validity of presented results.

\subsubsection{Derivation of self-consistent equations}

\begin{figure}[t]
%% \vspace{2mm}
\centering{
\includegraphics[angle=0,width=\columnwidth]{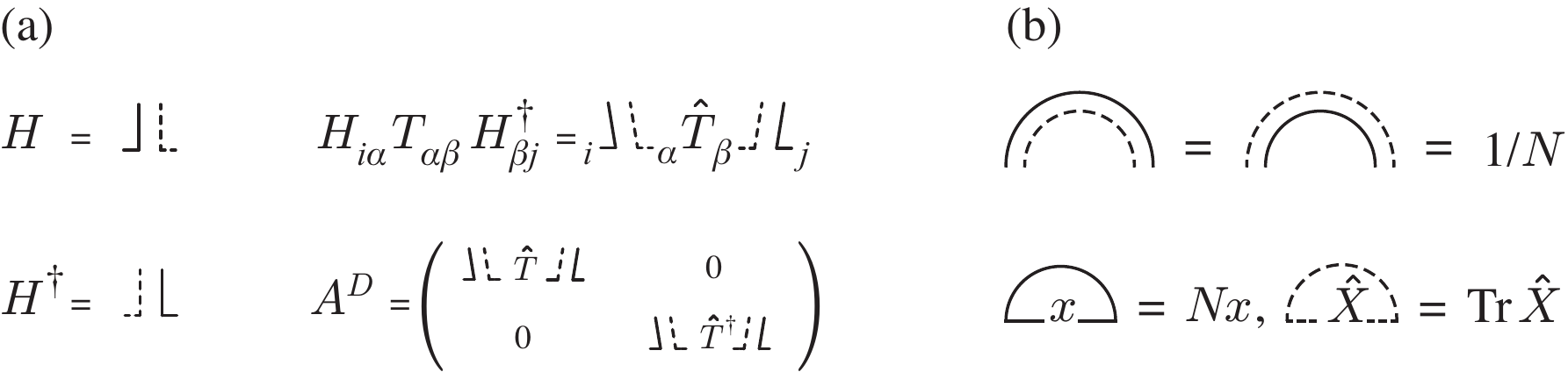}
\caption{
\label{HHAandTracediag2}
(a) Diagrammatic representations of the matrices $H$, $H^{\dagger}$, $A=HTH^{\dagger}$, and $A^D$. Full and dashed lines propagate in the bases $\{\mathbf{r}_i\}$ and $\{\psi_\alpha\}$, respectively (see Sec.\ \ref{SubsecERM}); $\hat{T}=\rho\hat{A}$.
(b) Diagrammatic notation for pairwise contractions (\ref{Contraction0}) and loop diagrams for any scalar $x$ in the basis $\{\mathbf{r}_i\}$, and for any operator $\hat{X}$ in an arbitrary basis $\{\psi_\alpha\}$.}}
\end{figure}

We start by expanding the $2\times 2$ resolvent matrix  $G(Z_{\epsilon})$ defined by Eqs.~(\ref{resolventQuaternion}) and (\ref{ze}) in series in $1/\mathcal{Z}_{\epsilon} = (1/Z_{\epsilon}) \otimes I_N $:
\begin{align}
\label{Gseries}
G(Z_{\epsilon})
&=  \frac{1}{N}\left\langle
 \textrm{Tr}_{N} \left[
\frac{1}{\mathcal{Z}_{\epsilon}}+
\frac{1}{\mathcal{Z}_{\epsilon}}\,A^D\,
\frac{1}{\mathcal{Z}_{\epsilon}}+\ldots
\right]
\right\rangle.
\end{align}
Inasmuch as $H_{i\alpha}$ are i.i.d. Gaussian random variables, the result of averaging $\moy{\dots}$ over the ensemble of matrices $H$ can be expressed through pairwise contractions (\ref{Contraction0}) only. Diagrammatic notations, already introduced in Sec.\ \ref{SecERMSelfConsistent} to evaluate efficiently the weight of different terms arising in the calculation, are reproduced in Fig.\ \ref{HHAandTracediag2}(a) for clarity.
The `propagator' $1/Z_{\epsilon}$ will be depicted by
\be
\label{InvZ}
\frac{1}{Z_\epsilon}
=
\left( \begin{array}{cc}
\frac{1}{z} & -\frac{i\epsilon}{\vert z\vert^2}   \\
 -\frac{i\epsilon}{\vert z\vert^2}&\frac{1}{z^*}
\end{array} \right)
=
\left( \begin{array}{cc}
\overline{\mbox{\scriptsize 1\;\;\;\;1}} &\overline{\mbox{\scriptsize 1\;\;\;\;2}}   \\
\overline{\mbox{\scriptsize 2\;\;\;\;1}}&\overline{\mbox{\scriptsize 2\;\;\;\;2}}
\end{array} \right).
\ee
%%%%%FIG%%%%%
%%%%%%%%%%%%%

\begin{figure}[b]
\centering{
\includegraphics[angle=0,width=0.9\columnwidth]{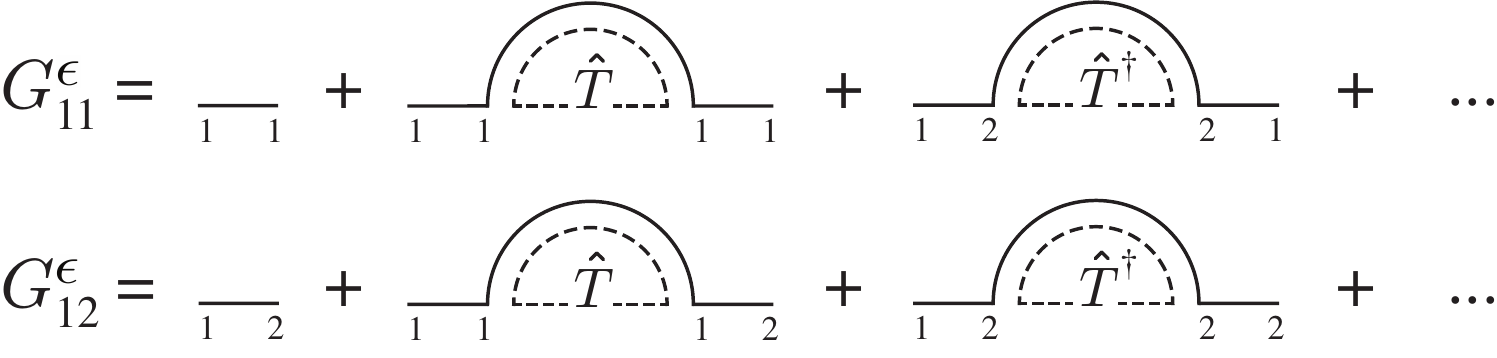}
\caption{
\label{Gdiag}
Diagrammatic expansion of the two independent elements of the matrix $G(Z_\epsilon)$. }}
\end{figure}

Since each contraction (\ref{Contraction0}) brings a factor $1/N$, and each loop corresponding to taking the trace of a matrix brings a factor $N$ [see Fig.\ \ref{HHAandTracediag2}(b)], only the planar rainbow-like diagrams, that contain as many loops as contractions, survive in the limit $N\to \infty$. Such diagrams appear, for example, in Fig.\ \ref{Gdiag}, where we show the beginning of the expansion of the two independent elements of $G(Z_\epsilon)$ defined by Eq.~(\ref{Gas2el}).

%%%%%FIG%%%%%
%%%%%%%%%%%%%

%%%%%FIG%%%%%
\begin{figure*}[t]
\centering{
\includegraphics[angle=0,width=0.8\textwidth]{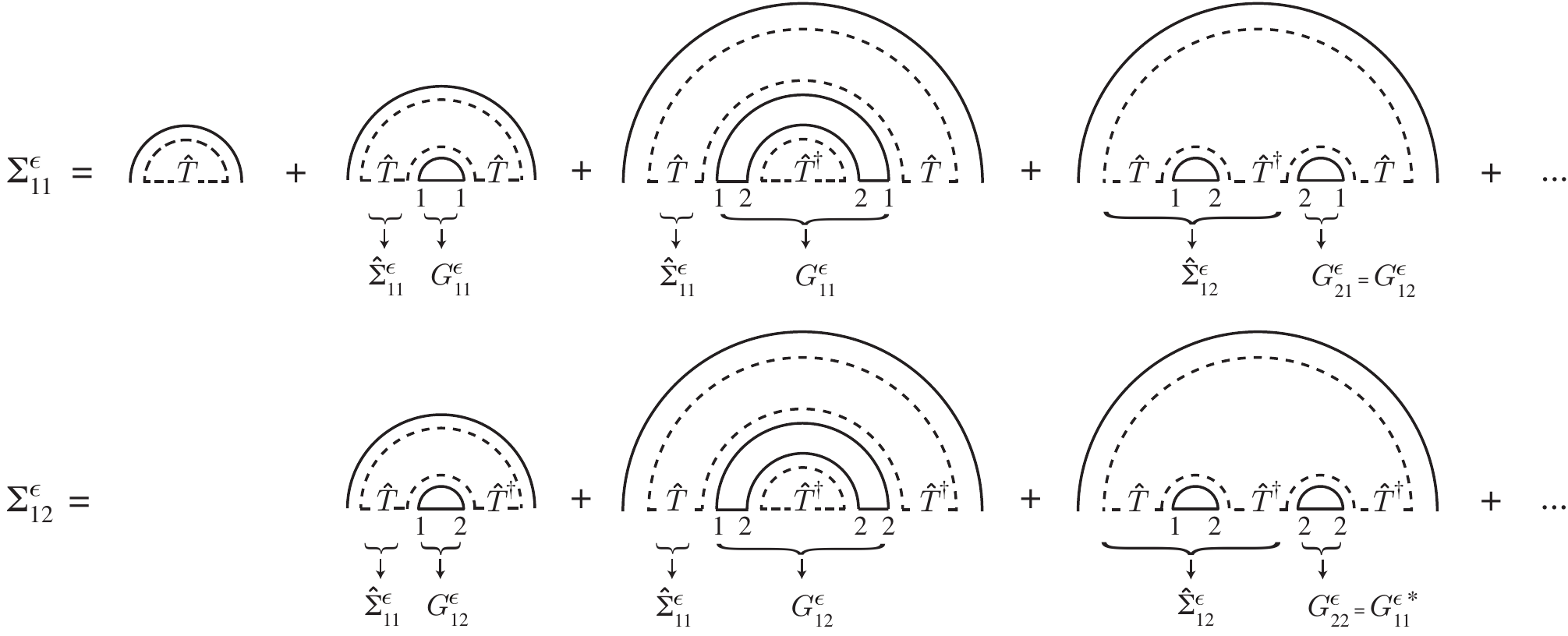}
\caption{
\label{Sigmadiag}
Diagrammatic expansion of the two independent elements of the self-energy $\Sigma(Z_\epsilon)$. Braces with arrows denote parts of diagrams that are beginning of diagrammatic expansions of the quantities which the arrows point to.}}
\end{figure*}
%%\end{widetext}
%%%%%%%%%%%%%

\begin{figure}[t]
\centering{
\includegraphics[angle=0,width=\columnwidth]{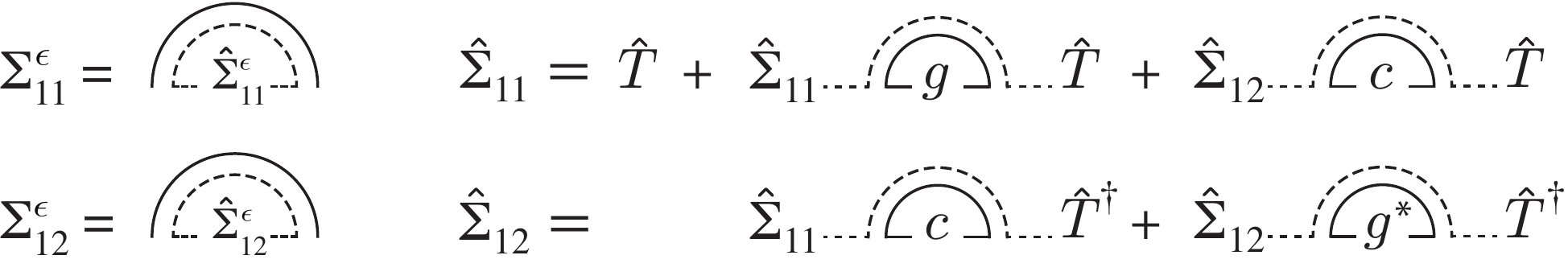}
\caption{
\label{SigmaSelfconsistentNH2}
The elements $\Sigma_{11}^\epsilon$ and $ \Sigma_{12}^\epsilon$  of the matrix $\Sigma(Z_{\epsilon})$ can be written as traces of operators $\hat{\Sigma}_{11}^\epsilon$ and $\hat{\Sigma}_{12}^\epsilon$: $\Sigma_{11}^\epsilon=\textrm{Tr}\hat{\Sigma}_{11}^\epsilon/N$ and $\Sigma_{12}^\epsilon=\textrm{Tr}\hat{\Sigma}_{12}^\epsilon/N$.
Operators $\hat{\Sigma}_{11}=\lim_{\epsilon \to 0^+ }\hat{\Sigma}_{11}^\epsilon$ and $\hat{\Sigma}_{12}=\lim_{\epsilon \to 0^+ }\hat{\Sigma}_{12}^\epsilon$ obey coupled equations, where $g=\lim_{\epsilon \to 0^+ }G_{11}^\epsilon$ and $c=\lim_{\epsilon \to 0^+ }G_{12}^\epsilon$ [see Eq.~(\ref{LimitResolvent})].}}
\end{figure}

In the standard way, rather than summing the diagrams for the resolvent, we introduce the $2\times 2$ self-energy matrix
\be
\label{DefSigmaMatrix}
\Sigma(Z_\epsilon) = Z_\epsilon - G(Z_\epsilon)^{-1}=\left( \begin{array}{cc}
\Sigma_{11}^\epsilon & \Sigma_{12}^\epsilon  \\
\Sigma_{12}^\epsilon & \Sigma_{11}^{\epsilon *}
\end{array} \right).
 \ee
It is equal to the sum of all one-particle irreducible diagrams contained in
\be
\label{Selfseries}
Z_\epsilon G(Z_\epsilon)Z_\epsilon=\frac{1}{N}  \left\langle \textrm{Tr}_N \left[
A^D+A^D\frac{1}{\mathcal{Z}_{\epsilon}}A^D+\ldots
\right]
\right\rangle.
\ee
The first dominant terms that appear in the expansion of the two matrix elements $\Sigma_{11}^\epsilon$ and $\Sigma_{12}^\epsilon$ are represented in Fig.\ \ref{Sigmadiag}.
In the two series of Fig.\ \ref{Sigmadiag} we recognize, under a pairwise contraction, the matrix elements $G_{11}^\epsilon$ and $G_{12}^\epsilon$ depicted in Fig.\ \ref{Gdiag}, as well as the two operators $\hat{\Sigma}_{11}^\epsilon$ and  $\hat{\Sigma}_{12}^\epsilon$ defined in Fig.\ \ref{SigmaSelfconsistentNH2}.
Equations obeyed by the operators
$\hat{\Sigma}_{11}=\lim_{\epsilon \to 0^+ }\hat{\Sigma}_{11}^\epsilon$ and $\hat{\Sigma}_{12}=\lim_{\epsilon \to 0}\hat{\Sigma}_{12}^\epsilon$
are obtained after summation of all planar rainbow diagrams in the expansion of Fig.\ \ref{Sigmadiag} and taking the limit $\epsilon \to 0^+$.\footnote{As usual in such a procedure, summation must be performed before taking the limit $\epsilon \to 0$. Hence, the off-diagonal element of the propagator $1/Z_{\epsilon}$ gives rise to non-vanishing terms after summation, although it is zero in the limit $\epsilon \to 0$.} The diagrammatic representation of these equations is shown in Fig.\ \ref{SigmaSelfconsistentNH2}.
%%%%%FIGA4%%%%%
%%%%%%%%%%%%%
Applying `Feynman' rules defined in Fig.\ \ref{HHAandTracediag2}(b), we obtain:
\begin{align}
\hat{\Sigma}_{11}&=(1+g\,\hat{\Sigma}_{11}+c\,\hat{\Sigma}_{12})\hat{T},
\label{self1NH}
\\
\hat{\Sigma}_{12}&=(c\,\hat{\Sigma}_{11}+g^*\,\hat{\Sigma}_{12})\hat{T}^\dagger,
\label{self2NH}
\end{align}
where $\hat{T}=\rho\hat{A}$. After some algebra,\footnote{ Although $[\hat{T}, \hat{T}^\dagger]\neq 0$, this calculation is easily performed by making cyclic permutations of operators under the trace operator.} $\Sigma_{11}=\textrm{Tr}\hat{\Sigma}_{11}/N$ and $\Sigma_{12}=\textrm{Tr}\hat{\Sigma}_{12}/N$ can be expressed as:
\begin{align}
\label{SolSigma11}
\Sigma_{11}&=\frac{1}{N}\textrm{Tr} \frac{(1-
g^*\hat{T}^{\dagger})\hat{T}}{(1-
g^*\hat{T}^{\dagger})(1- g\hat{T}) - c^2\hat{T}^{\dagger}\hat{T}},
\\
\label{SolSigma12}
\Sigma_{12}&= \frac{c}{N}\textrm{Tr}
\frac{\hat{T}^{\dagger}\hat{T}}{(1-
g^*\hat{T}^{\dagger})(1- g\hat{T}) - c^2\hat{T}^{\dagger}\hat{T}}.
\end{align}

Furthermore, as follows from Eq.\ (\ref{LimitResolvent}) and the definition (\ref{DefSigmaMatrix}) of the self-energy matrix, $g$ and $c$ are simply related to  $\Sigma_{11}$ and $\Sigma_{12}$ by
\be
\label{LinkGSigmaNH}
\left[ \begin{array}{cc}
g(z) &c(z)   \\
c(z)& g(z)^*
\end{array} \right]=\left( \begin{array}{cc}
z-\Sigma_{11} & -\Sigma_{12}   \\
-\Sigma_{12}& z^*-\Sigma_{11}^*
\end{array} \right)^{-1}.
\ee
Elimination of the self-energy from Eqs.~(\ref{SolSigma11}), (\ref{SolSigma12}) and (\ref{LinkGSigmaNH}) leads to two self-consistent equations for the resolvent $g(z)$ and the eigenvector correlator $c(z)$:
\begin{align}
&z = \frac{g^*}{|g|^2-c^2}+
\frac{1}{N}\textrm{Tr} \frac{(1-
g^*\hat{T}^{\dagger})\hat{T}}{(1-
g^*\hat{T}^{\dagger})(1- g\hat{T}) - c^2\hat{T}^{\dagger}\hat{T}},\;\;\;\;\;\;
\label{sc1NH}
\\
&\frac{1}{|g|^2-c^2} = \frac{1}{N}\textrm{Tr}
\frac{\hat{T}^{\dagger}\hat{T}}{(1-
g^*\hat{T}^{\dagger})(1- g\hat{T}) - c^2\hat{T}^{\dagger}\hat{T}}.
\label{sc2NH}
\end{align}

At this final stage, it is convenient to define the following operators
\begin{align}
\label{DefOpS0}
\hat{S}_0&=\hat{S}(g)=\frac{\hat{T}}{1-g\,\hat{T}},
\\
\label{DefOpS1}
\hat{S}_1&=\hat{S}(g+c^2\hat{S}_0^\dagger)=\frac{(1-
g^*\hat{T}^{\dagger})\hat{T}}{(1-
g^*\hat{T}^{\dagger})(1- g\hat{T}) - c^2\hat{T}^{\dagger}\hat{T}},
\end{align}
in terms of which Eqs.~(\ref{sc1NH}) and (\ref{sc2NH}) become
\begin{align}
&z = \frac{g^*}{|g|^2-c^2}+
\frac{1}{N}\textrm{Tr}\hat{S}_1,
\label{sc1NHv2}
\\
&\frac{1}{|g|^2-c^2} = \frac{1}{N}\textrm{Tr}
\hat{S}_1\hat{S}_0^\dagger.
\label{sc2NHv2}
\end{align}

Because $c(z)$ must vanish on the boundary $\delta\mathcal{D}$ of the support of the eigenvalue density $\mathcal{D}$, equations for $z\in\delta\mathcal{D}$ follow:
\begin{align}
&z = \frac{1}{g}+\frac{1}{N} \textrm{Tr} \hat{S}_0,
\label{contoura}
\\
&\frac{1}{|g|^2} = \frac{1}{N} \textrm{Tr} \hat{S}_0\hat{S}^{\dagger}_0.
\label{contourb}
\end{align}

\subsubsection{Analysis of self-consistent equations}

Equations (\ref{sc1NH}), (\ref{sc2NH}), (\ref{contoura}) and (\ref{contourb}) are the main results of this section. An equation for the borderline of the support of the eigenvalue density of a non-Hermitian ERM on the complex plane $z = \Lambda$ follows from Eqs.\ (\ref{contoura}) and (\ref{contourb}) upon elimination of $g$. The density of eigenvalues $\Lambda$ inside its support $\mathcal{D}$ can be found by solving Eqs.\ (\ref{sc1NH}) and (\ref{sc2NH}) with respect to $g(z)$ and then applying Eq.\ (\ref{LinkpgNH}).

At low density and in the framework of representation $A=HTH^\dagger$, the eigenvalue density of any Hermitian ERM obeys the Wigner semicircle law (\ref{WignerERM}).\footnote{We exclude rare ERMs for which the operator $\hat{T}$ has a small number of non-zero eigenvalues.} An analogous result exists for non-Hermitian ERMs as well. Indeed, using the approximation
\be
\hat{S}_1\simeq \hat{S}_0\simeq\hat{T}
\ee
valid at low densities, Eqs.~(\ref{contoura}) and (\ref{contourb}) for the borderline of the eigenvalue domain reduce to
\be
\label{contourApprox0}
\left\vert
z-\frac{1}{N} \textrm{Tr} \hat{T}
\right\vert^2
= \frac{1}{N} \textrm{Tr} \hat{T}\hat{T}^{\dagger},
\ee
and Eqs.~(\ref{sc1NHv2}) and  (\ref{sc2NHv2}) for $g(z)$ and $c(z)$ with $z\in\mc{D}$ become
\begin{align}
\label{ResolventApprox0}
g(z)& = \frac{z^* - \frac{1}{N} \mathrm{Tr} \hat{T}^{\dagger}}{\frac{1}{N} \mathrm{Tr} \hat{T} \hat{T}^{\dagger}},
\\
\label{CorrelatorApprox0}
c(z)&=\frac{1}{\frac{1}{N} \mathrm{Tr} \hat{T} \hat{T}^{\dagger}}
\left[
 \frac{\vert z - \frac{1}{N} \mathrm{Tr} \hat{T}\vert^2}{\frac{1}{N} \mathrm{Tr} \hat{T} \hat{T}^{\dagger}}
 -1
\right].
\end{align}
The term $ \textrm{Tr} \hat{T}/N$ that appears in Eqs.~(\ref{contourApprox0}), (\ref{ResolventApprox0}), and (\ref{CorrelatorApprox0}), leads to a shift of the eigenvalue distribution equal to
\be
\frac{ \textrm{Tr} \hat{T}}{N}=\frac{ \textrm{Tr} \hat{A}}{V}=\frac{\moy{\textrm{Tr}_NA}}{N}=\moy{\Lambda}.
\ee
We assume from here on that $A_{ii}=0$ ($i=1, \dots, N$), so that, in particular, $\moy{\Lambda}=0$. With this assumption, the term $ \textrm{Tr} \hat{T}\hat{T}^\dagger$ reads:
\begin{align}
\label{TTNH}
\textrm{Tr}(\hat{T}\hat{T}^\dagger) &= \rho^2\textrm{Tr}(\hat{A}\hat{A}^\dagger)=\rho^2 \iint_{V}  \mathrm{d}^d\vec{r}\; \mathrm{d}^d\vec{r'} \left|f(\vec{r},\vec{r'})\right|^2
\\
\label{TTNH2}
&= \left \langle \textrm{Tr}_N(AA^\dagger) \right \rangle
\nonumber \\
&= \left\langle
\sum_{n=1}^N\sum_{m=1}^N
\Lambda_n\Lambda_m^*\scp{L_n}{L_m}\scp{R_m}{R_n}
\right\rangle
\\
\label{TTNH3}
&\simeq
\left\langle
\sum_{n=1}^N\vert\Lambda_n\vert^2 \scp{L_n}{L_n}\scp{R_n}{R_n}
\right\rangle,
\\
\label{TTNH4}
&\simeq 2
\left\langle
\sum_{n=1}^N\vert\Lambda_n\vert^2
\right\rangle
= 2N
\left\langle
\vert\Lambda\vert^2
\right\rangle.
\end{align}
In Eqs.~(\ref{TTNH3}) and (\ref{TTNH4}) we assumed that, at low densities, $\scp{\vec{r_i}}{L_n}$ and $\scp{\vec{r_i}}{R_n}$ behave as Gaussian random variables. Introducing the shorthand notation
\be
\label{gammaNH}
\gamma=\frac{\textrm{Tr}(\hat{T}\hat{T}^\dagger)}{2N}\simeq \langle
\vert\Lambda\vert^2
\rangle,
\ee
we rewrite Eqs.~(\ref{contourApprox0}), (\ref{ResolventApprox0}), and (\ref{CorrelatorApprox0}) as
\begin{align}
\vert z \vert^2&=2\gamma \;\;\;\;(z\in\delta\mc{D}),
\\
g(z)&=\frac{z^*}{2\gamma} \;\;\;\;(z\in\mc{D}),
\\
c(z)&=\frac{1}{2\gamma}\left[\frac{\vert z\vert^2}{2\gamma}-1\right] \;\;\;\;(z\in\mc{D}).
\end{align}
This shows that, in the limit $N\to \infty$ and $\rho \to 0$ at fixed $\gamma$, the eigenvalues of an arbitrary traceless non-Hermitian ERM are uniformly distributed within a disk of radius $\sqrt{2\gamma}$. Within the disk, $p(\Lambda)=1/2\pi\gamma$. This is the famous Girko's law \cite{girko85}, first discovered by \citet{ginibre65} for the complex Gaussian ensemble. We recover this law because in the limit $N\to \infty$ and $\rho \to 0$, elements of $A$ essentially behave as i.i.d. variables and hence $\Sigma_{11}=0$ and  $\Sigma_{12}=c(z)/2\gamma$.

As it was the case for Hermitian ERMs, the solution of Eqs.~(\ref{sc1NH}), (\ref{sc2NH}), (\ref{contoura}) and (\ref{contourb}) for a given matrix $A$ can be greatly facilitated by a suitable choice of the basis in which traces appearing in these equations are expressed. In addition to $\{ \vec{r} \}$ and $\{ \vec{k}_{\alpha} \}$, a biorthogonal basis of right $\ket{\mathcal{R}_{\alpha}}$ and left $\ket{\mathcal{L}_{\alpha}}$ eigenvectors of $\hat{T}$ can be quite convenient. We recall that the right eigenvector $\ket{\mathcal{R}_{\alpha}}$ obeys
\begin{eqnarray}
\bra{\vec{r}} \hat{T} \ket{\mathcal{R}_{\alpha}} =
\rho \int_V\textrm{d}^d \vec{r'}f(\vec{r}, \vec{r'}) \mathcal{R}_\alpha(\vec{r'}) = \mu_\alpha \mathcal{R}_\alpha(\vec{r}),
\label{right}
\end{eqnarray}
where $\mu_{\alpha}$ is the eigenvalue corresponding to the eigenvector $\ket{\mathcal{R}_{\alpha}}$. The traces appearing in Eqs.\ (\ref{contoura}) and (\ref{contourb}) can be expressed as
\begin{align}
\label{trs}
\textrm{Tr} \hat{S}_0 &=
\sum_{\alpha} \braket{\mathcal{L}_{\alpha}}{\hat{S}_0}{\mathcal{R}_{\alpha}} = \sum_{\alpha} \frac{\mu_\alpha}{1-g\mu_\alpha},
\\
\label{trss}
\textrm{Tr} \hat{S}_0\hat{S}_0^{\dagger} &=
\sum_{\alpha,\beta}
\frac{\mu_\alpha \mu_\beta^* \scp{\mathcal{L}_\alpha}{\mathcal{L}_\beta}
\scp{\mathcal{R}_\beta}
{\mathcal{R}_\alpha}}{(1-g\mu_\alpha)(1-g\mu_\beta)^*},
\end{align}
respectively. Technically, the main difference with the study of Hermitian ERMs is that we now have to know the eigenvectors of $\hat{T}$ explicitly [compare Eqs.~(\ref{trss}) and (\ref{SolgERMEv})].

\subsection{Other approaches}
\label{SecOthernhERM}

\subsubsection{Dyson gas picture}
\label{SecDysonNH}

In this section, we extend the Dyson gas picture introduced for Hermitian matrices in Sec.\ \ref{MappingDyson}, to the non-Hermitian case. Let us first consider the ensemble of non-Hermitian random matrices with Gaussian probability density defined by Eq.\ (\ref{ComplexGaussianNH}) and originally introduced by \citet{ginibre65}. The probability density $P(A)$ of this ensemble is invariant under all unitary transformations, but not under the similarity transformation $A \to SAS^{-1}$ used to diagonalize $A=S^{-1}DS$ ($D$ denotes a diagonal matrix with elements $\Lambda_n$, $n=1 \dots N$). Hence, $P(A)$ depends explicitly on $S$ and not only on the eigenvalues of $A$. This feature is the main difference with the Gaussian ensemble (\ref{GaussianProba}) or the Wigner-Dyson ensemble (\ref{WignerDyson}). Hermitian matrices drawn from these two latter ensembles can be diagonalized by unitary matrices, so that $P(A)$ depends on $\{ \Lambda_n \}$ only. In order to obtain the joint probability density $P(\{\Lambda_n\})$ from Eq.~(\ref{ComplexGaussianNH}), we must change variables from $A_{ij}$ to parameters related to eigenvalues and eigenvectors of $A$. Since $\textrm{Tr}AA^\dagger$ depends on eigenvectors, the new variables have to be chosen carefully to facilitate further manipulations. The result is the following \cite{ginibre65, mehta04}:
\begin{align}
\label{GibbsDGNH}
P(\{\Lambda_n\})&=C'_Ne^{- \beta H^g(\{ \Lambda_n\})},
\\
\label{HgDGNH}
H^g(\{\lambda_n\})&=N\sum_{n=1}^NV^g(\Lambda_n)-\sum_{n<m}\ln|\Lambda_n-\Lambda_m|,
\\
\label{VgDGNH}
V^g(z)&=\frac{\vert z \vert^2}{2},
\end{align}
and $\beta = 2$.
We recognize in Eq.\ (\ref{GibbsDGNH}) the Boltzmann-Gibbs distribution of a Coulomb gas in thermal equilibrium at a temperature $T=1/\beta$. Equations (\ref{GibbsDGNH})--(\ref{VgDGNH}) have exactly the same form as Eqs.~(\ref{GibbsDG}) and (\ref{HgDG}). As for Hermitian matrices, the logarithmic pairwise repulsion comes from the Vandermonde-type Jacobian $\vert \mc{V}(\{\Lambda_n\})\vert^\beta$.

In the limit $N\to \infty$, we can perform coarse-graining of the energy functional $H^g$ [see Eq.~(\ref{energyCG})], and minimize it to obtain the equality
\be
\label{LinkVintVg2}
-\partial_z V^{\mathrm{int}}(z)=N\partial_z V^g(z),
\ee
where $V^{\mathrm{int}}(z)$ is the logarithmic pairwise repulsion (\ref{VCoulRep}). Eq.~(\ref{LinkVintVg2}) means that the force $N\partial_z V^g(z)$ experienced by each particle of the gas is compensated by the Coulomb repulsion by all other particles. The eigenvalue distribution (\ref{LinkpVSVintNH}) reads now:
\be
\label{LinkpVSVgNH}
p(\Lambda)= \left. \frac{1}{2\pi}\Delta_{x,y}V^{g}(z)\right|_{z=\Lambda}.
\ee
Note the difference with Eq.\ (\ref{pVsVg}) for Hermitian matrices: Eq.~(\ref{LinkpVSVgNH}) is local; the shape of the distribution at $\Lambda$ depends on the profile of $V^g$ in the vicinity of $\Lambda$ only, while in Eq.~(\ref{pVsVg}) the shape of $p(\Lambda)$ strongly depends on the boundaries of the distribution, meaning that the influence of $V^g$ on $p(\Lambda)$ is nonlocal. Note also that Eq.~(\ref{LinkpVSVgNH}) contains no information about the borderline of the support of eigenvalues. If $V^g$ is simple enough, the borderline can be obtained from the normalization constraint $\int \textrm{d}\Lambda p(\Lambda)=1$.\footnote{For complicated cases, the borderline may be found by inspection of Eq.~(\ref{Wignerequation}) that is still valid for non-Hermitian matrices. } For $V^g(z)=\vert z \vert^2/2$, we find that the eigenvalues are uniformly distributed inside a disk of radius $1$. This is the celebrated Ginibre's result \cite{ginibre65}.

Obviously, Eqs.~(\ref{GibbsDGNH}) and (\ref{HgDGNH}) also apply to any normal matrix $A$ ($[A, A^\dagger]=0$) with probability density
\be
\label{WignerDysonNormal}
P(A)=C_Ne^{- N\textrm{Tr}\mc{V}^g(AA^\dagger)},
\ee
where $\mc{V}^g$ is arbitrary, for $A$ can be diagonalized by a unitary matrix. The one-body potential appearing in Eq.~(\ref{HgDGNH}) is then given by $V^g(z)=\mc{V}^g(\vert z \vert^2)$. A counter-intuitive result is that solutions (\ref{GibbsDGNH}) and (\ref{HgDGNH}) may completely break down for most of random non-Hermitian matrices---i.e. for non-Hermitian matrices that are not normal or partially normal \cite{feinberg01, feinberg06}---distributed according to Eq.\ (\ref{WignerDysonNormal}). \citet{feinberg97a} proved the `single-ring theorem'. It stipulates that the shape of the eigenvalue distribution is either a disk or an annulus, whatever polynomial the potential $\mc{V}^g$ is. This is clearly in contradiction with what we could expect from Eqs.~(\ref{GibbsDGNH}) and (\ref{HgDGNH}), that tell us that the number of domains occupied by the eigenvalues on the complex plain should grow with the number of minima of $V^g(z)=\mc{V}^g(\vert z \vert^2)$. The polynomial $\mc{V}^g\sim A A^\dagger$ that corresponds to the complex Gaussian ensemble (\ref{ComplexGaussianNH}) is actually the only polynomial for which Eqs.~(\ref{GibbsDGNH}) and (\ref{HgDGNH}) are valid whatever the matrix $A$ obeying (\ref{WignerDysonNormal}) is. Remarkably, \citet{feinberg97a} also showed that the eigenvalue distribution of $A$ can nevertheless be found from the resolvent of the Hermitian matrix $AA^\dagger$. This resolvent has already been known in the literature for an arbitrary polynomial $\mc{V}^g$ \cite{feinberg97a}.

Although the Dyson gas picture was not rigorously justified for ERMs, it may be helpful to qualitatively understand the eigenvalue distributions obtained by other methods. For example, in the study of the Green's matrix $G$ in Sec.\ \ref{SecEigenvalueGreen}, we observe that the support $\mc{D}$ of $p(\Lambda)$ deforms when the density is increased, going through a transition from a disk-like to an annulus-like shape, and eventually splitting into multiple disconnected domains at high density (see Fig.\ \ref{FigGreencontours}). It is difficult to refrain from interpreting such transitions as phase transitions for the Dyson gas due to modifications of a hypothetic one-body potential $V^g$.

\subsubsection{Field representation}
\label{SecFieldNH}

Let us now briefly explain how to compute the eigenvalue distribution of a non-Hermitian matrix $A$ in the field-theoretical approach. We start with Eq.\ (\ref{pAAdagger3}) rewritten as
\be
\label{pAAdagger3v2}
p(\Lambda)
=-\left. \frac{1}{\pi N}
\lim_{\epsilon\to 0}
\partial_z\partial_{z^*}
\left\langle
\textrm{ln}
\mc{Z}^\epsilon(z)
\right\rangle
\right|_{z=\Lambda},
\ee
where we introduced the partition function
\begin{align}
\mc{Z}^\epsilon&=\textrm{det}
\left[ \begin{array}{cc}
\epsilon I_n & \textrm{i}(z-A)  \\
\textrm{i}(z^*-A^\dagger)& \epsilon I_n
\end{array} \right].
\end{align}
In order to evaluate $\moy{\textrm{ln}
\mc{Z}^\epsilon(z)}$, we follow the same procedure as in Sec.\ \ref{Fieldreprersentation}, namely, we apply the replica trick,
\be
\moy{\textrm{ln}\mc{Z}^\epsilon(z)}=\lim_{n\to 0}\frac{\moy{\mc{Z}^\epsilon(z)^n}-1}{n},
\ee
together with the representation
\begin{align}
\mc{Z}^\epsilon(z) &\varpropto \int \textrm{d}\phi_1 \dots \textrm{d}\phi_N e^{-\mc{H}(\Phi,z,\epsilon)},
\\
\mc{H}(\Phi,z,\epsilon) &= \sum_{i=1}^N\phi_i^\dagger\left(\epsilon I_2+\textrm{i}x\sigma_x-\textrm{i} y\sigma_y \right)\phi_i
\nonumber \\
&- \textrm{i} \sum_{i,j=1}^N\phi_i^\dagger\left(A^h_{ij}\sigma_x-A^s_{ij}\sigma_y \right)\phi_i,
\end{align}
where the $N$ fields $\phi_i$ are pairs of complex variables, $\sigma_x$ and $\sigma_y$ are Pauli matrices, $z=x+\textrm{i}y$, and $A=A^h+\textrm{i}A^s$, with $A^h$ and $A^s$ Hermitian matrices. This representation combined with the cavity method was used by \citet{rogers09} to analyze the eigenvalue distribution of sparse non-Hermitian matrices. A slightly different representation of $p(\Lambda)$, also based on the replica trick, can be found in a nice review of RMT by \citet{stephanov01}, where a derivation of Girko's law is also given.

For non-Hermitian ERMs of the form $A_{ij} = f(\vec{r}_i - \vec{r}_j)$ it seems feasible to generalize the field method proposed by \citet{mezard99} for Hermitian ERMs. Basically, it amounts to make the same approximations in the calculation of $\moy{\mc{Z}^\epsilon(z)^n}$ as those presented in details in Sec.\ \ref{Fieldreprersentation}. This leads to equations that have the same degree of validity as Eq.\ (\ref{SelfconsistentMezard2}). For example, the equations for the borderline of the eigenvalue domain are
\begin{align}
\label{contouraMezard}
&z = \frac{1}{g(z)}+\int\frac{\textrm{d}^d\mathbf{k}}{(2\pi)^d} \frac{f_0(\mathbf{k})}{1-\rho f_0(\mathbf{k})g(z)},
\\
\label{contourbMezard}
&\frac{1}{\vert g(z) \vert^2} = \int\frac{\textrm{d}^d\mathbf{k}}{(2\pi)^d} \frac{\rho \vert f_0(\mathbf{k}) \vert^2}{\vert 1-\rho f_0(\mathbf{k} )g(z)\vert^2},
\end{align}
where $f_0(\mathbf{k})$ is the Fourier transform of $f(\vec{r})$. These equation can also be obtained from Eqs.~(\ref{contoura}) and (\ref{contourb}) by using the approximation $\braket{\vec{k}}{\hat{A}}{\vec{k}'}\simeq \braket{\vec{k}}{\hat{A}}{\vec{k}} \delta_{\vec{k} \vec{k}'} \simeq f_0(\vec{k}) \delta_{\vec{k} \vec{k}'}$. Contrary to Eqs.~(\ref{contoura}) and (\ref{contourb}), Eqs.~(\ref{contouraMezard}) and (\ref{contourbMezard}) are parameterized only by the density $\rho=N/V$.

\subsubsection{Free probability}
\label{SecFreeNH}

The extension of free probability theory and, in particular, the generalization of the concept of Blue function, to non-Hermitian matrices is natural in quaternion space \cite{jarosz04}. The quaternion Blue matrix $B_X(Q)$ of any random matrix $X$ is the functional inverse of the quaternion resolvent matrix (\ref{resolventQuaternion}):
\be
\label{DefBQuat}
G_X[B_X(Q)]=B_X[G_X(Q)]=Q,
\ee
where $Q$ is a quaternion defined by Eq.~(\ref{DefQuaternion}). For convenience, we also introduce the quaternion $R$-transform:
\be
\label{DefRQuat}
R_X(Q)=B_X(Q)-\frac{1}{Q}.
\ee
For $Q=Z_\epsilon$ given by Eq.~(\ref{ze}), $R_X(Z_\epsilon)$ is simply related to the self-energy matrix (\ref{DefSigmaMatrix}) by
\be
R_X(Z_\epsilon)=\Sigma_X[B_X(Z_\epsilon)],
\ee
and therefore $R_X(z)=\lim_{\epsilon \to 0}R_X(Z_\epsilon)$ and $\Sigma_X(z)=\lim_{\epsilon \to 0}\Sigma_X(Z_\epsilon)$ are related through\footnote{To obtain Eq.~(\ref{LinkRBNH}), we use $B[\textrm{diag}(z,z^*)]=\textrm{diag}[\mc{B}(z), \mc{B}(z^*)]$. }
\be
\label{LinkRBNH}
R_X(z)=\Sigma_X[\mc{B}_X(z)],
\ee
where $\mc{B}_X(z)$ is the usual Blue function (\ref{DefBlue}).
We now mention two important properties of the matrices $G_X(Q)$ and $R_X(Q)$. First, $G_X(Q)$ and $R_X(Q)$ obey the following scaling relations \cite{jarosz04, jarosz06}:
\begin{align}
\label{ScalingGNH}
G_{\alpha X}(Q)&=G_{X}\left[\left(\begin{array}{cc}
1/\alpha & 0  \\
0& 1/\alpha^*
\end{array} \right) Q \right] \left(\begin{array}{cc}
1/\alpha & 0  \\
0& 1/\alpha^*
\end{array} \right),
\\
\label{ScalingRNH}
R_{\alpha X}(Q)&=\left(\begin{array}{cc}
\alpha & 0  \\
0& \alpha^*
\end{array} \right) R_{X}\left[Q \left(\begin{array}{cc}
\alpha & 0  \\
0& \alpha^*
\end{array} \right)  \right] ,
\end{align}
where $\alpha \in \mathbb{C}^*$. Second, both $G_X(Q)$ and $R_X(Q)$ can be expressed in terms of the resolvent $g_X(z)$ and the $\mc{R}$-transform $\mc{R}_X(z)$ only \cite{jarosz04, jarosz06}:
\begin{align}
\label{GXQvsg}
G_X(Q) &= \frac{1}{q-q^*}\left\{
\left[
qg_X(q)-q^*g_X(q^*)
\right]I_2 \right.
\nonumber \\
&- \left. \left[
g_X(q)-g_X(q^*)
\right]Q^\dagger
\right\},
\\
\label{RXQvsr}
R_X(Q)&=\frac{1}{q-q^*}\left\{
\left[
q\mc{R}_X(q)-q^*\mc{R}_X(q^*)
\right]I_2
\right.
\nonumber \\
&- \left.
\left[
\mc{R}_X(q)-\mc{R}_X(q^*)
\right]Q^\dagger
\right\},
\end{align}
where $q=x_0+\textrm{i}\vert \vec{x} \vert$ and $q^*$ are two complex conjugated eigenvalues of $Q$.\footnote{The relation (\ref{RXQvsr}) between $R_X(Q)$ and $\mc{R}_X(q)$ holds also between $B_X(Q)$ and $\mc{B}_X(q)$ because $Q^{-1}=Q^\dagger/qq^*$.}

For arbitrary $Q$, we can use algebraic properties of quaternions to show that the following addition law holds \cite{jarosz04, jarosz06}:
\be
\label{SumRNH}
R_{X_1+X_2}(Q)=R_{X_1}(Q)+R_{X_2}(Q),
\ee
where $X_1$ and $X_2$ are two non-Hermitian, asymptotically free random matrices. Therefore, applying successively Eqs.~(\ref{DefBlue}), (\ref{DefRed}), (\ref{RXQvsr}), (\ref{DefRQuat}) and (\ref{DefBQuat}) for $Q=Z_\epsilon$, we can infer $G_{X_1+X_2}(Z_\epsilon)$ from $g_{X_1}(z)$ and $g_{X_2}(z)$. The steps of the algorithm are:
\begin{align}
\label{algosum2}
g_{X_i} &\to \mc{B}_{X_i} \to \mc{R}_{X_i} \to R_{X_i} \to R_{X_1+X_2} \to
\nonumber \\
&\to B_{X_1+X_2} \to G_{X_1+X_2} .
\end{align}
The resolvent $g_{X_1+X_2} (z)$ and the eigenvector correlator $c_{X_1+X_2}(z)$ are finally found from  Eq.~(\ref{LimitResolvent}).

This algorithm is greatly simplified when we look for the eigenvalue distribution of a non-Hermitian matrix $X_1+\textrm{i}X_2$, where $X_1$ and $X_2$ are free Hermitian matrices with known $\mc{R}$-transforms. \citet{jarosz04,jarosz06} showed that the problem reduces to solving a simple system of three equations with three unknown variables, complex $u$, $v$, and real $t$:
\begin{align}
\nonumber
\mc{R}_{X1}(u)&=x+\frac{t-1}{u},
\\
\nonumber
\mc{R}_{X2}(v)&=y-\frac{t}{v},
\\
\vert u\vert&=\vert v\vert,
\label{Systemfree}
\end{align}
where $z=x+\textrm{i}y$. We express $u$ and $v$ via $t$ from the first two equations, substitute the results into the third equation, and then solve for $t$. The resolvent and the correlator are then given by
\begin{align}
\label{gFree}
g_{X_1+\textrm{i}X_2} (z)&=\textrm{Re}\,u-\textrm{i}\textrm{Re}\,v,
\\
\label{cFree}
c_{X_1+\textrm{i}X_2} (z)&=\left(\textrm{Re}\,u\right)^2+\left(\textrm{Re}\,v\right)^2-\vert u\vert^2.
\end{align}
Equation for the borderline $z\in\delta \mc{D}$ of the eigenvalue domain follows from $c_{X_1+\textrm{i}X_2} (z)=0$.\footnote{If we are interested only in the borderline $\delta \mc{D}$, i.e. the boundary between the holomorphic and nonholomorphic domains of $g_{X_1+\textrm{i}X_2} (z)$, it is also possible to use a conformal transformation that maps the cuts $t\in \mathbb{R}$ of $g_{X_1+X_2} (t)$ onto $\delta\mc{D}$. The equation $z=f(t)$ follows from $g_{X_1+\textrm{i}X_2} (z)=g_{X_1+X_2} (t)$ \cite{janik97a,jarosz06}.} From this simplified algorithm it is straightforward to recover the Girko's law for Gaussian Hermitian matrices $X_1$ and $X_2$ [$\mc{R}_1(z)=\mc{R}_2(z)=z$, see Eq.~(\ref{SelfenergyGaussian})]. A less trivial example of application of this algorithm is given in Sec.\ \ref{SecIndepnhERM}.

Finally, a generalization of the concept of $\mc{S}$-transform (\ref{DefSRMT}) to non-Hermitian matrices was presented by \citet{burda11}. These authors derived the multiplication law analogous to Eq.\ (\ref{Sproduct}) for Hermitian matrices, that allows to compute the eigenvalue distribution of a product $X_1 X_2$ of two random non-Hermitian matrices $X_1$ and $X_2$ from the known properties of the latter.

\subsection{Application of the general theory to specific Euclidean random matrix ensembles}
\label{applnonherm}

\subsubsection{Cardinal cosine $+ \rm{i} \times$(cardinal sine) matrix}
\label{SecIndepnhERM}

We start our study of non-Hermitian ERMs by the case of a $N\times N$ matrix
\begin{align}
\label{GERMIndep}
&X_{ij} = f( \mathbf{r}_i- \mathbf{r}_j )
\nonumber \\
&=
(1 - \delta_{ij})\left[
\frac{\cos(k_0 |\mathbf{r}_i - \mathbf{r}_j|)}{
k_0 |\mathbf{r}_i - \mathbf{r}_j|}
+\textrm{i}
 \frac{\sin(k_0 |\mathbf{r}'_i - \mathbf{r}'_j|)}{
k_0 |\mathbf{r}'_i - \mathbf{r}'_j|}
\right],
\end{align}
where $\{\vec{r}_i\}$ and $\{\vec{r}'_i\}$ are two different and independent sets of points.
We recognize in the real and imaginary parts of $X$ the two Hermitian ERMs $C$ and $S$ studied independently in Secs.\ \ref{ERMCosc} and \ref{ERMSinc}, respectively. The matrix $X=C+\textrm{i}[S'-I_N]$ is similar to the three-dimensional free-space Green's matrix $G$ to be studied in Sec.\ \ref{SecEigenvalueGreen}, except that its real and imaginary parts are not correlated. Using the definition (\ref{freeness}) of asymptotic freeness, it is easy to check that the matrices $C$ and $S'$ are asymptotically free, in agreement with the intuitive definition of freeness as statistical independence.

%%%%%%%%%%%%FIG%%%%%%%%%%%%%
\begin{figure*}[t]
\centering{
\includegraphics[width=0.8\textwidth]{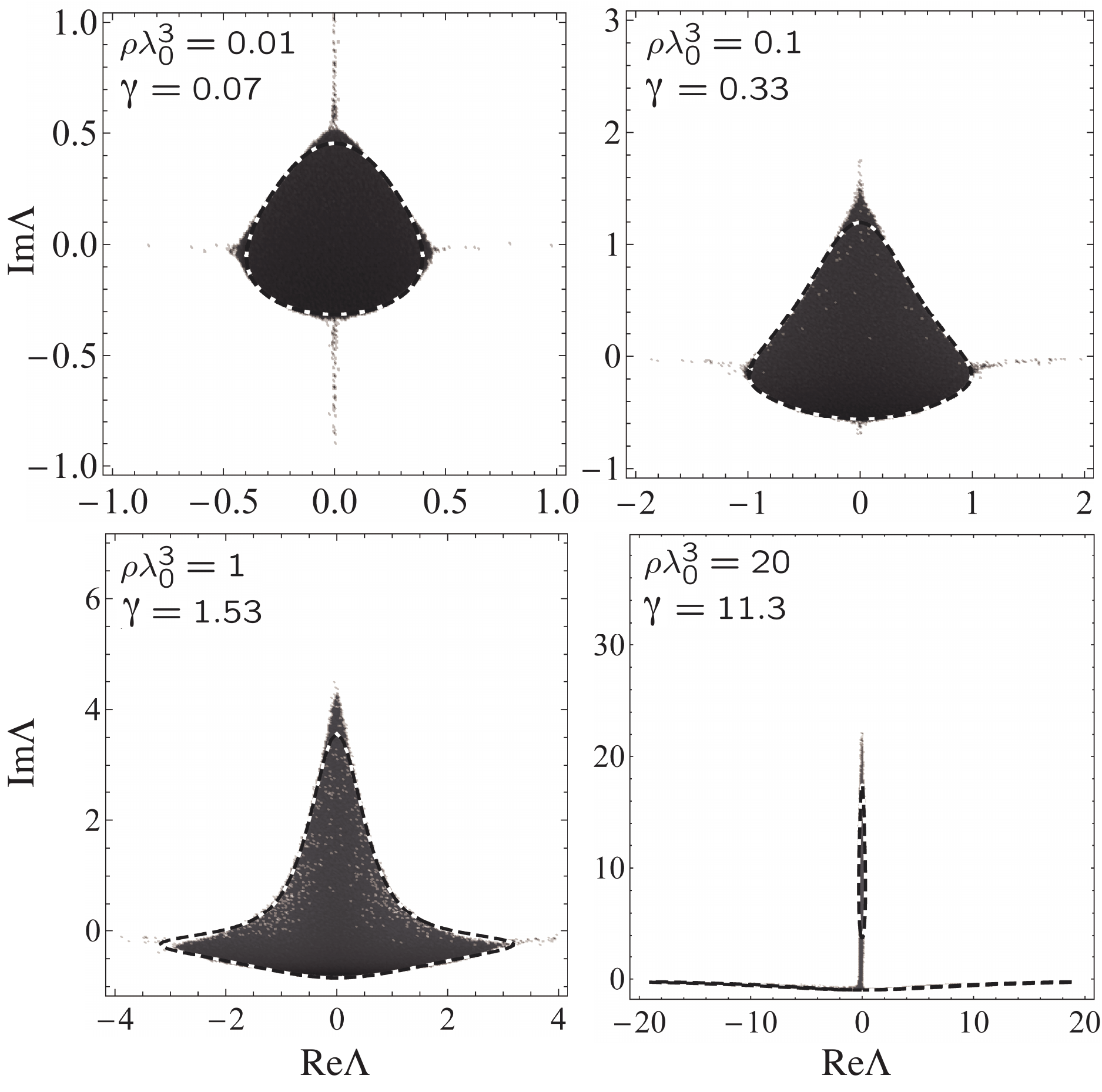}
\caption{\label{figfree}
Density plot of the logarithm of the probability density of eigenvalues $\Lambda_n$ of a square $N \times N$ Euclidean matrix $X$ defined by Eq.\ (\ref{GERMIndep})
at 4 different densities $\rho$ of points $\vec{r}_i$, $\vec{r}_i'$ per wavelength $\lambda_0 = 2\pi/k_0$ cube. $2N = 2 \times 10^4$ points $\vec{r}_i$ and $\vec{r}_i'$ ($i = 1, \dots, N$) are randomly chosen inside a 3D cube of side $L$; $\gamma=2.8N/(k_0L)^2$. The probability distributions are estimated from 10 realizations of $\{\mathbf{r}_i \}$ and $\{\mathbf{r}_i' \}$. Dashed lines show the domain of existence of eigenvalues following from the free probability theory. [Reproduced from \cite{skipetrov11}.]}}
\end{figure*}
%%%%%%%%%%%%%%%%%%%%%%%%%%%%%%

Since $X$ is of the form $X_1+\textrm{i} X_2$, where $X_1 = C$ and $X_2 = S'-I_N$ are two asymptotically free Hermitian matrices, we can make use of Eqs.~(\ref{Systemfree}), (\ref{gFree}) and (\ref{cFree}) to calculate the resolvent $g(z)$ and the eigenvector correlator $c(z)$ of $X$. In the limit of $\gamma \ll 1$, the $\mc{R}$-transforms of $X_1$ and $X_2$ are those of Gaussian and Wishart matrices, respectively: $\mc{R}_{X_1}(z)=\gamma z$ and $\mc{R}_{X_2}(z)=1/(1-\gamma z)$ (see Sec.\ \ref{revisited}). Solving Eqs.~(\ref{Systemfree}), (\ref{gFree}) and (\ref{cFree}), we find:
\begin{align}
\label{resolvantfree}
g(z=x+\textrm{i} y) &= \frac{x}{2\gamma}-
\frac{\textrm{i}}{2}\left[\frac{y}{\gamma(1+y)}+\frac{1}{2+y}\right],
 \\
 \label{corrfree}
c(z=x+\textrm{i} y) &= \left(\frac{x}{2\gamma}\right)^2 + \frac{1}{4}\left[ \frac{y}{\gamma(1+y)} - \frac{1}{2 + y} \right]^2
\nonumber \\
&- \frac{1}{\gamma(1 + y)(2 + y)}.
\end{align}
The correlator (\ref{corrfree}) must vanish on the borderline $\delta\mc{D}$ of the eigenvalue domain. We therefore readily obtain an equation for the borderline on the complex plane:
\be
\label{domainfree}
x^2 + \left( \frac{y}{1 + y} - \frac{\gamma}{2 + y} \right)^2 - \frac{4 \gamma}{(1 + y)(2 + y)} = 0.
\ee
The probability density inside the domain delimited by Eq.\ (\ref{domainfree}) is
\begin{align}
p(x,y) &=
\frac{1}{2\pi}\left[\partial_x\mathrm{Re}\,g(z)-\partial_y\mathrm{Im}\,g(z)\right]
\nonumber
\\
\label{pfree}
&= \frac{1}{4 \pi} \left[
\frac{1}{\gamma} + \frac{1}{\gamma (1+y)^2} - \frac{1}{(2+y)^2} \right].
\end{align}

A better model for the $\mc{R}$-transform of the matrix $X_1=C$ is given by Eq.\ (\ref{RCoscBox1}). If we use this equation instead of $\mc{R}_{X_1}(z) = \gamma z$ above, analytic calculation becomes impossible but we can still compute $g(z)$ and $c(z)$ numerically. The resulting borderline of the eigenvalue domain is shown in Fig.\ \ref{figfree} (dashed lines) together with the eigenvalue distribution of the matrix $X=C+\textrm{i}[S'-I_N]$ found by the numerical diagonalization of a set of $10^4 \times 10^4$ random matrices. At the lowest density considered $\rho \lambda_0^3 = 0.01$, the borderline following from Eq.\ (\ref{RCoscBox1}) is very close to Eq.~(\ref{domainfree}). At higher densities, the former describes numerical results much better than Eq.~(\ref{domainfree}).

Equation (\ref{domainfree}) predicts a splitting of the eigenvalue domain in two parts at $\gamma = 8$. The more accurate calculation using Eq.\ (\ref{RCoscBox1}) leads to a similar prediction (see the lower right panel of Fig.\ \ref{figfree}). However, the eigenvalues of the matrix $X$ do not show such a splitting and form an `inverted T' distribution on the complex plane instead. This is due to the fact that the Marchenko-Pastur law (\ref{pmp}) fails to describe the eigenvalue distribution of the matrix $S'$ at $\gamma > 1$ and hence the $\mc{R}$-transform $1/(1-\gamma z)$ that we assumed for $S'$ is not a good approximation anymore.

\subsubsection{Random Green's matrix}
\label{SecEigenvalueGreen}

Let us now illustrate the power of Eqs.  (\ref{sc1NH}), (\ref{sc2NH}), (\ref{contoura}), and (\ref{contourb}) on the example of the $N \times N$ random Green's matrix
\be
\label{ERMGreen}
G_{ij}=(1-\delta_{ij})\frac{\exp(\textrm{i} k_0| \vec{r}_i-\vec{r}_j|)}{k_0|\vec{r}_i-\vec{r}_j|}.
\ee
This non-Hermitian ERM is of special importance in the context of wave propagation in disordered media because its elements are proportional to the Green's function of Helmholtz equation, with $\vec{r}_i$ that may be thought of as positions of point-like scattering centers. It appeared in works of \citet{skipetrov11,rusek00, pinheiro04,massignan06,ernst69,svidzinsky10,pinheiro06,gremaud10,goetschy11a,goetschy11b}, but was studied only by extensive numerical simulations, except in the paper by \citet{svidzinsky10} where analytic results were obtained in the infinite density limit, and our works \cite{goetschy11a,goetschy11b} where an analytic theory applicable at any density was developed.

Very generally, the eigenvalue density of $G$ depends on two dimensionless parameters: the number of points per wavelength cubed $\rho \lambda_0^3$ and the second moment of $|\Lambda|$ calculated in the limit of low density: $\langle |\Lambda|^2 \rangle = \gamma=\textrm{Tr}(\hat{T}\hat{T}^\dagger)/N$ [Eq.~(\ref{gammaNH})]. Even though the latter result for $\langle |\Lambda|^2 \rangle$ can be rigourously justified only in the limit of low density $\rho \lambda_0^3 \ll 1$ [see Eq.~(\ref{TTNH4})], we checked numerically that it holds approximately up to densities as high as $\rho \lambda_0^3 \sim 100$. Equations (\ref{TTNH4}), (\ref{gammaNH}) and (\ref{Moment2HERM}) show that the second moment of $\vert\Lambda\vert$ is related to the second moments of the eigenvalues of $\textrm{Re}G$ and $\textrm{Im}G$:
\begin{align}
\gamma &= \langle\vert \Lambda_{G} \vert^2 \rangle =
\langle(\textrm{Re}\Lambda_{G})^2 \rangle
+
\langle(\textrm{Im}\Lambda_{G})^2 \rangle
\nonumber
\\
\label{LinkGammaReIm}
&=\frac{1}{2} \langle\Lambda_{\textrm{Re}G}^2\rangle
+
\frac{1}{2} \langle \Lambda_{\textrm{Im}G}^2 \rangle
\\
\label{LinkGammaReIm2}
&= \langle\Lambda_{\textrm{Re}G}^2\rangle
= \langle \Lambda_{\textrm{Im}G}^2\rangle.
\end{align}
Eq.~(\ref{LinkGammaReIm2}) holds for $k_0R\gg1$ and $\rho\lambda_0^3\lesssim 100$, whereas  Eq.~(\ref{LinkGammaReIm}) is also valid for any $k_0R$ and arbitrary non-Hermitian traceless ERM [provided that Eq.~(\ref{TTNH4}) holds]. We will see from the following that the two parameters $\rho \lambda_0^3$ and $\gamma$ control different properties of the eigenvalue density.

\paragraph{Borderline of the eigenvalue domain: approximate solution at low density.}
\label{SecSupportGreen}

%%%%%%%%%%%%%  FIG %%%%%%%%%
\begin{figure*}[t]
\centering{
\includegraphics[angle=0,width=0.8\textwidth]{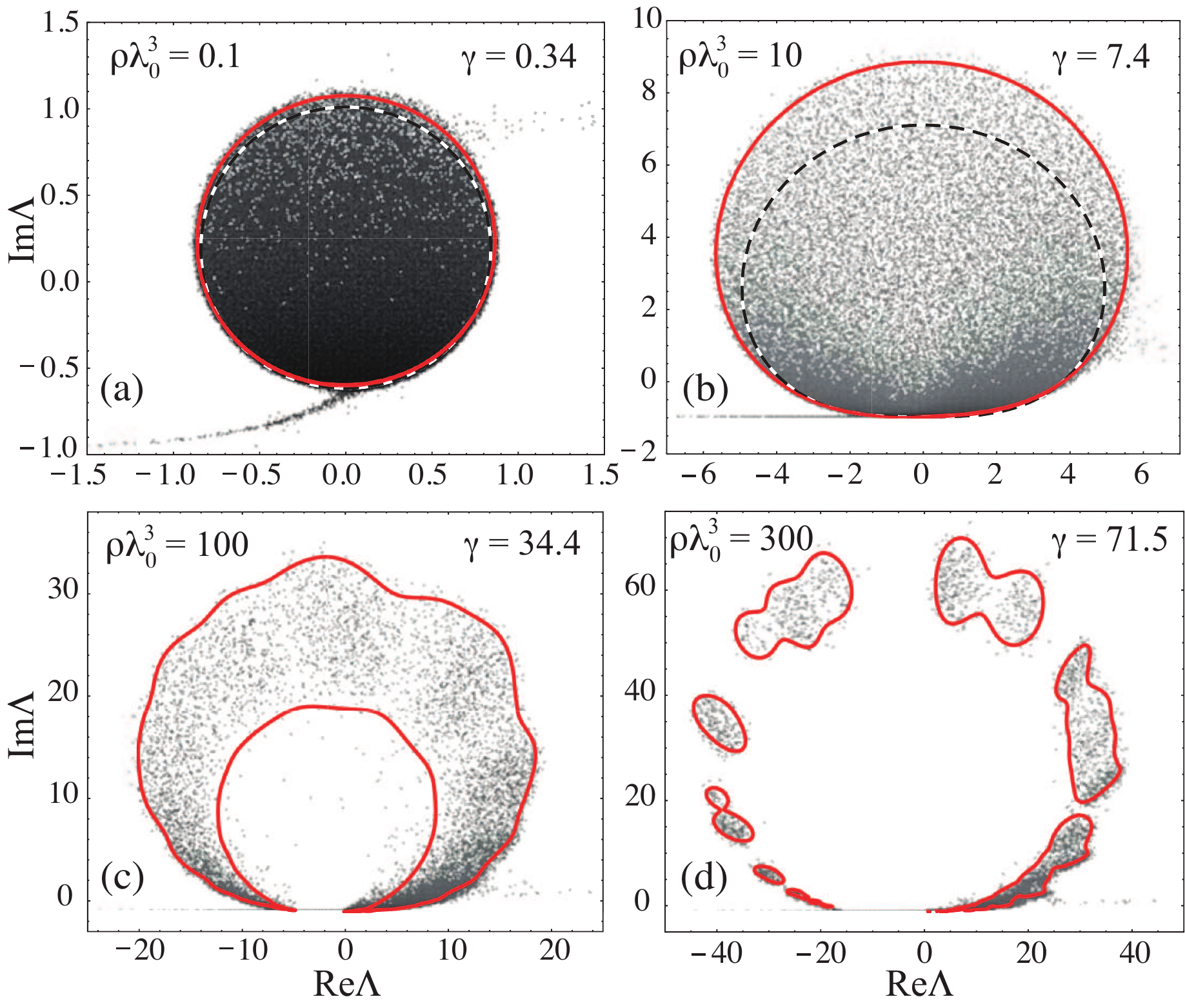}
\caption{
\label{FigGreencontours}
Density plots of the logarithm of eigenvalue density of the $N \times N$ random Green's matrix $G$ obtained by numerical diagonalization of 10 realizations of the matrix for $N = 10^4$. Points $\vec{r}_i$ are randomly chosen inside a sphere of radius $R$. The solid red lines represent the borderlines of the support of eigenvalue density following from Eq.\ (\ref{contour3}) in panels (a) and (b) and from Eqs.\ (\ref{contourT3}) and (\ref{contourT4}) in panels (c) and (d). The dashed lines show the diffusion approximation (\ref{diffusion}). [Reproduced from \cite{goetschy11a}.]}}
\end{figure*}
%%%%%%%%%%%%%%%%%%%%%%%%%%

We first focus on the borderline of the support of eigenvalues which is easier to visualize. We assume that the $N$ points are chosen inside a sphere of radius $R$. For arbitrary $k_0R$, the parameter $\gamma$ is then given by $\gamma =9N/8(k_0R)^2$. Traces appearing in Eqs.\ (\ref{contoura}) and (\ref{contourb}) in the $\ket{\vec{r}}$-representation read
\begin{align}
\mathrm{Tr} \hat{S}_0 &=
\mathrm{Tr} \left( \frac{\hat{T}}{1 - g \hat {T}} \right) =
\mathrm{Tr} \left( \hat{T} + g \hat{T} \hat{S}_0 \right)
\nonumber
\\
\label{contourR1}
&= g\iint_{V} \mathrm{d}^3\vec{r}\; \mathrm{d}^3\vec{r'}\, T(\vec{r},\vec{r'})S_0(\vec{r'},\vec{r}),
\\
\label{contourR2}
\mathrm{Tr} \hat{S}_0 \hat{S}_0^{\dagger}& = \iint_{V}  \mathrm{d}^3\vec{r}\; \mathrm{d}^3\vec{r'} \left|S_0(\vec{r},\vec{r'})\right|^2,
\end{align}
where  $T(\vec{r},\vec{r'})=\rho \bra{ \vec{r}}\hat{A}\ket{ \vec{r'}}= \rho \exp(\textrm{i} k_0 | \vec{r}-\vec{r'}|)/k_0 | \vec{r}-\vec{r'}|$ and in Eq.\ (\ref{contourR1}) we used the fact that  $\textrm{Tr} \hat{T} = \rho \textrm{Tr} \hat{A} = 0$, as follows from Eq.\ (\ref{ERMGreen}). $S_0(\vec{r},\vec{r'})=\bra{ \vec{r}}\hat{S}_0\ket{ \vec{r'}}$ obeys
\be
\label{Sinspace}
S_0(\vec{r},\vec{r'})=T(\vec{r},\vec{r'})+g\int_{V} \mathrm{d}^3\vec{r''} T(\vec{r},\vec{r''})S_0(\vec{r''},\vec{r'}),
\ee
as follows from the definition of $\hat{S}_0$.
Noting that
\be
\label{THelmholtz}
\left(\Delta_{\vec{r}}+k_0^2+\textrm{i}\epsilon\right)T(\vec{r},\vec{r'})
=-\frac{4\pi\rho}{k_0} \delta^{(3)}(\vec{r}-\vec{r}'),
\ee
where $\epsilon \to 0^+$, we apply the operator $\Delta_{\vec{r}}+k_0^2+\textrm{i}\epsilon$ to Eq.\ (\ref{Sinspace}) and obtain
\begin{align}
\label{SHemholtz}
\Delta_{\vec{r}}S_0(\vec{r},\vec{r'}) &+
k_0^2\left[1+g\frac{\rho\lambda_0^3}{2\pi^2}\Pi_V(\vec{r})+
\textrm{i}\epsilon\right]S_0(\vec{r},\vec{r'})
\nonumber \\
&= -\frac{4\pi\rho}{k_0}\delta^{(3)}(\vec{r}-\vec{r}'),
\end{align}
where $\Pi_V(\vec{r})=1$ for $\vec{r} \in V$ and $0$ elsewhere. In the limit of low density $\rho\lambda_0^3 \to 0$, an approximate solution of this equation is obtained by neglecting `reflections' of the `wave' $S_0(\vec{r},\vec{r'})$ on the boundaries of the volume $V$ and thus setting $\Pi_V(\vec{r}) = 1$ everywhere. This yields
\begin{align}
S_0(\vec{r},\vec{r'})&\simeq \rho \frac{\exp\left[\textrm{i} \kappa(g) | \vec{r}-\vec{r'}|\right]}{k_0 | \vec{r}-\vec{r'}|},
\\
\label{DefKappag}
\kappa(g)&=k_0\sqrt{1+\frac{g\rho\lambda_0^3}{2\pi^2}}.
\end{align}

We now plug the explicit expressions for $T(\vec{r},\vec{r'})$ and $S_0(\vec{r},\vec{r'})$ into Eqs.\ (\ref{contourR1}) and (\ref{contourR2}). This yields
\begin{align}
\label{contourR3}
&\mathrm{Tr} \hat{S}_0 = 2\gamma N g h[-\textrm{i}\kappa(g)R-\textrm{i}k_0R],&
\\
\label{contourR4}
&\mathrm{Tr} \hat{S}_0 \hat{S}_0^{\dagger} = 2\gamma N h[2\textrm{Im} \kappa(g)R],
\end{align}
with
\begin{align}
\label{defh}
h(x) &= \frac{1}{6x^4}\left[3-6x^2+8x^3-3(1+2x)e^{-2x}\right].
\end{align}
In the low-density limit, $g$ can be eliminated from Eqs.\ (\ref{contoura}) and (\ref{contourb}) by neglecting $\mathrm{Tr} \hat{S}_0/N$ in Eq.\ (\ref{contoura}) and substituting $g = 1/z$ into Eq.\ (\ref{contourR4}). This gives
\be
\label{contour2}
|\Lambda|^2 = 2 \gamma h \left[
2\textrm{Im} \kappa\left( 1/\Lambda \right) R
\right].
\ee
If the argument of the function $h$ in Eq.\ (\ref{contour2}) is expanded in series in $\rho \lambda_0^3$, Eq.\ (\ref{contour2}) becomes:
\be
\label{contour3}
|\Lambda|^2 \simeq 2 \gamma h \left( -8\gamma \frac{\textrm{Im}\Lambda}{3|\Lambda|^2} \right).
\ee
By comparing Eq.\ (\ref{contour3}) with the exact solution (see Sec.\ \ref{SecExactSolNH} and Fig.\ \ref{FigGreencontours}), we conclude that it is valid up to densities as high as $\rho \lambda_0^3 \simeq 10$.

For $\gamma \ll 1$, the density of eigenvalues is roughly uniform within a circular domain of radius $\sqrt{2\gamma}$, see Fig.\ \ref{FigGreencontours}(a). The domain grows in size and shifts up upon increasing $\gamma$. At $\gamma \gtrsim 1$ it starts to `feel' the `wall' $\textrm{Im} \Lambda = -1$ and deforms [Fig.\ \ref{FigGreencontours}(b)]. Before considering the shape of the eigenvalue domain at higher densities, we would like to show how the link with scattering theory can be used to derive an equation for its borderline.

\paragraph{Borderline of the eigenvalue domain from the scattering theory.}
\label{SecMapST}

Because the elements $G_{ij}$ of the Green's matrix $G$ describe propagation of a scalar wave between points $\vec{r}_i$ and $\vec{r}_j$ in space, it is quite natural that a link exists between some of the properties of the eigenvalue density of $G$ and the theory of wave scattering in an ensemble of point scatterers. In particular, if we define $I_{ij}$ as the intensity of a wave at $\vec{r}_j$ due to a source at $\vec{r}_i$, $I(t) = \sum_{i \neq  j} I_{ij}$ can be written as \cite{goetschy11a}
\be
I(t) = \textrm{Tr} \frac{1}{[t-\mathcal{G}^{-1}][t-\mathcal{G}^{-1}]^{\dagger}},
\ee
where $t$ is the scattering matrix of an individual scatterer and $\mathcal{G} = -k_0G/4\pi$. This is to be compared with the expression for the correlator of right and left eigenvectors of an arbitrary matrix $A$, $c(z)=\lim_{\epsilon \to 0^+}G_{12}^\epsilon$, following from Eq.\ (\ref{G12explicit}):
\be
\label{corr2}
c(z) = -\lim_{\epsilon \to 0^+} \frac{\mathrm{i} \epsilon}{N}
 \left\langle \textrm{Tr} \frac{1}{(z-A)(z-A)^{\dagger}+\epsilon^2} \right\rangle.
\ee
For $A = \mathcal{G}^{-1}$ and $z = t$ we thus have
\be
\label{ci}
c(t) = -\lim_{\epsilon \to 0^+} \frac{\mathrm{i} \epsilon}{N}
\langle I(t) \rangle.
\ee
This should become different from zero when $t$ enters the support of the eigenvalue density of $\mathcal{G}^{-1}$ or, equivalently, when $1/t$ enters the support of the eigenvalue density of $\mathcal{G}$. The only way to obtain $c(t) \neq 0$ for $\epsilon \to 0^+$ is to make $\moy{I(t)}$ diverge. In the framework of the linear model of scattering, this can be achieved by realizing a random laser \cite{goetschy11b}. We thus come to the conclusion that finding the borderline of the support of the eigenvalue density $p(\Lambda)$ of the $N \times N$ Green's matrix (\ref{ERMGreen}) is mathematically equivalent to calculating the random lasing threshold in an ensemble of $N$ identical point-like scatterers with $t = -4\pi/k_0 \Lambda$. Application of this link to the study of random lasers will be discussed in Sec.\ \ref{laser}. In the diffusion approximation, for example, the threshold of such a random laser was found by \citet{froufeperez09} and leads to the following equation for the borderline:
\be
\label{diffusion}
|\Lambda|^2 = \frac{8\gamma}{\sqrt{3}\pi} \sqrt{1+\textrm{Im} \Lambda}
\left(1 + \frac{|\Lambda|^2}{|\Lambda|^2+4 \gamma} \right).
\ee
We show this equation in Figs.\ \ref{FigGreencontours}(a) and (b) by dashed lines. It gives satisfactory results only in the weak scattering regime $\rho \lambda_0^3 \lesssim 10$. However, it should be noted that in this regime, it describes the lower part of the borderline, corresponding to small $\mathrm{Im} \Lambda$, better than Eq.\ (\ref{contour3}).

\paragraph{Borderline of the eigenvalue domain: exact solution of self-consistent equations.}
\label{SecExactSolNH}

The approximate equation (\ref{contour2}) for the borderline of the support of eigenvalue density yields a closed line on the complex plane until $\rho\lambda_0^3 \simeq 30$, after which the line opens from below. This opening is reminiscent of the gap predicted by the analytic theory  for the eigenvalue distribution of the matrix $C = \textrm{Re}G$ in Sec.\ \ref{ERMCosc}.  This signals that an important change in behavior might be expected at this density. And indeed, we observe that a `hole' opens in the eigenvalue density for $\rho\lambda_0^3 \gtrsim 30$. As we see in Fig.\ \ref{FigGreencontours}(c), this hole is perfectly described by Eqs.\ (\ref{contoura}) and (\ref{contourb}) which we now solve in the biorthogonal basis of right $\ket{\mathcal{R}_{\alpha}}$ and left $\ket{\mathcal{L}_{\alpha}}$ eigenvectors of the operator $\hat{T}$. These eigenvectors obey $\hat{T}\ket{\mathcal{R}_{\alpha}}=\mu_{\alpha}\ket{\mathcal{R}_{\alpha}}$ and $\hat{T}^\dagger\ket{\mathcal{L}_{\alpha}}=\mu_{\alpha}^*\ket{\mathcal{L}_{\alpha}}$. In this basis, Eqs.\ (\ref{contoura}) and (\ref{contourb}) read
\begin{align}
&z = \frac{1}{g} + \frac{g}{N} \sum_{\alpha}\frac{\mu_\alpha^2}{1-g\mu_\alpha},&
\label{contourT1}
\\
&\frac{1}{|g|^2} = \frac{1}{N} \sum_{\alpha,\beta}
\frac{\mu_\alpha \mu_\beta^*
\scp{\mathcal{L}_\alpha}{\mathcal{L}_\beta}
\scp{\mathcal{R}_\beta}{\mathcal{R}_\alpha}
}
{(1-g\mu_\alpha)(1-g\mu_\beta)^*},
\label{contourT2}
\end{align}
where we made use of the fact that $\mathrm{Tr} \hat{T} = 0$ and therefore $\mathrm{Tr} \hat{S}_0 = g \mathrm{Tr} \hat{T} \hat{S}_0$ [see Eq.~(\ref{contourR1})].
The problem essentially reduces to solving the eigenvalue equation
\be
\label{rightreal}
\rho\int_V \mathrm{d}^3 \vec{r'} \frac{\exp(\textrm{i} k_0 | \vec{r}-\vec{r'}|)}{k_0 | \vec{r}-\vec{r'}|}\mathcal{R}_{\alpha}(\vec{r'})=
\mu_\alpha\mathcal{R}_{\alpha}(\vec{r}),
\ee
where $\vec{r} \in V$. As follows from Eq.\ (\ref{THelmholtz}), $\mathcal{R}_{\alpha}(\vec{r})$ is also an eigenvector of the Laplacian, $\Delta_{\vec{r}}\mathcal{R}_{\alpha}(\vec{r})=-\kappa_{\alpha}^2\mathcal{R}_{\alpha}(\vec{r})$, with $\kappa_{\alpha}=\kappa(1/\mu_\alpha)$. In a sphere of radius $R$, using the decomposition of the kernel of Eq.\ (\ref{rightreal}) in spherical harmonics \cite{gradshteyn80}:
\begin{align}
\label{DecompoExp}
&\frac{\exp(\textrm{i}k_0 |\mathbf{r} - \mathbf{r}'|)}{
k_0 |\mathbf{r} - \mathbf{r}'|} = 4\textrm{i}\pi \sum_{l=0}^\infty\sum_{m=-l}^l
j_l\left[k_0\textrm{min}(r,r')\right]
\nonumber \\
&\hspace{5mm}\times h^{(1)}_l\left[k_0\textrm{max}(r,r')\right]Y_{lm}(\theta,\phi)
Y_{lm}(\theta',\phi')^*,
\end{align}
 it is quite easy to find that \cite{svidzinsky10}
\be
\mathcal{R}_{\alpha}(\vec{r}) = \mathcal{R}_{lmp}(\vec{r})=\mathcal{A}_{lp}j_l(\kappa_{lp}r)Y_{lm}(\theta,\phi),
\ee
where $\theta$ and $\phi$ are the polar and azimuthal angles of the vector $\vec{r}$, respectively, $j_l$ are spherical Bessel functions of the first kind,  $h^{(1)}_l$ are spherical Hankel functions, $Y_{lm}$ are spherical harmonics, $\mathcal{A}_{lp}$ are normalization coefficients, and $\alpha=\{ l,m,p \}$. Furthermore, coefficients $\kappa_{lp}$ obey \cite{svidzinsky10}
\be
\label{modesk}
\frac{\kappa_{lp}}{k_0}=\frac{j_l(\kappa_{lp}R)}{j_{l-1}(\kappa_{lp}R)}\frac{h^{(1)}_{l-1}(k_0R)}{h^{(1)}_l(k_0R)}.
\ee
Integer $p$ labels the different solutions of this equation for a given $l$. Hence, the eigenvalues
\be
\label{mulpGreen}
\mu_{lp}=\frac{\rho\lambda_0^3}{2\pi^2}\frac{1}{(\kappa_{lp}/k_0)^2-1}
\ee
are $(2l+1)$-times degenerate ($m\in [-l,l]$).

In the limit $k_0R \to \infty$, for $l\ll k_0R$ and $l\ll \kappa_{lp}R$, we can use asymptotic expressions for the spherical functions in Eq.\ (\ref{modesk}) to obtain
\be
\frac{\mathrm{i}}{2}\textrm{ln}\left(\frac{\kappa_{lp}+k_0}{\kappa_{lp}-k_0}\right)=-\kappa_{lp}R+\left(\frac{l}{2}+p\right)\pi.
\ee
In this limit, the eigenvalues $\mu_{lp}$ are therefore localized in the vicinity of a roughly circular line\footnote{An equation of a circle can be found from Eq.~(\ref{largerho}) by expanding $\kappa(1/\mu)$ in series in $1/\rho\lambda_0^3$. The resulting equation is $(x+\rho\lambda_0^3/8\pi^2)^2+(y-{\cal R}/2+1)^2 = {\cal R}^2$ with ${\cal R} = 4\gamma/3W(4k_0R)$, $\mu=x+\textrm{i}y$, and $W(t)$ the Lambert function (a function inverse of $f(W)=We^W$); $W(t)\simeq \textrm{ln} t$ for $\vert t \vert\gg1$. Hence, ${\cal R} \simeq 4\gamma/3\textrm{ln}(4k_0R)$ for $k_0R\gg1$.} in the complex plane given by
\be
\label{largerho}
\left|\frac{\kappa(1/\mu)-k_0}{\kappa(1/\mu)+k_0} \right|^2  \left|e^{4 \mathrm{i} \kappa(1/\mu)R} \right|=1.
\ee

Let us now study the eigenvectors. Using standard properties of spherical harmonics and spherical Bessel functions \cite{morse53}, we can show that
\begin{align}
&\scp{\mathcal{R}^*_{lmp}}{\mathcal{R}_{l'm'p'}}
= (-1)^m\mathcal{A}_{lp}^2\frac{R^3}{2}\Big[j_l(\kappa_{lp}R)^2
\nonumber \\
&\hspace{5mm}- j_{l-1}(\kappa_{lp}R)j_{l+1}(\kappa_{lp}R)\Big]
\delta_{l,l'}\delta_{m,-m'}\delta_{p,p'}.
\end{align}
From the normalization condition $\scp{\mathcal{L}_{lmp}}{\mathcal{R}_{l'm'p'}}
=\delta_{l,l'}\delta_{m,m'}\delta_{p,p'}$, we find that $\mathcal{L}_{lmp}(\vec{r})=(-1)^m\mathcal{R}_{l(-m)p}(\vec{r})^*$ and
\be
\mathcal{A}_{lp}=\sqrt{\frac{2}{R^3}}\frac{1}{\sqrt{j_l(\kappa_{lp}R)^2-j_{l-1}(\kappa_{lp}R)j_{l+1}(\kappa_{lp}R)}}.
\ee
On the other hand, we also have
\begin{align}
&\scp{\mathcal{R}_{lmp}}{\mathcal{R}_{l'm'p'}} =
\frac{R^2\mathcal{A}^*_{lp}\mathcal{A}_{lp'}}{\kappa_{lp'}^2-
\kappa_{lp}^{*2}}\Big[\kappa_{lp}^* j_{l-1}(\kappa_{lp}^*R) j_{l}(\kappa_{lp'}R)
\nonumber \\
&\hspace{5mm}- \kappa_{lp'} j_{l-1}(\kappa_{lp'}R) j_{l}(\kappa_{lp}^*R)\Big]\delta_{l,l'}\delta_{m,m'},
\end{align}
and $\scp{\mathcal{L}_{lmp}}{\mathcal{L}_{l'm'p'}}=\scp{\mathcal{R}_{lmp}}{\mathcal{R}_{lmp'}}\delta_{l,l'}\delta_{m,m'}$. It is now convenient to introduce a new coefficient
\begin{align}
\label{coeffC}
\footnotesize{C_{lpp'}} &=
4
\bigg[\kappa_{lp}^*R j_{l-1}(\kappa_{lp}^*R) j_{l}(\kappa_{lp'}R)
\nonumber \\
&-\kappa_{lp'}R j_{l-1}(\kappa_{lp'}R) j_{l}(\kappa_{lp}^*R)\bigg]^2
\bigg[\kappa_{lp'}^2R^2-\kappa_{lp}^{*2}R^2\bigg]^{-2}
\nonumber \\
&\times \bigg[j_l(\kappa^*_{lp}R)^2-j_{l-1}(\kappa^*_{lp}R)
j_{l+1}(\kappa^*_{lp}R)\bigg]^{-1}
\nonumber \\
&\times \bigg[j_l(\kappa_{lp'}R)^2-j_{l-1}(\kappa_{lp'}R)
j_{l+1}(\kappa_{lp'}R)\bigg]^{-1},
\end{align}
in terms of which Eqs.~(\ref{contourT1}) and (\ref{contourT2}) become
\begin{align}
&z =\frac{1}{g}+\frac{g}{N} \sum_l \sum_p\frac{(2l+1)\mu_{lp}^2}{1-g\mu_{lp}},&
\label{contourT3}
\\
&\frac{1}{|g|^2} = \frac{1}{N}\sum_l \sum_p\sum_{p'}
\frac{(2l+1)\mu_{lp'}\mu_{lp}^*C_{lpp'}}
{(1-g\mu_{lp'})(1-g\mu_{lp})^*}.
\label{contourT4}
\end{align}

To find the borderline of the support of eigenvalue density of the matrix (\ref{ERMGreen}) shown in Figs.\ \ref{FigGreencontours}(c) and  \ref{FigGreencontours}(d), we apply the following recipe. (1) Find solutions $\kappa_{lp}$ of Eq.\ (\ref{modesk}) numerically and then compute the corresponding $\mu_{lp}$.  (2) Compute the coefficients $C_{lpp'}$ using Eq.\ (\ref{coeffC}). (3) Find lines on the complex plane $1/g$ defined by Eq.\ (\ref{contourT4}). (4) Transform the lines on the complex plane $1/g$ into contours on the complex plane $z$ using Eq.\ (\ref{contourT3}). The latter contours are the borderlines of the support of eigenvalue density $p(\Lambda)$.

%%%%%%%%%%%%%  FIG %%%%%%%%%
\begin{figure}[t]
\centering{
\includegraphics[angle=0,width=0.9\columnwidth]{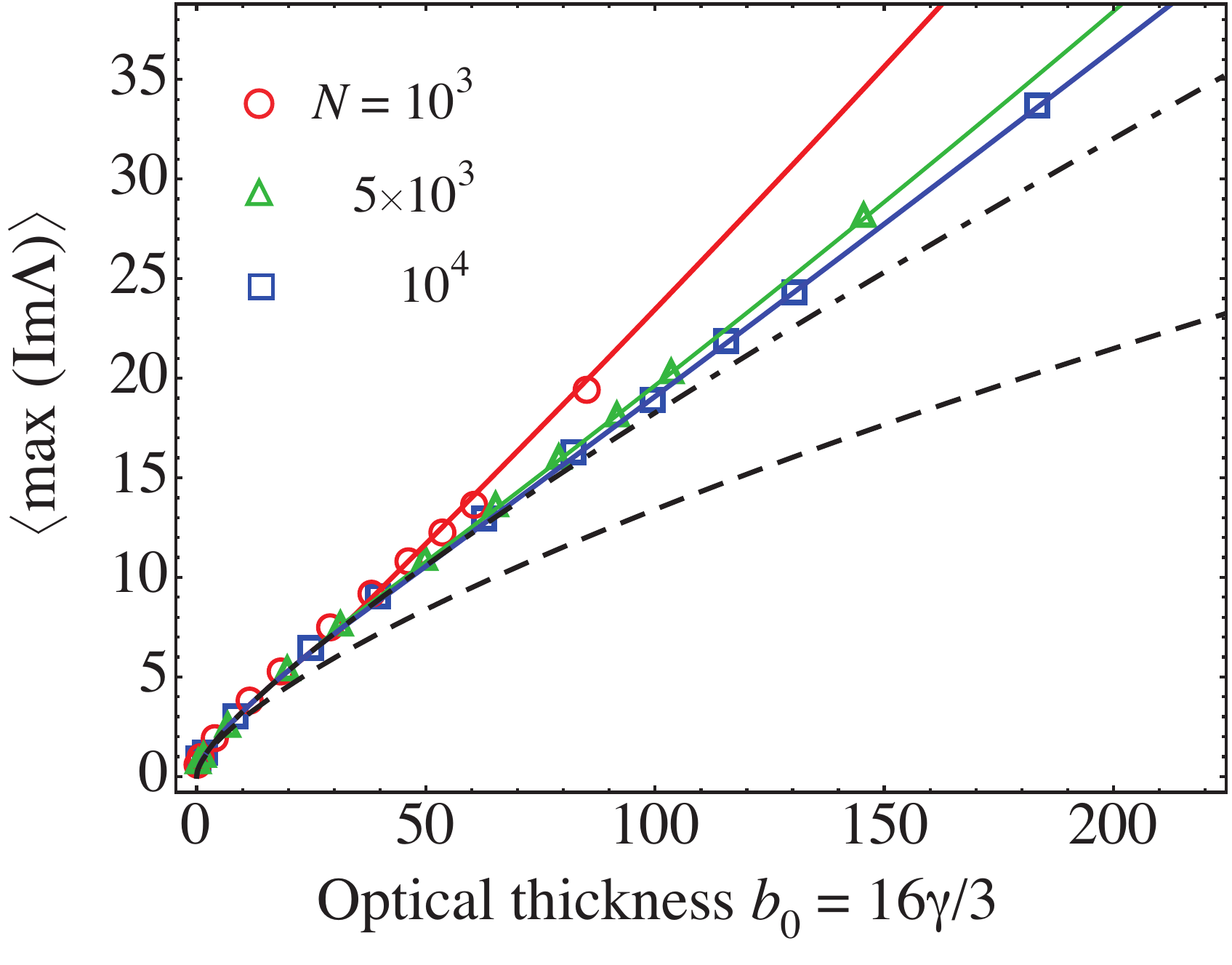}
\caption{
\label{MeanMaxIm}
Mean maximum value of the imaginary part of eigenvalues $\Lambda$ of the $N \times N$ random Green's matrix $G$. Analytic results (solid lines) following from Eqs.~(\ref{contourT3}) and (\ref{contourT4}) are compared with the results of numerical diagonalization for three different matrix sizes $N$ (symbols). Analytic results depend both on $\gamma$ and $\rho\lambda_0^3$, except for $\rho\lambda_0^3\lesssim 10$ when they reduce to Eq.~(\ref{contour3}) (dot-dashed line). The dashed line represents the prediction of the diffusion approximation (\ref{diffusion}). The horizontal axis is the on-resonance optical thickness of a cloud of cold atoms, relevant for the problem of random lasing discussed in Sec.\ \ref{laser}.
}}
\end{figure}
%%%%%%%%%%%%%%%%%%%%%%%%%%

At high density, the crown formed by the eigenvalues blows up in spots centered around $\mu_\alpha-\mathrm{i}$, where $\mu_\alpha$ are the eigenvalues of $\hat{T}$, as we show in Fig.\ \ref{FigGreencontours}(d). When the density is further increased, the clouds of eigenvalues of $A$ turn clockwise along the circular line given by Eq.\ (\ref{largerho}) and shrink in size. The eigenvalues $\Lambda$ eventually become equal to $\mu_\alpha-\mathrm{i}$. They then fall on the circular line (\ref{largerho}) and the problem looses its statistical nature. As follows from our analysis, the parameter $\gamma$ controls the overall extent of the support of eigenvalue density $\mathcal{D}$ on the complex plane, whereas its structure depends also on the density $\rho\lambda_0^3$. At fixed $\gamma$, $\mathcal{D}$ goes through a transition from a disk-like to an annulus-like shape, and eventually splits into multiple disconnected spots upon increasing $\rho\lambda_0^3$.
The transition from disk-like to the annulus-like shape is reminiscent of the disk-annulus transition in the eigenvalue distribution of rotationally invariant non-Hermitian random matrix ensembles \cite{feinberg97a} (see the discussion at the end of Sec.\ \ref{SecDysonNH}).

Quite remarkably, Eqs.\ (\ref{contourT3}) and (\ref{contourT4}) properly capture the transition to the continuous medium regime (high density) and to the small sample regime (small $k_0R$). To illustrate this point, we calculated $\moy{\textrm{max}(\textrm{Im}\Lambda)}$ from Eqs.~(\ref{contourT3}) and (\ref{contourT4}), and found excellent agreement with numerical results at all values of parameters, including high densities $\rho\lambda_0^3$, see Fig.\ \ref{MeanMaxIm}. In contrast, the analysis of the lower part of the spectrum in Fig.\ \ref{FigGreencontours} shows that the theory fails to describe it with sufficient accuracy, especially at high densities $\rho\lambda_0^3 \gtrsim 10$.

\paragraph{Super- and subradiant states.}
\label{SecBranches}

An important additional feature of the numerical results in Fig.\ \ref{FigGreencontours} that is not described by Eqs.\ (\ref{contoura}) and (\ref{contourb}) is the eigenvalues that concentrate around the two hyperbolic spirals, $|\Lambda|=1/\arg{\Lambda}$ and its reflection through the origin. These spirals correspond to the two eigenvalues $\pm G_{12}$ of the matrix (\ref{ERMGreen}) for $N = 2$. The eigenvectors corresponding to these eigenvalues are localized on pairs of very close points $\vert \vec{r}_i -\vec{r}_j\vert \ll \lambda_0$. They are analogous to super- and subradiant states of a pair of atoms \cite{gross82} and correspond to the so-called proximity resonances \cite{heller96,rusek002}: the superradiant state corresponds to $\mathrm{Re} \Lambda$, $\mathrm{Im} \Lambda > 0$, whereas the subradiant---to  $\mathrm{Re} \Lambda$, $\mathrm{Im} \Lambda < 0$. Numerical results show that in the limit of $\rho\lambda_0^3 \to \infty$, the lower branch is much more populated than the upper one.

A rough model that partially mimics this  behavior is given by the $N\times N$ matrix:
\be
\tilde{G} = G_{12}\begin{pmatrix}
0&  1& \dots & 1  \\
1& \ddots& \ddots &\vdots\\
\vdots & \ddots & \ddots &1 \\
1 & \dots & 1 & 0
\end{pmatrix},
\ee
where $G_{12}=e^{\textrm{i} k_0\vert \vec{r}_1 -\vec{r}_2\vert}/k_0\vert\vec{r}_1 -\vec{r}_2\vert$, and $\vec{r}_1$ and $\vec{r}_2$ are randomly chosen points inside the sphere of radius $R$. This matrix has two different eigenvalues: the non-degenerate eigenvalue $\Lambda=(N-1)G_{12}$ corresponds to the superradiant state $(1, \dots, 1)/\sqrt{N}$; the $(N-1)$-degenerate eigenvalues $\Lambda=-G_{12}$ correspond to subradiant states localized on pairs of points $(1, 0, \dots, 0, -1, 0, \dots, 0)/\sqrt{2}$. In the limit $N\to \infty$, only subradiant states contribute significantly to the eigenvalue distribution of $\tilde{G}$. Using the definition (\ref{DefpNH}) we can show that the latter is then given by:
\be
\label{densityLowerBrach}
p(\Lambda)=\frac{3}{(k_0R)^3}\frac{1}{\vert\Lambda\vert^2}\,s\left(\frac{1}{2k_0R\vert\Lambda \vert}\right)\delta\left(\textrm{arg}\Lambda+\frac{1}{\vert \Lambda \vert}\right),
\ee
where $s(x) = 1 - 3 x/2 + x^3/2$. Loosely speaking, the true eigenvalue distribution of the Green's matrix $G$ is a superposition of Eqs.~(\ref{sc1NH}), (\ref{sc2NH}) and (\ref{densityLowerBrach}). With the qualitative picture of the Dyson gas in mind, we could say that the lower `branch' $|\Lambda|=-1/\arg{\Lambda}$ plays the role of a channel for the gas of eigenvalues, through which the latter can escape from the eigenvalue domain predicted by Eqs.~(\ref{contoura}) and (\ref{contourb}). This effect is more pronounced at high density because the eigenvalues accumulate near the line $\textrm{Im}\Lambda=-1$, so that the repulsive interaction between eigenvalues forces the latter to flow into the lower branch. From numerical results for $N \leq 10^4$, we estimate the statistical weight of subradiant states to be important at large densities, of the order of $1 - \mathrm{const}/(\rho \lambda_0^3)^p$ with $p \sim 1$ \cite{goetschy11a}. This is consistent with the estimation of the number of subradiant states in a large atomic cloud by \citet{ernst69}. As explained earlier, the lack of the spiral branches corresponding to super- and subradiant states in the theory can be traced back to the assumption of statistical independence of the elements of the matrix $H$ in the representation $A=HTH^\dagger$ [see Eq.\ (\ref{AHTH})].

An important implication of the existence of the hyperbolic spiral branches in the eigenvalue distribution of the Green's matrix $G$ is that, at least at low density, quantities such as $\moy{\textrm{min}( \textrm{Re} \Lambda)}$ or $\moy{\textrm{min}( \textrm{Im} \Lambda)}$, that are $\textit{a priori}$ difficult to calculate, can be readily found from $2$-body interactions \cite{skipetrov11,goetschy11c}.

\paragraph{Density of eigenvalues.}
\label{SecDensitygreen}

\begin{figure*}[t]
\centering{
\includegraphics[angle=0,width=0.8\textwidth]{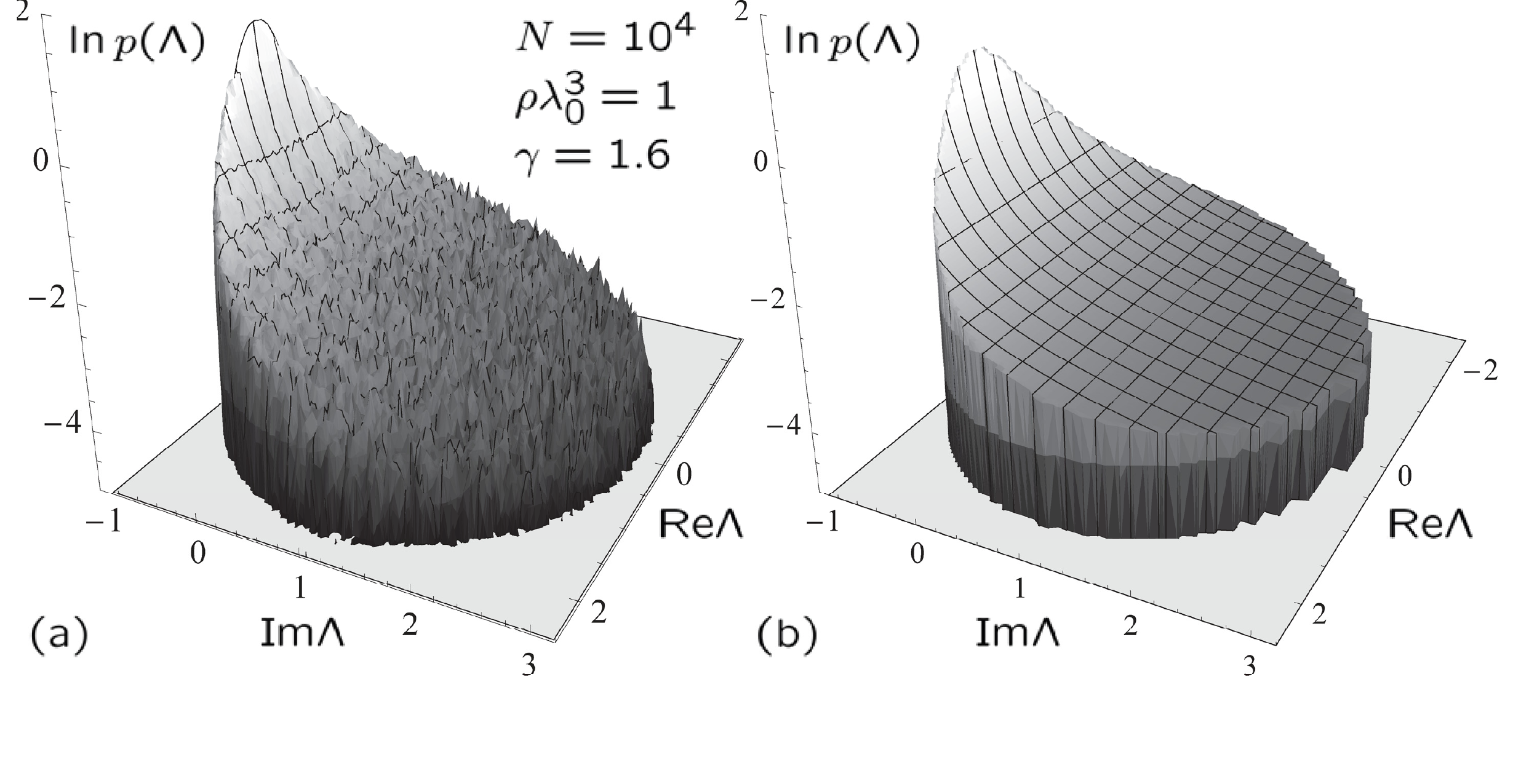}
\vspace{-1cm}
\caption{
\label{plotGreenFull}
Logarithm of the eigenvalue density of the $N \times N$ random Green's matrix $G$. Numerical results obtained by diagonalizing 20 realizations of the matrix for $N = 10^4$ (a) are compared with the solution of Eq.\ (\ref{gLowDvsS02}) (b). Points $\vec{r}_i$ are chosen randomly inside a sphere of radius $R$; $\gamma=9N/8(k_0R)^2$.
}}
\end{figure*}

Equations (\ref{sc1NHv2}) and (\ref{sc2NHv2}) allow us to analyze not only the borderline of the eigenvalue domain, but also the eigenvalue density $p(\Lambda)$ inside it. Very generally, $p(\Lambda)$ is roughly symmetric with respect to the line $\textrm{Re}\Lambda =0$ and decays with $\textrm{Im}\Lambda$. In the regime of low densities $\rho\lambda_0^3 \lesssim 1$, an approximation of Eqs.~(\ref{sc1NHv2}) and (\ref{sc2NHv2}) can be obtained by replacing the operator $\hat{S}_1$ by $\hat{S}_0$. This amounts to neglecting the term $c^2\hat{T}\hat{T}^\dagger$ in the denominator of Eq.~(\ref{DefOpS1}). Then, Eqs.~(\ref{sc1NHv2}) and (\ref{sc2NHv2}) reduce to two equations in which the resolvent $g(z)$ and the eigenvector correlator $c(z)$ are decoupled:
\begin{align}
\label{gLowDvsS0}
g(z)& = \frac{z^* -\frac{1}{N} \mathrm{Tr} \hat{S}_0^{\dagger}}{\frac{1}{N} \mathrm{Tr} \hat{S}_0 \hat{S}_0^{\dagger}},
\\
\label{cLowDvsS0}
c(z)^2&=\vert g(z) \vert^2-\frac{N}{ \mathrm{Tr} \hat{S}_0\hat{S}_0^{\dagger}}.
\end{align}
Assuming explicitly that the $N$ points are distributed in a sphere of radius $R$, we can make use of the results of Sec.\ \ref{SecSupportGreen} to compute traces in these equations, so that Eqs.~(\ref{gLowDvsS0}) and (\ref{cLowDvsS0}) become
\begin{eqnarray}
\label{gLowDvsS02}
g(z) &=& \frac{z^* - 2\gamma  g(z)^* h\left(\textrm{i}\kappa[g(z)]^*R+\textrm{i}k_0R\right)}{2\gamma  h\left(2\,\textrm{Im} \kappa [g(z)] R\right)},
\\
\label{cLowDvsS02}
c(z)^2 &=& \vert g(z) \vert^2-\frac{1}{2\gamma  h\left(2\,\textrm{Im} \kappa [g(z)] R\right)},
\end{eqnarray}
where the functions $\kappa(g)$  and $h(x)$ are defined by Eqs.~(\ref{DefKappag}) and (\ref{defh}), respectively. We find the resolvent $g(z)$ by solving Eq.~(\ref{gLowDvsS02}) numerically and then evaluate the eigenvalue density $p(\Lambda)$ with the help of Eq.~(\ref{LinkpgNH2}). Note that Eq.~(\ref{gLowDvsS02}) applies only within the eigenvalue domain $\mc{D}$ given by Eq.~(\ref{contour2}). Fig.\ \ref{plotGreenFull} shows the full distribution $p(\Lambda)$ obtained in this way for $N=10^4$ and $\rho\lambda_0^3=1$, together with the result of numerical diagonalization.

%%%%%%%%%%%%%  FIG %%%%%%%%%
%%%%%%%%%%%%%%%%%%%%%%%%%%

%%%%%%%%%%%%%  FIG %%%%%%%%%
\begin{figure*}
\centering{
\includegraphics[angle=0,width=0.8\textwidth]{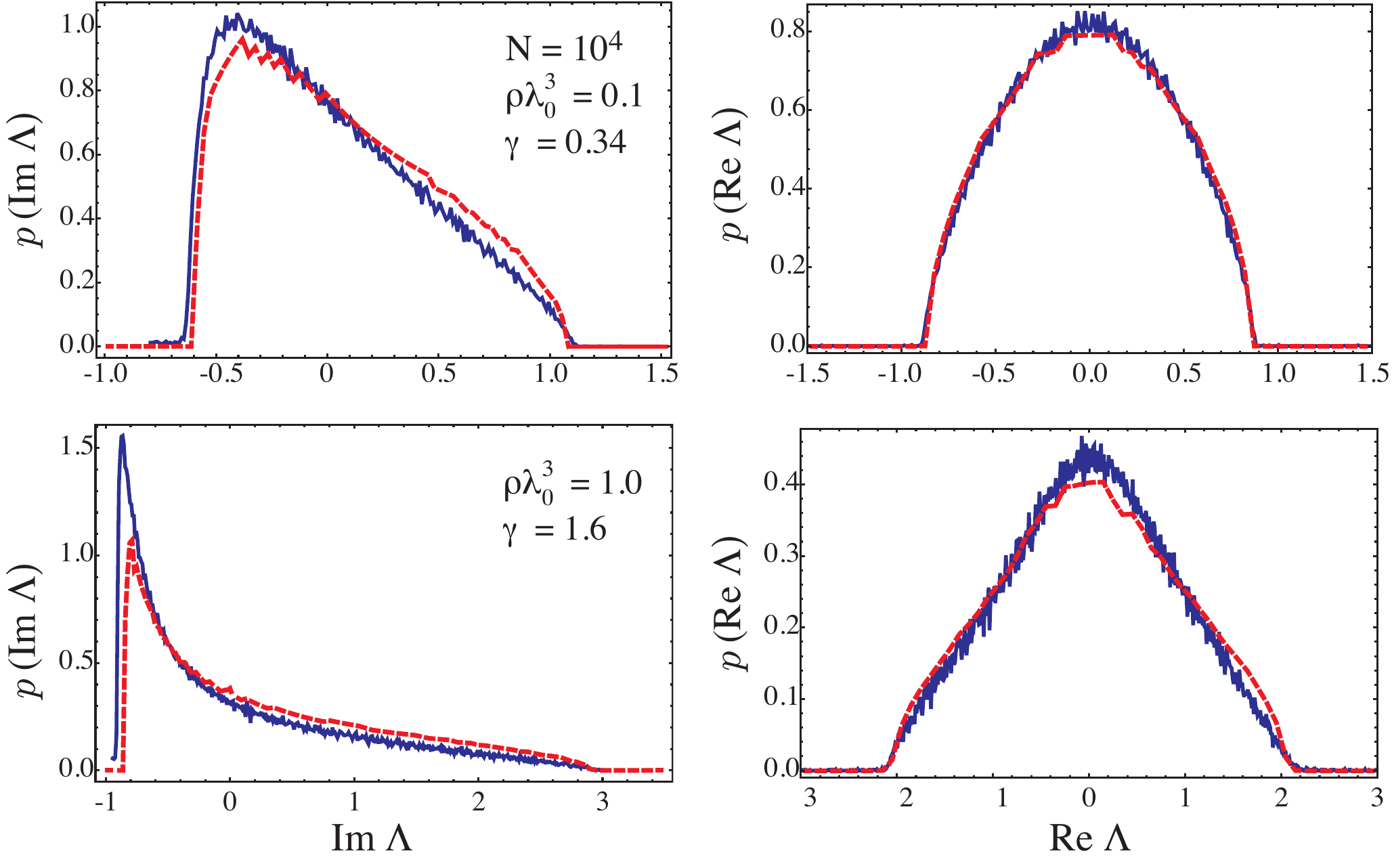}
\caption{
\label{ReandImProj}
Marginal probability density of the imaginary (left column) and real (right column) parts of eigenvalues $\Lambda$ of the $N \times N$ random Green's matrix (\ref{ERMGreen}) at two different densities $\rho$ of
$N$ points $\vec{r}_i$ randomly chosen inside a sphere of radius $R$. $\gamma=9N/8(k_0R)^2$. Results of numerical diagonalization (blue solid lines) obtained for $N=10^4$ after averaging over $10$ realizations are compared to the solution of Eq.~(\ref{gLowDvsS02}) (red dashed line).
}}
\end{figure*}
%%%%%%%%%%%%%%%%%%%%%%%%%%

\paragraph{Marginal distributions of real and imaginary parts of $\Lambda$.}
\label{SecMarginalGreen}

An interesting link exists between the probability densities of the eigenvalues of matrices $C$ and $S$ studied in Sec.\ \ref{ERMCosc} and \ref{ERMSinc}, respectively, and the probability densities of real and imaginary parts of the eigenvalues of the Green's matrix $G$. Remember that $C$ and $S$ define the real and imaginary parts of $G$. We note that at low densities $\rho\lambda_0^3\lesssim 1$,  the eigenvalue distributions of $G$, $\mathrm{Re}G$ and $\mathrm{Im}G$ are parameterized by a single parameter, their second moment $\gamma = \moy{\vert \Lambda \vert^2}$. It is thus also the case for the marginal distributions $p(\textrm{Re}\Lambda_{G})$ and $p(\textrm{Im}\Lambda_{G})$. In addition, Eqs.~(\ref{LinkGammaReIm}) and  (\ref{LinkGammaReIm2}) suggest that
$\gamma = \langle\Lambda_{\textrm{Re}G}^2\rangle
= 2 \langle ( \textrm{Re}\Lambda_{G})^2 \rangle
= \langle \Lambda_{\textrm{Im}G}^2 \rangle
= 2 \langle (\textrm{Im}\Lambda_{G})^2 \rangle$,
as long as $k_0R\gg 1$ and the density is not too high. It is therefore reasonable to conjecture that $p(\textrm{Im}\Lambda_{G})$ and $p(\textrm{Re}\Lambda_{G})$ may be described by equations for $p(\Lambda_{\textrm{Im}G})$ and $p(\Lambda_{\textrm{Re}G})$ with $\gamma$ replaced by $\gamma/2$:
\begin{eqnarray}
p(\textrm{Re}\Lambda_{G}, \gamma) &\simeq& p(\Lambda_{\textrm{Re}G}, \gamma/2),
\label{marginalre}
\\
p(\textrm{Im}\Lambda_{G}, \gamma) &\simeq& p(\Lambda_{\textrm{Im}G}, \gamma/2).
\label{marginalim}
\end{eqnarray}
This conjecture was verified by comparing analytic results for $\mathrm{Re}G$ and $\mathrm{Im}G$ with numerical simulations for $G$ and turns out to hold with satisfactory accuracy, as long as $\gamma \lesssim 2$ \cite{skipetrov11}. For $\gamma > 2$, the distributions of $\textrm{Im}\Lambda_{G}$ and $\Lambda_{\textrm{Im}G}$ were found to be very different, whereas the distributions of $\textrm{Re}\Lambda_{G}$ and $\Lambda_{\textrm{Re}G}$ remain similar.

The marginal probability distributions $p(\textrm{Im}\Lambda_{G})$ and $p(\textrm{Re}\Lambda_{G})$ can be obtained by projecting $p(\Lambda_G)$ following from Eq.\ (\ref{gLowDvsS02}) on the real and imaginary axes. As we show in Fig.\ \ref{ReandImProj}, a good quantitative agreement is found with numerical simulations for $\rho\lambda_0^3 \lesssim 1$. At higher densities, Eq.~(\ref{gLowDvsS02}) is not a good approximation for Eqs.~(\ref{sc1NHv2}) and (\ref{sc2NHv2}) anymore. At the same time, Eqs.~(\ref{sc1NHv2}) and (\ref{sc2NHv2}) are difficult to solve exactly because $g(z)$ and $c(z)$ are coupled and $\textrm{Tr}\hat{S}_1\hat{S}_0^\dagger$ has no `simple' expression in the biorthogonal basis of eigenvectors of the operator $\hat{T}$, in contrast to $\textrm{Tr}\hat{S}_0\hat{S}_0^\dagger$ [see Eqs.~(\ref{trs}), (\ref{trss}), (\ref{contourT3}), and (\ref{contourT4})].

\section{Applications}
\label{Applications}

Let us now see how the mathematical results reviewed in the previous sections can be applied to understand real physical systems.

\subsection{Vibrations in topologically disordered systems}
\label{SecTopo}

Amorphous solids, glasses and supercooled liquids exhibit a number of interesting vibrational properties that still lack a satisfactory explanation agreed upon by all specialists \cite{philips81,klinger10}. Euclidean random matrices can be used to model the behavior of these topologically disordered systems and thus to propose possible explanations of such properties \cite{grigera01a,grigera01b,grigera02,grigera03,ciliberti03,ganter11,grigera11}. Following \citet{grigera01a}, let us model a topologically disordered three-dimensional system,---say, an amorphous solid,---as an ensemble of $N \gg 1$ identical particles that harmonically oscillate around their equilibrium positions $\vec{r}_i$ ($i = 1, \ldots, N$). The latter are randomly distributed in a large volume $V$ with an average density $\rho = N/V$. For simplicity, we assume that the displacements $\Delta x_i(t)$ of all particles are collinear and take place along a fixed direction that we choose to be the $x$ axis of the reference frame. The instantaneous position of the particle $i$ is then $\vec{R}_i(t) = \vec{r}_i + \vec{e}_x \Delta x_i(t)$, where $\vec{e}_x$ is the unit vector codirectional with the $x$ axis. The vector nature of displacements can be taken into account but does not influence the qualitative conclusions of the scalar analysis \cite{ciliberti03}. The energy of the system is $U = -\frac12 \sum_{i,j} f(\vec{r}_i - \vec{r}_j) [\Delta x_i(t) - \Delta x_j(t)]^2$, where the second derivative of the pair potential $-f(\vec{r}_i - \vec{r}_j)$ is supposed to be a rapidly decaying function of its argument. Vibrations in this model can be characterized by studying the Hessian matrix $A$ defined by Eq.\ (\ref{DefinitionERMu}) with $u = 1$.
%% \begin{eqnarray}
%% A_{ij} = \delta_{ij} \sum\limits_{l = 1}^N f(\vec{r}_i - %% \vec{r}_l) - f(\vec{r}_i - \vec{r}_j).
%% \label{hessian}
%% \end{eqnarray}

In the limit of infinite density $\rho \to \infty$, this model describes an elastic continuum medium. We deduce from the analysis of Sec.\ \ref{SecERMHighDensity} that the resolvent $g_{\mathbf{k}}(z)$ defined in Eq.\ (\ref{Defgk}) is given by
\begin{eqnarray}
g_{\mathbf{k}}^0(z) = \frac{1}{z - \varepsilon(\vec{k})},
\label{res0}
\end{eqnarray}
with the dispersion relation $\varepsilon(\vec{k}) = \rho [f_0(\vec{k}) - f_0(\vec{0})]$. In particular, for spherically symmetric smooth $f(\vec{r})$ at small $k$ we have $\varepsilon(\vec{k}) = c^2 k^2$. This describes the propagation of sound with a speed $c$ and a linear dispersion relation $\omega = \sqrt{\varepsilon(\vec{k})} = c k$. Interesting and still not fully understood properties arise when the density $\rho$ is high but finite, i.e. when a certain amount of disorder is introduced in an otherwise homogeneous medium. Attempts were undertaken to use the Euclidean RMT to explain some of these properties in the framework of the high-density expansion developed in Sec.\ \ref{SecERMHighDensity}.

\subsubsection{Brillouin peak in the dynamic structure factor}
\label{brillouin}

One of the main quantities measured in inelastic x-ray and neutron scattering experiments designed to characterize the topologically disordered systems is the dynamic structure factor (DSF) $S(k, \omega)$. For a system of $N$ identical particles it is \cite{hansen86}
\begin{eqnarray}
S(k, \omega) = \frac{1}{N} \sum\limits_{i,j = 1}^N
\int_{-\infty}^{\infty} dt e^{\textrm{i} \omega t} \left\langle e^{\textrm{i} \vec{k} \cdot [\vec{R}_i(t) - \vec{R}_j(0)]} \right\rangle.
\label{dsf}
\end{eqnarray}
Taking into account only the vibrational modes of the system, we can express $S(k, \omega)$ through the resolvent $g_{\mathbf{k}}(z)$ defined in Eq.\ (\ref{Defgk}) as
\begin{eqnarray}
S(k, \omega) = -\frac{2 k_{\mathrm{B}} T k^2}{\omega \pi} \lim\limits_{\epsilon \to 0^+} \mathrm{Im} g_{\vec{k}}(\omega^2 + \textrm{i} \epsilon).
\label{dsf2}
\end{eqnarray}

Early experiments [see, e.g., \cite{benassi96,monaco98}] showed that at $k \lesssim k_0$, where $k_0$ is the position of the first maximum of the static structure factor, DSF exhibited a low-frequency peak with a width $\Gamma$ that scaled approximately as $\Gamma \propto k^2$. This was very surprising because naive arguments suggest that Rayleigh scattering in the low-frequency (and long-wavelength) limit should lead to $\Gamma \propto k^4$. \citet{grigera01a} used the self-consistent equations (\ref{sc1u1}) and (\ref{sc2u1}) to find that $\mathrm{Im} \sigma_{\vec{k}}(z) \propto z^{(d-2)/2} k^2$ and $\Gamma = \mathrm{Im} \sigma_{\vec{k}}(\omega^2)/\omega \propto k^2$ for $d = 3$, in contradiction with the expectation for Rayleigh scattering but in agreement with experimental observations. This result was confirmed by a perturbative calculation up to the second order in $1/\rho$ \cite{grigera01b}. However, more recently \citet{ganter11} pointed out the incompleteness of the previous analysis and presented a refined calculation of the resolvent $g_{\vec{k}}(z)$ up to the second order in $1/\rho$ leading to $\mathrm{Im} \sigma_{\vec{k}}(z) \propto z^{d/2} k^2$ and hence to $\Gamma \propto k^4$. Shortly after that, \citet{grigera11} demonstrated the exact cancelation of terms $\propto z^{(d-2)/2} k^2$ in all orders in $1/\rho$ and also found an additional contribution $\propto z^{(d-2)/2} k^4$ to $\mathrm{Im} \sigma_{\vec{k}}(z)$. This again leads to $\Gamma \propto k^4$ and shows that the simple Euclidean RMT model that we presented above does not seem to be sufficient to explain the anomalous scaling of $\Gamma$ with $k$. Recent experimental findings \cite{monaco09} suggest that $\Gamma \propto k^2$ is characteristic for the region of relatively high frequencies, where strong temperature dependence of $\Gamma$ is observed and the simple model of harmonically oscillating particles that we consider in this section does not apply. The model becomes justified at lower frequencies, where $\Gamma$ is independent of temperature, and its prediction $\Gamma \propto k^4$, which is in agreement with the behavior expected for Rayleigh scattering, is verified by the experiment \cite{monaco09}.

\subsubsection{The boson peak}
\label{bosonpeak}

\begin{figure}
\includegraphics[angle=0,width=0.9\columnwidth]{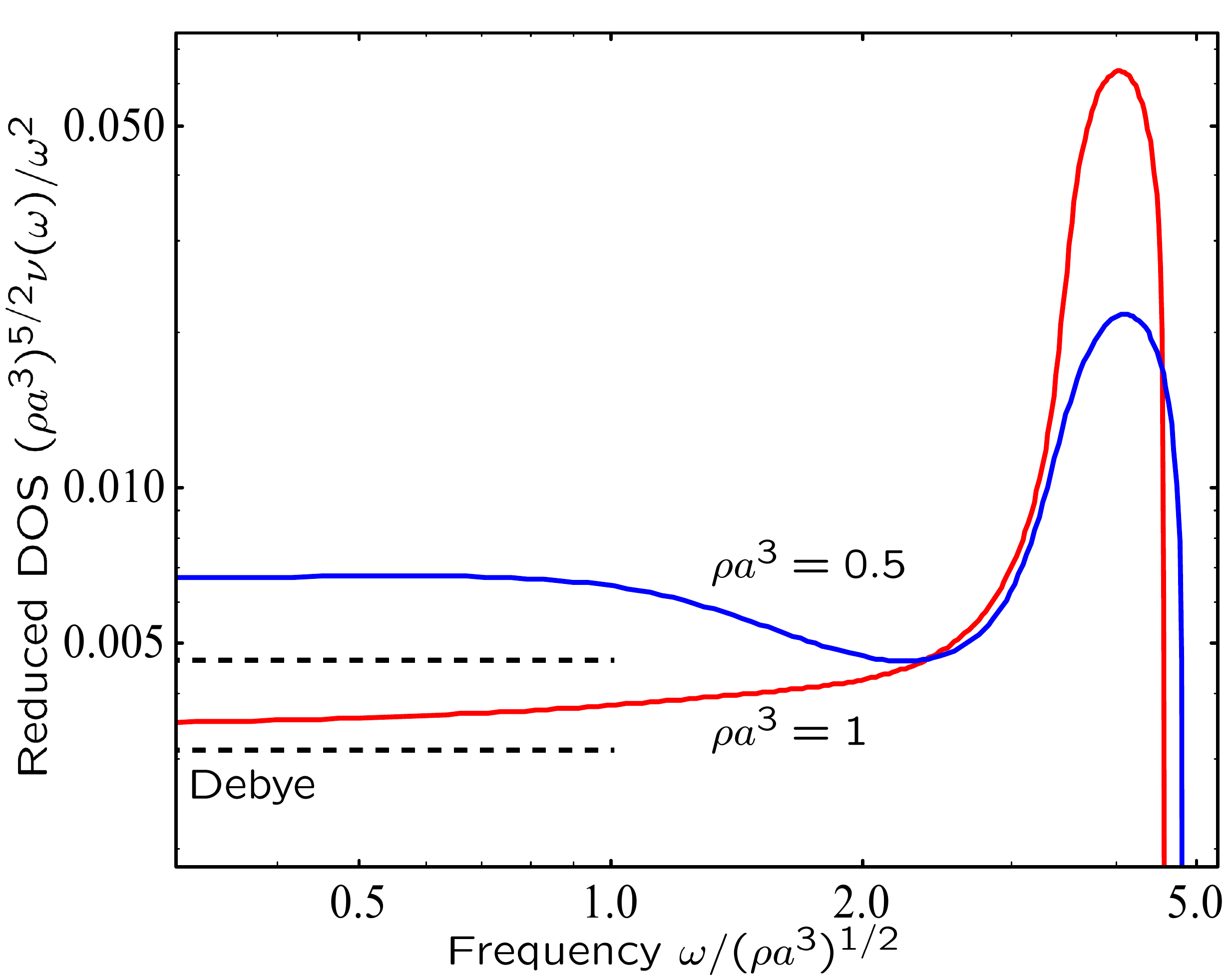}
\caption{
\label{figbp}
Reduced density of states $\nu(\omega)/\omega^2$ of a topologically disordered system at two different densities $\rho a^3$ (log-log plot). The solid lines are obtained by numerical solution of Eqs.\ (\ref{sc1u1}), (\ref{sc2u1}) and (\ref{eqbp}) for the ERM defined by $f(\vec{r}) = -\exp(-r^2/2 a^2)$ at two densities $\rho a^3 = 1$ and $0.5$. The dashed lines follow from Eq.\ (\ref{eqbp}) with $g_{\vec{k}}(z)$ replaced by $g_{\vec{k}}^0(z)$, in the limit of $\omega \to 0$.}
\end{figure}

Another interesting property of topologically disordered systems is the so-called `boson peak' in the density of states (DOS) $\nu(\omega)$. For our model of a topologically disordered system, DOS can be expressed through the resolvent (\ref{Defresolvent}) as
\begin{eqnarray}
\nu(\omega) = -\frac{2 \omega}{\pi} \lim\limits_{\epsilon \to 0^+} \mathrm{Im} g(\omega^2 + \mathrm{i} \epsilon).
\label{dos}
\end{eqnarray}

An equation for $g(z)$ can be obtained from Eqs.\ (\ref{sc1u1}) and (\ref{sc2u1}) using Eq.\ (\ref{LinkggkERM}) \cite{grigera01a}:
\begin{eqnarray}
\frac{1}{\rho g(z)} &=& \frac{z}{\rho} + f_0(\vec{0}) - A g(z)
\nonumber \\
&-& \int\frac{\textrm{d}^3 \vec{q}}{(2\pi)^3} f_0(\vec{q})^2
g_{\vec{k}}(z),
\label{eqbp}
\end{eqnarray}
where $A = \int \textrm{d}^3 \vec{q} f_0(\vec{q})^2/(2\pi)^3$. In the limit of high density ($\rho \to \infty$) and low frequencies ($\mathrm{Re} z = \omega^2 \to 0$), we can substitute $g_{\vec{k}}(z)$ in the last term of Eq.\ (\ref{eqbp}) by $g_{\vec{k}}^0(z)$ defined in Eq.\ (\ref{res0}) and obtain the Debye spectrum  $\nu(\omega) \propto \omega^2/\rho^{5/2}$ \cite{grigera01a,grigera02}. The quadratic scaling of DOS with frequency is precisely the behavior that follows from the so-called Debye model for DOS in crystals. The disordered nature of an amorphous solid manifests itself in an important correction to the above scaling at finite $\rho$. Experimentally, one observes states in excess of the Debye law that give rise to a peak in the reduced DOS $\nu(\omega)/\omega^2$, commonly known as the boson peak \cite{klinger10}. The origin of the boson peak is still a subject of active research activity that we do not intend to review here [see, e.g., \cite{chumakov11} for a recent work and a selection of references]. \citet{grigera01a,grigera02} argued that the peak in DOS can be obtained from the self-consistent equations (\ref{sc1u1}) and (\ref{sc2u1}). Indeed, solving these equations for $g_{\vec{k}}(z)$ numerically, substituting the solution into Eq.\ (\ref{eqbp}), solving the latter for $g(z)$, and then using Eq.\ (\ref{dos}), we obtain the reduced DOS shown in Fig.\ \ref{figbp}. At the lower density, we observe an excess of low-frequency states with respect to the Debye limit shown by dashed lines, even though no well-defined peak structure can be identified. Such an excess is less important or even absent at the higher density.

The appearance of the boson peak in DOS is believed to be generic for a certain class of ERM ensembles, independent of the specific function $f(\vec{r})$ that generates the ensemble, as far as the latter decays sufficiently fast with $r$. \citet{grigera02} studied it for $f(\vec{r}) = (1-\alpha r^2/a^2) \exp(-r^2/2 a^2)$, where the parameter $\alpha$ were varied. They also checked that computation of DOS using the method of moments yields similar results, thus proving that the appearance of the boson peak in ERM theory is not an artifact of self-consistent equations (\ref{sc1u1}) and (\ref{sc2u1}). In general, low-frequency vibrations in topologically disordered systems are still a subject of intense theoretical research, as witnessed, for example, by the recent works of \citet{amir10,amir13}.

\subsubsection{Anderson localization}
\label{SecAndersonTopo}

A universal phenomenon that is expected to take place for waves in disordered systems is Anderson localization \cite{anderson58}. In 3D infinite systems, Anderson localization manifests itself by a transition from extended to localized states when varying either the strength of disorder or the energy of the state.

\citet{ciliberti05,huang09,krich11,amir13} studied Anderson localization in variants of the ERM model considered in this section. Focusing on liquid instantaneous normal modes (INMs), \citet{ciliberti05} generated random equilibrium configurations of points $\{ \vec{r}_i \}$ in a realistic system using a previously developed Monte-Carlo algorithm at a fixed density and several temperatures. These random configurations were used to obtain the pair correlation function $g(r)$ that allows to compute the probability density function (PDF) of elements of the ERM $A$ under assumption that higher-order correlations can be neglected. In its turn, this distribution allows one to obtain an equation for the PDF of diagonal elements of a matrix resolvent ${\cal G}(z) = (z - A)^{-1}$. The latter equation was solved numerically using a population dynamics algorithm and the stability of a purely real `population' of ${\cal G}_{nn}$ was analyzed. The idea behind is that when a small imaginary part is added to ${\cal G}_{nn}$ and the latter is evolved according to the above-mentioned algorithm, $\mathrm{Im} {\cal G}_{nn}$ grows in time in the extended phase or decays in the localized phase. Indeed, in terms of the eigenvectors $R_n$ and eigenvalues $\Lambda_{n}$ of the matrix $A$ we can write
\be
%% \mathrm{Im} {\cal G}_{ii}(z) = \mathrm{Im} \sum_{\alpha}
%%\frac{|\scp{i}{\alpha}|^2}{z - \Lambda_{\alpha}},
\mathrm{Im} {\cal G}_{nn}(z) = \mathrm{Im} \sum_{m}
\frac{|\sum_{i=1}^{N} R_n^*(\vec{r}_i) R_m(\vec{r}_i)|^2}{z - \Lambda_{m}},
\label{imr}
\ee
where $z = \Lambda + \mathrm{i} \epsilon$ and $\epsilon \to 0^+$. When $\Lambda$ belongs to the part of the spectrum for which eigenstates are extended, $|\sum_{i=1}^{N} R_n^*(\vec{r}_i) R_m(\vec{r}_i)|^2 \sim 1/N$ and $\mathrm{Im} {\cal G}_{nn} \sim 1$. In contrast, for $\Lambda$ in the part of the spectrum corresponding to localized states, $R_n^*(\vec{r}_i) R_m(\vec{r}_i)$ will be sizable only for a small fraction of sites $\vec{r}_i$ and $\mathrm{Im} {\cal G}_{nn} = 0$ for most $n$. This approach allowed \citet{ciliberti05} to show that two mobility edges exist in the spectrum of the matrix $A$ (one at negative and another one at positive $\Lambda$) and to roughly determine their positions.

A different approach to essentially the same problem was employed by \citet{huang09}. They were able to differentiate between the parts of the eigenvalue spectrum corresponding to extended and localized states by analyzing the level spacing statistics, i.e. the statistical distribution $P(s)$ of normalized spacing $s = (\Lambda_{n+1} - \Lambda_{n})/\langle \Lambda_{n+1} - \Lambda_{n} \rangle$ between ordered eigenvalues $\Lambda_n$. This distribution is expected to have different shapes in the extended part of the spectrum, where repulsion of eigenvalues is expected [$P(s) \propto s$ for small $s$], in its localized part, where $P(s) \propto \exp(-s)$, and at the mobility edge, where $P(s)$ takes an intermediate form. Using extensive numerical simulations, \citet{huang09} not only located the two mobility edges in the spectrum of ERM $A$ quite precisely, but was also able to determine the critical exponent of the localization transition from the finite-size scaling of the second moment of $P(s)$. The value of the critical exponent $\nu \simeq 1.6$ is in agreement with the one found for different models of the same universality class in 3D.

A similar estimate of $\nu$ was obtained by \citet{krich11} who studied ERMs generated by an exponentially decaying function $f(\vec{r})$ (see Sec.\ \ref{expERM}). They performed a finite-size scaling analysis using powerful numerical methods. An analytic treatment of Anderson localization in this model was developed very recently by \citet{amir13} who found the critical frequency of the localization transition as a function of density of randomly distributed points $\vec{r}_i$ and the dimensionality $d$ of the system.

\subsection{Electron glass dynamics}
\label{ElectronGlass}

An electron glass is a highly disordered solid in which most electronic states are localized and where long-range Coulomb interactions play an important role. It is referred to as a glass because it exhibits glassy behaviors, such as slow relaxation of the conductance and dependence of the relaxation on the time of application of a perturbation. The latter property is called `aging' \cite{amir11}.

A simple model of the electron glass is an ensemble of exponentially localized electronic states randomly distributed in space and weakly coupled by phonons. At equilibrium, the hopping rate between the states localized at $\vec{r}_i$ and $\vec{r}_j$ is given by the Fermi's golden rule
\be
\label{gammaEq}
\gamma_{ij}^0\varpropto f(E_i)[1-f(E_j)]e^{-r_{ij}/\xi}[1+b(E_i-E_j)],
\ee
where $f$ and $b$ are the Fermi-Dirac and Bose-Einstein distributions that characterize the statistics of electrons and phonons, respectively. $E_i=\epsilon_i+\sum_{j\neq i}e^2f(E_j)/r_{ij}>E_j$ is the potential energy of the state $i$ and $\epsilon_i$ is the energy in the absence of Coulomb interactions. If $E_i<E_j$, the expression in square brackets in Eq.\ (\ref{gammaEq}) have to be replaced by $b(E_i-E_j)$.
Within the local mean-field approach, the average occupation number $n_i$ evolves, under a small perturbation, as $\textrm{d}n_i/\textrm{d}t = \sum_j (\gamma_{ji}-\gamma_{ij})$, where $\gamma_{ij}$ is obtained from Eq.\ (\ref{gammaEq}) upon replacing $f(E_i)$ by $n_i$  \cite{amir08}. In the vicinity of a metastable state characterized by the electronic configuration $\vec{n_0}$, the linearized equation of motion for $\delta \vec{n}=\vec{n}-\vec{n}_0$ takes the form of a master equation $\textrm{d}\delta\vec{n}/\textrm{d}t = \tilde{A}\delta\vec{n}$, where the off-diagonal elements of the matrix $\tilde{A}$ are
\be
\label{AElectronGlass}
\tilde{A}_{ij}=\frac{\gamma_{ij}^0}{n_j^0(1-n_j^0)}-\sum_{k\neq j,i}\frac{e^2\gamma_{ik}^0}{k_BT}\left(\frac{1}{r_{ij}}-\frac{1}{r_{jk}}\right)
\ee
and the diagonal elements $\tilde{A}_{ii} = -\sum_{j\neq i}\tilde{A}_{ij}$ guarantee the particle number conservation.

Near a stable point, the eigenvalues of $\tilde{A}$ are negative and their statistical distribution $p(\Lambda)$ determines the relaxation properties of the glass. A numerical study performed by \citet{amir08} shows that for parameters relevant to experiments, $p(\Lambda)$ is similar to the distribution obtained by keeping only the first term in Eq.\ (\ref{AElectronGlass}) and neglecting the remaining energy dependence, such that $\tilde{A}_{ij}$ is replaced by $A_{ij} = e^{-r_{ij}/\xi}$. Therefore, the exponential ERM $A$ studied in Sec.\ \ref{expERM} appears to be of fundamental importance for the study of slow relaxation processes in an electron glass. In the regime of low density $\rho\xi^d \ll 1$, where almost all eigenvectors of $A$ are localized, one readily finds from Eq.\ (\ref{PexpLowDensity}) that $p(\Lambda) \propto 1/\vert\Lambda\vert$. This behavior gives rise to a logarithmic relaxation of a small deviation $\delta \vec{n} (t)$ from the metastable state. Indeed, expanding $\delta \vec{n} (t)$ over the basis $\{\vec{R}_k\}$ composed of eigenvectors $\vec{R}_k$ of the matrix $A$: $\delta \vec{n} (t)=\sum_{k} c_k \vec{R}_k e^{-\vert \Lambda_k \vert t}$, and assuming that the eigenvectors are uniformly excited, we get
\be
\label{DynDerN}
\vert \delta \vec{n} (t)  \vert \sim\int_{\Lambda_{\mathrm{min}}}^{\Lambda_{\mathrm{max}}} \textrm{d} \Lambda \frac{e^{-\Lambda t}}{\Lambda} \sim -\gamma_E-\textrm{ln}(\Lambda_{\mathrm{min}}t)
\ee
for $1/\Lambda_{\mathrm{max}} < t < 1/\Lambda_{\mathrm{min}}$, where $\gamma_E$ is the Euler constant \cite{amir08}. It is worth noting that the Coulomb interaction does not play any role in this slow relaxation, since the latter has been neglected in the replacement of $\tilde{A}$ by $A$.

In real experiments, a logarithmic decay is observed not directly for the deviation  $\vert \delta \vec{n}\vert$ of the occupation number but for the conductance $\delta \sigma $ of the electron-glass \cite{vaknin00}. It is believed that the relation $\delta \sigma(\vert \delta \vec{n}\vert)$ is linear \cite{amir08}, which implies that a kick out of equilibrium increases the conductance before the latter starts relaxing slowly. One possible explanation for this effect is related to the long-range Coulomb interaction as follows. At low temperature and at equilibrium, the interaction creates a soft gap in the density of states at Fermi energy, known as the Coulomb gap, that reduces the conductance. When a small perturbation is applied,
the Coulomb gap is suppressed, causing the conductance to increase before relaxing back to equilibrium \cite{amir11}. Hence, $\delta \sigma (t)$ evolves according to Eq.\ (\ref{DynDerN}). This result allows us to understand aging experiments, in which a gate voltage is applied during a time $t_w$ to a sample initially at equilibrium. By taking into account the fact that not all the modes have the time to relax during $t_w$, it is easy to show that at time $t$ (measured from the end of the excitation), the conductance relaxes as
\be
\label{AgingLn}
\delta \sigma (t, t_w)=C\,\textrm{ln}(1+t_w/t),
\ee
where $C$ is a nonuniversal constant \cite{amir09}. The fact that $\delta \sigma (t, t_w)$ depends on the ratio $t_w/t$ is known as `full aging' and is a direct consequence of $p(\Lambda)\sim 1/\vert \Lambda \vert$. The behavior predicted by Eq.\ (\ref{AgingLn}) was observed in electron glasses such as InO \cite{vaknin00} or granular aluminium \cite{grenet07}, as well as in structural glasses such as the plastic Mylar \cite{ludwig03}. This suggests that the ERM model of relaxation described in the present section might be a general paradigm for aging in glasses of different physical origins \cite{amir09}.

\subsection{Waves in open chaotic systems}
\label{chaos}

In classical mechanics, a chaotic system is a deterministic, often quite simple system that exhibits extreme sensitivity to initial conditions. This means that in practice, it is impossible to make any predictions for the state of such a system after a sufficiently long time. Billiards having irregular shapes are examples of classically chaotic systems. Interestingly enough, when a quantum (or wave) problem is solved in a system that is classically chaotic, the chaotic behavior is suppressed and the evolution of the wave function of a quantum particle (or of the amplitude of a classical wave) can be calculated without problems. However, the resulting solution bears subtle signatures of the chaotic behavior that the system would exhibit if it were classical. These signatures are subject of the field of `quantum chaos' \cite{haake10}.

Particularly interesting physics takes place in open chaotic systems that couple to the exterior world through $M$ `channels'. The statistical properties of the matrix $X$ considered in Sec.\ \ref{SecIndepnhERM} are strongly reminiscent of those of effective Hamiltonians used to characterize open chaotic systems \cite{fyodorov97,fyodorov05,mahaux69,mitchell10}. If we recall the physical meaning of the matrix $X$, we understand that this analogy is not an accident. \citet{mahaux69} introduced the random matrix model for the $M\times M$ scattering matrix of an open chaotic system [see also \cite{mitchell10}]:
\begin{align}
\label{SChaos}
\mc{S}(E)&=I_M-\textrm{i}aH^\dagger\frac{1}{E - H_{\mathrm{eff}}}H,
\\
\label{HeRMT}
H_{\mathrm{eff}}&=H_0-\frac{\textrm{i}a}{2}HH^\dagger,
\end{align}
where $H_0$ is a Hermitian matrix that describes the closed part of the system under consideration, $E$ is the energy of the incoming wave, $H$ is a $N\times M$ matrix that contains entries coupling the $N$ internal states to the $M$ external channels, and $a>0$ is an overall coupling parameter controlling the `degree of non-Hermiticity' of the effective Hamiltonian $H_{\mathrm{eff}}$. $H_0$ is commonly drawn from the Gaussian ensemble (\ref{GaussianProba}), and $H$ is chosen such that $HH^\dagger$ is a Wishart matrix. Randomness in $H_0$ and $HH^\dagger$ is assumed to be independent, meaning that $H_0$ and $HH^\dagger$ are asymptotically free matrices. The eigenvalue distribution of $H_{\mathrm{eff}}$ was considered previously by \citet{haake92} (with the help of the replica trick), \citet{lehmann95} (using the supersymmetry method) and \citet{janik97a} (using the free probability theory). The splitting of the domain of existence of eigenvalues in two parts was observed when $a$ was increased. This is slightly different from the matrix $X$ studied in Sec.\ \ref{SecIndepnhERM} that has elements with equal variances $\gamma/N$ of real and imaginary parts (hence always the same degree of non-Hermiticity).

To fully understand the similarities and the differences between the two models,  let us consider a realistic $M \times M$ scattering matrix $\mc{S}(\omega)$ describing the scattering of light by $N$ randomly placed atoms with polarizability $\alpha(\omega)$ \cite{goetschy11c}:
\begin{eqnarray}
\label{ScaterringHTH}
\mc{S}(\omega) &=& I_M+TH^\dagger
\frac{1}{\mc{A}(\omega)^{-1}-G}H,
\\
G &=& HTH^\dagger,
\end{eqnarray}
where $\omega$ is the frequency of light, $\mc{A}(\omega) = \mathrm{diag}[k_0^3 \alpha(\omega)/4\pi]$ is the polarizability matrix, $H$ is defined by Eq.~(\ref{DefHinA}), and $G$ is the Green's matrix considered in Sec.\ \ref{SecEigenvalueGreen}. By choosing a simple model for the atomic polarizability $\alpha(\omega) = -(4\pi/k_0^3) (\Gamma_0/2)/(\omega- \omega_0 + \mathrm{i} \Gamma_0/2)$, we obtain
\begin{eqnarray}
\label{ScaterringHTH2}
\mc{S}(\omega) &=& I_M - TH^\dagger
\frac{\Gamma_0/2}{\omega-H_{\mathrm{eff}}} H,
\\
\label{HeRMT2}
H_{\mathrm{eff}} &=& \omega_0 I_N - \frac{\Gamma_0}{2}\textrm{Re}G - \frac{\textrm{i}\Gamma_0}{2}\left[
\textrm{Im}G+I_N
\right].
\end{eqnarray}

The main difference between the effective Hamiltonian (\ref{HeRMT2}) with respect to Eq.\ (\ref{HeRMT}) is that its Hermitian and anti-Hermitian parts are correlated because they both depend on the matrix $G$. A model with uncorrelated Hermitian and anti-Hermitian parts can be obtained by replacing $G$ by the matrix $X$ studied in Sec.\ \ref{SecIndepnhERM}. However, we see from the comparison of the eigenvalue densities of matrices $X$ (Fig.\ \ref{figfree}) and $G$ (Fig.\ \ref{FigGreencontours}) that such a replacement can be justified only at very low densities and $\gamma \ll 1$, when eigenvalues of both $X$ and $G$ are uniformly distributed within a circle on the complex plane.\footnote{Note that even at low densities and $\gamma \ll 1$, the eigenvalue distributions of matrices $X$ and $G$ exhibit differences due to the eigenvalues that fall outside this circle: they follow two hyperbolic spirals for $G$ but form a `cross' for $X$.} If we restrict our consideration to low densities and $\gamma \ll 1$ and accept to replace $G$ by $X$ in the effective Hamiltonian (\ref{HeRMT2}), the latter becomes very close to Eq.\ (\ref{HeRMT}): $\textrm{Re}X$ is well approximated by a Gaussian matrix, mimicking therefore the random Hamiltonian $H_0$ of Eq.\ (\ref{HeRMT}), and the anti-Hermitian part of Eq.~(\ref{HeRMT2}) $S=\textrm{Im}X + I_N$ is well approximated by the Wishart matrix $\gamma HH^\dagger$, which coincides with the anti-Hermitian part of Eq.~(\ref{HeRMT}) with $a=\gamma=N/M$. The eigenvalue domain of $X$ [and hence of $H_{\mathrm{eff}}$ of Eq.\ (\ref{HeRMT2})] splits in two parts upon increasing $\gamma$ (see dashed contours in Fig.\ \ref{figfree}), exactly in the same way as does the eigenvalue domain of $H_{\mathrm{eff}}$ defined by Eq.\ (\ref{HeRMT}) \cite{haake92,lehmann95,janik97a}. However, it is important to keep in mind that this similarity stems from the approximations made for the matrix $X$ that are not justified at large $\gamma$ at which the splitting takes place. Indeed, we see from Fig.\ \ref{figfree} that the numerically computed eigenvalues do not split in two groups, in contradiction with the prediction of the approximate theory shown by the dashed line. The splitting of the eigenvalue domain is therefore an artefact of approximations. The similar splitting discussed by \citet{haake92,lehmann95,janik97a,mitchell10} might also be an artefact of the model (\ref{HeRMT}) that fails to describe the correct effective Hamiltonian of the system at large $a$.\footnote{Note that splittings of the eigenvalue domains are, however, observed for ERMs $S$ (Sec.\ \ref{ERMSinc}), $C$ (Sec.\ \ref{ERMCosc}) and $G$ (Sec.\  \ref{SecEigenvalueGreen}). The crucial point is that they do not occur at $\gamma \sim N/(k_0L)^2 \sim 1$, but at $k_0 L \sim 1$. In particular, in the limit of very small sample $k_0L \ll 1$, the cloud of eigenvalues of $G$ with the largest imaginary part describes the phenomenon of superradiance \cite{gross82}.}

\subsection{Waves in random media}
\label{random}

The effective Hamiltonian (\ref{HeRMT2}) for a scalar wave in an ensemble of randomly distributed, resonant point scatterers (atoms) can be rewritten as
\begin{eqnarray}
\label{HeRMT3}
H_{\mathrm{eff}} &=& \left( \omega_0 - \mathrm{i} \frac{\Gamma_0}{2} \right) I_N - \frac{\Gamma_0}{2} G.
\end{eqnarray}
We thus see that the Green's matrix $G$ defined by Eq.\ (\ref{ERMGreen}) essentially plays the role of the effective Hamiltonian for light scattering in an ensemble of point scatterers. An eigenvector of $G$ associated with an eigenvalue $\Lambda_k$  corresponds to a quasi-mode of the system that has an eigenfrequency $\omega_k = \omega_0 -(\Gamma_0/2) \mathrm{Re} \Lambda_k$ and decays with a rate $\Gamma_k/2 = (\Gamma_0/2) (1 + \mathrm{Im} \Lambda_k)$. Hence, the study of the statistical properties of $\Lambda_k$ allows us to better understand scattering of waves among a large ensemble of scattering centers. In this context, a few particular problems will be discussed below.

\subsubsection{Collective spontaneous emission of large atomic ensembles}
\label{collective}

An interesting problem of modern quantum optics concerns relaxation of elementary excitation in large atomic systems. To put it simple, a photon is stored in an ensemble of (cold, meaning immobile) atoms, and one is interested in the properties (frequency, direction of propagation, etc.) of the photon re-emitted by the atoms at a later time.  It is a specific case of the superradiance protocol, with only one photon and no restriction concerning the size of the system. Theoretically, this problem was addressed a long time ago by \citet{ernst69}, but has been popularized only very recently by the group of Scully \cite{scully06, das08, svidzinsky08, svidzinsky208, svidzinsky09, scully09, scully209, scully10, svidzinsky10}, as well as by Manassah and Friedberg [see, for example, \cite{manassah10} and references therein]. The reason for this renewed interest is probably the recent development of experimental setups free from effects that may obscure cooperative phenomena (e.g., Doppler effect or near field atom-atom interactions), using either cold atoms or ultrathin solid-state samples \cite{rohlsberger10}.

Consider $N$ immobile two-level atoms at random positions $\{ \vec{r}_i \}$ in a three-dimensional space. Each atom has the ground state $\ket{g}$ and the excited state $\ket{e}$. The atoms can interact with a single quasi-resonant photon of frequency $\omega$ close to the frequency $\omega_0$ of the atomic transition. The quantum state of the whole system can be  written as \cite{svidzinsky09,svidzinsky10}
\begin{eqnarray}
\ket{\Psi(t)} &=& \sum\limits_{j=1}^{N} \beta_j(t)
\ket{(N-1):g, j:e}\ket{0}
\nonumber \\
&+&
\sum_{\mathbf{k}} \gamma_{\mathbf{k}}(t)
\ket{N:g}\ket{\mathbf{k}}
\nonumber \\
&+&
\sum_{i<j}^N \sum_{\mathbf{k}}
\alpha_{ij\mathbf{k}}
\ket{(N-2):g,i:e,j:e}
\ket{\mathbf{k}}, \;\;\;\;\;\;
\label{state1}
\end{eqnarray}
where the first sum is over states with one atom (atom $j$) in the excited state and no photons, the second one is over states with all atoms in the ground state and a photon in the mode $\ket{\vec{k}}$, and the last sum is over states with two atoms ($i$ and $j$) excited and a photon in the mode $\ket{\vec{k}}$. The last term---in which the photon is virtual---is necessary to describe non-resonant processes and recover the proper dynamics of the system as a whole. The evolution equation for the vector $\bm\beta = (\beta_1,\ldots, \beta_N)$ reads \cite{svidzinsky09, svidzinsky10}:
\be
\label{betadyn}
\frac{\textrm{d}\bm \beta}{\textrm{d}t} = - \bm\beta(t) +\textrm{i}G \bm\beta(t),
\ee
where the time is in units of the lifetime of the excited state $\Gamma_0^{-1}$  and $G$  is the Green's matrix (\ref{ERMGreen}). According to this equation, and as we already obtained above from the effective Hamiltonian (\ref{HeRMT3}), the real and imaginary parts of an eigenvalue $\Lambda_k$ of $G$ yield the frequency shift and the modification of the decay rate of the corresponding eigenstate. The frequency shift originates from non-resonant contributions (so-called `off-shell processes') in the light-matter interaction, and is sometimes called `collective Lamb shift' \cite{scully09, svidzinsky09, scully10, svidzinsky10}. Both the decay rate and the frequency shift were studied by \citet{svidzinsky09} and \citet{svidzinsky10} in the limit of a very dense atomic cloud ($\rho \lambda_0^3 \rightarrow \infty$) by replacing summation by integration in the last term of Eq.~(\ref{betadyn}), $[G \bm \beta(t)]_{i} = \sum_{j=1}^{N} G_{ij} \beta_j(t)$.
This is equivalent to averaging Eq.\ (\ref{betadyn}) over all possible configurations $\{ \mathbf{r}_i \}$ of atoms and amounts to the neglect of the statistical nature of the initial problem in which both the decay rates and frequency shifts were random quantities, depending on the positions $\{ \vec{r}_i \}$ of atoms. As a consequence, \citet{svidzinsky09} and \citet{svidzinsky10} find deterministic eigenvalues $\Lambda_k$. Besides, all subradiant states of the Green's matrix, the importance of which was already pointed out by \citet{ernst69}, are lost.

\begin{figure}[t]
\centering{
\includegraphics[angle=0,width=0.9\columnwidth]{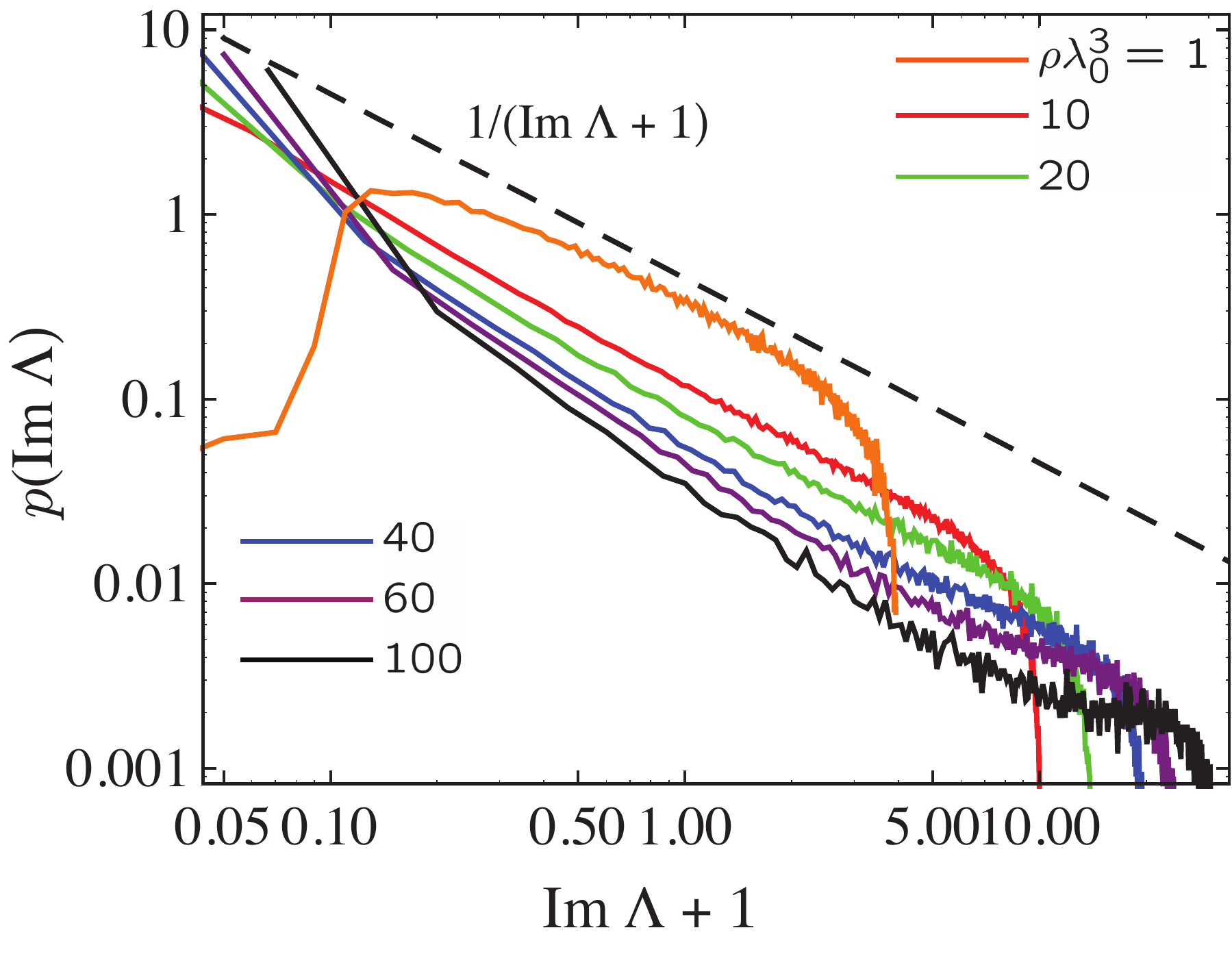}
\caption{
\label{ImProjAll}
Distribution of dimensionless decay rates $y = \mathrm{Im} \Lambda + 1$ of eigenstates of an ensemble of $N = 10^4$ identical two-level atoms in a cube of side $L$. Results at densities $\rho\lambda_0^3=1, 10, 20, 40, 60$ and $100$ (curves from top to bottom) are compared with the asymptotic law $1/y$ shown by the dashed line. [Reproduced from \cite{skipetrov11}.]
}}
\end{figure}

In this context, the study of the random Green's matrix reviewed in Sec.\ \ref{SecEigenvalueGreen} appears to be quite useful. In fact, it allows us to immediately obtain analytical results for the statistical distributions of the dimensionless decay rate $y = 1 + \mathrm{Im} \Lambda$ and frequency shift $x = \mathrm{Re} \Lambda$ at low density and small on-resonance optical thickness $b_0 = 16 \gamma/3$ \cite{skipetrov11}. According to Eqs.\ (\ref{marginalre}) and (\ref{marginalim}), $p(y)$ is given by the Marchenko-Pastur law (\ref{pmp}) with $\Lambda$ replaced by $y$, while $p(x)$ follows from the more involved Eq.\ (\ref{RCoscBox1}) after replacing $\Lambda$ by $x$. In both cases, $\gamma$ should be replaced by $\gamma/2$ and $b_0$ is the only parameter of the distribution. At higher densities or for $b_0 \gtrsim 1$, the more advanced results presented in Fig.\ \ref{ReandImProj} apply. For $\rho \lambda_0^3 \gtrsim 10$, no analytical results for the distributions of $x$ and $y$ are available. Numerical simulations show that the distribution of dimensionless decay rates $y = \mathrm{Im} \Lambda + 1$ tends to $1/y$ as the density is increased \cite{skipetrov11}, see Fig.\ \ref{ImProjAll}.

In a recent work, \citet{akkermans08} claimed that statistical properties of decay rates $y$ can be understood, at least qualitatively, by dropping the real part of the Green's matrix, inasmuch as the latter is expected to be responsible for the collective Lamb shift. This picture might not be entirely correct, because the decay rates and frequency shifts are related to the imaginary and real parts of the eigenvalues of the matrix $G$, and not to the imaginary and real parts of the matrix itself. Nevertheless, the results of Sec.\ \ref{SecMarginalGreen} suggest that under certain conditions, the distributions of imaginary parts of eigenvalues of matrix $G$ and of eigenvalues of $\mathrm{Im}G$ have indeed some similarities.

\subsubsection{Random laser}
\label{laser}

The emission of light by a conventional laser requires two important ingredients: an active medium that amplifies light and a feedback mechanism that ensures that light passes through the active medium many times \cite{siegman86}. The feedback is typically provided by an optical cavity that also plays a major role in the selection of modes in which the laser emits. In a random laser, the cavity is absent and the feedback is ensured by the multiple scattering of light \cite{cao05,wiersma08}.

%%%%%%%%%%%% FIGURE %%%%%%%%
\begin{figure}[t!]
\centering{
\hspace{0mm}
\includegraphics[angle=0,width=\columnwidth]{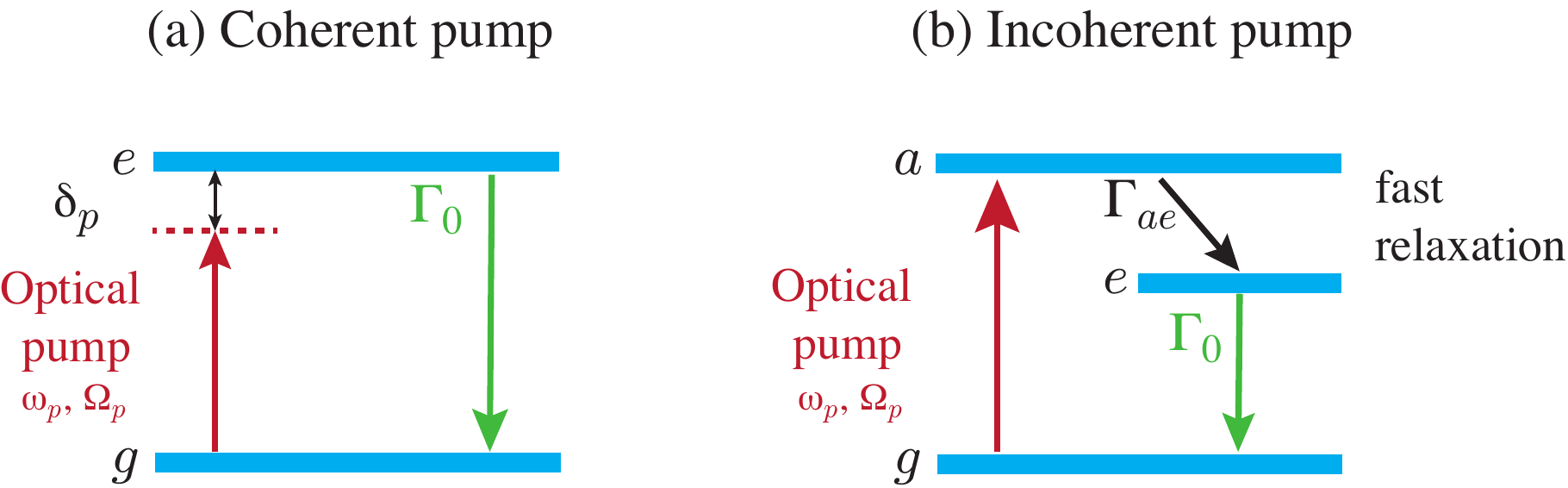}
\caption{ (a) Energy levels of a two-level atom in the field of a quasi-resonant pump beam (amplitude $\Omega_p$, frequency $\omega_p$).  (b) Energy levels of a three-level atom pumped incoherently through the auxiliary level $a$; $\Gamma_{ae}\gg \Gamma_0, \Gamma_0\Omega_p$.
}
\label{PumpScheme}
}
\end{figure}
%%%%%%%%%%%%%%%%%%%%%%%%%

Euclidean RMT was used to describe the simplest random laser---a random ensemble of atoms under an external pump \cite{goetschy11b}. Consider a gas of $N$ three-level atoms at random positions $\mathbf{r}_i$ ($i = 1, \ldots ,N$) in free three-dimensional space. The atoms are illuminated by a strong external pump field resonant with the transition from the ground state $\ket{g_i}$ to the upper auxiliary level $\ket{a_i}$ of each atom. The atoms rapidly decay to the upper level $\ket{e_i}$ of the laser transition at a rate $\Gamma_{ae} \gg \Gamma_{eg} = \Gamma_0 \gg \Gamma_{ag}$, as we illustrate in Fig.\ \ref{PumpScheme}(b). Interaction of atoms with the electromagnetic field which is near-resonant with the transition from $\ket{e_i}$ to $\ket{g_i}$ (energy difference $\hbar \omega_0$) is described by $5N$  equations of motion for atomic operators that are coupled to the quantum propagation equation for the electric field \cite{goetschy11c}. After elimination of the electric field, and in the semiclassical approximation, these equations can be reduced to a system of $2N$ equations for the expectation values $S^+_i$ and $\Pi_i$ of atomic raising and population imbalance operators $\hat{S}^+_i = \ket{e_i}\bra{g_i}$ and $\hat{\Pi}_i = \ket{e_i}\bra{e_i}-\ket{g_i}\bra{g_i}$, respectively \cite{goetschy11b}:
\begin{eqnarray}
\label{DdynLaser2}
\frac{\textrm{d}S_i^+}{\textrm{d}t} &=& \left[\textrm{i}\frac{\omega_0}{\Gamma_0}
-\frac{1}{2}(1+W_i)\right]S_i^+
\nonumber \\
&+& \frac{\textrm{i}}{2}\Pi_i \sum_{j}G_{ij}^* S_j^+,
\\
\label{PidynLaser2}
\frac{\textrm{d}\Pi_i}{\textrm{d}t} &=& -\left(1+W_i\right)\Pi_i+W_i-1
\nonumber \\
&-& 2\textrm{Im}\left[S_i^+\sum_{j}G_{ij} S_j^-\right].
\end{eqnarray}
Here the time $t$ is in units of $\Gamma_0^{-1}$and $W_i$ is the pumping rate that determines the equilibrium population imbalance $\Pi^{\mathrm{eq}}_i = (W_i-1)/(W_i+1)$. We also recognize the random Green's matrix $G$ that couples different atoms.

\begin{figure}[t]
\centering{
%\vspace{2mm}
\includegraphics[angle=0,width=\columnwidth]{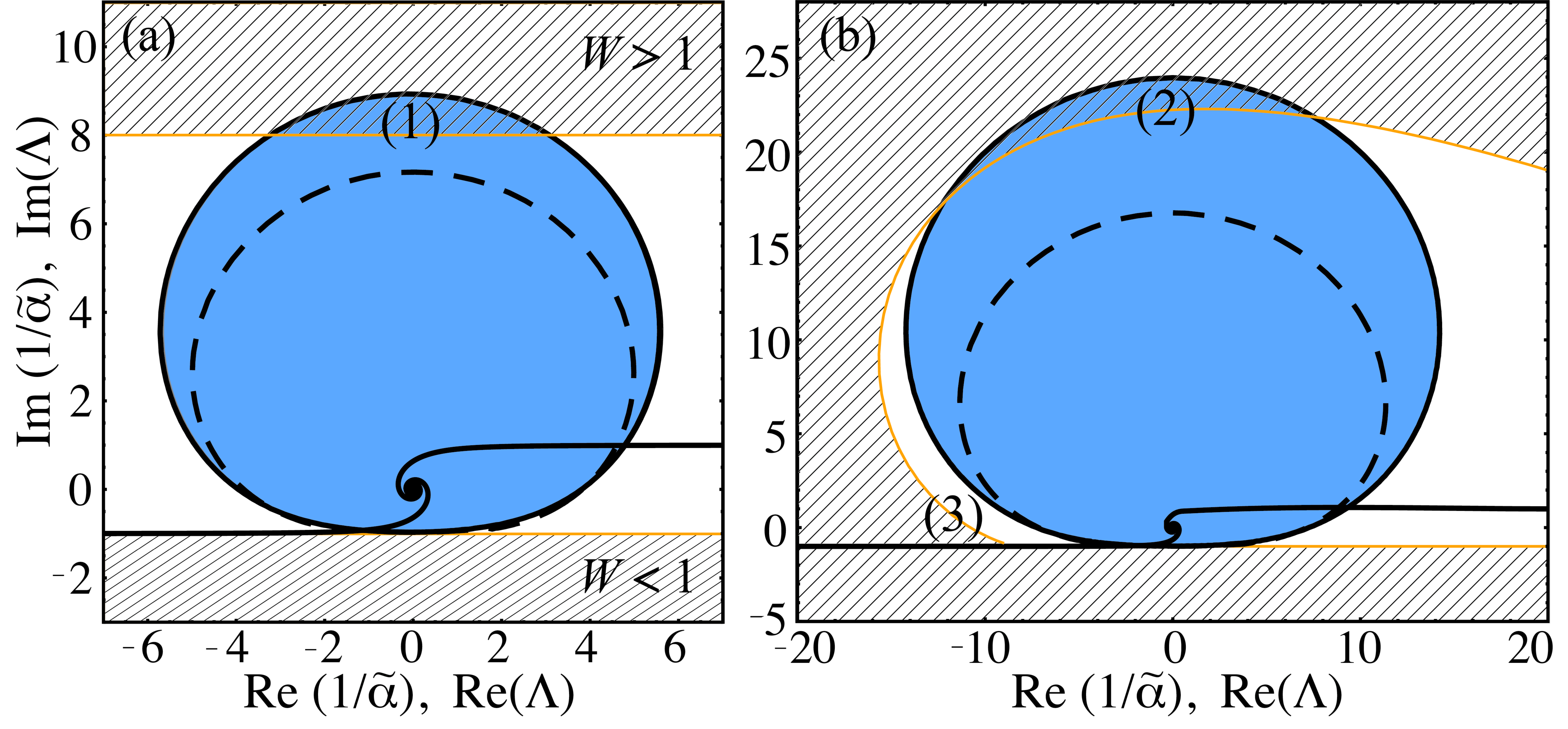}
\caption{
\label{ContourIllustration}
The domain $\mathcal{D}_\alpha$ (hatched) spanned by
$1/\tilde{\alpha}$ and the domain $\mathcal{D}_\Lambda$ (blue area delimited by the solid line) occupied by the eigenvalues $\Lambda$ of the random Green's matrix (\ref{ERMGreen}). (a) Incoherent gain, corresponding to the atomic level structure of Fig.\ \ref{PumpScheme}(b). (b) Coherent Mollow gain, corresponding to the atomic level structure of Fig.\ \ref{PumpScheme}(a) with $\Delta_p = (\omega_p - \omega_0)/\Gamma_0 = 1$.  Lasing starts when $\mathcal{D}_\alpha$ and $\mathcal{D}_\Lambda$ overlap: regions (1), (2), (3). The borderline of $\mathcal{D}_\Lambda$ is given by Eq.\ (\ref{thresholdLowD}) with the optical thickness $b_0=40$ in (a) and $b_0=140$ in (b). The dashed lines show the borderline of $\mathcal{D}_\Lambda$ following from the diffusion approximation [Eq.\ (\ref{diffusion})].
[Reproduced from \cite{goetschy11b}.]
}}
\end{figure}

The lasing threshold can be found by analyzing the linear stability of the stationary solution $\Pi_i = \Pi_i^{\mathrm{eq}}$, $S_i^{\pm} = 0$ of Eqs.\ (\ref{DdynLaser2}) and (\ref{PidynLaser2}). For a uniform pump $W_i = W$, lasing starts whenever
\be
\label{thresholdUniform}
\frac{2W}{1+W} \mathrm{Im}\Lambda_n > (1+W)+\mathrm{Im} \Lambda_n,
\ee
where $\Lambda_n$ is an eigenvalue of $G$. This condition can be rewritten in a more general form, applicable to any point-like scatterers, as
\be
\label{thresholdUniform2}
\Lambda_n = \frac{1}{\tilde{\alpha}(\omega_L)},
\ee
where $\tilde{\alpha}(\omega) = (k_0^3/4\pi) \alpha(\omega)$ and $\alpha(\omega)$ is the polarizability of the scatterer (atom). The lasing frequency $\omega_L$ is to be determined from this equation. In Fig.\ \ref{ContourIllustration} we illustrate the solution of this equation for a three-level atom that we considered above [Fig.\ \ref{ContourIllustration}(a)], but also for a two-level atom under a quasi-resonant pump [Fig.\ \ref{ContourIllustration}(a), see the level structure in Fig.\ \ref{PumpScheme}(a)]. In the latter case, the polarizability $\alpha(\omega)$ is given by the famous Mollow result \cite{mollow72}. According to Eq.\ (\ref{thresholdUniform2}), to find the lasing threshold, one has to find an overlap between the eigenvalue domain ${\cal D}_{\Lambda}$ of the Green's matrix and the domain ${\cal D}_{\alpha}$ spanned by $1/\tilde{\alpha}(\omega)$ when the parameters of $\tilde{\alpha}(\omega)$ are varied. The two domains ${\cal D}_{\Lambda}$ and ${\cal D}_{\alpha}$ are shown in Fig.\ \ref{ContourIllustration} in blue and gray hatched, respectively, for the parameters that correspond to the random laser just above the threshold in both cases.

%%%%%%%%%%%%FIG%%%%%%%%%%%%
%%%%%%%%%%%%%%%%%%%%%%%%%%%%%

As we discussed in Sec.\ \ref{SecEigenvalueGreen}, at a moderate density $\rho\lambda_0^3\lesssim 10$, the eigenvalue domain $\mathcal{D}_\Lambda$ consists of two parts: a (roughly circular) `bulk' and a pair of spiral branches. Depending on the particular model of atomic polarizability $\tilde{\alpha}$, either the bulk or the branches may touch ${\cal D}_{\alpha}$ and give rise to the laser effect. The lasing threshold due to the bulk of eigenvalues is readily obtained by combining the analytic equation (\ref{contour3}) for the borderline of $\mathcal{D}_{\Lambda}$ at low density $\rho\lambda_0^3 \lesssim 10$ and Eq.\ (\ref{thresholdUniform2}):
\be
\label{thresholdLowD}
\frac38 b_0 |\tilde{\alpha}|^2 h\left( \frac12 b_0\textrm{Im}\tilde{\alpha} \right) = 1,
\ee
where $h(x)$ is given by Eq.~(\ref{defh}). Note that the threshold depends only on the on-resonance optical thickness $b_0$ but not on the density of atoms $\rho \lambda_0^3$. Interestingly enough, for both gain mechanisms represented in Fig.\ \ref{ContourIllustration}, the threshold condition (\ref{thresholdLowD}) involves the eigenvalue with the largest imaginary part. The analytic calculation of $\langle \max(\textrm{Im}\Lambda) \rangle$ presented in Sec.\ \ref{SecExactSolNH} is in excellent agreement with numerical results, see Fig.\ \ref{MeanMaxIm}. It is quite remarkable that the agreement is present at all values of parameters, including high densities $\rho \lambda_0^3 \gg 1$ that were necessary to reach large optical thicknesses $b_0 \gg 1$ in numerical calculations with moderate $N \leq 10^4$. Because it is $\langle \max(\textrm{Im}\Lambda) \rangle$ that controls the laser threshold, we conclude that ERM theory applies to random lasing in ensembles of atoms all the way from weak ($\rho\lambda_0^3 \ll 1$) to strong ($\rho\lambda_0^3 \gg 1$) scattering regime.

\citet{goetschy11b} and \citet{goetschy11c} have shown that Eq.\ (\ref{thresholdLowD}) predicts a lower threshold for lasing than the diffusion approximation, based on Eq.\ (\ref{diffusion}). They also analyzed lasing due to the eigenvalues that belong to spiral branches shown by solid black lines in Fig.\ \ref{ContourIllustration} and that may enter into play in the case of Mollow gain, corresponding to the level structure of Fig.\ \ref{PumpScheme}(a). ERM theory of random lasing seems to be particularly adapted to the description of lasing in cold atomic systems and may be the method of choice for the description of recent experiments \cite{baudouin13}.

%%%%%%%%%%%%%  FIG %%%%%%%%%
\begin{figure*}
\centering{
\includegraphics[angle=0,width=0.8\textwidth]{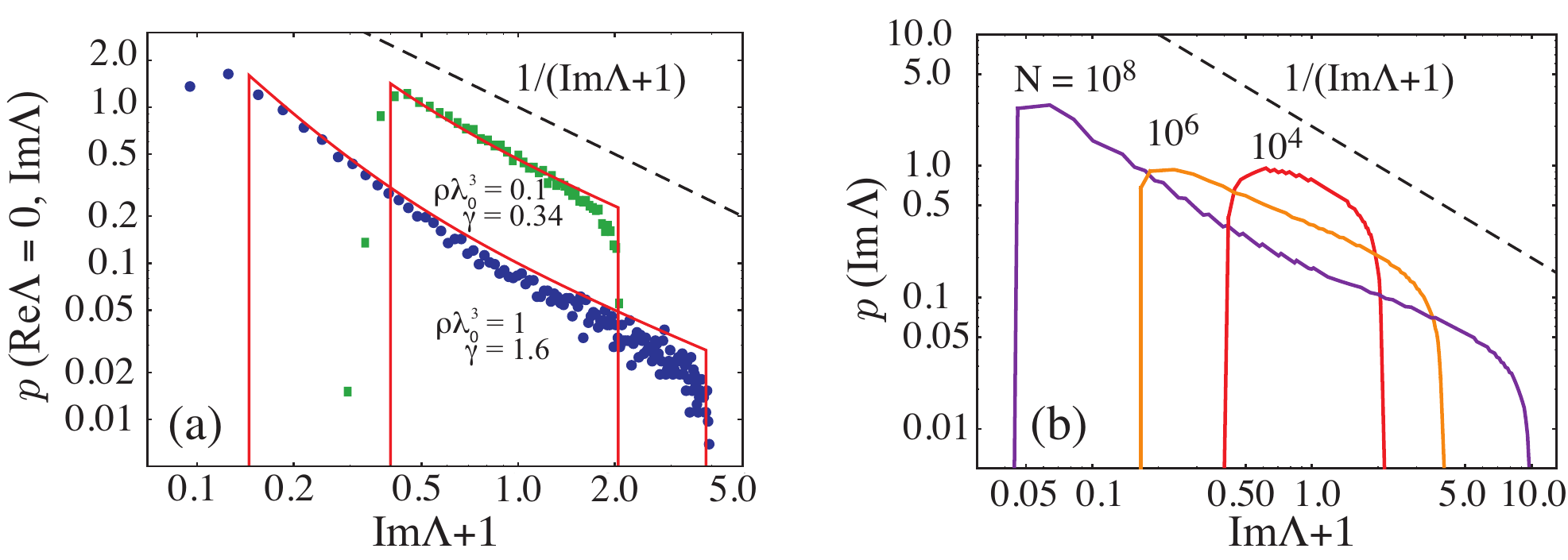}
\caption{
\label{CutAndMarginal}
(a) Cuts of the eigenvalue density $p(\Lambda)$ of the $N\times N$ Green's  matrix (\ref{ERMGreen}) along the imaginary axis $\textrm{Re}\Lambda=0$. $N=10^4$ points $\vec{r}_i$ are randomly chosen inside a sphere of radius $R$; $\gamma=9N/8(k_0R)^2$ . Results of numerical diagonalization (symbols) are compared with the solution of Eq.~(\ref{gLowDvsS02}) (solid red lines). (b) Marginal probability density of the imaginary part of eigenvalues of Eq. (\ref{ERMGreen}). Solutions of Eq.~(\ref{gLowDvsS02}) (solid lines) at $N=10^4$ ($\gamma=0.34$), $10^6$ ($\gamma=1.6$), and $10^8$ ($\gamma=7.4$) for $\rho\lambda_0^3=0.1$ are compared with the asymptotic law $1/(\textrm{Im} \Lambda +1)$ (dashed line).
}}
\end{figure*}
%%%%%%%%%%%%%%%%%%%%%%%%%%

\subsubsection{Anderson localization in open random media}
\label{SecAndersonGreen}

The methods described in Sec.\ \ref{SecAndersonTopo} and aimed at the study of Anderson localization in an infinite medium cannot be applied in open random media, where excitations can leak outside the medium. This is the case, for example, for light or sound in random ensembles of small scatterers. As follows from Eq.\ (\ref{HeRMT3}), for point-like scatterers the study of Anderson localization reduces to the study of statistical properties of eigenvectors of the random Green's matrix $G$. According to the Ioffe-Regel criterium of localization, in a 3D medium the localization transition is expected for $k \ell \simeq 1$. Here $k$ is the wavenumber of the wave and $\ell$ is the scattering mean free path due to disorder. None of these is actually known when the scattering is strong, but we can estimate $k \ell$ by replacing $k$ by $k_0$ and $\ell$ by $\ell_0 = k_0^2/4 \pi \rho$, the on-resonance mean free path in the independent scattering approximation \cite{lagendijk96}. This leads to the conclusion that in a system of point scatterers, Anderson localization can be expected at $\rho \lambda_0^3 \gtrsim 20$.

\paragraph{Distribution of decay rates.}
\label{SecDecayRates}

First attempts to use random Green's matrices to study Anderson localization were undertaken by \citet{rusek00} and \citet{pinheiro04}. In particular, in the latter work the authors looked for signatures of Anderson localization in the probability distribution $p(y)$ of dimensionless decay rates $y = \mathrm{Im} \Lambda + 1$ of eigenstates that we defined in Sec.\ \ref{collective}. Based on heuristic arguments and extensive numerical simulations, they concluded that $p(y)$ takes a universal form $p(y) \propto 1/y$ (for a certain intermediate range of $y$'s) when the regime of Anderson localization is reached. Although Fig.\ \ref{ImProjAll} seems to confirm this conclusion, one may note from this figure that $p(y)$ tends to decay as $1/y$ at densities as low as $\rho \lambda_0^3 = 1$, which is far from the mobility edge expected at  $\rho \lambda_0^3 \approx 20$. To get a deeper insight into this issue, we show cuts of $p(\Lambda)$ along the imaginary axis $\textrm{Re}\Lambda=0$ in Fig.\ \ref{CutAndMarginal}(a). We clearly observe that $p(\mathrm{Re} \Lambda = 0, \mathrm{Im} \Lambda)$ decays as  $1/(\textrm{Im} \Lambda+1)$, even though the density of points $\rho \lambda_0^3$ is too low to bring the system to the localization transition. For $\gamma \lesssim 1$, although $p(\Lambda) \varpropto 1/(\textrm{Im}\Lambda +1)$, the marginal distribution $p(\textrm{Im}\Lambda)$ follows the Marchenko-Pastur law [see Fig.\ \ref{ReandImProj}] due to the circular shape of the support of $p(\Lambda)$. Incidentally, we see that the Marchenko-Pastur law (\ref{pmp}) for $\gamma \lesssim 1$ can be seen as a projection of a two-dimensional distribution $p(\Lambda)$  on the imaginary axis, provided that $p(\Lambda) \varpropto 1/(\textrm{Im}\Lambda +1)$ inside a circle of radius $\sqrt{2\gamma}$ centered at $(0,\gamma/2)$ and $p(\Lambda) = 0$ elsewhere. The power-law decay becomes visible in the marginal distribution $p(\textrm{Im} \Lambda)$ (see Fig.\ \ref{ImProjAll}) only when the support of $p(\textrm{Im} \Lambda)$ is sufficiently wide, i.e. for $\gamma \gtrsim 1$.  Because the condition $\gamma \gtrsim 1$ can be obeyed at any, even very low density by just increasing the number of scatterers $N$, it seems that no direct link can be established between the power-law decay of $p(\textrm{Im} \Lambda)$ and Anderson localization. This also seems to be confirmed by the calculation using the analytic Eq.\ (\ref{gLowDvsS02}) at large values of $N$ that are inaccessible for numerical simulations and low density $\rho\lambda_0^3 = 0.1$ [Fig.\ \ref{CutAndMarginal}(b)].

\paragraph{Properties of eigenvectors.}
\label{SecIPR}

%%%%%%%%%%%%FIG%%%%%%%%%%%%
\begin{figure}
\centering{
%\vspace{2mm}
\includegraphics[width=0.8\columnwidth]{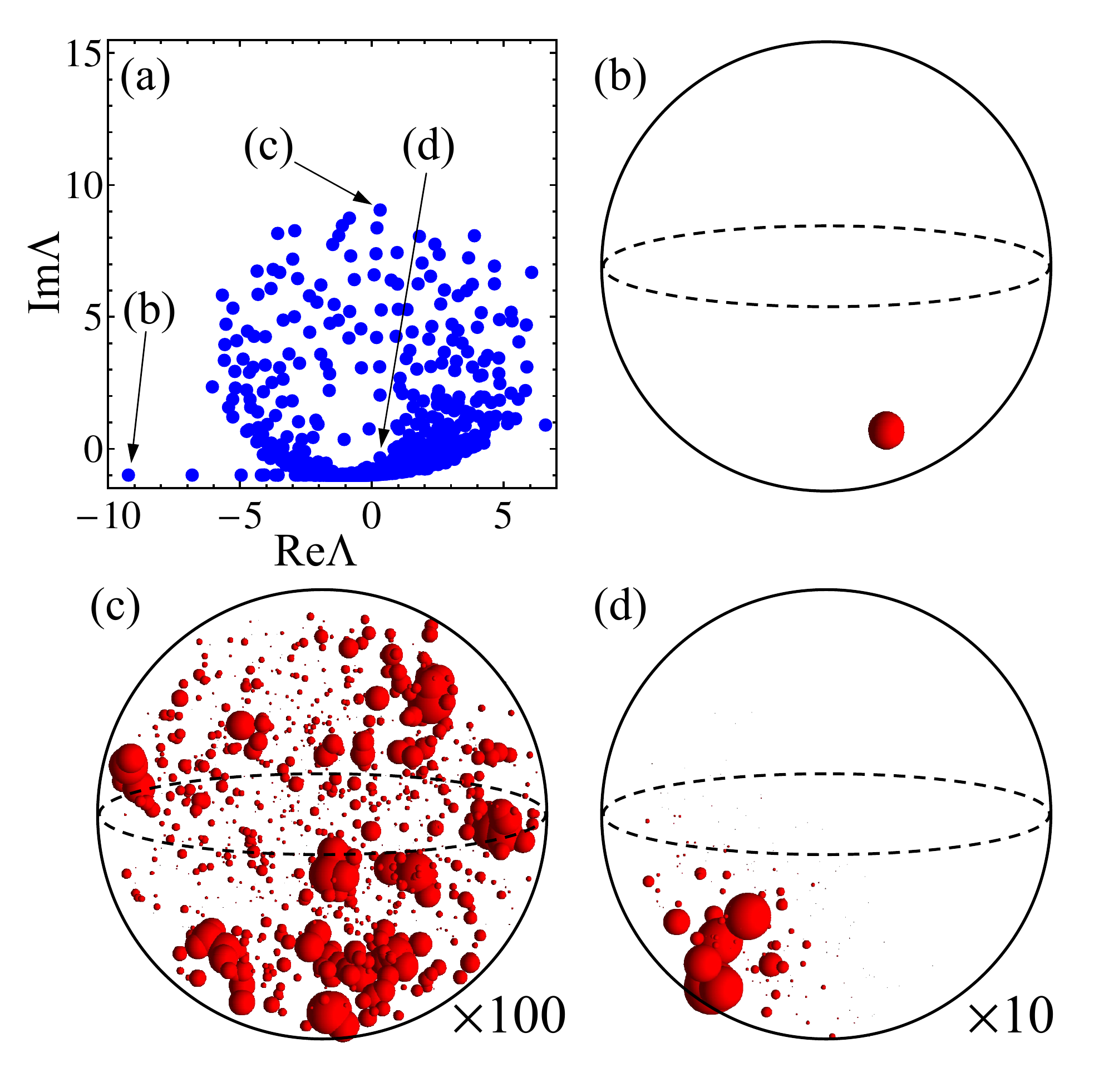}
\caption{
\label{ModesOneShot}
(a) Eigenvalues $\Lambda$ of a single random realization of the Green's matrix $G$ (dots) for a cloud of optical thickness $b_0 = 40$, composed of $N = 10^3$ atoms. (b)--(d) Intensities $|R_n^i|^2$ corresponding to the mode in the subradiant branch, localized on a pair of atoms (b), the mode with the largest $\mathrm{Im} \Lambda$ (c) and the mode corresponding to the smallest $|\Lambda|$ (d). A mode $\mathbf{R}_n = \{ R_n^1, R_n^2, \ldots, R_n^N \}$ is represented by spheres centered at positions of atoms
$\mathbf{r}_i$ and having radii equal to $1 \times$ (b), $100 \times$ (c), and $10 \times |R_n^i|^2$ (d). [Reproduced from \citet{goetschy11b}.]
}}
\end{figure}
%%%%%%%%%%%%%%%%%%%%%%%%%%%

%%%%%%%%%%%%%  FIG %%%%%%%%%
\begin{figure}
\centering{
\includegraphics[angle=0,width=0.95\columnwidth]{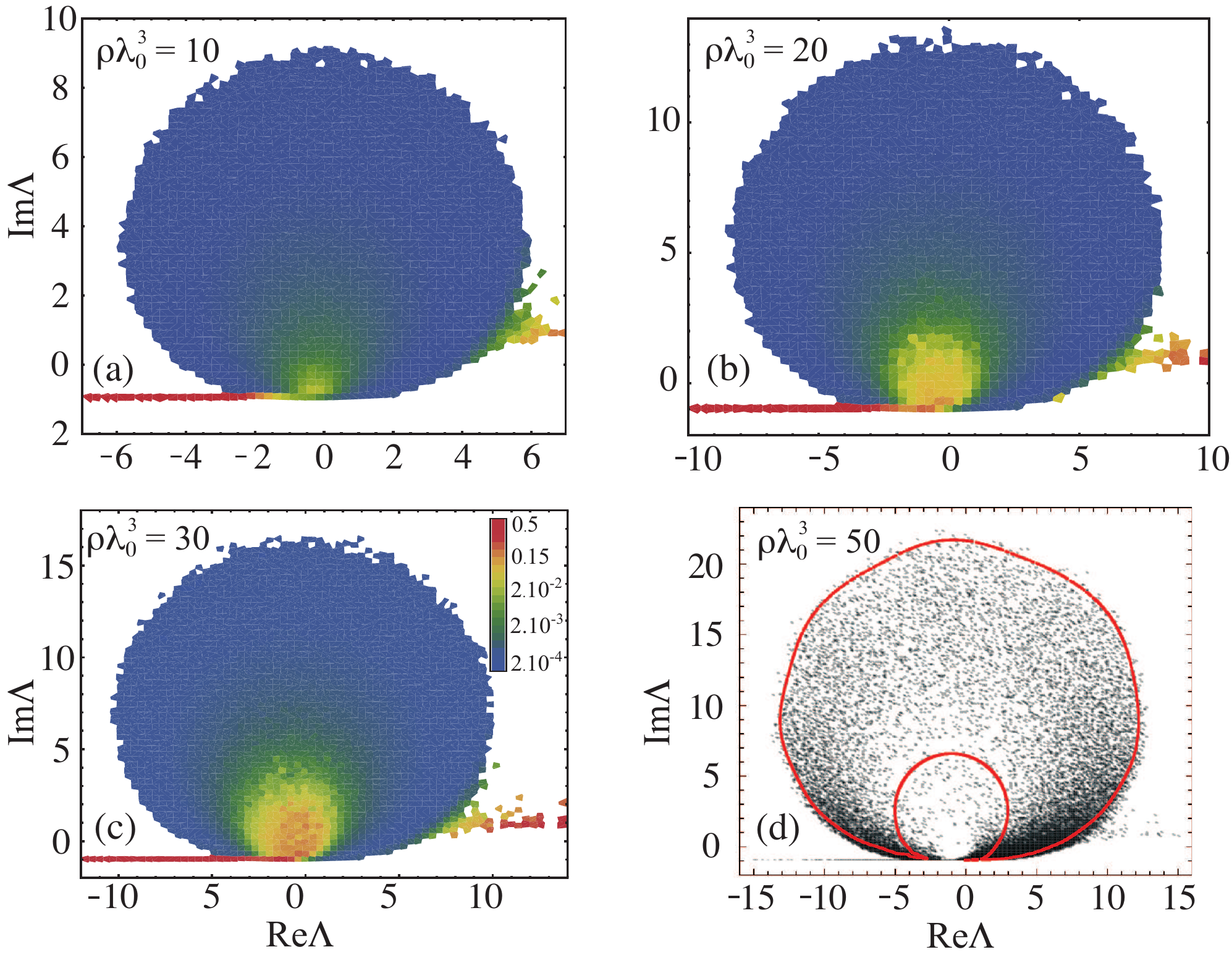}
\caption{
\label{figIPRGreen}
(a)--(c) Density plots of the logarithm of the average inverse participation ratio of eigenvectors of the Green's matrix (\ref{ERMGreen}). For each of these plots, we found eigenvalues of 10 different random realizations of $10^4 \times 10^4$ Green's matrix numerically (with points $\vec{r}_i$ randomly chosen inside a sphere of radius $R$), computed their IPRs using Eq.\ (\ref{iprNH}), and then determined $\mathrm{IPR}(\Lambda)$ by integrating Eq.\ (\ref{iprNH2}) over a small area $(\Delta \Lambda)^2$ around $\Lambda$, for a grid of $\Lambda$'s on the complex plane. (d) Density plot of the logarithm of the eigenvalue density of the Green's matrix (\ref{ERMGreen}). The solid red line represents the borderline of the support of eigenvalue density following from Eqs.~(\ref{contourT3}) and (\ref{contourT4}).
}}
\end{figure}
%%%%%%%%%%%%%%%%%%%%%%%%%%

A more direct way of studying Anderson localization is to look at the eigenvectors of the Green's matrix. At high enough density $\rho \lambda_0^3$, the eigenvectors of three different types coexist \cite{goetschy11b}, as we illustrate in Fig.\ \ref{ModesOneShot}. The eigenvectors of type (d) appear only at $\rho \lambda_0^3 \gtrsim 20$ and are localized due to disorder. A quantitative measure of degree of localization of an eigenvector $R_n=\{R_{n}(\vec{r}_1), \dots, R_{n}(\vec{r}_N)\}$ can be obtained by computing its inverse participation ratio (IPR)
\be
\textrm{IPR}_{n} = \frac{\sum_{i=1}^N |R_{n}(\vec{r}_i)|^4}{\left[\sum_{i=1}^N |R_{n}(\vec{r}_i)|^2 \right]^2}.
\label{iprNH}
\ee
An eigenvector extended over $M$ points is characterized by $\mathrm{IPR} \sim 1/M$. The average value of IPR corresponding to eigenvectors with eigenvalues in the vicinity of $\Lambda$ can be defined as
\be
\textrm{IPR}(\Lambda) = \frac{1}{Np(\Lambda)}
\left\langle
\sum\limits_{n=1}^{N} \mathrm{IPR}_{n}\;
\delta^2(\Lambda - \Lambda_{n})
\right \rangle,
\label{iprNH2}
\ee
where averaging is over all possible configurations of $N$ points in a sphere. The numerical analysis of the average IPR defined by this equation reveals the following scenario \cite{goetschy11a}. At low density $\rho \lambda_0^3 \lesssim 10$, $\mathrm{IPR} \simeq 2/N$ for all eigenvectors except those corresponding to the eigenvalues that belong to spiral branches  [see Fig.\ \ref{FigGreencontours}(a) and (b) and section \ref{SecBranches}] for which $\mathrm{IPR} \simeq \frac12$. These states are localized on pairs of points that are very close together and correspond to proximity resonances \cite{rusek00}. The prefactor $2$ in the result for $\mathrm{IPR}$ of extended eigenvectors is due to the Gaussian statistics of eigenvectors at low densities. For $\rho \lambda_0^3 \gtrsim 10$ [Fig.\ \ref{figIPRGreen}(a) and (b)], IPR starts to grow in the vicinity of $\Lambda = 0$ and reaches maximum values $\sim 0.1$ at $\rho \lambda_0^3 \simeq 30$ [Fig.\ \ref{figIPRGreen}(c)]. Contrary to common belief \cite{rusek00}, neither localized states necessarily have $\mathrm{Im} \Lambda$ close to $-1$, nor states with $\mathrm{Im} \Lambda \simeq -1$ are always localized, as can be seen from Fig.\ \ref{figIPRGreen}(c). For $\rho \lambda_0^3 > 30$, the localized states start to disappear and a hole opens in the eigenvalue density. As can be seen from the comparison of Fig.\ \ref{figIPRGreen}(c) and (d), it is quite remarkable that the opening of the hole in $p(\Lambda)$ [Fig.\ \ref{figIPRGreen}(d)] proceeds by disappearance of localized states [i.e. of states with $\mathrm{IPR} \gg 1/N$ in Fig.\ \ref{figIPRGreen}(c)]. Although this suggests a link between the hole in $p(\Lambda)$ and Anderson localization, further work is needed to arrive at any definitive conclusions \cite{goetschy11a}.

\section{Conclusion and perspectives}
\label{Conclusion}

In this paper, we tried to give an overview of the current state of the art of the Euclidean random matrix theory. We presented a number of approaches that can be used to deal with ERMs and gave examples of the practical use of ERM theory to understand real physical systems. It is clear, however, that much remains to be done in this field. Our understanding of mathematical properties of ERMs is far from being complete and their applications could certainly be much wider than the few examples that we considered.

One possible direction of research in which progress is needed concerns the representation of an arbitrary Euclidean matrix as $A = H T H^{\dagger}$ that we heavily exploited in this review. The assumption that the matrix $H$ has i.i.d. elements $H_{i \alpha}$ allowed us to progress in the calculation but it is certainly not sufficient. Correlations between $H_{i \alpha}$ could be taken into account by analogy with the work on correlated Wishart matrices $HH^\dagger$ \cite{marchenko67, sengupta99, tulino04}. Another way of accounting for these correlations is to use the Dyson gas picture in which the possibility for two points $\vec{r}_i$ to be very close to each other can be taken into account when obtaining the one-body potential $V^g$, leading to correlated $H_{i \alpha}$.

Another way of extending the analysis of ERMs reviewed here is to use the representation $A = HTH^\dagger$ to access quantities that are more advanced than the eigenvalue density $p(\Lambda)$. For Hermitian matrices, the connected part of the two-point correlation function
\begin{eqnarray}
p_c(\Lambda, \Lambda') &=& \frac{1}{N^2} \langle\sum_{n=1}^N \sum_{n'=1}^N \delta(\Lambda-\Lambda_n) \delta(\Lambda'-\Lambda_{n'}) \rangle_c,\;\;\;\;\;\;\;\;
\end{eqnarray}
where $\moy{xy}_c=\moy{xy}-\moy{x}\moy{y}$, can be obtained from the two-point resolvent
\begin{align}
g_c(z, z')&=\frac{1}{N^2}\left\langle
\textrm{Tr}\frac{1}{z-A}\textrm{Tr}\frac{1}{z'-A}
\right\rangle_c
\\
&=\frac{1}{N^2}\partial_z\partial_{z'}\left\langle
\textrm{Tr}\log(z-A)\textrm{Tr}\log(z'-A)
\right\rangle_c
\end{align}
by invoking the relation \cite{brezin94}:
\begin{eqnarray}
\label{pcVSgc}
p_c(\Lambda, \Lambda') &=& -\frac{1}{4\pi^2}[g_c(+,+)+g_c(-,-)
\nonumber \\
&-& g_c(+,-)-g_c(-,+)],
\end{eqnarray}
where we introduced a shorthand notation $g_c(\pm,\pm)=g_c(\Lambda\pm i\epsilon, \Lambda'\pm i\epsilon')$. Using an elegant diagrammatic approach, \citet{brezin95} showed that $g_c(z,z')$ can be expressed as
\be
\label{gcVsU}
g_c(z, z')=-\frac{1}{N^2}\partial_z\partial_{z'}\log\left[1-U(z,z') g(z)g(z')\right],
\ee
where $U(z,z')$ is the irreducible vertex that contains the sum of all irreducible diagrams contained in the expansion of $g_c(z,z')$. For ERMs $A=HTH^\dagger$, the diagrammatic method developed in Sec.\ \ref{SecERMSelfConsistent} can be used to obtain
\be
\label{UVertxERM}
U(z, z')=\frac{1}{N^2}\textrm{Tr}\left\{
\frac{\hat{T}^2}{\left[1-g(z)\hat{T}\right]\left[1-g(z')\hat{T}\right]}
\right\}
\ee
in the limit $N \to \infty$, with $g(z)$ the solution of Eq.~(\ref{SolgERM}). Inserting Eq.~(\ref{UVertxERM}) into Eq.~(\ref{gcVsU}) yields the two-point correlation function (\ref{pcVSgc}).

Finally, the application of the theory of Euclidean random matrices may be quite fruitful in those branches of physics where they naturally appear but were treated only numerically until now. In a metal, magnetic moments localized at random positions $\vec{r}_i$ interact indirectly via polarization of conduction electrons through the Ruderman-Kittel-Kasuya-Yosida (RKKY) potential. In the three-dimensional space, this interaction can be described by an ERM \cite{aristov97}
\be
\label{RKKYERM}
J_{ij} \propto \frac{\cos\left(2k_F\vert\mathbf{r}_i
-\mathbf{r}_j\vert\right)}{\vert\mathbf{r}_i-\mathbf{r}_j\vert^3},
\ee
where $k_F$ is the Fermi wave vector. Another example is the problem of light scattering in an ensemble of point scatterers at positions $\vec{r}_i$ that one wants to analyze with account for the vector character of light. It can be studied using an ERM composed of $3 \times 3$ blocks \cite{rusek96,goetschy11c}
\begin{eqnarray}
\label{DyadicERM}
G_{ij} &=& \frac{3}{2}(1-\delta_{ij})\frac{\exp(\textrm{i} k_0r_{ij})}{k_0r_{ij}}\left\{
\left[ 1 + \frac{\mathrm{i}}{k r_{ij}} - \frac{1}{(k r_{ij})^2}  \right] I_3 \right.
\nonumber \\
&-& \left. \left[ 1 + \frac{\mathrm{3i}}{k r_{ij}} - \frac{3}{(k r_{ij})^2} \right] \frac{\vec{r}_{ij} \otimes \vec{r}_{ij}}{r_{ij}^2}
\right\},
\end{eqnarray}
where $\vec{r}_{ij}=\vec{r}_i-\vec{r}_j$. We are sure that, besides these two examples to which the methods described in the present review can be readily applied, other physical problems await to be studied in the framework of ERM models.

\acknowledgements
SES acknowledges support by the Federal Program for Scientific
and Scientific-Pedagogical Personnel of Innovative Russia for 2009--2013 (contract No. 14.B37.21.1938).

\bibliographystyle{apsrmp}

%\bibliography{nc,cds,sw,connes}
\bibliography{erm_references}

\begin{thebibliography}{133}
\expandafter\ifx\csname natexlab\endcsname\relax\def\natexlab#1{#1}\fi
\expandafter\ifx\csname bibnamefont\endcsname\relax
  \def\bibnamefont#1{#1}\fi
\expandafter\ifx\csname bibfnamefont\endcsname\relax
  \def\bibfnamefont#1{#1}\fi
\expandafter\ifx\csname citenamefont\endcsname\relax
  \def\citenamefont#1{#1}\fi
\expandafter\ifx\csname url\endcsname\relax
  \def\url#1{\texttt{#1}}\fi
\expandafter\ifx\csname urlprefix\endcsname\relax\def\urlprefix{URL }\fi
\providecommand{\bibinfo}[2]{#2}
\providecommand{\eprint}[2][]{\url{#2}}

\bibitem[{\citenamefont{Akkermans} \emph{et~al.}(2008)\citenamefont{Akkermans,
  Gero, and Kaiser}}]{akkermans08}
\bibinfo{author}{\bibnamefont{Akkermans}, \bibfnamefont{E.}},
  \bibinfo{author}{\bibfnamefont{A.}~\bibnamefont{Gero}}, and
  \bibinfo{author}{\bibfnamefont{R.}~\bibnamefont{Kaiser}},
  \bibinfo{year}{2008}, \bibinfo{journal}{Phys. Rev. Lett.}
  \textbf{\bibinfo{volume}{101}}, \bibinfo{pages}{103602}.

\bibitem[{\citenamefont{Amir} \emph{et~al.}(2012)\citenamefont{Amir, Krich,
  Vitelli, Oreg, and Imry}}]{amir13}
\bibinfo{author}{\bibnamefont{Amir}, \bibfnamefont{A.}},
  \bibinfo{author}{\bibfnamefont{J.}~\bibnamefont{Krich}},
  \bibinfo{author}{\bibfnamefont{V.}~\bibnamefont{Vitelli}},
  \bibinfo{author}{\bibfnamefont{Y.}~\bibnamefont{Oreg}}, and
  \bibinfo{author}{\bibfnamefont{Y.}~\bibnamefont{Imry}}, \bibinfo{year}{2012},
  \eprint{arXiv:1209.2169}.

\bibitem[{\citenamefont{Amir} \emph{et~al.}(2008)\citenamefont{Amir, Oreg, and
  Imry}}]{amir08}
\bibinfo{author}{\bibnamefont{Amir}, \bibfnamefont{A.}},
  \bibinfo{author}{\bibfnamefont{Y.}~\bibnamefont{Oreg}}, and
  \bibinfo{author}{\bibfnamefont{Y.}~\bibnamefont{Imry}}, \bibinfo{year}{2008},
  \bibinfo{journal}{Phys. Rev. B} \textbf{\bibinfo{volume}{77}},
  \bibinfo{pages}{165207}.

\bibitem[{\citenamefont{Amir} \emph{et~al.}(2009)\citenamefont{Amir, Oreg, and
  Imry}}]{amir09}
\bibinfo{author}{\bibnamefont{Amir}, \bibfnamefont{A.}},
  \bibinfo{author}{\bibfnamefont{Y.}~\bibnamefont{Oreg}}, and
  \bibinfo{author}{\bibfnamefont{Y.}~\bibnamefont{Imry}}, \bibinfo{year}{2009},
  \bibinfo{journal}{Phys. Rev. Lett.} \textbf{\bibinfo{volume}{103}},
  \bibinfo{pages}{126403}.

\bibitem[{\citenamefont{Amir} \emph{et~al.}(2010)\citenamefont{Amir, Oreg, and
  Imry}}]{amir10}
\bibinfo{author}{\bibnamefont{Amir}, \bibfnamefont{A.}},
  \bibinfo{author}{\bibfnamefont{Y.}~\bibnamefont{Oreg}}, and
  \bibinfo{author}{\bibfnamefont{Y.}~\bibnamefont{Imry}}, \bibinfo{year}{2010},
  \bibinfo{journal}{Phys. Rev. Lett.} \textbf{\bibinfo{volume}{105}},
  \bibinfo{pages}{070601}.

\bibitem[{\citenamefont{Amir} \emph{et~al.}(2011)\citenamefont{Amir, Oreg, and
  Imry}}]{amir11}
\bibinfo{author}{\bibnamefont{Amir}, \bibfnamefont{A.}},
  \bibinfo{author}{\bibfnamefont{Y.}~\bibnamefont{Oreg}}, and
  \bibinfo{author}{\bibfnamefont{Y.}~\bibnamefont{Imry}}, \bibinfo{year}{2011},
  \bibinfo{journal}{Annu. Rev. Condens. Matter Phys.}
  \textbf{\bibinfo{volume}{2}}, \bibinfo{pages}{235}.

\bibitem[{\citenamefont{Anderson}(1958)}]{anderson58}
\bibinfo{author}{\bibnamefont{Anderson}, \bibfnamefont{P.~W.}},
  \bibinfo{year}{1958}, \bibinfo{journal}{Phys. Rev.}
  \textbf{\bibinfo{volume}{109}}, \bibinfo{pages}{1492}.

\bibitem[{\citenamefont{Antezza} \emph{et~al.}(2010)\citenamefont{Antezza,
  Castin, and Hutchinson}}]{antezza10}
\bibinfo{author}{\bibnamefont{Antezza}, \bibfnamefont{M.}},
  \bibinfo{author}{\bibfnamefont{Y.}~\bibnamefont{Castin}}, and
  \bibinfo{author}{\bibfnamefont{D.~A.~W.} \bibnamefont{Hutchinson}},
  \bibinfo{year}{2010}, \bibinfo{journal}{Phys. Rev. A}
  \textbf{\bibinfo{volume}{82}}, \bibinfo{pages}{043602}.

\bibitem[{\citenamefont{Aristov}(1997)}]{aristov97}
\bibinfo{author}{\bibnamefont{Aristov}, \bibfnamefont{D.~N.}},
  \bibinfo{year}{1997}, \bibinfo{journal}{Phys. Rev. B}
  \textbf{\bibinfo{volume}{55}}, \bibinfo{pages}{8064}.

\bibitem[{\citenamefont{Bai}(1999)}]{bai99}
\bibinfo{author}{\bibnamefont{Bai}, \bibfnamefont{Z.~D.}},
  \bibinfo{year}{1999}, \bibinfo{journal}{Stat. Sinica}
  \textbf{\bibinfo{volume}{9}}, \bibinfo{pages}{611}.

\bibitem[{\citenamefont{Baudouin} \emph{et~al.}(2013)\citenamefont{Baudouin,
  Mercadier, Guarrera, Guerin, and Kaiser}}]{baudouin13}
\bibinfo{author}{\bibnamefont{Baudouin}, \bibfnamefont{Q.}},
  \bibinfo{author}{\bibfnamefont{N.}~\bibnamefont{Mercadier}},
  \bibinfo{author}{\bibfnamefont{V.}~\bibnamefont{Guarrera}},
  \bibinfo{author}{\bibfnamefont{W.}~\bibnamefont{Guerin}}, and
  \bibinfo{author}{\bibfnamefont{R.}~\bibnamefont{Kaiser}},
  \bibinfo{year}{2013}, \eprint{arXiv:1301.0522}.

\bibitem[{\citenamefont{Beenakker}(1997)}]{beenakker97}
\bibinfo{author}{\bibnamefont{Beenakker}, \bibfnamefont{C.~W.~J.}},
  \bibinfo{year}{1997}, \bibinfo{journal}{Rev. Mod. Phys.}
  \textbf{\bibinfo{volume}{69}}, \bibinfo{pages}{731}.

\bibitem[{\citenamefont{Benassi} \emph{et~al.}(1996)\citenamefont{Benassi,
  Krisch, Masciovecchio, Mazzacurati, Monaco, Ruocco, Sette, and
  Verbeni}}]{benassi96}
\bibinfo{author}{\bibnamefont{Benassi}, \bibfnamefont{P.}},
  \bibinfo{author}{\bibfnamefont{M.}~\bibnamefont{Krisch}},
  \bibinfo{author}{\bibfnamefont{C.}~\bibnamefont{Masciovecchio}},
  \bibinfo{author}{\bibfnamefont{V.}~\bibnamefont{Mazzacurati}},
  \bibinfo{author}{\bibfnamefont{G.}~\bibnamefont{Monaco}},
  \bibinfo{author}{\bibfnamefont{G.}~\bibnamefont{Ruocco}},
  \bibinfo{author}{\bibfnamefont{F.}~\bibnamefont{Sette}}, and
  \bibinfo{author}{\bibfnamefont{R.}~\bibnamefont{Verbeni}},
  \bibinfo{year}{1996}, \bibinfo{journal}{Phys. Rev. Lett.}
  \textbf{\bibinfo{volume}{77}}, \bibinfo{pages}{3835}.

\bibitem[{\citenamefont{Bogomolny} \emph{et~al.}(2003)\citenamefont{Bogomolny,
  Bohigas, and Schmit}}]{bogomolny03}
\bibinfo{author}{\bibnamefont{Bogomolny}, \bibfnamefont{E.}},
  \bibinfo{author}{\bibfnamefont{O.}~\bibnamefont{Bohigas}}, and
  \bibinfo{author}{\bibfnamefont{C.}~\bibnamefont{Schmit}},
  \bibinfo{year}{2003}, \bibinfo{journal}{J. Phys. A: Math. Gen.}
  \textbf{\bibinfo{volume}{36}}, \bibinfo{pages}{3595}.

\bibitem[{\citenamefont{Br{\'e}zin and Zee}(1994)}]{brezin94}
\bibinfo{author}{\bibnamefont{Br{\'e}zin}, \bibfnamefont{E.}}, and
  \bibinfo{author}{\bibfnamefont{A.}~\bibnamefont{Zee}}, \bibinfo{year}{1994},
  \bibinfo{journal}{Phys. Rev. E} \textbf{\bibinfo{volume}{49}},
  \bibinfo{pages}{2588}.

\bibitem[{\citenamefont{Br{\'e}zin and Zee}(1995)}]{brezin95}
\bibinfo{author}{\bibnamefont{Br{\'e}zin}, \bibfnamefont{E.}}, and
  \bibinfo{author}{\bibfnamefont{A.}~\bibnamefont{Zee}}, \bibinfo{year}{1995},
  \bibinfo{journal}{Nucl. Phys. B} \textbf{\bibinfo{volume}{453}},
  \bibinfo{pages}{531}.

\bibitem[{\citenamefont{Brody} \emph{et~al.}(1981)\citenamefont{Brody, Flores,
  French, Mello, Pandey, and Wong}}]{brody81}
\bibinfo{author}{\bibnamefont{Brody}, \bibfnamefont{T.~A.}},
  \bibinfo{author}{\bibfnamefont{J.}~\bibnamefont{Flores}},
  \bibinfo{author}{\bibfnamefont{J.~B.} \bibnamefont{French}},
  \bibinfo{author}{\bibfnamefont{P.~A.} \bibnamefont{Mello}},
  \bibinfo{author}{\bibfnamefont{A.}~\bibnamefont{Pandey}}, and
  \bibinfo{author}{\bibfnamefont{S.~S.~M.} \bibnamefont{Wong}},
  \bibinfo{year}{1981}, \bibinfo{journal}{Rev. Mod. Phys.}
  \textbf{\bibinfo{volume}{53}}, \bibinfo{pages}{385}.

\bibitem[{\citenamefont{Burda} \emph{et~al.}(2011)\citenamefont{Burda, Janik,
  and Nowak}}]{burda11}
\bibinfo{author}{\bibnamefont{Burda}, \bibfnamefont{Z.}},
  \bibinfo{author}{\bibfnamefont{R.~A.} \bibnamefont{Janik}}, and
  \bibinfo{author}{\bibfnamefont{M.~A.} \bibnamefont{Nowak}},
  \bibinfo{year}{2011}, \bibinfo{journal}{Phys. Rev. E}
  \textbf{\bibinfo{volume}{84}}, \bibinfo{pages}{061125}.

\bibitem[{\citenamefont{Cao}(2005)}]{cao05}
\bibinfo{author}{\bibnamefont{Cao}, \bibfnamefont{H.}}, \bibinfo{year}{2005},
  \bibinfo{journal}{J. Phys. A: Math. Gen.} \textbf{\bibinfo{volume}{38}},
  \bibinfo{pages}{10497}.

\bibitem[{\citenamefont{Chalker and Mehlig}(1998)}]{chalker98}
\bibinfo{author}{\bibnamefont{Chalker}, \bibfnamefont{J.~T.}}, and
  \bibinfo{author}{\bibfnamefont{B.}~\bibnamefont{Mehlig}},
  \bibinfo{year}{1998}, \bibinfo{journal}{Phys. Rev. Lett.}
  \textbf{\bibinfo{volume}{81}}, \bibinfo{pages}{3367}.

\bibitem[{\citenamefont{Chalker and Wang}(1997)}]{chalker97}
\bibinfo{author}{\bibnamefont{Chalker}, \bibfnamefont{J.~T.}}, and
  \bibinfo{author}{\bibfnamefont{Z.~J.} \bibnamefont{Wang}},
  \bibinfo{year}{1997}, \bibinfo{journal}{Phys. Rev. Lett.}
  \textbf{\bibinfo{volume}{79}}, \bibinfo{pages}{1797}.

\bibitem[{\citenamefont{Chamon and Mudry}(2001)}]{chamon01}
\bibinfo{author}{\bibnamefont{Chamon}, \bibfnamefont{C.}}, and
  \bibinfo{author}{\bibfnamefont{C.}~\bibnamefont{Mudry}},
  \bibinfo{year}{2001}, \bibinfo{journal}{Phys. Rev. B}
  \textbf{\bibinfo{volume}{63}}, \bibinfo{pages}{100503(R)}.

\bibitem[{\citenamefont{Chumakov} \emph{et~al.}(2011)\citenamefont{Chumakov,
  Monaco, Monaco, Crichton, Bosak, R\"uffer, Meyer, Kargl, Comez, Fioretto,
  Giefers, Roitsch} \emph{et~al.}}]{chumakov11}
\bibinfo{author}{\bibnamefont{Chumakov}, \bibfnamefont{A.~I.}},
  \bibinfo{author}{\bibfnamefont{G.}~\bibnamefont{Monaco}},
  \bibinfo{author}{\bibfnamefont{A.}~\bibnamefont{Monaco}},
  \bibinfo{author}{\bibfnamefont{W.~A.} \bibnamefont{Crichton}},
  \bibinfo{author}{\bibfnamefont{A.}~\bibnamefont{Bosak}},
  \bibinfo{author}{\bibfnamefont{R.}~\bibnamefont{R\"uffer}},
  \bibinfo{author}{\bibfnamefont{A.}~\bibnamefont{Meyer}},
  \bibinfo{author}{\bibfnamefont{F.}~\bibnamefont{Kargl}},
  \bibinfo{author}{\bibfnamefont{L.}~\bibnamefont{Comez}},
  \bibinfo{author}{\bibfnamefont{D.}~\bibnamefont{Fioretto}},
  \bibinfo{author}{\bibfnamefont{H.}~\bibnamefont{Giefers}},
  \bibinfo{author}{\bibfnamefont{S.}~\bibnamefont{Roitsch}}, \emph{et~al.},
  \bibinfo{year}{2011}, \bibinfo{journal}{Phys. Rev. Lett.}
  \textbf{\bibinfo{volume}{106}}, \bibinfo{pages}{225501}.

\bibitem[{\citenamefont{Ciliberti} \emph{et~al.}(2003)\citenamefont{Ciliberti,
  Grigera, Martin-Mayor, Parisi, and Verrocchio}}]{ciliberti03}
\bibinfo{author}{\bibnamefont{Ciliberti}, \bibfnamefont{S.}},
  \bibinfo{author}{\bibfnamefont{T.}~\bibnamefont{Grigera}},
  \bibinfo{author}{\bibfnamefont{V.}~\bibnamefont{Martin-Mayor}},
  \bibinfo{author}{\bibfnamefont{G.}~\bibnamefont{Parisi}}, and
  \bibinfo{author}{\bibfnamefont{P.}~\bibnamefont{Verrocchio}},
  \bibinfo{year}{2003}, \bibinfo{journal}{J. Chem. Phys}
  \textbf{\bibinfo{volume}{119}}(\bibinfo{number}{16}), \bibinfo{pages}{8577}.

\bibitem[{\citenamefont{Ciliberti} \emph{et~al.}(2005)\citenamefont{Ciliberti,
  Grigera, Martin-Mayor, Parisi, and Verrocchio}}]{ciliberti05}
\bibinfo{author}{\bibnamefont{Ciliberti}, \bibfnamefont{S.}},
  \bibinfo{author}{\bibfnamefont{T.~S.} \bibnamefont{Grigera}},
  \bibinfo{author}{\bibfnamefont{V.}~\bibnamefont{Martin-Mayor}},
  \bibinfo{author}{\bibfnamefont{G.}~\bibnamefont{Parisi}}, and
  \bibinfo{author}{\bibfnamefont{P.}~\bibnamefont{Verrocchio}},
  \bibinfo{year}{2005}, \bibinfo{journal}{Phys. Rev. B}
  \textbf{\bibinfo{volume}{71}}, \bibinfo{pages}{153104}.

\bibitem[{\citenamefont{Das} \emph{et~al.}(2008)\citenamefont{Das, Agarwal, and
  Scully}}]{das08}
\bibinfo{author}{\bibnamefont{Das}, \bibfnamefont{S.}},
  \bibinfo{author}{\bibfnamefont{G.~S.} \bibnamefont{Agarwal}}, and
  \bibinfo{author}{\bibfnamefont{M.~O.} \bibnamefont{Scully}},
  \bibinfo{year}{2008}, \bibinfo{journal}{Phys. Rev. Lett.}
  \textbf{\bibinfo{volume}{101}}, \bibinfo{pages}{153601}.

\bibitem[{\citenamefont{Dasgupta and Ma}(1980)}]{dasgupta80}
\bibinfo{author}{\bibnamefont{Dasgupta}, \bibfnamefont{C.}}, and
  \bibinfo{author}{\bibfnamefont{S.~K.} \bibnamefont{Ma}},
  \bibinfo{year}{1980}, \bibinfo{journal}{Phys. Rev. B}
  \textbf{\bibinfo{volume}{22}}, \bibinfo{pages}{1305}.

\bibitem[{\citenamefont{Dyson}(1962{\natexlab{a}})}]{dyson62a}
\bibinfo{author}{\bibnamefont{Dyson}, \bibfnamefont{F.~J.}},
  \bibinfo{year}{1962}{\natexlab{a}}, \bibinfo{journal}{J. Math. Phys.}
  \textbf{\bibinfo{volume}{3}}, \bibinfo{pages}{140}.

\bibitem[{\citenamefont{Dyson}(1962{\natexlab{b}})}]{dyson62b}
\bibinfo{author}{\bibnamefont{Dyson}, \bibfnamefont{F.~J.}},
  \bibinfo{year}{1962}{\natexlab{b}}, \bibinfo{journal}{J. Math. Phys.}
  \textbf{\bibinfo{volume}{3}}, \bibinfo{pages}{157}.

\bibitem[{\citenamefont{Dyson}(1962{\natexlab{c}})}]{dyson62c}
\bibinfo{author}{\bibnamefont{Dyson}, \bibfnamefont{F.~J.}},
  \bibinfo{year}{1962}{\natexlab{c}}, \bibinfo{journal}{J. Math. Phys.}
  \textbf{\bibinfo{volume}{3}}, \bibinfo{pages}{166}.

\bibitem[{\citenamefont{Dyson}(1972)}]{dyson72}
\bibinfo{author}{\bibnamefont{Dyson}, \bibfnamefont{F.~J.}},
  \bibinfo{year}{1972}, \bibinfo{journal}{J. Math. Phys.}
  \textbf{\bibinfo{volume}{13}}, \bibinfo{pages}{90}.

\bibitem[{\citenamefont{Edwards and Jones}(1976)}]{edwards79}
\bibinfo{author}{\bibnamefont{Edwards}, \bibfnamefont{S.~F.}}, and
  \bibinfo{author}{\bibfnamefont{R.~C.} \bibnamefont{Jones}},
  \bibinfo{year}{1976}, \bibinfo{journal}{J. Phys. A: Math. Gen.}
  \textbf{\bibinfo{volume}{9}}, \bibinfo{pages}{1595}.

\bibitem[{\citenamefont{Efetov}(1997)}]{efetov97}
\bibinfo{author}{\bibnamefont{Efetov}, \bibfnamefont{K.~B.}},
  \bibinfo{year}{1997}, \emph{\bibinfo{title}{Supersymmetry in Disorder and
  Chaos}} (\bibinfo{publisher}{Cambridge University Press, Cambridge}).

\bibitem[{\citenamefont{Ernst}(1968)}]{ernst69}
\bibinfo{author}{\bibnamefont{Ernst}, \bibfnamefont{V.}}, \bibinfo{year}{1968},
  \bibinfo{journal}{Z. Phys.} \textbf{\bibinfo{volume}{218}},
  \bibinfo{pages}{111}.

\bibitem[{\citenamefont{Feinberg}(2006)}]{feinberg06}
\bibinfo{author}{\bibnamefont{Feinberg}, \bibfnamefont{J.}},
  \bibinfo{year}{2006}, \bibinfo{journal}{J. Phys. A: Math. Gen.}
  \textbf{\bibinfo{volume}{39}}, \bibinfo{pages}{10029}.

\bibitem[{\citenamefont{Feinberg} \emph{et~al.}(2001)\citenamefont{Feinberg,
  Scalettar, and Zee}}]{feinberg01}
\bibinfo{author}{\bibnamefont{Feinberg}, \bibfnamefont{J.}},
  \bibinfo{author}{\bibfnamefont{R.}~\bibnamefont{Scalettar}}, and
  \bibinfo{author}{\bibfnamefont{A.}~\bibnamefont{Zee}}, \bibinfo{year}{2001},
  \bibinfo{journal}{J. Math. Phys.} \textbf{\bibinfo{volume}{42}},
  \bibinfo{pages}{5712}.

\bibitem[{\citenamefont{Feinberg and Zee}(1997{\natexlab{a}})}]{feinberg97a}
\bibinfo{author}{\bibnamefont{Feinberg}, \bibfnamefont{J.}}, and
  \bibinfo{author}{\bibfnamefont{A.}~\bibnamefont{Zee}},
  \bibinfo{year}{1997}{\natexlab{a}}, \bibinfo{journal}{Nucl. Phys. B}
  \textbf{\bibinfo{volume}{501}}, \bibinfo{pages}{643}.

\bibitem[{\citenamefont{Feinberg and Zee}(1997{\natexlab{b}})}]{feinberg97b}
\bibinfo{author}{\bibnamefont{Feinberg}, \bibfnamefont{J.}}, and
  \bibinfo{author}{\bibfnamefont{A.}~\bibnamefont{Zee}},
  \bibinfo{year}{1997}{\natexlab{b}}, \bibinfo{journal}{Nucl. Phys. B}
  \textbf{\bibinfo{volume}{504}}, \bibinfo{pages}{579}.

\bibitem[{\citenamefont{Feinberg and Zee}(1999{\natexlab{a}})}]{feinberg99}
\bibinfo{author}{\bibnamefont{Feinberg}, \bibfnamefont{J.}}, and
  \bibinfo{author}{\bibfnamefont{A.}~\bibnamefont{Zee}},
  \bibinfo{year}{1999}{\natexlab{a}}, \bibinfo{journal}{Phys. Rev. E}
  \textbf{\bibinfo{volume}{59}}, \bibinfo{pages}{6433}.

\bibitem[{\citenamefont{Feinberg and Zee}(1999{\natexlab{b}})}]{feinberg99b}
\bibinfo{author}{\bibnamefont{Feinberg}, \bibfnamefont{J.}}, and
  \bibinfo{author}{\bibfnamefont{A.}~\bibnamefont{Zee}},
  \bibinfo{year}{1999}{\natexlab{b}}, \bibinfo{journal}{Nucl. Phys. B}
  \textbf{\bibinfo{volume}{552}}, \bibinfo{pages}{599}.

\bibitem[{\citenamefont{Fisher}(1995)}]{fisher95}
\bibinfo{author}{\bibnamefont{Fisher}, \bibfnamefont{D.~S.}},
  \bibinfo{year}{1995}, \bibinfo{journal}{Phys. Rev. B}
  \textbf{\bibinfo{volume}{51}}, \bibinfo{pages}{6411}.

\bibitem[{\citenamefont{Froufe-P\'erez}
  \emph{et~al.}(2009)\citenamefont{Froufe-P\'erez, Gu\'erin, Carminati, , and
  Kaiser}}]{froufeperez09}
\bibinfo{author}{\bibnamefont{Froufe-P\'erez}, \bibfnamefont{L.~S.}},
  \bibinfo{author}{\bibfnamefont{W.}~\bibnamefont{Gu\'erin}},
  \bibinfo{author}{\bibfnamefont{R.}~\bibnamefont{Carminati}}, , and
  \bibinfo{author}{\bibfnamefont{R.}~\bibnamefont{Kaiser}},
  \bibinfo{year}{2009}, \bibinfo{journal}{Phys. Rev. Lett.}
  \textbf{\bibinfo{volume}{102}}, \bibinfo{pages}{173903}.

\bibitem[{\citenamefont{Fyodorov} \emph{et~al.}(2005)\citenamefont{Fyodorov,
  Savin, and Sommers}}]{fyodorov05}
\bibinfo{author}{\bibnamefont{Fyodorov}, \bibfnamefont{Y.~V.}},
  \bibinfo{author}{\bibfnamefont{D.~V.} \bibnamefont{Savin}}, and
  \bibinfo{author}{\bibfnamefont{H.~J.} \bibnamefont{Sommers}},
  \bibinfo{year}{2005}, \bibinfo{journal}{J. Phys. A: Math. Gen.}
  \textbf{\bibinfo{volume}{38}}, \bibinfo{pages}{10731}.

\bibitem[{\citenamefont{Fyodorov and Sommers}(1997)}]{fyodorov97}
\bibinfo{author}{\bibnamefont{Fyodorov}, \bibfnamefont{Y.~V.}}, and
  \bibinfo{author}{\bibfnamefont{H.~J.} \bibnamefont{Sommers}},
  \bibinfo{year}{1997}, \bibinfo{journal}{J. Math. Phys.}
  \textbf{\bibinfo{volume}{38}}, \bibinfo{pages}{1918}.

\bibitem[{\citenamefont{Fyodorov and Sommers}(2003)}]{fyodorov03}
\bibinfo{author}{\bibnamefont{Fyodorov}, \bibfnamefont{Y.~V.}}, and
  \bibinfo{author}{\bibfnamefont{H.~J.} \bibnamefont{Sommers}},
  \bibinfo{year}{2003}, \bibinfo{journal}{J. Phys. A: Math. Gen.}
  \textbf{\bibinfo{volume}{36}}, \bibinfo{pages}{3303}.

\bibitem[{\citenamefont{Ganter and Schirmacher}(2011)}]{ganter11}
\bibinfo{author}{\bibnamefont{Ganter}, \bibfnamefont{C.}}, and
  \bibinfo{author}{\bibfnamefont{W.}~\bibnamefont{Schirmacher}},
  \bibinfo{year}{2011}, \bibinfo{journal}{Phil. Mag.}
  \textbf{\bibinfo{volume}{91}}, \bibinfo{pages}{1894}.

\bibitem[{\citenamefont{Ginibre}(1965)}]{ginibre65}
\bibinfo{author}{\bibnamefont{Ginibre}, \bibfnamefont{J.}},
  \bibinfo{year}{1965}, \bibinfo{journal}{J. Math. Phys.}
  \textbf{\bibinfo{volume}{6}}, \bibinfo{pages}{440}.

\bibitem[{\citenamefont{Girko}(1985)}]{girko85}
\bibinfo{author}{\bibnamefont{Girko}, \bibfnamefont{V.~L.}},
  \bibinfo{year}{1985}, \bibinfo{journal}{Theory Probab. Appl.}
  \textbf{\bibinfo{volume}{29}}, \bibinfo{pages}{694}.

\bibitem[{\citenamefont{Goetschy}(2011)}]{goetschy11c}
\bibinfo{author}{\bibnamefont{Goetschy}, \bibfnamefont{A.}},
  \bibinfo{year}{2011}, \emph{\bibinfo{title}{Light in Disordered Atomic
  Systems: Euclidean Matrix Theory of Random Lasing. PhD Thesis}}
  (\bibinfo{publisher}{J. Fourier University--Grenoble 1, France}).

\bibitem[{\citenamefont{Goetschy and
  Skipetrov}(2011{\natexlab{a}})}]{goetschy11b}
\bibinfo{author}{\bibnamefont{Goetschy}, \bibfnamefont{A.}}, and
  \bibinfo{author}{\bibfnamefont{S.~E.} \bibnamefont{Skipetrov}},
  \bibinfo{year}{2011}{\natexlab{a}}, \bibinfo{journal}{Europhys. Lett.}
  \textbf{\bibinfo{volume}{96}}, \bibinfo{pages}{34005}.

\bibitem[{\citenamefont{Goetschy and
  Skipetrov}(2011{\natexlab{b}})}]{goetschy11a}
\bibinfo{author}{\bibnamefont{Goetschy}, \bibfnamefont{A.}}, and
  \bibinfo{author}{\bibfnamefont{S.~E.} \bibnamefont{Skipetrov}},
  \bibinfo{year}{2011}{\natexlab{b}}, \bibinfo{journal}{Phys. Rev. E}
  \textbf{\bibinfo{volume}{84}}, \bibinfo{pages}{011150}.

\bibitem[{\citenamefont{Gradshteyn and Ryzhik}(1980)}]{gradshteyn80}
\bibinfo{author}{\bibnamefont{Gradshteyn}, \bibfnamefont{I.~S.}}, and
  \bibinfo{author}{\bibfnamefont{I.~M.} \bibnamefont{Ryzhik}},
  \bibinfo{year}{1980}, \emph{\bibinfo{title}{Table of Integrals, Series and
  Products}} (\bibinfo{publisher}{Academic Press, Inc., London}),
  \bibinfo{edition}{fourth} edition.

\bibitem[{\citenamefont{Gray}(2006)}]{gray06}
\bibinfo{author}{\bibnamefont{Gray}, \bibfnamefont{R.}}, \bibinfo{year}{2006},
  \emph{\bibinfo{title}{Toeplitz and Circulant Matrices: A review}}
  (\bibinfo{publisher}{Now Publishers, Delft}).

\bibitem[{\citenamefont{Gr\'{e}maud and Wellens}(2010)}]{gremaud10}
\bibinfo{author}{\bibnamefont{Gr\'{e}maud}, \bibfnamefont{B.}}, and
  \bibinfo{author}{\bibfnamefont{T.}~\bibnamefont{Wellens}},
  \bibinfo{year}{2010}, \bibinfo{journal}{Phys. Rev. Lett.}
  \textbf{\bibinfo{volume}{104}}, \bibinfo{pages}{133901}.

\bibitem[{\citenamefont{Grenander and Szeg{\"{o}}}(1958)}]{grenander58}
\bibinfo{author}{\bibnamefont{Grenander}, \bibfnamefont{U.}}, and
  \bibinfo{author}{\bibfnamefont{G.}~\bibnamefont{Szeg{\"{o}}}},
  \bibinfo{year}{1958}, \emph{\bibinfo{title}{Toeplitz Forms and Their
  Applications}} (\bibinfo{publisher}{University of California Press, Berkeley
  and Los Angeles}).

\bibitem[{\citenamefont{Grenet} \emph{et~al.}(2007)\citenamefont{Grenet,
  Delahaye, Sabra, and Gay}}]{grenet07}
\bibinfo{author}{\bibnamefont{Grenet}, \bibfnamefont{T.}},
  \bibinfo{author}{\bibfnamefont{J.}~\bibnamefont{Delahaye}},
  \bibinfo{author}{\bibfnamefont{M.}~\bibnamefont{Sabra}}, and
  \bibinfo{author}{\bibfnamefont{F.}~\bibnamefont{Gay}}, \bibinfo{year}{2007},
  \bibinfo{journal}{Eur. Phys. J. B} \textbf{\bibinfo{volume}{56}},
  \bibinfo{pages}{183}.

\bibitem[{\citenamefont{Grigera} \emph{et~al.}(2011)\citenamefont{Grigera,
  Martin-Mayor, Parisi, Urbani, and Verrocchio}}]{grigera11}
\bibinfo{author}{\bibnamefont{Grigera}, \bibfnamefont{T.~S.}},
  \bibinfo{author}{\bibfnamefont{V.}~\bibnamefont{Martin-Mayor}},
  \bibinfo{author}{\bibfnamefont{G.}~\bibnamefont{Parisi}},
  \bibinfo{author}{\bibfnamefont{P.}~\bibnamefont{Urbani}}, and
  \bibinfo{author}{\bibfnamefont{P.}~\bibnamefont{Verrocchio}},
  \bibinfo{year}{2011}, \bibinfo{journal}{J. Stat. Mech.}
  \textbf{\bibinfo{volume}{11}}, \bibinfo{pages}{P02015}.

\bibitem[{\citenamefont{Grigera}
  \emph{et~al.}(2001{\natexlab{a}})\citenamefont{Grigera, Martin-Mayor, Parisi,
  and Verrocchio}}]{grigera01a}
\bibinfo{author}{\bibnamefont{Grigera}, \bibfnamefont{T.~S.}},
  \bibinfo{author}{\bibfnamefont{V.}~\bibnamefont{Martin-Mayor}},
  \bibinfo{author}{\bibfnamefont{G.}~\bibnamefont{Parisi}}, and
  \bibinfo{author}{\bibfnamefont{P.}~\bibnamefont{Verrocchio}},
  \bibinfo{year}{2001}{\natexlab{a}}, \bibinfo{journal}{Phys. Rev. Lett.}
  \textbf{\bibinfo{volume}{87}}, \bibinfo{pages}{085502}.

\bibitem[{\citenamefont{Grigera} \emph{et~al.}(2002)\citenamefont{Grigera,
  Martin-Mayor, Parisi, and Verrocchio}}]{grigera02}
\bibinfo{author}{\bibnamefont{Grigera}, \bibfnamefont{T.~S.}},
  \bibinfo{author}{\bibfnamefont{V.}~\bibnamefont{Martin-Mayor}},
  \bibinfo{author}{\bibfnamefont{G.}~\bibnamefont{Parisi}}, and
  \bibinfo{author}{\bibfnamefont{P.}~\bibnamefont{Verrocchio}},
  \bibinfo{year}{2002}, \bibinfo{journal}{J. Phys.: Condens. Matter}
  \textbf{\bibinfo{volume}{14}}, \bibinfo{pages}{2167}.

\bibitem[{\citenamefont{Grigera} \emph{et~al.}(2003)\citenamefont{Grigera,
  Martin-Mayor, Parisi, and Verrocchio}}]{grigera03}
\bibinfo{author}{\bibnamefont{Grigera}, \bibfnamefont{T.~S.}},
  \bibinfo{author}{\bibfnamefont{V.}~\bibnamefont{Martin-Mayor}},
  \bibinfo{author}{\bibfnamefont{G.}~\bibnamefont{Parisi}}, and
  \bibinfo{author}{\bibfnamefont{P.}~\bibnamefont{Verrocchio}},
  \bibinfo{year}{2003}, \bibinfo{journal}{Nature}
  \textbf{\bibinfo{volume}{422}}, \bibinfo{pages}{289}.

\bibitem[{\citenamefont{Grigera}
  \emph{et~al.}(2001{\natexlab{b}})\citenamefont{Grigera, M\'{e}zard, Parisi,
  and Verrocchio}}]{grigera01b}
\bibinfo{author}{\bibnamefont{Grigera}, \bibfnamefont{T.~S.}},
  \bibinfo{author}{\bibfnamefont{M.}~\bibnamefont{M\'{e}zard}},
  \bibinfo{author}{\bibfnamefont{G.}~\bibnamefont{Parisi}}, and
  \bibinfo{author}{\bibfnamefont{P.}~\bibnamefont{Verrocchio}},
  \bibinfo{year}{2001}{\natexlab{b}}, \bibinfo{journal}{J. Chem. Phys.}
  \textbf{\bibinfo{volume}{114}}, \bibinfo{pages}{8068}.

\bibitem[{\citenamefont{Gross and Haroche}(1982)}]{gross82}
\bibinfo{author}{\bibnamefont{Gross}, \bibfnamefont{M.}}, and
  \bibinfo{author}{\bibfnamefont{S.}~\bibnamefont{Haroche}},
  \bibinfo{year}{1982}, \bibinfo{journal}{Phys. Rep.}
  \textbf{\bibinfo{volume}{93}}, \bibinfo{pages}{301}.

\bibitem[{\citenamefont{Guhr} \emph{et~al.}(1998)\citenamefont{Guhr,
  M{\"u}ller-Groeling, and Weidenm{\"u}ller}}]{guhr98}
\bibinfo{author}{\bibnamefont{Guhr}, \bibfnamefont{T.}},
  \bibinfo{author}{\bibfnamefont{A.}~\bibnamefont{M{\"u}ller-Groeling}}, and
  \bibinfo{author}{\bibfnamefont{H.~A.} \bibnamefont{Weidenm{\"u}ller}},
  \bibinfo{year}{1998}, \bibinfo{journal}{Phys. Rep.}
  \textbf{\bibinfo{volume}{299}}, \bibinfo{pages}{189}.

\bibitem[{\citenamefont{Haake}(2010)}]{haake10}
\bibinfo{author}{\bibnamefont{Haake}, \bibfnamefont{F.}}, \bibinfo{year}{2010},
  \emph{\bibinfo{title}{Quantum Signatures of Chaos}}
  (\bibinfo{publisher}{Springer, Heidelberg}), \bibinfo{edition}{third}
  edition.

\bibitem[{\citenamefont{Haake} \emph{et~al.}(1992)\citenamefont{Haake,
  Izrailev, Lehmann, Saher, and Sommers}}]{haake92}
\bibinfo{author}{\bibnamefont{Haake}, \bibfnamefont{F.}},
  \bibinfo{author}{\bibfnamefont{F.}~\bibnamefont{Izrailev}},
  \bibinfo{author}{\bibfnamefont{N.}~\bibnamefont{Lehmann}},
  \bibinfo{author}{\bibfnamefont{D.}~\bibnamefont{Saher}}, and
  \bibinfo{author}{\bibfnamefont{H.~J.} \bibnamefont{Sommers}},
  \bibinfo{year}{1992}, \bibinfo{journal}{Z. Phys. B}
  \textbf{\bibinfo{volume}{88}}, \bibinfo{pages}{359}.

\bibitem[{\citenamefont{Hansen and McDonald}(1986)}]{hansen86}
\bibinfo{author}{\bibnamefont{Hansen}, \bibfnamefont{J.~P.}}, and
  \bibinfo{author}{\bibfnamefont{I.~R.} \bibnamefont{McDonald}},
  \bibinfo{year}{1986}, \emph{\bibinfo{title}{Theory of Simple Liquids}}
  (\bibinfo{publisher}{Academic, London}).

\bibitem[{\citenamefont{Hatano and Nelson}(1996)}]{hatano96}
\bibinfo{author}{\bibnamefont{Hatano}, \bibfnamefont{N.}}, and
  \bibinfo{author}{\bibfnamefont{D.~R.} \bibnamefont{Nelson}},
  \bibinfo{year}{1996}, \bibinfo{journal}{Phys. Rev. Lett.}
  \textbf{\bibinfo{volume}{77}}, \bibinfo{pages}{570}.

\bibitem[{\citenamefont{Heller}(1996)}]{heller96}
\bibinfo{author}{\bibnamefont{Heller}, \bibfnamefont{E.~J.}},
  \bibinfo{year}{1996}, \bibinfo{journal}{Phys. Rev. Lett.}
  \textbf{\bibinfo{volume}{77}}, \bibinfo{pages}{4122}.

\bibitem[{\citenamefont{Huang and Wu}(2009)}]{huang09}
\bibinfo{author}{\bibnamefont{Huang}, \bibfnamefont{B.}}, and
  \bibinfo{author}{\bibfnamefont{T.-M.} \bibnamefont{Wu}},
  \bibinfo{year}{2009}, \bibinfo{journal}{Phys. Rev. E}
  \textbf{\bibinfo{volume}{79}}, \bibinfo{pages}{041105}.

\bibitem[{\citenamefont{Igloi and Monthus}(2005)}]{igloi05}
\bibinfo{author}{\bibnamefont{Igloi}, \bibfnamefont{F.}}, and
  \bibinfo{author}{\bibfnamefont{C.}~\bibnamefont{Monthus}},
  \bibinfo{year}{2005}, \bibinfo{journal}{Phys. Rep.}
  \textbf{\bibinfo{volume}{412}}, \bibinfo{pages}{277}.

\bibitem[{\citenamefont{Janik} \emph{et~al.}(1999)\citenamefont{Janik,
  N{\"o}renberg, Nowak, Papp, and Zahed}}]{janik99}
\bibinfo{author}{\bibnamefont{Janik}, \bibfnamefont{R.~A.}},
  \bibinfo{author}{\bibfnamefont{W.}~\bibnamefont{N{\"o}renberg}},
  \bibinfo{author}{\bibfnamefont{M.~A.} \bibnamefont{Nowak}},
  \bibinfo{author}{\bibfnamefont{G.}~\bibnamefont{Papp}}, and
  \bibinfo{author}{\bibfnamefont{I.}~\bibnamefont{Zahed}},
  \bibinfo{year}{1999}, \bibinfo{journal}{Phys. Rev. E}
  \textbf{\bibinfo{volume}{60}}, \bibinfo{pages}{2699}.

\bibitem[{\citenamefont{Janik and Nowak}(2003)}]{janik03}
\bibinfo{author}{\bibnamefont{Janik}, \bibfnamefont{R.~A.}}, and
  \bibinfo{author}{\bibfnamefont{M.~A.} \bibnamefont{Nowak}},
  \bibinfo{year}{2003}, \bibinfo{journal}{J. Phys. A: Math Gen.}
  \textbf{\bibinfo{volume}{36}}, \bibinfo{pages}{3629}.

\bibitem[{\citenamefont{Janik}
  \emph{et~al.}(1997{\natexlab{a}})\citenamefont{Janik, Nowak, Papp, Wambach,
  and Zahed}}]{janik97a}
\bibinfo{author}{\bibnamefont{Janik}, \bibfnamefont{R.~A.}},
  \bibinfo{author}{\bibfnamefont{M.~A.} \bibnamefont{Nowak}},
  \bibinfo{author}{\bibfnamefont{G.}~\bibnamefont{Papp}},
  \bibinfo{author}{\bibfnamefont{J.}~\bibnamefont{Wambach}}, and
  \bibinfo{author}{\bibfnamefont{I.}~\bibnamefont{Zahed}},
  \bibinfo{year}{1997}{\natexlab{a}}, \bibinfo{journal}{Phys. Rev. E}
  \textbf{\bibinfo{volume}{55}}, \bibinfo{pages}{4100}.

\bibitem[{\citenamefont{Janik}
  \emph{et~al.}(1997{\natexlab{b}})\citenamefont{Janik, Nowak, Papp, and
  Zahed}}]{janik97b}
\bibinfo{author}{\bibnamefont{Janik}, \bibfnamefont{R.~A.}},
  \bibinfo{author}{\bibfnamefont{M.~A.} \bibnamefont{Nowak}},
  \bibinfo{author}{\bibfnamefont{G.}~\bibnamefont{Papp}}, and
  \bibinfo{author}{\bibfnamefont{I.}~\bibnamefont{Zahed}},
  \bibinfo{year}{1997}{\natexlab{b}}, \bibinfo{journal}{Nucl. Phys. B}
  \textbf{\bibinfo{volume}{501}}, \bibinfo{pages}{603}.

\bibitem[{\citenamefont{Janik} \emph{et~al.}(2001)\citenamefont{Janik, Nowak,
  Papp, and Zahed}}]{janik01}
\bibinfo{author}{\bibnamefont{Janik}, \bibfnamefont{R.~A.}},
  \bibinfo{author}{\bibfnamefont{M.~A.} \bibnamefont{Nowak}},
  \bibinfo{author}{\bibfnamefont{G.}~\bibnamefont{Papp}}, and
  \bibinfo{author}{\bibfnamefont{I.}~\bibnamefont{Zahed}},
  \bibinfo{year}{2001}, \bibinfo{journal}{Physica E}
  \textbf{\bibinfo{volume}{9}}, \bibinfo{pages}{456}.

\bibitem[{\citenamefont{Jarosz and Nowak}(2004)}]{jarosz04}
\bibinfo{author}{\bibnamefont{Jarosz}, \bibfnamefont{A.}}, and
  \bibinfo{author}{\bibfnamefont{M.~A.} \bibnamefont{Nowak}},
  \bibinfo{year}{2004}, \eprint{arXiv:math-ph/0402057}.

\bibitem[{\citenamefont{Jarosz and Nowak}(2006)}]{jarosz06}
\bibinfo{author}{\bibnamefont{Jarosz}, \bibfnamefont{A.}}, and
  \bibinfo{author}{\bibfnamefont{M.~A.} \bibnamefont{Nowak}},
  \bibinfo{year}{2006}, \bibinfo{journal}{J. Phys. A: Math. Gen.}
  \textbf{\bibinfo{volume}{39}}, \bibinfo{pages}{10107}.

\bibitem[{\citenamefont{Jurkiewicz}
  \emph{et~al.}(2008)\citenamefont{Jurkiewicz, Lukaszewski, and
  Nowak}}]{jurkiewicz08}
\bibinfo{author}{\bibnamefont{Jurkiewicz}, \bibfnamefont{J.}},
  \bibinfo{author}{\bibfnamefont{G.}~\bibnamefont{Lukaszewski}}, and
  \bibinfo{author}{\bibfnamefont{M.~A.} \bibnamefont{Nowak}},
  \bibinfo{year}{2008}, \bibinfo{journal}{Acta Phys. Pol. B}
  \textbf{\bibinfo{volume}{39}}, \bibinfo{pages}{799}.

\bibitem[{\citenamefont{Klinger}(2010)}]{klinger10}
\bibinfo{author}{\bibnamefont{Klinger}, \bibfnamefont{M.}},
  \bibinfo{year}{2010}, \bibinfo{journal}{Physics Reports}
  \textbf{\bibinfo{volume}{492}}(\bibinfo{number}{4–5}), \bibinfo{pages}{111}.

\bibitem[{\citenamefont{Krich and Aspuru-Guzik}(2011)}]{krich11}
\bibinfo{author}{\bibnamefont{Krich}, \bibfnamefont{J.~J.}}, and
  \bibinfo{author}{\bibfnamefont{A.}~\bibnamefont{Aspuru-Guzik}},
  \bibinfo{year}{2011}, \bibinfo{journal}{Phys. Rev. Lett.}
  \textbf{\bibinfo{volume}{106}}, \bibinfo{pages}{156405}.

\bibitem[{\citenamefont{Lagendijk and van Tiggelen}(1996)}]{lagendijk96}
\bibinfo{author}{\bibnamefont{Lagendijk}, \bibfnamefont{A.}}, and
  \bibinfo{author}{\bibfnamefont{B.~A.} \bibnamefont{van Tiggelen}},
  \bibinfo{year}{1996}, \bibinfo{journal}{Phys. Rep.}
  \textbf{\bibinfo{volume}{270}}, \bibinfo{pages}{143}.

\bibitem[{\citenamefont{Lehmann} \emph{et~al.}(1995)\citenamefont{Lehmann,
  Saher, Sokolov, and Sommers}}]{lehmann95}
\bibinfo{author}{\bibnamefont{Lehmann}, \bibfnamefont{N.}},
  \bibinfo{author}{\bibfnamefont{D.}~\bibnamefont{Saher}},
  \bibinfo{author}{\bibfnamefont{V.~V.} \bibnamefont{Sokolov}}, and
  \bibinfo{author}{\bibfnamefont{H.~J.} \bibnamefont{Sommers}},
  \bibinfo{year}{1995}, \bibinfo{journal}{Nucl. Phys. A}
  \textbf{\bibinfo{volume}{582}}, \bibinfo{pages}{223}.

\bibitem[{\citenamefont{Ludwig} \emph{et~al.}(2003)\citenamefont{Ludwig,
  Nalbach, Rosenberg, and Osheroff}}]{ludwig03}
\bibinfo{author}{\bibnamefont{Ludwig}, \bibfnamefont{S.}},
  \bibinfo{author}{\bibfnamefont{P.}~\bibnamefont{Nalbach}},
  \bibinfo{author}{\bibfnamefont{D.}~\bibnamefont{Rosenberg}}, and
  \bibinfo{author}{\bibfnamefont{D.}~\bibnamefont{Osheroff}},
  \bibinfo{year}{2003}, \bibinfo{journal}{Phys. Rev. Lett.}
  \textbf{\bibinfo{volume}{90}}, \bibinfo{pages}{105501}.

\bibitem[{\citenamefont{Ma} \emph{et~al.}(1979)\citenamefont{Ma, Dasgupta, and
  Hu}}]{ma79}
\bibinfo{author}{\bibnamefont{Ma}, \bibfnamefont{S.}},
  \bibinfo{author}{\bibfnamefont{C.}~\bibnamefont{Dasgupta}}, and
  \bibinfo{author}{\bibfnamefont{C.~K.} \bibnamefont{Hu}},
  \bibinfo{year}{1979}, \bibinfo{journal}{Phys. Rev. Lett.}
  \textbf{\bibinfo{volume}{43}}, \bibinfo{pages}{1434}.

\bibitem[{\citenamefont{Mahaux and Weidenm{\"u}ller}(1969)}]{mahaux69}
\bibinfo{author}{\bibnamefont{Mahaux}, \bibfnamefont{C.}}, and
  \bibinfo{author}{\bibfnamefont{H.~A.} \bibnamefont{Weidenm{\"u}ller}},
  \bibinfo{year}{1969}, \emph{\bibinfo{title}{Shell-Model Approach to Nuclear
  Reactions}} (\bibinfo{publisher}{North-Holland, Amsterdam}).

\bibitem[{\citenamefont{Manassah}(2010)}]{manassah10}
\bibinfo{author}{\bibnamefont{Manassah}, \bibfnamefont{J.~T.}},
  \bibinfo{year}{2010}, \bibinfo{journal}{Phys. Rev. A}
  \textbf{\bibinfo{volume}{82}}, \bibinfo{pages}{053816}.

\bibitem[{\citenamefont{Marchenko and Pastur}(1967)}]{marchenko67}
\bibinfo{author}{\bibnamefont{Marchenko}, \bibfnamefont{V.~A.}}, and
  \bibinfo{author}{\bibfnamefont{L.~A.} \bibnamefont{Pastur}},
  \bibinfo{year}{1967}, \bibinfo{journal}{Math. USSR-Sb}
  \textbf{\bibinfo{volume}{1}}, \bibinfo{pages}{457}.

\bibitem[{\citenamefont{Martin-Mayor}
  \emph{et~al.}(2001)\citenamefont{Martin-Mayor, M\'{e}zard, Parisi, and
  Verrocchio}}]{martin01}
\bibinfo{author}{\bibnamefont{Martin-Mayor}, \bibfnamefont{V.}},
  \bibinfo{author}{\bibfnamefont{M.}~\bibnamefont{M\'{e}zard}},
  \bibinfo{author}{\bibfnamefont{G.}~\bibnamefont{Parisi}}, and
  \bibinfo{author}{\bibfnamefont{P.}~\bibnamefont{Verrocchio}},
  \bibinfo{year}{2001}, \bibinfo{journal}{J. Chem. Phys.}
  \textbf{\bibinfo{volume}{114}}, \bibinfo{pages}{8068}.

\bibitem[{\citenamefont{Massignan and Castin}(2006)}]{massignan06}
\bibinfo{author}{\bibnamefont{Massignan}, \bibfnamefont{P.}}, and
  \bibinfo{author}{\bibfnamefont{Y.}~\bibnamefont{Castin}},
  \bibinfo{year}{2006}, \bibinfo{journal}{Phys. Rev. A}
  \textbf{\bibinfo{volume}{74}}, \bibinfo{pages}{013616}.

\bibitem[{\citenamefont{Mehta}(2004)}]{mehta04}
\bibinfo{author}{\bibnamefont{Mehta}, \bibfnamefont{M.~L.}},
  \bibinfo{year}{2004}, \emph{\bibinfo{title}{Random Matrices}}
  (\bibinfo{publisher}{Elsevier, Amsterdam}).

\bibitem[{\citenamefont{M\'{e}zard}
  \emph{et~al.}(1999)\citenamefont{M\'{e}zard, Parisi, and Zee}}]{mezard99}
\bibinfo{author}{\bibnamefont{M\'{e}zard}, \bibfnamefont{M.}},
  \bibinfo{author}{\bibfnamefont{G.}~\bibnamefont{Parisi}}, and
  \bibinfo{author}{\bibfnamefont{A.}~\bibnamefont{Zee}}, \bibinfo{year}{1999},
  \bibinfo{journal}{Nucl. Phys. B} \textbf{\bibinfo{volume}{559}},
  \bibinfo{pages}{689}.

\bibitem[{\citenamefont{Mitchell} \emph{et~al.}(2010)\citenamefont{Mitchell,
  Richter, and Weidenm{\"u}ller}}]{mitchell10}
\bibinfo{author}{\bibnamefont{Mitchell}, \bibfnamefont{G.~E.}},
  \bibinfo{author}{\bibfnamefont{A.}~\bibnamefont{Richter}}, and
  \bibinfo{author}{\bibfnamefont{H.~A.} \bibnamefont{Weidenm{\"u}ller}},
  \bibinfo{year}{2010}, \bibinfo{journal}{Rev. Mod. Phys.}
  \textbf{\bibinfo{volume}{82}}, \bibinfo{pages}{2845}.

\bibitem[{\citenamefont{Mollow}(1972)}]{mollow72}
\bibinfo{author}{\bibnamefont{Mollow}, \bibfnamefont{B.~R.}},
  \bibinfo{year}{1972}, \bibinfo{journal}{Phys. Rev. A}
  \textbf{\bibinfo{volume}{5}}, \bibinfo{pages}{2217}.

\bibitem[{\citenamefont{Monaco} \emph{et~al.}(1998)\citenamefont{Monaco,
  Masciovecchio, Ruocco, and Sette}}]{monaco98}
\bibinfo{author}{\bibnamefont{Monaco}, \bibfnamefont{G.}},
  \bibinfo{author}{\bibfnamefont{C.}~\bibnamefont{Masciovecchio}},
  \bibinfo{author}{\bibfnamefont{G.}~\bibnamefont{Ruocco}}, and
  \bibinfo{author}{\bibfnamefont{F.}~\bibnamefont{Sette}},
  \bibinfo{year}{1998}, \bibinfo{journal}{Phys. Rev. Lett.}
  \textbf{\bibinfo{volume}{80}}, \bibinfo{pages}{2161}.

\bibitem[{\citenamefont{Monaco and Mossa}(2009)}]{monaco09}
\bibinfo{author}{\bibnamefont{Monaco}, \bibfnamefont{G.}}, and
  \bibinfo{author}{\bibfnamefont{S.}~\bibnamefont{Mossa}},
  \bibinfo{year}{2009}, \bibinfo{journal}{PNAS}
  \textbf{\bibinfo{volume}{106}}(\bibinfo{number}{40}), \bibinfo{pages}{16907}.

\bibitem[{\citenamefont{Monthus and Garel}(2011)}]{monthus11}
\bibinfo{author}{\bibnamefont{Monthus}, \bibfnamefont{C.}}, and
  \bibinfo{author}{\bibfnamefont{T.}~\bibnamefont{Garel}},
  \bibinfo{year}{2011}, \bibinfo{journal}{J. Phys. A: Math. Theor.}
  \textbf{\bibinfo{volume}{44}}, \bibinfo{pages}{085001}.

\bibitem[{\citenamefont{Morse and Feshbach}(1953)}]{morse53}
\bibinfo{author}{\bibnamefont{Morse}, \bibfnamefont{P.~M.}}, and
  \bibinfo{author}{\bibfnamefont{H.}~\bibnamefont{Feshbach}},
  \bibinfo{year}{1953}, \emph{\bibinfo{title}{Methods of Theoretical Physics}}
  (\bibinfo{publisher}{McGraw-Hill, New York}).

\bibitem[{\citenamefont{Philips}(1981)}]{philips81}
\bibinfo{editor}{\bibnamefont{Philips}, \bibfnamefont{W.~A.}} (ed.),
  \bibinfo{year}{1981}, \emph{\bibinfo{title}{Amorphous Solids: Low-Temperature
  Properties}} (\bibinfo{publisher}{Springer, Berlin}).

\bibitem[{\citenamefont{Pinheiro} \emph{et~al.}(2004)\citenamefont{Pinheiro,
  Rusek, Orlowski, and van Tiggelen}}]{pinheiro04}
\bibinfo{author}{\bibnamefont{Pinheiro}, \bibfnamefont{F.~A.}},
  \bibinfo{author}{\bibfnamefont{M.}~\bibnamefont{Rusek}},
  \bibinfo{author}{\bibfnamefont{A.}~\bibnamefont{Orlowski}}, and
  \bibinfo{author}{\bibfnamefont{B.~A.} \bibnamefont{van Tiggelen}},
  \bibinfo{year}{2004}, \bibinfo{journal}{Phys. Rev. E}
  \textbf{\bibinfo{volume}{69}}, \bibinfo{pages}{026605}.

\bibitem[{\citenamefont{Pinheiro and Sampaio}(2006)}]{pinheiro06}
\bibinfo{author}{\bibnamefont{Pinheiro}, \bibfnamefont{F.~A.}}, and
  \bibinfo{author}{\bibfnamefont{L.~C.} \bibnamefont{Sampaio}},
  \bibinfo{year}{2006}, \bibinfo{journal}{Phys. Rev. A}
  \textbf{\bibinfo{volume}{73}}, \bibinfo{pages}{013826}.

\bibitem[{\citenamefont{Prudnikov} \emph{et~al.}(1990)\citenamefont{Prudnikov,
  Brychkov, and Marichev}}]{prudnikov90}
\bibinfo{author}{\bibnamefont{Prudnikov}, \bibfnamefont{A.}},
  \bibinfo{author}{\bibfnamefont{Y.}~\bibnamefont{Brychkov}}, and
  \bibinfo{author}{\bibfnamefont{O.}~\bibnamefont{Marichev}},
  \bibinfo{year}{1990}, \emph{\bibinfo{title}{Integrals and Series. Vol.3: More
  Special Functions}} (\bibinfo{publisher}{Gordon and Breach, Newark, NJ}).

\bibitem[{\citenamefont{Rajan and Abbott}(2006)}]{rajan06}
\bibinfo{author}{\bibnamefont{Rajan}, \bibfnamefont{K.}}, and
  \bibinfo{author}{\bibfnamefont{L.~F.} \bibnamefont{Abbott}},
  \bibinfo{year}{2006}, \bibinfo{journal}{Phys. Rev. Lett.}
  \textbf{\bibinfo{volume}{97}}, \bibinfo{pages}{188104}.

\bibitem[{\citenamefont{Rogers and Castillo}(2009)}]{rogers09}
\bibinfo{author}{\bibnamefont{Rogers}, \bibfnamefont{T.}}, and
  \bibinfo{author}{\bibfnamefont{I.~P.} \bibnamefont{Castillo}},
  \bibinfo{year}{2009}, \bibinfo{journal}{Phys. Rev. E}
  \textbf{\bibinfo{volume}{79}}, \bibinfo{pages}{012101}.

\bibitem[{\citenamefont{R{\"o}hlsberger}
  \emph{et~al.}(2010)\citenamefont{R{\"o}hlsberger, Schlage, Sahoo, Couet, and
  R{\"u}ffer}}]{rohlsberger10}
\bibinfo{author}{\bibnamefont{R{\"o}hlsberger}, \bibfnamefont{R.}},
  \bibinfo{author}{\bibfnamefont{K.}~\bibnamefont{Schlage}},
  \bibinfo{author}{\bibfnamefont{B.}~\bibnamefont{Sahoo}},
  \bibinfo{author}{\bibfnamefont{S.}~\bibnamefont{Couet}}, and
  \bibinfo{author}{\bibfnamefont{R.}~\bibnamefont{R{\"u}ffer}},
  \bibinfo{year}{2010}, \bibinfo{journal}{Science}
  \textbf{\bibinfo{volume}{328}}, \bibinfo{pages}{1248}.

\bibitem[{\citenamefont{Rusek}
  \emph{et~al.}(2000{\natexlab{a}})\citenamefont{Rusek, Mostowski, and
  Orlowski}}]{rusek00}
\bibinfo{author}{\bibnamefont{Rusek}, \bibfnamefont{M.}},
  \bibinfo{author}{\bibfnamefont{J.}~\bibnamefont{Mostowski}}, and
  \bibinfo{author}{\bibfnamefont{A.}~\bibnamefont{Orlowski}},
  \bibinfo{year}{2000}{\natexlab{a}}, \bibinfo{journal}{Phys. Rev. A}
  \textbf{\bibinfo{volume}{61}}, \bibinfo{pages}{022704}.

\bibitem[{\citenamefont{Rusek}
  \emph{et~al.}(2000{\natexlab{b}})\citenamefont{Rusek, Mostowski, and
  Orlowski}}]{rusek002}
\bibinfo{author}{\bibnamefont{Rusek}, \bibfnamefont{M.}},
  \bibinfo{author}{\bibfnamefont{J.}~\bibnamefont{Mostowski}}, and
  \bibinfo{author}{\bibfnamefont{A.}~\bibnamefont{Orlowski}},
  \bibinfo{year}{2000}{\natexlab{b}}, \bibinfo{journal}{Phys. Rev. A}
  \textbf{\bibinfo{volume}{61}}, \bibinfo{pages}{022704}.

\bibitem[{\citenamefont{Rusek} \emph{et~al.}(1996)\citenamefont{Rusek,
  Or\l{}owski, and Mostowski}}]{rusek96}
\bibinfo{author}{\bibnamefont{Rusek}, \bibfnamefont{M.}},
  \bibinfo{author}{\bibfnamefont{A.}~\bibnamefont{Or\l{}owski}}, and
  \bibinfo{author}{\bibfnamefont{J.}~\bibnamefont{Mostowski}},
  \bibinfo{year}{1996}, \bibinfo{journal}{Phys. Rev. E}
  \textbf{\bibinfo{volume}{53}}, \bibinfo{pages}{4122}.

\bibitem[{\citenamefont{Scully}(2009)}]{scully09}
\bibinfo{author}{\bibnamefont{Scully}, \bibfnamefont{M.~O.}},
  \bibinfo{year}{2009}, \bibinfo{journal}{Phys. Rev. Lett.}
  \textbf{\bibinfo{volume}{102}}, \bibinfo{pages}{143601}.

\bibitem[{\citenamefont{Scully} \emph{et~al.}(2006)\citenamefont{Scully, Fry,
  Ooi, and W\'{o}dkiewicz}}]{scully06}
\bibinfo{author}{\bibnamefont{Scully}, \bibfnamefont{M.~O.}},
  \bibinfo{author}{\bibfnamefont{E.~S.} \bibnamefont{Fry}},
  \bibinfo{author}{\bibfnamefont{C.~H.~R.} \bibnamefont{Ooi}}, and
  \bibinfo{author}{\bibfnamefont{K.}~\bibnamefont{W\'{o}dkiewicz}},
  \bibinfo{year}{2006}, \bibinfo{journal}{Phys. Rev. Lett.}
  \textbf{\bibinfo{volume}{96}}, \bibinfo{pages}{010501}.

\bibitem[{\citenamefont{Scully and Svidzinsky}(2009)}]{scully209}
\bibinfo{author}{\bibnamefont{Scully}, \bibfnamefont{M.~O.}}, and
  \bibinfo{author}{\bibfnamefont{A.~A.} \bibnamefont{Svidzinsky}},
  \bibinfo{year}{2009}, \bibinfo{journal}{Science}
  \textbf{\bibinfo{volume}{325}}, \bibinfo{pages}{1510}.

\bibitem[{\citenamefont{Scully and Svidzinsky}(2010)}]{scully10}
\bibinfo{author}{\bibnamefont{Scully}, \bibfnamefont{M.~O.}}, and
  \bibinfo{author}{\bibfnamefont{A.~A.} \bibnamefont{Svidzinsky}},
  \bibinfo{year}{2010}, \bibinfo{journal}{Science}
  \textbf{\bibinfo{volume}{328}}, \bibinfo{pages}{1239}.

\bibitem[{\citenamefont{Sengupta and Mitra}(1999)}]{sengupta99}
\bibinfo{author}{\bibnamefont{Sengupta}, \bibfnamefont{A.~M.}}, and
  \bibinfo{author}{\bibfnamefont{P.~P.} \bibnamefont{Mitra}},
  \bibinfo{year}{1999}, \bibinfo{journal}{Phys. Rev. E}
  \textbf{\bibinfo{volume}{60}}, \bibinfo{pages}{3389}.

\bibitem[{\citenamefont{Siegman}(1986)}]{siegman86}
\bibinfo{author}{\bibnamefont{Siegman}, \bibfnamefont{A.~E.}},
  \bibinfo{year}{1986}, \emph{\bibinfo{title}{Lasers}}
  (\bibinfo{publisher}{University Science Books}).

\bibitem[{\citenamefont{Skipetrov and Goetschy}(2011)}]{skipetrov11}
\bibinfo{author}{\bibnamefont{Skipetrov}, \bibfnamefont{S.~E.}}, and
  \bibinfo{author}{\bibfnamefont{A.}~\bibnamefont{Goetschy}},
  \bibinfo{year}{2011}, \bibinfo{journal}{J. Phys. A}
  \textbf{\bibinfo{volume}{44}}, \bibinfo{pages}{065102}.

\bibitem[{\citenamefont{Sommers} \emph{et~al.}(1988)\citenamefont{Sommers,
  Crisanti, Sompolinsky, and Stein}}]{sommers88}
\bibinfo{author}{\bibnamefont{Sommers}, \bibfnamefont{H.~J.}},
  \bibinfo{author}{\bibfnamefont{A.}~\bibnamefont{Crisanti}},
  \bibinfo{author}{\bibfnamefont{H.}~\bibnamefont{Sompolinsky}}, and
  \bibinfo{author}{\bibfnamefont{Y.}~\bibnamefont{Stein}},
  \bibinfo{year}{1988}, \bibinfo{journal}{Phys. Rev. Lett.}
  \textbf{\bibinfo{volume}{60}}, \bibinfo{pages}{1895}.

\bibitem[{\citenamefont{Stephanov}(1996)}]{stephanov96}
\bibinfo{author}{\bibnamefont{Stephanov}, \bibfnamefont{M.~A.}},
  \bibinfo{year}{1996}, \bibinfo{journal}{Phys. Rev. Lett.}
  \textbf{\bibinfo{volume}{76}}, \bibinfo{pages}{4472}.

\bibitem[{\citenamefont{Stephanov} \emph{et~al.}(2001)\citenamefont{Stephanov,
  Verbaarschot, and Wettig}}]{stephanov01}
\bibinfo{author}{\bibnamefont{Stephanov}, \bibfnamefont{M.~A.}},
  \bibinfo{author}{\bibfnamefont{J.~J.~M.} \bibnamefont{Verbaarschot}}, and
  \bibinfo{author}{\bibfnamefont{T.}~\bibnamefont{Wettig}},
  \bibinfo{year}{2001}, \emph{\bibinfo{title}{Random Matrices}}
  (\bibinfo{publisher}{Wiley Encyclopedia of Electrical and Electronics
  Engineering supplement I, New York}).

\bibitem[{\citenamefont{Svidzinsky and Chang}(2008)}]{svidzinsky208}
\bibinfo{author}{\bibnamefont{Svidzinsky}, \bibfnamefont{A.~A.}}, and
  \bibinfo{author}{\bibfnamefont{J.-T.} \bibnamefont{Chang}},
  \bibinfo{year}{2008}, \bibinfo{journal}{Phys. Rev. A}
  \textbf{\bibinfo{volume}{77}}, \bibinfo{pages}{043833}.

\bibitem[{\citenamefont{Svidzinsky}
  \emph{et~al.}(2008)\citenamefont{Svidzinsky, Chang, and
  Scully}}]{svidzinsky08}
\bibinfo{author}{\bibnamefont{Svidzinsky}, \bibfnamefont{A.~A.}},
  \bibinfo{author}{\bibfnamefont{J.-T.} \bibnamefont{Chang}}, and
  \bibinfo{author}{\bibfnamefont{M.~O.} \bibnamefont{Scully}},
  \bibinfo{year}{2008}, \bibinfo{journal}{Phys. Rev. Lett.}
  \textbf{\bibinfo{volume}{100}}, \bibinfo{pages}{160504}.

\bibitem[{\citenamefont{Svidzinsky}
  \emph{et~al.}(2010)\citenamefont{Svidzinsky, Chang, and
  Scully}}]{svidzinsky10}
\bibinfo{author}{\bibnamefont{Svidzinsky}, \bibfnamefont{A.~A.}},
  \bibinfo{author}{\bibfnamefont{J.-T.} \bibnamefont{Chang}}, and
  \bibinfo{author}{\bibfnamefont{M.~O.} \bibnamefont{Scully}},
  \bibinfo{year}{2010}, \bibinfo{journal}{Phys. Rev. A}
  \textbf{\bibinfo{volume}{81}}, \bibinfo{pages}{053821}.

\bibitem[{\citenamefont{Svidzinsky and Scully}(2009)}]{svidzinsky09}
\bibinfo{author}{\bibnamefont{Svidzinsky}, \bibfnamefont{A.~A.}}, and
  \bibinfo{author}{\bibfnamefont{M.~O.} \bibnamefont{Scully}},
  \bibinfo{year}{2009}, \bibinfo{journal}{Opt. Commun.}
  \textbf{\bibinfo{volume}{282}}, \bibinfo{pages}{2894}.

\bibitem[{\citenamefont{Tierz}(2003)}]{tierz03}
\bibinfo{author}{\bibnamefont{Tierz}, \bibfnamefont{M.}}, \bibinfo{year}{2003},
  \eprint{arXiv:cond-math/0106485}.

\bibitem[{\citenamefont{Tricomi}(1957)}]{tricomi57}
\bibinfo{author}{\bibnamefont{Tricomi}, \bibfnamefont{F.~G.}},
  \bibinfo{year}{1957}, \emph{\bibinfo{title}{Integral Equations}}
  (\bibinfo{publisher}{Interscience, London}).

\bibitem[{\citenamefont{Tulino and Verd{\'u}}(2004)}]{tulino04}
\bibinfo{author}{\bibnamefont{Tulino}, \bibfnamefont{A.~M.}}, and
  \bibinfo{author}{\bibfnamefont{S.}~\bibnamefont{Verd{\'u}}},
  \bibinfo{year}{2004}, \emph{\bibinfo{title}{Random Matrix Theory and Wireless
  Communications}} (\bibinfo{publisher}{Now Publishers, Delft}).

\bibitem[{\citenamefont{Vaknin} \emph{et~al.}(2000)\citenamefont{Vaknin,
  Ovadyahu, and Pollak}}]{vaknin00}
\bibinfo{author}{\bibnamefont{Vaknin}, \bibfnamefont{A.}},
  \bibinfo{author}{\bibfnamefont{Z.}~\bibnamefont{Ovadyahu}}, and
  \bibinfo{author}{\bibfnamefont{M.}~\bibnamefont{Pollak}},
  \bibinfo{year}{2000}, \bibinfo{journal}{Phys. Rev. Lett.}
  \textbf{\bibinfo{volume}{84}}, \bibinfo{pages}{3402}.

\bibitem[{\citenamefont{Verbaarschot and Wettig}(2000)}]{verbaarschot00}
\bibinfo{author}{\bibnamefont{Verbaarschot}, \bibfnamefont{J.~J.~M.}}, and
  \bibinfo{author}{\bibfnamefont{T.}~\bibnamefont{Wettig}},
  \bibinfo{year}{2000}, \bibinfo{journal}{Annu. Rev. Nucl. Sci.}
  \textbf{\bibinfo{volume}{50}}, \bibinfo{pages}{343}.

\bibitem[{\citenamefont{Voiculescu}(1983)}]{voiculescu83}
\bibinfo{author}{\bibnamefont{Voiculescu}, \bibfnamefont{D.~V.}},
  \bibinfo{year}{1983}, \bibinfo{journal}{Acta. Sci. Math.}
  \textbf{\bibinfo{volume}{45}}, \bibinfo{pages}{429}.

\bibitem[{\citenamefont{Voiculescu}
  \emph{et~al.}(1992)\citenamefont{Voiculescu, Dykema, and
  Nica}}]{voiculescu92}
\bibinfo{author}{\bibnamefont{Voiculescu}, \bibfnamefont{D.~V.}},
  \bibinfo{author}{\bibfnamefont{K.~J.} \bibnamefont{Dykema}}, and
  \bibinfo{author}{\bibfnamefont{A.}~\bibnamefont{Nica}}, \bibinfo{year}{1992},
  \emph{\bibinfo{title}{Free Random Variables}} (\bibinfo{publisher}{American
  Mathematical Society}).

\bibitem[{\citenamefont{Wiersma}(2008)}]{wiersma08}
\bibinfo{author}{\bibnamefont{Wiersma}, \bibfnamefont{D.~S.}},
  \bibinfo{year}{2008}, \bibinfo{journal}{Nature Phys.}
  \textbf{\bibinfo{volume}{4}}, \bibinfo{pages}{359}.

\bibitem[{\citenamefont{Wigner}(1955)}]{wigner55}
\bibinfo{author}{\bibnamefont{Wigner}, \bibfnamefont{E.~P.}},
  \bibinfo{year}{1955}, \bibinfo{journal}{Ann. Math.}
  \textbf{\bibinfo{volume}{62}}, \bibinfo{pages}{548}.

\bibitem[{\citenamefont{Wigner}(1958)}]{wigner58}
\bibinfo{author}{\bibnamefont{Wigner}, \bibfnamefont{E.~P.}},
  \bibinfo{year}{1958}, \bibinfo{journal}{Ann. Math.}
  \textbf{\bibinfo{volume}{67}}, \bibinfo{pages}{325}.

\bibitem[{\citenamefont{Wishart}(1928)}]{wishart28}
\bibinfo{author}{\bibnamefont{Wishart}, \bibfnamefont{J.}},
  \bibinfo{year}{1928}, \bibinfo{journal}{Biometrika A}
  \textbf{\bibinfo{volume}{20}}, \bibinfo{pages}{32}.

\bibitem[{\citenamefont{Zee}(1996)}]{zee96}
\bibinfo{author}{\bibnamefont{Zee}, \bibfnamefont{A.}}, \bibinfo{year}{1996},
  \bibinfo{journal}{Nucl. Phys. B} \textbf{\bibinfo{volume}{474}},
  \bibinfo{pages}{726}.

\end{thebibliography}

\end{document}